\documentclass[usenatbib]{mnras}
\usepackage[figuresright]{rotating}
\usepackage{times}
\usepackage{geometry}
\usepackage{txfonts}

\usepackage[T1]{fontenc}
\usepackage{ae,aecompl}

\usepackage{caption}
\usepackage{array,footnote,color}
\usepackage{lscape}
\usepackage{graphicx}

\newcommand\nodata{ ~$\cdots$~ }
\newcolumntype{x}[1]{>{\centering\let\newline\\\arraybackslash\hspace{0pt}}p{#1}}
\newcommand{\atlas}{A{\small TLAS}$^{\rm 3D}$}

\DeclareRobustCommand{\VAN}[3]{#2}

\title[Star Formation in ETGs]{Star Formation in Nearby Early-Type Galaxies: The Radio Continuum Perspective}

\author[Kristina Nyland et al.]{\parbox{\textwidth}{
Kristina Nyland,$^{1,2}$\thanks{E-mail: knyland@nrao.edu}
Lisa M. Young,$^{3}$ 
Joan M. Wrobel,$^{4}$
Timothy A. Davis,$^{5}$
Martin Bureau,$^{6}$
Katherine Alatalo,$^{7}$\thanks{Hubble fellow}
Raffaella Morganti$^{2,8}$,
Pierre-Alain Duc,$^{9}$
P. T. de Zeeuw,$^{10,11}$
Richard M. McDermid,$^{12,13}$
Alison F. Crocker,$^{14}$
and Tom Oosterloo$^{2,8}$}
\vspace{0.4cm}\\ 
\parbox{\textwidth}{
$^{1}$National Radio Astronomy Observatory, Charlottesville, VA 22903, USA\\
$^{2}$Netherlands Institute for Radio Astronomy (ASTRON), Postbus 2, 7990 AA Dwingeloo, The Netherlands\\
$^{3}$Physics Department, New Mexico Institute of Mining and Technology, Socorro, NM 87801, USA\\
$^{4}$National Radio Astronomy Observatory, Socorro, NM 87801, USA\\
$^{5}$School of Physics \&\ Astronomy, Cardiff University, Queens Buildings, The Parade, Cardiff, CF24 3AA, UK\\
$^{6}$Sub-Dept. of Astrophysics, Dept. of Physics, University of Oxford, Denys Wilkinson Building, Keble Road, Oxford, OX1 3RH, UK\\
$^{7}$Observatories of the Carnegie Institution of Washington, 813 Santa Barbara Street, Pasadena, CA 91101, USA\\
$^{8}$Kapteyn Astronomical Institute, University of Groningen, Postbus 800, 9700 AV Groningen, The Netherlands\\
$^{9}$Observatoire de Paris, LERMA and CNRS, 61 Av. de l`Observatoire, F-75014 Paris, France\\
$^{10}$European Southern Observatory, Karl-Schwarzschild-Str. 2, 85748 Garching, Germany\\
$^{11}$Sterrewacht Leiden, Leiden University, Postbus 9513, 2300 RA Leiden, the Netherlands\\
$^{12}$Department of Physics and Astronomy, Macquarie University, Sydney NSW 2109,  Australia\\
$^{13}$Australian Astronomical Observatory, PO Box 915, North Ryde, NSW 1670, Australia\\
$^{14}$Physics Department, Reed College, Portland, OR 97202, USA}}

\date{Accepted YEAR MONTH DAY.  Received YEAR MONTH DAY; in original form YEAR MONTH DAY}

\pagerange{\pageref{firstpage}--\pageref{lastpage}} 
\pubyear{2016}

\begin{document}
\label{firstpage}
\maketitle

\begin{abstract}
We present a 1.4~GHz Karl G. Jansky Very Large Array (VLA) study of a sample of early-type galaxies (ETGs) from the volume- and magnitude-limited \atlas\ survey.  
The radio morphologies of these ETGs at a resolution of $\theta_{\mathrm{FWHM}} \approx$ 5$^{\prime \prime}$ are diverse and include sources that are compact on sub-kpc scales, resolved structures similar to those seen in star-forming spiral galaxies, and kpc-scale radio jets/lobes associated with active nuclei.  We compare the 1.4~GHz, molecular gas, and infrared (IR) properties of these ETGs.  The most CO-rich \atlas\ ETGs have radio luminosities consistent with extrapolations from H$_{2}$ mass-derived star formation rates from studies of late-type galaxies.  These ETGs also follow the radio-IR correlation.  However, ETGs with lower molecular gas masses tend to have less radio emission relative to their CO and IR emission compared to spirals.  
The fraction of galaxies in our sample with high IR-radio ratios is much higher than in previous studies, and cannot be explained by a systematic underestimation of the radio luminosity due to the presence extended, low-surface-brightness emission that was resolved-out in our VLA observations.  
In addition, we find that the high IR-radio ratios tend to occur at low IR luminosities, but are not associated with low dynamical mass or metallicity.
Thus, we have identified a population of ETGs that have a genuine shortfall of radio emission relative to both their IR and molecular gas emission.  
A number of mechanisms may conspire to cause this radio deficiency, including a bottom-heavy stellar initial mass function, weak magnetic fields, a higher prevalence of environmental effects compared to spirals and enhanced cosmic ray losses.
\end{abstract}


\clearpage
\section{Introduction and Motivation}
Early-type (elliptical and lenticular) galaxies (ETGs) were once considered a homogeneous class of ``red and dead" systems devoid of cold gas and young stars, archetypes of the end point of hierarchical galaxy formation and evolution.  However, evidence is mounting that a significant fraction of nearby ETGs are in fact still continuing to form stars.  We now know that ETGs commonly host neutral hydrogen (H{\tt I}) distributed in discs, rings, or disturbed structures, with masses ranging from $\sim10^{6} - 10^{8}$ M$_{\odot}$ (e.g., \citealt{morganti+06, oosterloo+10}).  Recent statistical searches for H{\tt I} have reported detection rates of $\sim$40\% in field ETGs, and $\sim$10\% in ETGs in more densely populated environments \citep{serra+14}.

In addition to cold atomic gas, CO studies have found that many ETGs also harbor substantial reservoirs of molecular gas (e.g., \citealt{knapp+96, welch+03, combes+07}).  Recently, the first statistically-complete single-dish CO survey of molecular gas in the \atlas\ galaxies quantified the prevalence of molecular gas in ETGs, reporting a detection rate of 22\% $\pm$ 3\% \citep{young+11}.  Interferometric molecular gas imaging studies have shown that ETG molecular gas reservoirs span a range of diverse morphologies and kinematics \citep{young+08, crocker+11, alatalo+13, davis+13}.  While secular processes such as stellar mass loss from asymptotic giant branch (AGB) or post-asymptotic giant branch (pAGB) stars may be responsible for the presence of the molecular gas in ETGs in some cases \citep{faber+76, knapp+92, mathews+03, temi+07}, the disturbed morphologies and kinematics of the gas in other cases point to an external origin (i.e., mergers; \citealt{sarzi+06, young+08, duc+15, davis+11, davis+16}).  Other authors have suggested that molecular gas in massive ETGs galaxies may originate from cooled gas from the hot X-ray halos in which these galaxies typically reside \citep{werner+14}.

While it has become clear that many ETGs contain significant cold gas reservoirs, the ultimate fate of this gas has remained a subject of debate.  Whether the gas is actively engaged in star formation (SF), and the efficiency of that SF compared to spiral galaxies, is still unclear.  The difficultly in addressing these questions largely arises from the fact that common SF tracers, such as ultraviolet (UV) and infrared (IR) emission, may be contaminated by emission from the underlying evolved stellar population in ETGs \citep{jeong+09, temi+09, sarzi+10, davis+14}.  Emission from active galactic nuclei (AGNs) in ETGs can also contaminate many standard SF tracers.

Nevertheless, recent studies have argued in favor of the presence of ongoing SF in ETGs.  The detection of young stellar populations through UV observations with the {\it Galaxy Evolution Explorer} and the {\it Hubble Space Telescope}, especially in gas-rich ETGs, has provided support for this scenario \citep{yi+05, kaviraj+07, ford+13}.  UV emission re-processed by dust in star-forming galaxies and re-emitted in the IR provides another avenue for SF studies of ETGs, and is less susceptible to dust extinction compared to SFR tracers at shorter wavelengths.  Although the possibility of contamination from old stars complicates the use of IR emission as a SFR tracer in ETGs, techniques for isolating the portion of IR emission associated with SF have shown promising results (e.g., \citealt{davis+14}).  

Another potential ETG SFR tracer is radio continuum emission.  Unlike other tracers, such as optical or UV emission, centimeter-wave radio continuum emission is virtually unaffected by extinction or obscuration \citep{condon+92}.  Recent upgrades at the Karl G. Jansky Very Large Array (VLA) offer the ability to obtain sensitive measurements over relatively short timespans, making radio continuum observations an efficient means of detecting even weak SF in ETGs.  Although radio continuum emission may be contaminated by AGNs, strong AGNs can be readily identified based on their radio morphologies (e.g., \citealt{wrobel+91b}) and through comparisons with other SF and AGN diagnostics (e.g., \citealt{nyland+16}).

Radio continuum emission is well-established as a SF tracer in late-type galaxies.  Studies of the relationship between radio continuum and IR emission have demonstrated a tight correlation between these two quantities that extends over at least three orders of magnitude among ``normal" star-forming galaxies (e.g., \citealt{helou+85, condon+92, yun+01}).  This so-called ``radio-IR" relation is believed to be driven by SF in the host galaxy.  The radio continuum emission is generated by massive stars as they end their lives as supernovae, accelerating cosmic rays and subsequently producing non-thermal synchrotron emission.  Dusty H{\tt II} regions in turn re-radiate optical and UV light emitted by young stars at IR wavelengths.  

Numerous studies of the radio-IR relation for samples of star-forming spiral galaxies using IR data at both far-infrared (FIR) and mid-infrared (MIR) wavelengths (e.g., \citealt{yun+01, condon+02, appleton+04, sargent+10}) have been performed.  However, detailed studies of the radio-IR correlation in ETGs have been rare.  Some authors have reported that ETGs closely follow the same tight radio-IR correlation as spiral galaxies \citep{walsh+89, combes+07}, while others have found that ETGs as a class tend to be systematically ``radio faint" \citep{wrobel+91b, lucero+07, crocker+11}.  A large, sensitive study of the radio continuum emission on kpc-scales of a statistical sample of ETGs is therefore needed to improve our understanding of the incidence and efficiency of SF in bulge-dominated galaxies.

Here, we present new 1.4~GHz VLA observations at 5$^{\prime \prime}$ spatial resolution of a subset of the statistically-complete \atlas\ survey.  We combine these new VLA data with existing archival 1.4~GHz measurements to study the global relationship between the radio continuum and IR emission in ETGs.  We also compare the radio continuum emission properties to those of the molecular gas in our sample galaxies, all of which have single-dish CO observations available, to study the SF efficiency in ETGs.  In Section~\ref{sec:sample}, we describe the \atlas\ survey.  
We explain the selection, observations, data reduction, and results of our new VLA observations in Section~\ref{sec:data}.  Ancillary molecular and infrared data are discussed in Section~\ref{sec:multiwav}.
In Section~\ref{radio_corr}, we describe the radio-CO, radio-IR, and IR-CO relations and discuss potential explanations for the observed deficit of radio emission in Section~\ref{sec:discussion}.
We summarize our results and provide concluding remarks in Section~\ref{sec:summary}.

\section{Sample}
\label{sec:sample}
Our sample is drawn from the \atlas\ survey.  This volume- and magnitude-limited ($D < 42$ Mpc and $M_{\mathrm{K}} < -21.5$) survey of 260 ETGs uses multiwavelength data \citep{cappellari+11} and theoretical models \citep{bois+11, khochfar+11, naab+14} to characterize the local population of ETGs and study their formation histories.  The \atlas\ sample includes ETGs from a variety of environments with diverse kinematics, morphologies, and interstellar medium (ISM) properties.  The rich database of optical observations includes two-dimensional integral field spectroscopy (IFS) with the {\tt SAURON} instrument \citep{bacon+01} on the William Herschel Telescope.  This data is used to classify the \atlas\ galaxies on the basis of their stellar kinematics as ``slow rotators'' and ``fast rotators" \citep{emsellem+07, emsellem+11}.  Slow rotators are generally massive ellipticals and have little ordered rotation in their stellar velocity fields, while fast rotators are characterized by regular rotation.  The fast rotator class contains lenticulars and some lower-mass ellipticals whose discy nature was not previously recognized.

The \atlas\ survey also includes ground-based imaging from the Sloan Digital Sky Survey (\citealt{york+00}) or Isaac Newton Telescope (\citealt{scott+13}), as well as extremely deep optical observations with the MegaCam instrument at the Canada-France-Hawaii Telescope \citep{duc+11, duc+15}.  
Molecular gas observations are available for the full \atlas\ sample from single-dish $^{12}$CO(1-0) and (2-1) observations with the Institut de Radioastronomie Millim\'{e}trique (IRAM) 30-m telescope \citep{young+11}, and represent the first large, statistical search for molecular gas in a sample of ETGs.  A variety of other large datasets covering subsets of the full \atlas\ sample are also available and include H{\tt I} imaging from the Westerbork Radio Synthesis Telescope \citep{serra+12, serra+14}, interferometric $^{12}$CO(1-0) maps  \citep{alatalo+13} from the Combined Array for Research in Millimeter Astronomy (CARMA), and high-resolution ($\theta_{\mathrm{FWHM}} \sim 0.5^{\prime \prime}$) VLA observations of the nuclear radio emission at 5~GHz \citep{nyland+16}.

\begin{table*}
\begin{minipage}{12cm}
\caption{New VLA Observations.}
\label{tab:projects}
\begin{tabular*}{12cm}{lcccccc}
\hline
\hline
 Project & Dates & Time     & Galaxies & BW      & Spws & Frequency \\
                    &                                & (hours)  &                 & (MHz)  &            & (GHz)          \\
(1) & (2) & (3) & (4) & (5) & (6) & (7)\\
\hline 
 10C-173 &  March 13 - 31, 2011 & 10 & 20 &  256 & 2 & 1.39\\
12A-404 &  June 5 -  August 9, 2012 & 23 & 52 & 1024 & 16 & 1.50\\ 
\hline
\hline
\end{tabular*}
 
 \medskip
{Column 1: Project ID.  Column 2: Observing dates.  Column 3: Total project length.  Column 4: Number of galaxies.  Column 5: Total observing bandwidth per polarization.  Column 6: Number of spectral windows.  Column 7: Central observing frequency.}

 \end{minipage} 
 \end{table*}

\section{Radio Continuum Data}
\label{sec:data}

\subsection{VLA Sample Selection}
We obtained new 1.4~GHz VLA observations of 72 ETGs drawn from the \atlas\ survey \citep{cappellari+11}.  Since our primary goal was to study SF in ETGs, we included as many of the 56 CO-detected \atlas\ galaxies as possible in our new observations.  Of the 72 ETGs that we observed at 1.4~GHz, 52 have single-dish CO detections with IRAM at a spatial resolution of 22$^{\prime \prime}$ \citep{young+11}.  The 4 CO-detected \atlas\ ETGs that we did not observe are NGC4283, NGC4435, NGC4476, and NGC4477. 
These galaxies were included in the Faint Images of the Radio Sky at Twenty Centimeters (FIRST; \citealt{becker+95}) survey at 5$^{\prime \prime}$ spatial resolution, though none were detected.
In addition to the 52 CO-detected galaxies, we also observed 20 \atlas\ ETGs with CO upper limits only.  These new observations of 20 molecular gas-poor ETGs, combined with archival observations from FIRST, thus provide a comparative ``control" sample for the VLA observations of the CO-detected ETGs.

\subsection{Observations}
\label{obs_RC}
We observed during the VLA B configuration at $L$ band (1-2~GHz) over two projects, 10C-173 and 12A-404, spanning a total of 33 hours.  Our observational set-up is summarized in Table~\ref{tab:projects}.  Project 10C-173 was observed as part of the Open Shared Risk Observing program, which offered 256~MHz of total bandwidth.  The full bandwidth for this project was split into two 128~MHz-wide spectral windows (SPWs), each containing 64 channels.  We required 25 minutes of integration time per galaxy to achieve our desired RMS noise of 25~$\mu$Jy beam$^{-1}$.  For Project 12A-404, we were able to utilize the full $L$-band bandwidth of 1024~MHz.  We divided this bandwidth into 16 SPWs, each spanning 64~MHz and containing 64 channels.  The wider bandwidth of project 12A-404 allowed us to reach an RMS noise of 25~$\mu$Jy beam$^{-1}$ for each galaxy in about 15 minutes.  

We divided each project into independent scheduling blocks (SBs) for flexible dynamic scheduling.  We phase-referenced each galaxy to a nearby calibrator within 10 degrees, and chose calibrators with expected amplitude closure errors of no more than 10\% to ensure robust calibration solutions.  In addition, the positional accuracy of most of our phase calibrators was $<0.002^{\prime \prime}$.  In order to set the amplitude scale to an accuracy of 3\%, as well as calibrate the bandpass and instrumental delays, we observed the most conveniently-located standard flux calibrator (3C286, 3C48, 3C147, or 3C138) once per SB \citep{perley+13}.

\subsection{Calibration and Imaging}
\label{sec:cal_imaging}
Our data reduction strategy follows that of the higher-resolution 5~GHz VLA study of the \atlas\ galaxies presented in \citet{nyland+16}, and we refer readers there for details.  We flagged, calibrated, and imaged each SB using the Common Astronomy Software Applications (CASA) package (version 4.1.0) and the CASA VLA calibration pipeline version 1.2.0\footnote{https://science.nrao.edu/facilities/vla/data-processing/pipeline}.  All of our SBs were Hanning smoothed prior to the pipeline calibration to minimize Gibbs ringing due to bright radio frequency interference.  Nevertheless, typically one to three SPWs per SB in Project 12A-404 had to be flagged entirely from the dataset to improve the quality of our images.
  
We imaged our data in CASA using the CLEAN task in the Multi Frequency Synthesis mode \citep{conway+90}.  Due to the large fractional bandwidths ($\sim$67\% from 1-2 GHz), we imaged each galaxy with the parameter nterms = 2 \citep{rau+11}.  We chose Briggs weighting \citep{briggs+95} with a robustness parameter of 0.5 for the best compromise among sensitivity, sidelobe suppression, and spatial resolution.  To correct for the effects of non-coplanar baselines, we set the parameters gridmode = `widefield' and wprojplanes = 128.  We produced large images covering the full $L$-band primary beam (30$^{\prime}$) with a cell size of 0.75$^{\prime \prime}$.  Self calibration was performed when necessary following standard procedures.  Sources with evidence of extended structures were imaged using the multiscale algorithm \citep{cornwell+08}.

\subsection{Image Analysis}
\label{sec:image_analysis}
Measurements of source fluxes, sizes, and their corresponding uncertainties follow the detailed description in \citet{nyland+16}.  In brief, the RMS noise of each image was determined by averaging the flux densities in several source-free regions.  For detections, we required a peak flux density of $S_{\mathrm{peak}} >$ 5$\sigma$, where $\sigma$ is the RMS noise.  Upper limits for non-detections were set to $S_{\mathrm{peak}} <$ 5$\sigma$.  We also required radio sources to be spatially coincident with the host galaxy optical position to within 3$^{\prime \prime}$.  For each radio source with a Gaussian-like morphology, we determined the source parameters (peak flux density, integrated flux density, deconvolved major and minor axes, and deconvolved position angle) by fitting a single two-dimensional elliptical Gaussian model using the JMFIT task in the 31DEC15 release of the Astronomical Image Processing System ({\tt AIPS}).  

For sources with more complex/extended morphologies, we measured the spatial parameters by hand using the CASA Viewer and calculated the integrated flux density using the task IMSTAT.  
The image and source parameters are summarized in Tables~\ref{tab:radio_parms} and \ref{tab:gauss}.  Maps of our detected sources are provided in Figure~\ref{fig:radio_images} and relative contour levels are given in Table~\ref{tab:contours}.

\subsection{Detection Rate and Morphology}
The detection rates in projects 10C-173 and 12A-404 are 19/20 and 35/52, respectively, and the combined detection rate for both projects is 51/72 (71 $\pm$ 5\%).  Including the galaxies with archival data at comparable spatial resolution from FIRST (see Section~\ref{sec:first}), the total detection rate of \atlas\ ETGs with kpc-scale 1.4~GHz emission is 79/252 (31 $\pm$ 3\%).  This combined detection rate is likely a lower limit due to the poorer sensitivity of FIRST compared to our new observations.

Many of the detected source morphologies resemble the resolved, disc-like radio structures present in typical star-forming spirals and span scales of 200 to 900 pc for the nearest ($D = $ 11.1~Mpc) to the farthest ($D = $ 45.8~Mpc) ETGs, respectively.  The fraction of detected ETGs with resolved emission is 41/51 (80 $\pm$ 6\%; see Table~\ref{tab:gauss}).  There are 19/51 sources (37 $\pm$ 7\%) with distinct multiple components or extended morphologies on scales of $\sim$1~kpc or larger.  Optical images of these 19 sources overlaid with the radio contours are shown in Figure~\ref{fig:radio_overlays}.  The source with the largest spatial extent spans $\approx$18~kpc and is characterized by prominent twin radio jets launched by the active nucleus hosted by NGC3665.  In 8 galaxies, the 1.4~GHz emission is distributed among multiple components.  We summarize the flux and spatial properties of these multi-component sources in Tables~\ref{tab:multi_flux} and \ref{tab:multi_spatial}.  

\subsection{Comparison to Previous Studies}
\label{sec:previous_studies}

\subsubsection{NVSS}
All of the \atlas\ galaxies fall within the survey area of the 1.4~GHz NRAO VLA Sky Survey (NVSS; \citealt{condon+98}).  There are 54/260 (21 $\pm$ 3\%) \atlas\ ETGs detected in the NVSS catalog (within a search radius of 10$^{\prime \prime}$) at a detection threshold of 2.5 mJy beam$^{-1}$.  For most of these galaxies, the emission is unresolved at the low spatial resolution ($\theta_{\mathrm{FWHM}} \approx 45^{\prime \prime}$) of NVSS.  Nevertheless, for the 32 ETGs detected in both NVSS and our new VLA observations, the flux densities are generally in good agreement.  Accounting for the typical power-law dependence\footnote{$S \propto \nu^{\alpha}$, where $S$ is the radio continuum flux density, $\nu$ is the frequency, and $\alpha$ is the radio spectral index.  The radio spectral index is assumed to have a value of $\alpha \approx -0.7$ for unabsorbed, non-thermal, synchrotron emission \citep{condon+92, marvil+15}.} of radio flux density with frequency, the median ratio between the NVSS and VLA flux densities is 1.13.  We address the possibility of resolved-out radio emission and its influence on our analysis in Section~\ref{sec:resolved_out_radio}.

\subsubsection{FIRST}
\label{sec:first}
FIRST provides the largest compilation of 1.4~GHz images with spatial resolutions ($\theta_{\mathrm{FWHM}} \approx 5^{\prime \prime}$) comparable to the new VLA observations presented here.  Although 239 (92\%) of the \atlas\ galaxies are included in the FIRST survey footprint, only 57 (24 $\pm$ 3\%) have flux densities above the 5$\sigma$ detection threshold of 1 mJy beam$^{-1}$ (within a search radius of 5$^{\prime \prime}$).  Our new VLA data are typically a factor of 5 times more sensitive than FIRST, and this is reflected in our higher detection rate.  We detect 1.4~GHz emission in 15 galaxies that were previously undetected in FIRST.  

For ETGs detected in both our new 1.4~GHz data and FIRST, we find good agreement between the flux densities, with a median flux ratio of 0.98.  The single significant outlier is NGC3665, however, the Gaussian-fit integrated flux density reported in the FIRST catalog\footnote{http://sundog.stsci.edu/index.html} substantially underestimates the total 1.4~GHz emission in this extended radio source (see Figure~\ref{fig:radio_images}) by over an order of magnitude.  After re-measuring the integrated flux density in the NGC3665 FIRST image by hand, we found good agreement between the FIRST data and our new VLA observations.  

\subsubsection{Previous ETG Surveys}
\citet{sadler+89} and \citet{wrobel+91b} performed 5~GHz imaging studies of large samples of ETGs and concluded that the radio morphologies and multiwavelength source properties indicated that the radio emission in at least some ETGs is likely related to recent SF.  
The volume-limited study by \citet{wrobel+91b} is the most comparable ETG survey to the 1.4~GHz study of the \atlas\ ETGs presented here.  While sample sizes and spatial resolutions are similar, our new 1.4~GHz observations reach sensitivities nearly an order of magnitude deeper after adjusting the 5~GHz detection threshold of the \citet{wrobel+91b} study to 1.4~GHz assuming a standard radio spectral index of $\alpha = -0.7$.  The detection fraction in \citet{wrobel+91b} is 52/198 (26 $\pm$ 3\%) galaxies, 7/52 (13 $\pm$ 5\%) of which display extended, disc-like morphologies strongly suggestive of a SF origin.  

Forty ETGs are included in the 1.4~GHz study presented here and \citet{wrobel+91b}.  The detection rates for these ETGs are 21/40 (53 $\pm$ 8\%) and 28/40 (70 $\pm$ 7\%) for the 5~GHz \citet{wrobel+91b} observations and the 1.4~GHz observations presented here, respectively.  If the ETGs with archival FIRST data are included along with our new 1.4~GHz observations, the overlap between the \citet{wrobel+91b} and the \atlas\ samples increases to 143 galaxies.  Of these, only 36/143 (25 $\pm$ 4\%) were detected by \citet{wrobel+91b}.  The total (new + archival) 1.4~GHz detection rate of the ETGs common to both studies at 5$^{\prime \prime}$ resolution is 40/143 (28 $\pm$ 4\%).

We also compare our new 1.4~GHz data to a higher-resolution, complementary 5~GHz study of the nuclear emission in the \atlas\ ETGs \citep{nyland+16}.  There are 142 galaxies with both 1.4~GHz data at $\approx5^{\prime \prime}$ resolution (this work) and 5~GHz data at $\approx0.5^{\prime \prime}$ ($\sim25-100$~pc) resolution \citep{nyland+16}.  Of these 142 ETGs, 74 (52 $\pm$ 4\%) are detected at each band, with 60 (42 $\pm$ 4\%) detected in both datasets.  Fifty-four (38 $\pm$ 4\%) ETGs are non-detections in both our new 1.4~GHz data and the 5~GHz data from \citet{nyland+16}.  These galaxies may be genuinely quiescent ETGs with no measurable SF or AGN emission.

Fourteen (10 $\pm$ 3\%) ETGs (see Table~\ref{tab:summary_all}) were detected only in the high-resolution 5~GHz observations.  This could be due to the higher sensitivity of these 5~GHz data.  Another possibility is that the nuclear radio sources in these ETGs are associated primarily with low-luminosity AGNs \citep{ho+08} rather than SF.  

For a different set of 14 ETGs (see Table~\ref{tab:summary_all}), we detect emission in our lower-resolution 1.4~GHz data, but not in the high-resolution 5~GHz data presented in \citet{nyland+16}.  In these galaxies, the majority of the radio emission is likely distributed on scales larger than $\sim$100~pc, and may have been resolved-out in the higher-resolution data.  The dominance of radio continuum emission on larger scales in these galaxies suggests that their radio emission is primarily associated with SF.  This is supported by the fact that 11/14 (79 $\pm$ 11\%) of these galaxies also harbor molecular gas \citep{young+11}.  The three galaxies without molecular gas detections are NGC1023, NGC3193, and NGC6547, though NGC1023 does contain a large, disturbed H{\tt I} reservoir \citep{serra+12}.
  
\section{Multiwavelength Data}
\label{sec:multiwav}
A summary of the CO and IR data included in our analysis is provided in Table~\ref{tab:summary_all}.  In the remainder of this section, we describe the CO and IR data used to compute the CO-radio and IR-radio ratios.

\subsection{Molecular Gas Data}
\label{sec:molecular_gas}
As mentioned in Section~\ref{sec:sample}, one of the most unique aspects of the \atlas\ survey of ETGs is the availability of CO data for the full sample \citep{young+11}.  This allows a direct measurement of the amount of raw material available for future SF.  Nearly 25\% of the \atlas\ galaxies were detected in \citet{young+11}, with H$_2$ masses ranging from 1.3 $\times$ 10$^{7}$ to 1.9 $\times$ 10$^{9}$ M$_{\odot}$.  We use these CO data in concert with our 1.4~GHz VLA data to investigate the relationship between radio luminosity and molecular gas mass in Section~\ref{sec:radio_CO}.  

\subsection{Infrared Data}
\label{sec:IR_data}

\subsubsection{{\it IRAS}}
\label{sec:FIR_data}
The FIR luminosity provides an estimate of the integrated $42.5-122.5\mu$m emission \citep{helou+88}, and is commonly defined as follows:

\begin{equation}
\label{eq:LFIR}
L_{\mathrm{FIR}}(L_{\odot}) \equiv \left( 1 +  \frac{S_{100\mu \mathrm{m}} } {2.58\, S_{60\mu \mathrm{m}}} \right) L_{60 \mu \mathrm{m}},
\end{equation}

\noindent where $S_{60 \mu \mathrm{m}}$ and $S_{100 \mu \mathrm{m}}$ are the {\it Infrared Astronomical Satellite} ({\it IRAS}; \citealt{soifer+87}) 60 and 100$\mu$m band flux densities in Jy, respectively, and $L_{60 \mu \mathrm{m}}$ is measured in solar luminosities (\citealt{yun+01}).  

We obtained the FIR data at 60 and 100$\mu$m from NED.  FIR measurements from {\it IRAS} were available for 195 of the \atlas\ galaxies, however, only 96 galaxies were detected at both 60 and 100$\mu$m.  We discuss the FIR data further in Section~\ref{sec:FIR}, where we study the global FIR-radio relation.

\subsubsection{{\it WISE}}
\label{sec:MIR_data}
Sensitive MIR data from the {\it Wide-field Infrared Survey Explorer} ({\it WISE}; \citealt{wright+10}) are available for the full \atlas\ sample, and we utilize these data in Section~\ref{sec:MIR} to examine the relationship between the MIR and radio continuum emission.  All of the \atlas\ galaxies are detected in the 4 {\it WISE} bands.  In the W1, W2, and W3 bands at 3.4$\mu$m, 4.6$\mu$m, and 12$\mu$m, respectively, all of the \atlas\ galaxies are robustly detected.  In the W4 band at 22$\mu$m, 29 galaxies have signal-to-noise ratios less than 2 in their profile fits.  However, the aperture photometry fluxes measured within an area defined by the spatial properties of the near-infrared emission from the Two Micron All Sky Survey (2MASS; \citealt{skrutskie+06}) of each galaxy yields a measurement within the sensitivity limits of the W4 band.  Thus, we consider these 29 galaxies as genuine, albeit weak, detections.
  
We extracted {\it WISE} photometry from the AllWISE source catalog \citep{cutri+13}  and performed cross matching with the official \atlas\ positions \citep{cappellari+11} within a search radius of 5$^{\prime \prime}$.  The W4-band data provide a spatial resolution of $\theta_{\mathrm{FWHM}} \approx 11.8^{\prime \prime}$.  Although most of the \atlas\ galaxies are only marginally resolved at 22$\mu$m, we selected photometric measurements derived within the elliptical area of the 2MASS emission for each galaxy ({\it w4gmag}) when possible.  If {\it w4gmag} magnitudes were unavailable, we used the Gaussian profile fit magnitudes instead ({\it w4mpro}).

\section{Global Relationships}
\label{radio_corr}

\subsection{Radio-H$_{2}$ Relation}
\label{sec:radio_CO}
Previous studies have found a strong correlation between the radio luminosity and CO luminosity in samples of spiral galaxies (e.g., \citealt{adler+91, murgia+02, liu+10, liu+15}), with some studies reporting the correlation is as tight as the radio-FIR relation (e.g., \citealt{murgia+05}).  However, little information about whether molecular-gas-rich ETGs similarly follow this relationship is available.

\begin{figure}
\includegraphics[clip=true, trim=0cm 0cm 1.8cm 2cm, scale=0.4]{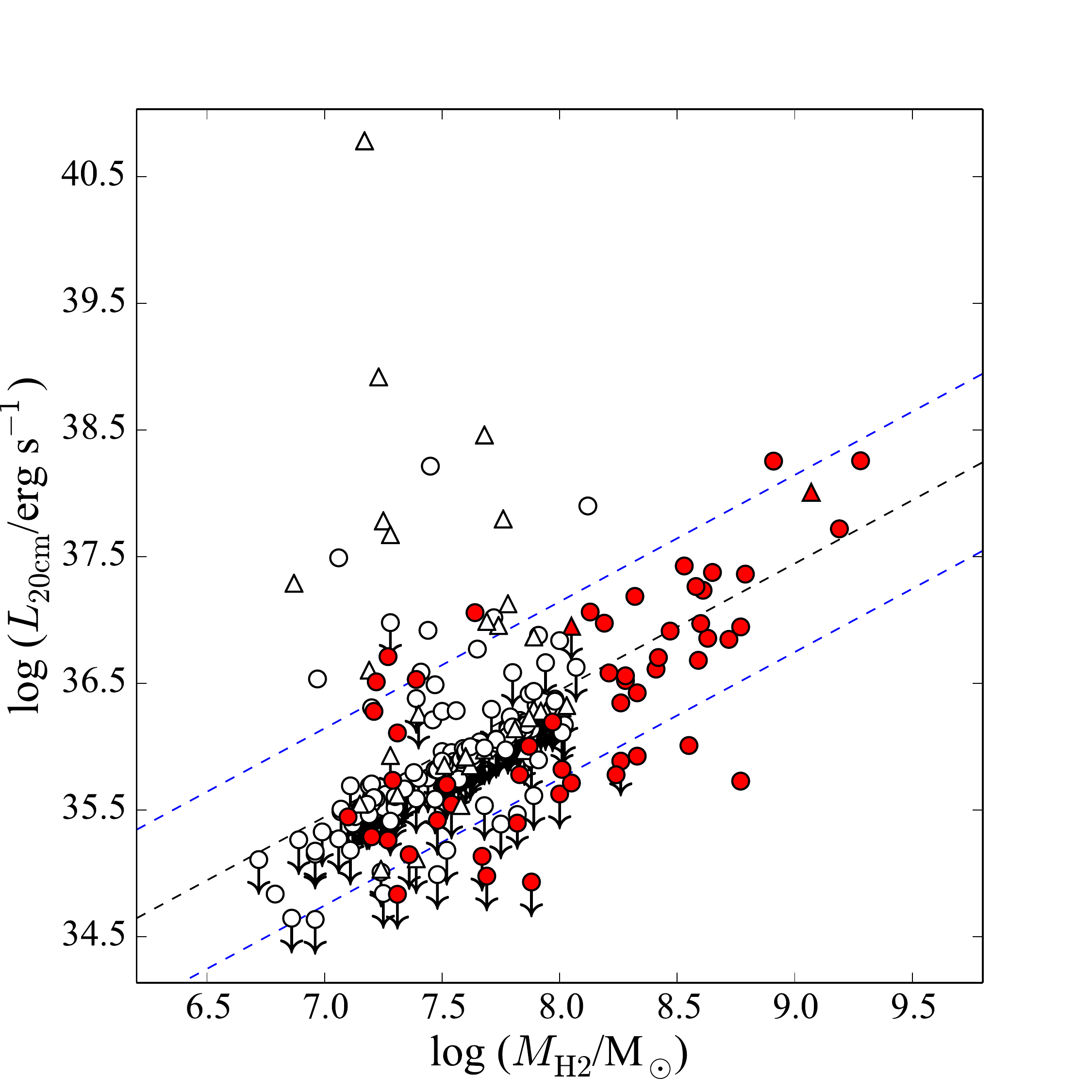}
\caption{Global radio-M$_{\mathrm{H2}}$ relation for the \atlas\ survey.  H$_2$ masses were derived from the single-dish IRAM CO measurements \citep{young+11}.  CO detections are highlighted by red symbols and CO upper limits are shown as white symbols.  Upper limits to the 1.4~GHz luminosity are shown as downward-pointing arrows.Circles represent fast rotators and triangles represent slow rotators \citep{emsellem+11}.  
The dashed black line represents the expected radio luminosity (Equation~15, \citealt{murphy+11}) if the SFRs of the \atlas\ ETGs agree with the SFRs predicted by the CO-derived H$_2$ mass.  
Assuming a conversion factor of $\alpha \equiv M_{\mathrm{gas}}/L_{\mathrm{CO}} 
= 4.6 \mathrm{M}_{\odot}$ (K~km~s$^{-1}$ pc$^{2})^{-1}$ \citep{solomon+05}, this SFR relation is SFR $= 1.43 \times 10^{-9}$ ($M_{\mathrm{H2}}$/M$_{\odot}$) M$_{\odot}$ yr$^{-1}$.  
The upper and lower dashed blue lines denote $L_{20\mathrm{cm}}/M_{\mathrm{H2}}$ ratios of factors of 5 above and below the expected radio luminosity at a given molecular gas mass for typical star-forming galaxies.}
\label{fig:radio_CO}
\end{figure}

In Figure~\ref{fig:radio_CO}, we investigate the relationship between the molecular gas mass and radio luminosity.  The dashed black line in this figure traces the expected 1.4~GHz luminosity based on the H$_2$-mass-derived SFR \citep{gao+04} and the calibration between the SFR and radio continuum luminosity from \citet{murphy+11}.  In other words, this line denotes the radio luminosity one would expect if the H$_2$-SFR and radio-SFR relationships previously established for star-forming spiral galaxies were also true for ETGs.
Some of the most molecular gas-rich ETGs shown in Figure~\ref{fig:radio_CO} have 1.4~GHz luminosities consistent with this extrapolation, suggesting they are forming new stars with efficiencies similar to those found in spiral galaxies.  However, other ETGs in Figure~\ref{fig:radio_CO}, particularly those with lower H$_2$ masses, appear to have less radio continuum emission than expected.  In these galaxies, the radio emission may be genuinely suppressed.  Alternatively, variations in the CO-to-H$_2$ conversion factor ($X_{\mathrm{CO}}$) could cause the H$_2$ masses to be overestimated (see Section~\ref{sec:Xco}).  Galaxies that are obvious outliers in Figure~\ref{fig:radio_CO}, with high radio luminosities and only upper limits to their molecular gas masses, are likely massive ETGs dominated by AGN emission (see Section~\ref{sec:AGNs}).

Of the 56 CO-detected and candidate star-forming ETGs shown in Figure~\ref{fig:radio_CO}, at least 18 (32 $\pm$ 6\%) have 1.4~GHz luminosities a factor of 5 above/below the predicted radio luminosity indicated by the dashed line.  The 5 CO-detected ETGs with radio emission exceeding the level expected from SF are NGC2768, NGC3245, NGC3665, NGC4111, and NGC4203.  The enhanced radio emission in these galaxies may be the result of nuclear activity.  A clear example of this is NGC3665, a low-power AGN host with prominent kpc-scale radio jets (see Figure~\ref{fig:radio_images}) that are responsible for the excess radio emission.  Two other galaxies, NGC2768 and NGC4203, are classified as LINERs based on their optical emission line ratios \citep{nyland+16}, and may also be contaminated by nuclear activity at 1.4~GHz.

There are 13 CO-detected ETGs with radio luminosities deficient by at least a factor of 5 from the level predicted by standard SF relations.  Of these, 7 have 1.4~GHz detections (NGC4150, NGC4429, NGC4459, NGC4753, NGC5273, NGC5379, and UGC09519), and 6 have upper limits only (NGC3156, NGC4119, NGC4324, NGC4596, PGC016060, and PGC061468).  For the ETGs with the most extreme radio deficiencies, NGC4119 and UGC09519, the radio emission is deficient by factors of about 25 and 30, respectively.  
An additional 6 galaxies (NGC0509, NGC3073, NGC3599, NGC4283, NGC4476, and NGC4477) have radio upper limits within a factor of 5 above/below the dashed line in Figure~\ref{fig:radio_CO}.  

If the radio deficiency relative to the H$_2$ mass genuinely exists and is not the result of a varying $X_{\mathrm{CO}}$, possible causes include reduced star formation efficiency (SFE), predominantly low-mass SF, weak galactic magnetic fields, and enhanced cosmic ray losses.  We further discuss these potential explanations in Section~\ref{sec:discussion}.  In the following section, we examine the relationship between the radio continuum and IR emission, another interesting proxy of the global SF conditions.

\subsection{Radio-Infrared Relation}
\label{sec:IR}

\subsubsection{Far-Infrared}
\label{sec:FIR}
Many previous studies have explored the FIR-radio relation for various samples of galaxies (e.g., \citealt{yun+01, condon+02}).  These studies have determined a range of average q-values characteristic of typical SF, where the q-value is defined as:

\begin{equation}
\label{eq:q}
q \equiv \log \left( \frac{\mathrm{FIR}}{3.75 \times 10^{12} \, \mathrm{W} \, \mathrm{m}^{-2}} \right) - \log \left(\frac{S_{1.4 \, \mathrm{GHz}}}{\mathrm{W} \, \mathrm{m}^{-2} \, \mathrm{Hz}^{-1}} \right),
\end{equation}

\medskip
\noindent and FIR is the standard FIR estimator defined as:

\begin{equation}
\label{eq:FIR}
\mathrm{FIR} \equiv 1.26 \times 10^{-14} \, (2.58\,S_{60 \mu \mathrm{m}} + S_{100 \mu \mathrm{m}})  \, \mathrm{W} \, \mathrm{m}^{-2}.
\end{equation}

\noindent One of the most widely-cited publications, \citet{yun+01}, reports an average q-value of 2.34, with q $<$ 1.64 and q $>$ 3.00 defining ``radio-excess" and ``FIR-excess" galaxies, respectively.

\begin{figure*}
\includegraphics[clip=true, trim=0cm 0cm 2.2cm 1.8cm, scale=0.4]{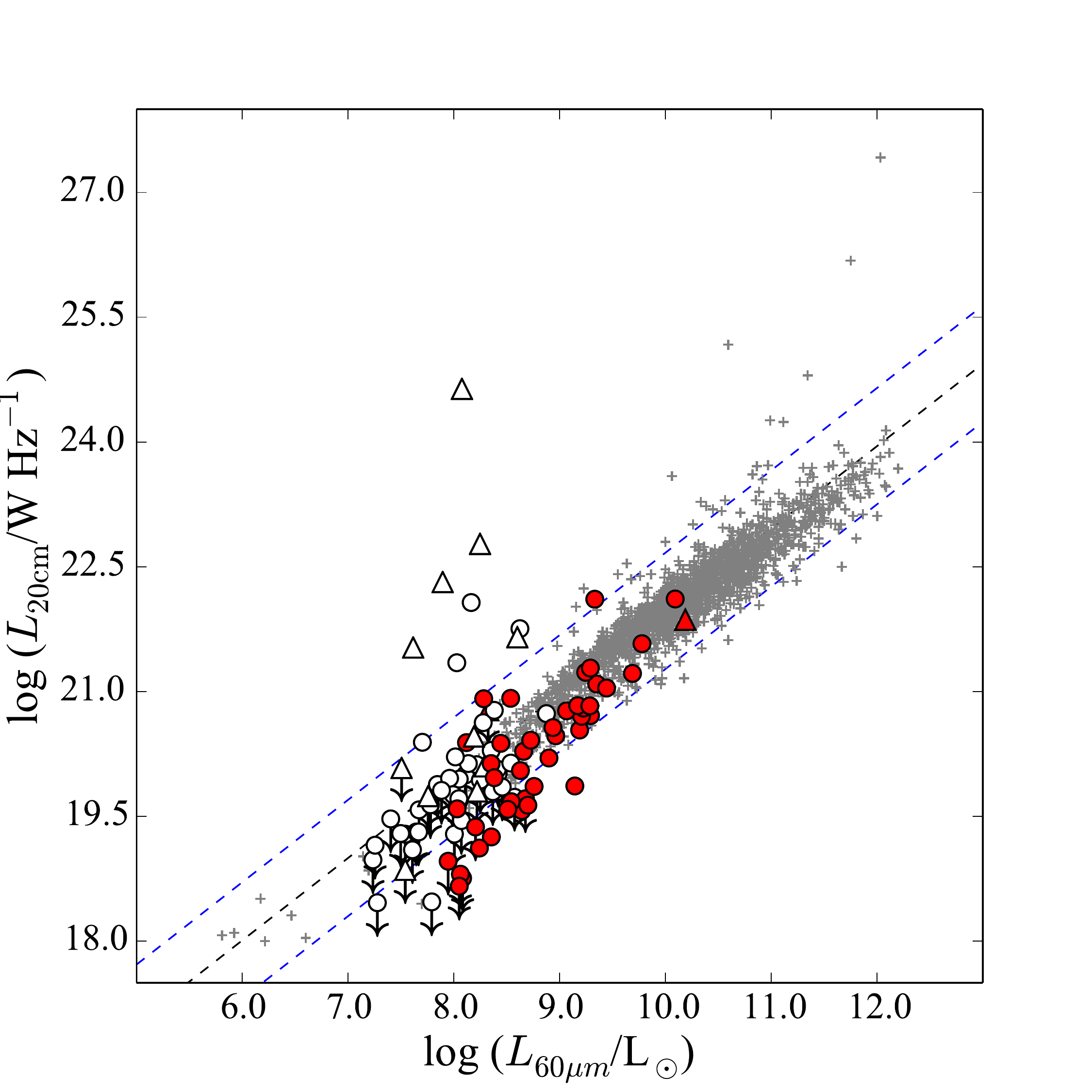}
\includegraphics[clip=true, trim=0cm 0cm 2.2cm 1.8cm, scale=0.4]{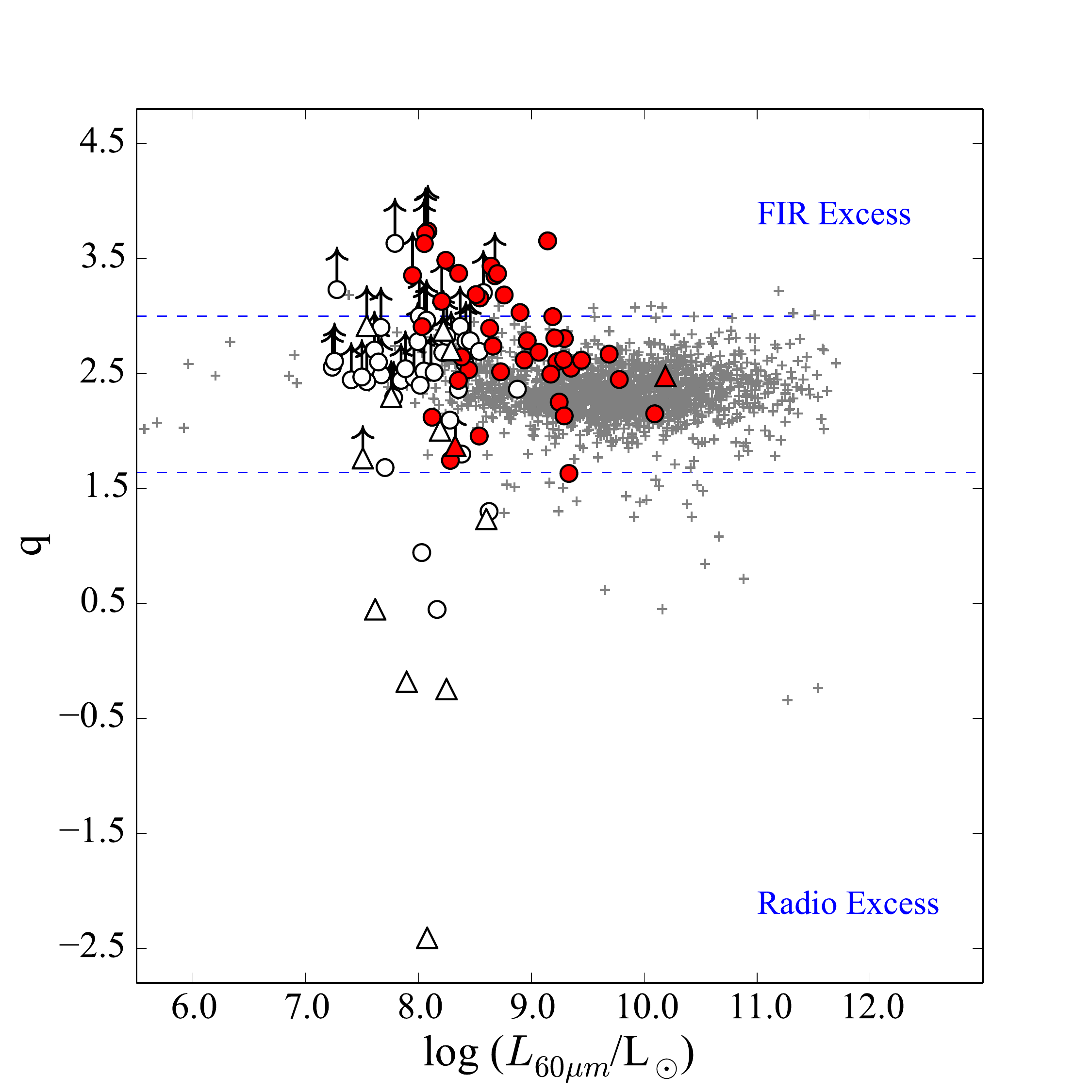}
\includegraphics[clip=true, trim=0cm 0cm 2.2cm 1.5cm, scale=0.4]{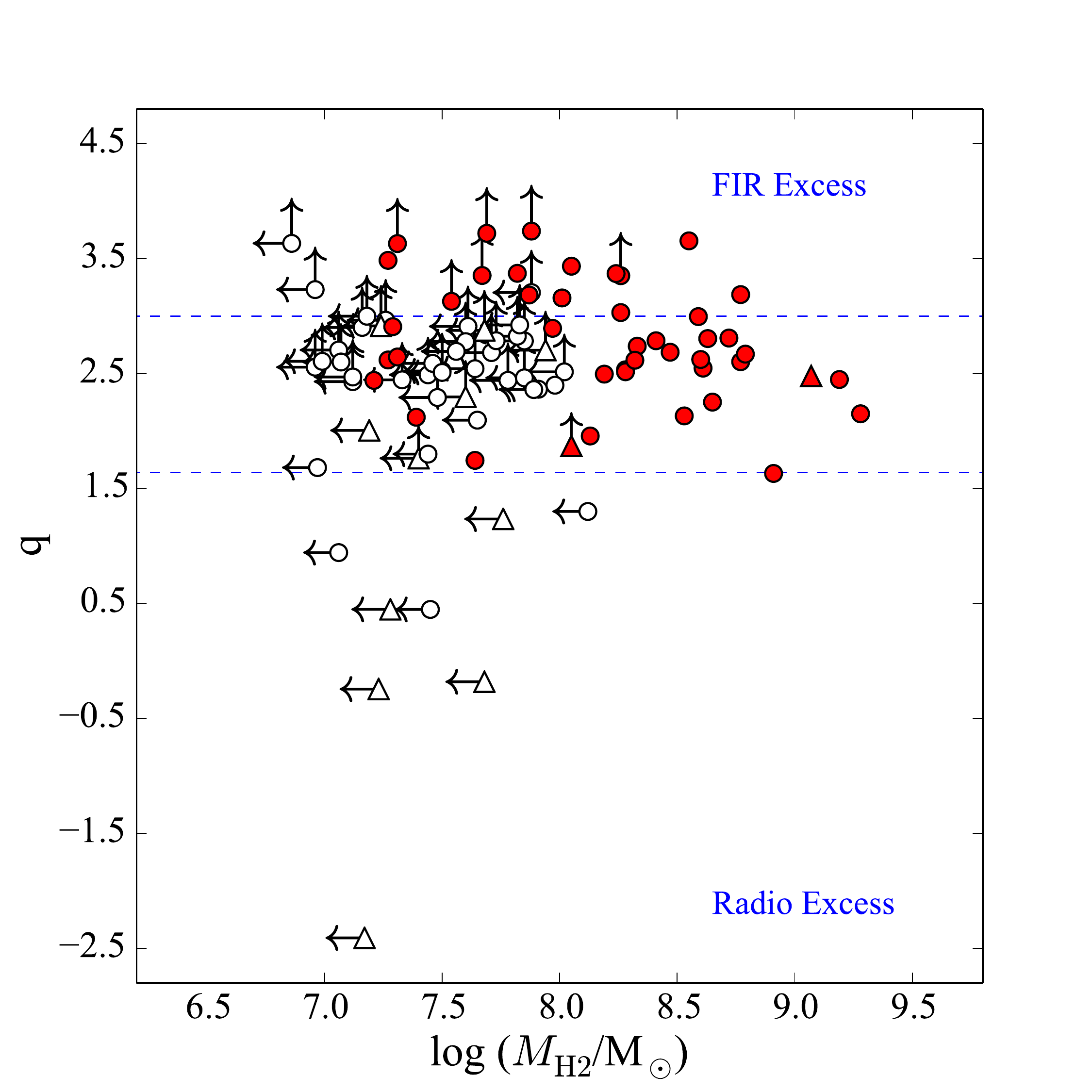}
\caption{FIR-radio relation of the \atlas\ galaxies.  Symbols filled in red represent the \atlas\ IRAM single-dish CO detections, while white symbols represent CO upper limits \citep{young+11}.  Circles represent fast rotators while triangles represent slow rotators \citep{emsellem+11}.  The grey symbols show the distribution of data points included in the analysis of the FIR-radio correlation presented in \citet{yun+01}.  
{\bf (Top Left:)} The global radio-60$\mu$m relation for the subset of the \atlas\ galaxies with {\it IRAS} 60$\mu$m detections.  The dashed black line is the formal fit to the relation defined in \citet{yun+01}.  The dashed blue lines denote factors of 5 above and below the fit to the 20cm-60$\mu$m relation.  Upper limits to the 1.4~GHz luminosity are shown as downward-pointing arrows.
{\bf (Top Right:)} The logarithmic FIR-radio flux density ratio, q, as a function of the 60$\mu$m luminosity.  The upper and lower dashed blue lines denote the classic divisions between sources with excess FIR (q $>$ 3.00) and radio (q $<$ 1.64) emission, respectively \citep{yun+01}.  Lower-limits to the q-value are shown as upward-pointing arrows.
{\bf (Bottom:)} Same as the top right panel, except here the q-value is shown as a function of H$_{2}$ mass \citep{young+11}.  Upper limits to the H$_{2}$ mass are shown as leftward-pointing arrows.
}
\label{fig:radio_FIR}
\end{figure*}

In the top left panel of Figure~\ref{fig:radio_FIR}, we have plotted the 20cm radio luminosity as a function of the FIR luminosity measured at 60$\mu$m for the 94 \atlas\ galaxies with {\it IRAS} detections at both 60 and 100$\mu$m.  A few galaxies have excess radio continuum emission well beyond what would be expected if they were dominated by SF alone.  These sources lie above the relationship for typical star-forming galaxies illustrated by the upper blue dashed line in the top left panel of Figure~\ref{fig:radio_FIR}, and include many well-known AGNs in our sample.  The top right and bottom panels of Figure~\ref{fig:radio_FIR} also clearly highlight these galaxies.  The 9 galaxies in the radio excess category are NGC3665, NGC3998, NGC4261, NGC4278, NGC4374, NGC4486, NGC4552, NGC5322, and NGC5353.  Only two of these galaxies, NGC3665 \citep{young+11, alatalo+13} and NGC3998 \citep{baldi+15}, are known to harbour any molecular gas.

Thirty-five ETGs detected at 1.4~GHz and have q-values consistent with typical star-forming galaxies, suggesting the presence of active SF in these systems (for alternative possibilities, see Section~\ref{sec:normal_alternative}).  These ETGs tend to have high FIR luminosities (top left and right panels of Figure~\ref{fig:radio_FIR}) and H$_{2}$ masses (bottom panel of Figure~\ref{fig:radio_FIR}).  However, even among the ETGs with ``normal" q-values consistent with SF, there is still a tendency towards higher q-values.  Most of our sample galaxies have systematically high FIR-radio ratios at a given 60$\mu$m luminosity and H$_{2}$ mass, suggesting that star-forming ETGs are either over-luminous in the FIR or under-luminous at radio frequencies compared to typical star-forming spirals.  This effect becomes more significant at low FIR luminosities, in-line with reports from previous studies of a possible steepening of the relation for galaxies with $L_{60\mu \mathrm{m}} < 10^{9}$ $L_{\odot}$ \citep{yun+01}.

As shown in the top right panel of Figure~\ref{fig:radio_FIR}, many of the ETGs in our study may be classified as FIR-excess sources based on their high FIR-radio ratios (q $>$ 3.00; \citealt{yun+01}).  A total of 18 galaxies (19\%) are characterized by FIR-radio ratios in the FIR-excess regime (see Table~\ref{tab:summary_all}).  To put this into perspective, less than 1\% of the galaxies included in the study by \citet{yun+01} fell into the FIR-excess category.  An additional 32 galaxies in our study with q-values in the range of normal star-forming galaxies only have 20cm upper limits, meaning their q-values are {\it lower limits} and may be even higher in reality.

The results of our FIR-radio analysis are generally consistent with previous studies.  
\citet{wrobel+91b} reported that, while ellipticals tended to lie above the FIR-radio relation due to excess radio emission likely originating from AGNs, lenticular galaxies generally conformed to the relation.  However, they also identified a population of FIR-excess lenticulars, most of which were non-detections in their 5~GHz radio continuum study.  These results are consistent with our study, in which many of the radio-excess sources are classified kinematically as slow rotators (massive ellipticals) and all of the FIR-excess sources are fast rotators (lower-mass ellipticals and lenticulars).  The fraction of sources in the FIR-excess category in \citet{wrobel+91b} is roughly 10\%, much more similar to the fraction found in our study (19\%) than in studies of normal star-forming spiral galaxies (e.g., $< 1$\%; \citealt{yun+01}).

More recently, \citet{combes+07} presented a study of the molecular gas and SF properties of the galaxies included in the SAURON survey \citep{dezeeuw+02}, a representative sample of 48 nearby ETGs with IFS observations.  They concluded that the ETGs typically follow the radio-FIR relation, especially those with high H$_2$ masses.  However, many of their FIR-radio ratio measurements were based on upper limits from FIRST, suggesting that some of the ETGs might actually reside in the FIR-excess regime.  Additional studies (e.g., \citealt{lucero+07, crocker+11}) have confirmed that, while some ETGs are characterized by FIR-radio ratios consistent with star-forming spiral galaxies, many ETGs not dominated by AGNs show enhancements in their FIR emission relative to their emission at radio frequencies.

\subsubsection{Mid-Infrared}
\label{sec:MIR}
FIR emission is a robust SF tracer since it is sensitive to cool dust embedded deep within dense molecular cores present in star-forming regions.  However, only $~\sim$36\% of the \atlas\ galaxies are detected in the FIR with {\it IRAS}.  Detection rates in the MIR at 22$\mu$m from the {\it WISE} All Sky Survey, on the other hand, are 100\%.  MIR emission in star-forming galaxies arises from re-radiation of optical/UV emission by interstellar dust associated with newly formed massive stars.  Unlike FIR emission, MIR emission traces warm dust, and as a consequence SFRs based on MIR data may be underestimated in purely star-forming galaxies (e.g., \citealt{calzetti+07, jarrett+13}).  MIR emission may also arise from AGNs (e.g., \citealt{xilouris+04}) and circumstellar dust associated with evolved stars that have passed through the (p)AGB phase \citep{knapp+92, athey+02, temi+09, madau+14}.  Thus, MIR emission may overestimate SFRs in ETGs hosting dusty AGNs and/or substantial circumstellar dust from an underlying evolved stellar population.

\begin{figure}
\includegraphics[clip=true, trim=0cm 0cm 2cm 1.5cm, scale=0.4]{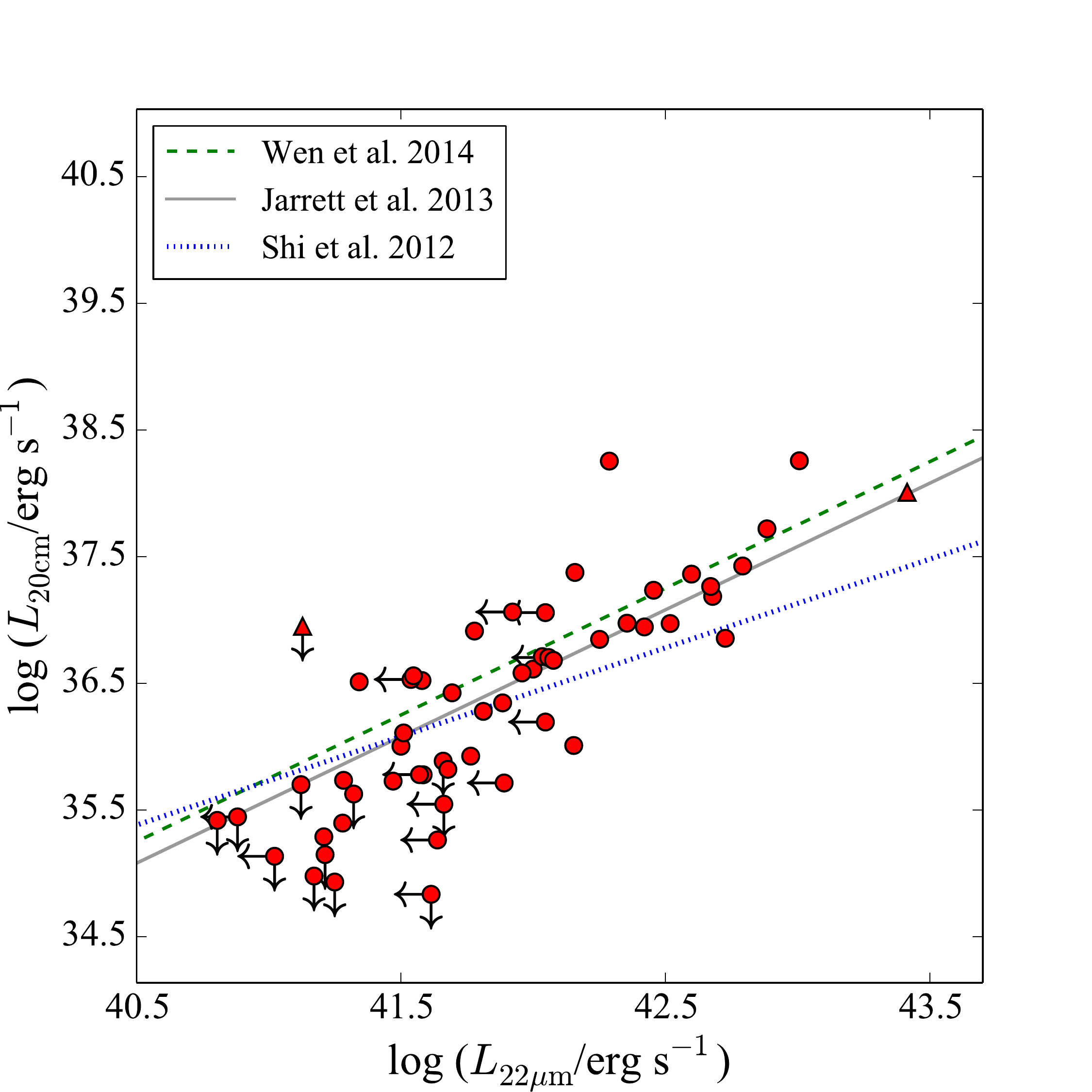}
\caption{Global radio-22$\mu$m relation for the molecular gas-rich \atlas\ ETGs.  The 22$\mu$m fluxes have been corrected for the contribution of pAGB stars using Equation~1 from \citet{davis+14}.  Symbols filled in red represent the \atlas\ IRAM single-dish CO detections, while white symbols represent CO upper limits \citep{young+11}.  Circles represent fast rotators while triangles represent slow rotators \citep{emsellem+11}.
The lines represent a series of linear fits to the radio-22$\mu$m relation from the literature (green dashed: \citealt{shi+12};  solid grey: \citealt{jarrett+13}; and blue dotted: \citealt{wen+14}).}
\label{fig:radio_w4}
\end{figure}

\begin{figure}
\includegraphics[clip=true, trim=0cm 0cm 2cm 1.5cm, scale=0.4]{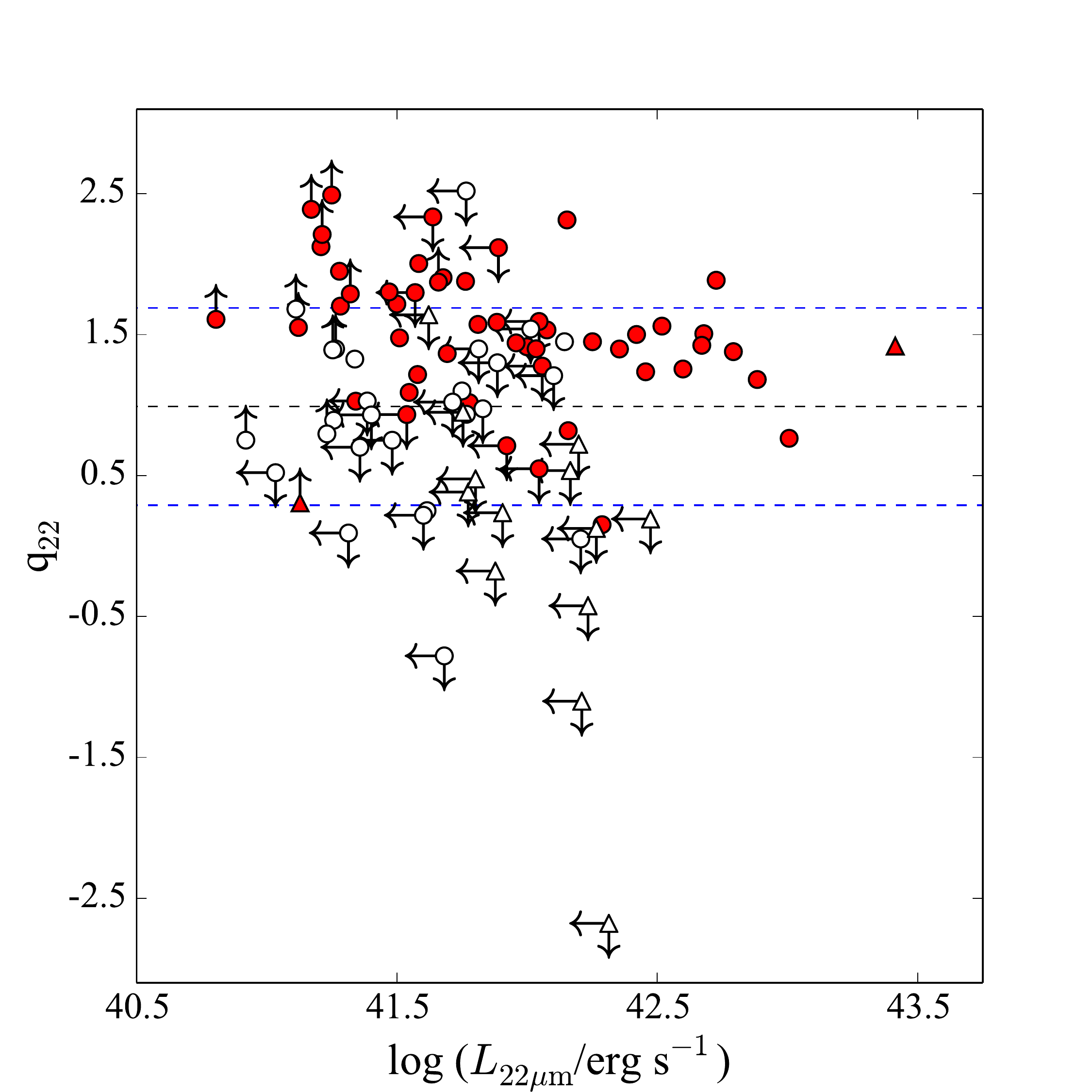}
\caption{Logarithmic 22$\mu$m-radio ratio (q$_{22}$) as a function of corrected 22$\mu$m luminosity.  Symbols filled in red represent the \atlas\ IRAM single-dish CO detections, while white symbols represent CO upper limits \citep{young+11}.  Circles represent fast rotators while triangles represent slow rotators \citep{emsellem+11}.  The upper and lower dashed blue lines denote q$_{22}$ values of factors of 5 above and below the median value of the \citet{yun+01} sample of q$_{22} = 0.99$ (black dashed line), respectively.}
\label{fig:q_MIR}
\end{figure}

While separating the SF/AGN contributions to the MIR is not possible given sensitivity and spatial resolution limitations, removing the contamination to the MIR due to evolved stars is more straightforward.  We use the relation between the 2MASS $K_{\mathrm{s}}$-band luminosity and the {\it WISE} 22$\mu$m luminosity from \citet{davis+14} to estimate the portion of the MIR emission produced by old, passively evolving stars.  We then subtract this ``passive'' 22$\mu$m component from the observed {\it WISE} 22$\mu$m luminosity to obtain the MIR component related to SF.  When the passive component of the MIR emission has been removed, we refer to the 22$\mu$m luminosity as ``corrected."  The empirical relation for the corrected 22$\mu$m luminosity used in this study can be found in Equation~1 of \citet{davis+14}.

Calibrations of the MIR SFR have been studied extensively in the literature with instruments such as {\em Spitzer} (e.g., \citealt{calzetti+07, rieke+09, rujopakarn+13b}) and {\it WISE} \citep{donoso+12, shi+12, lee+13, jarrett+13, cluver+14, wen+14}.  A number of studies have also analyzed the MIR-radio relation \citep{elbaz+02, gruppioni+03, appleton+04, beswick+08}.  The general consensus in the literature is that the radio and MIR emission are indeed correlated, albeit with somewhat increased scatter compared to the FIR-radio relation.  Likely reasons for the increased scatter in the MIR-radio relation include the higher susceptibility to dust extinction at MIR wavelengths, as well as stronger contamination associated with evolved stars and dusty AGNs.

We investigate the MIR-radio relation for the \atlas\ sample in Figure~\ref{fig:radio_w4}.   For the MIR measurements, we required that our corrected 22$\mu$m luminosities exceed the intrinsic scatter of the 22$\mu$m-2.2$\mu$m relation defined in \citet{davis+14} of $\approx$0.4 dex to be considered ``detections."  Most of the \atlas\ ETGs have only upper limits to their MIR and radio emission, and so we only show the 1.4~GHz luminosity as a function of the 22$\mu$m luminosity for the 56 \atlas\ ETGs with molecular gas detections.  The characteristics of the MIR-radio relation in these molecular gas-rich ETGs is particularly relevant since they are good SF candidates.  This figure shows similar behaviour to the radio-CO and radio-FIR relationships shown in Figures~\ref{fig:radio_CO} and \ref{fig:radio_FIR}.  However, we note that many of the CO-detected ETGs in Figure~\ref{fig:radio_w4} have high MIR-radio ratios even after the passive contribution to the 22$\mu$m emission has been subtracted.

Figure~\ref{fig:radio_w4} also shows a series of linear fits to the 22$\mu$m-20cm relation from the literature \citep{shi+12, jarrett+13, wen+14}.  The closest fit to our data above 22$\mu$m luminosities of 10$^{42}$ erg~s$^{-1}$ is that of \citet{jarrett+13}, who studied the MIR-radio relation for a small sample of local galaxies (including three ETGs) with SFRs ranging from 0$-$3 M$_{\odot}$ yr$^{-1}$.  Since the relationship between the radio and MIR emission in \citet{jarrett+13} was consistent with previous studies using 24$\mu$m data from {\em Spitzer} (e.g., \citealt{rieke+09}), the relationship between the 1.4~GHz and the {\it WISE} 22$\mu$m emission in our sample is also in good agreement with these studies.  For $L_{22\mu \mathrm{m}} <$ 10$^{42}$ erg~s$^{-1}$, the radio luminosities measured for the \atlas\ galaxies begin to decline sharply from the literature extrapolations of the 22$\mu$m-radio relations.  This observed steepening of the MIR-radio relation for less MIR-luminous ETGs is consistent with the behaviour of the FIR-radio relation discussed in Section~\ref{sec:FIR}.

We show the logarithmic 22$\,\mu$m-radio ratio, q$_{22} \equiv \log_{10}(S_{22\mu\mathrm{m}}/S_{20\mathrm{cm}})$, as a function of the corrected 22$\,\mu$m luminosity in Figure~\ref{fig:q_MIR}.  A few obvious outliers associated with active nuclei have extremely low q$_{22}$-values, while a number of other galaxies with high 22$\mu$m luminosities are consistent with normal SF.  The majority of the galaxies have only upper limits on one or both parameters or are consistent with high q$_{22}$-values.  The median q$_{22}$ value for the subset of CO-detected, star-forming \atlas\ galaxies shown in Figure~\ref{fig:q_MIR} is 1.52.  For comparison, we computed the median q$_{22}$ value of the sample of spirals studied in \citet{yun+01} and found a substantially lower value of 0.99.

\subsection{CO-Infrared Relation}
\label{sec:CO_IR}
So far we have considered the global relationships of radio luminosity vs.\ molecular gas mass and radio luminosity vs.\ IR luminosity.  In these relationships, there appears to be a relative deficiency in the radio continuum luminosity compared to normal, star-forming spirals.  Before we delve into a discussion of the possible causes of this observed deficiency, we first examine the relationship between the H$_2$ mass and IR luminosity to check if any of the radio-deficient ETGs have extra contributions to the IR from AGN activity.  

\begin{figure}
\includegraphics[clip=true, trim=0cm 0cm 2cm 2cm, scale=0.4]{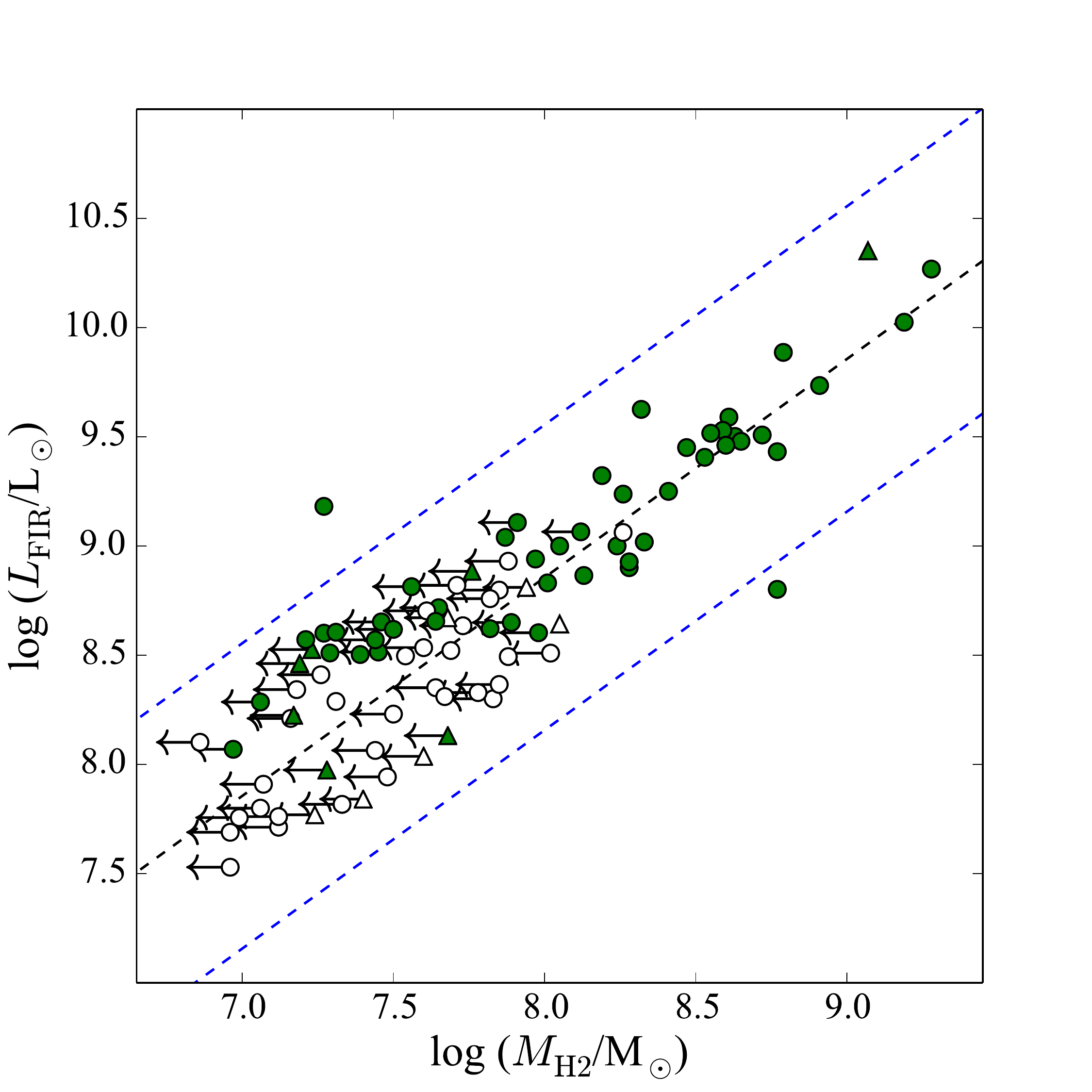}
\caption{Global $M_{\mathrm{H2}}-L_{\mathrm{FIR}}$ relation.  H$_2$ masses were derived from the single-dish IRAM CO measurements \citep{young+11}.  CO upper limits are represented by left-pointing arrows.  Green symbols are 1.4~GHz detections and white symbols are 1.4~GHz upper limits.  Circles represent fast rotators and triangles represent slow rotators \citep{emsellem+11}.  The dotted black line is an extrapolation of the IR-CO relation of spirals from \citet{gao+04}, $L_{\mathrm{FIR}}/L_{\mathrm{CO}}= 33 \Rightarrow \log L_{\mathrm{FIR}} = \log M_{\mathrm{H}2}$ + 0.86, where we have assumed a conversion factor of $\alpha \equiv$ $M_{\mathrm{gas}}$/$L_{\mathrm{CO}}$ = 4.6 M$_{\odot}$ (K km s$^{-1}$ pc$^{2}$)$^{-1}$ \citep{solomon+05}.  The upper and lower dashed blue lines denote $M_{\mathrm{H2}}/L_{\mathrm{FIR}}$ ratios of factors of 5 above and below the standard relation for spirals, respectively.
The two outliers to the extrapolated FIR-CO relation from \citet{gao+04} are NGC3245 (above) and UGC09519 (below).}
\label{fig:CO_IR}
\end{figure}

In Figure~\ref{fig:CO_IR}, we show the FIR luminosity as a function of the H$_2$ mass for the 94 \atlas\ galaxies in our sample with detections at both 60 and 100$\,\mu$m.  The H$_2$ mass and FIR luminosity are tightly related, consistent with the previous conclusions of \citet{combes+07}, who examined the H$_2$-FIR relationship for a smaller subset of the \atlas\ galaxies.  This suggests that inflation of the IR luminosities due to AGN contamination is likely not significant in our sample.  
Only two galaxies, NGC3245 and UGC09519, have H$_2$-FIR-ratios that lie slightly outside (above and below, respectively) a factor of 5 of the H$_2$-FIR relation from \citet{gao+04}.  NGC3245 may have some contribution in the IR due to AGN dust heating based on AGN evidence at other wavelengths \citep{filho+04, nyland+16}.  The low FIR luminosity of UGC09519, which is a candidate FIR-excess source, suggests the SF efficiency in this galaxy may be significantly reduced compared to that of spirals.

\section{Discussion}
\label{sec:discussion}
As mentioned previously, \citet{yun+01} reported that only 9 out of 1809 galaxies ($\approx$0.5\%) in their sample were characterized by q $>$ 3.00.    
However, we find that galaxies with high molecular gas-radio and IR-radio ratios are much more common in our sample, in agreement with the results of previous studies of the radio-IR correlation in ETGs (e.g., \citealt{wrobel+91b, lucero+07, crocker+11}).  As discussed in Section~\ref{sec:FIR}, 19\%, and perhaps as high as 53\%, of the \atlas\ galaxies with {\it IRAS} FIR measurements available are candidate FIR-excess sources.  The fact that 39\% of the CO-detected \atlas\ ETGs also have q $>$ 3.00 indicates that in some galaxies the FIR excess persists even in the presence of significant supplies of molecular gas.  These unusually high H$_2$-radio and IR-radio ratios could either be caused by enhanced CO and/or IR emission, or a relative deficiency of radio continuum emission compared to normal, star-forming galaxies.  Although it is difficult to definitively identify the foremost cause of the high q-values in the \atlas\ ETGs, we discuss a number of possibilities in the remainder of this section.

\subsection{Excess CO Emission}
\label{sec:excess_CO}

\subsubsection{{\bf $X_{\mathrm{CO}}$} Factor}
\label{sec:Xco}
The conversion factor used to derive the H$_2$ masses for the \atlas\ galaxies is $X_{\mathrm{CO}}$ = $N_{\mathrm{H2}}$/I$_{\mathrm{CO}}$ =  3 $\times$ 10$^{20}$ cm$^{-2}$ (K~km~s$^{-1}$)$^{-1}$ \citep{dickman+86} and is discussed in detail in \citet{young+11}.  However, if $X_{\mathrm{CO}}$ is in fact lower than this value, then the H$_2$ masses used in the analysis of Section~\ref{sec:radio_CO} would be overestimates.  It has long been suggested that the $X_{\mathrm{CO}}$ factor may depend on a variety of ISM parameters, such as metallicity and density (for a review, see \citealt{kennicutt+12} and \citealt{bolatto+13}).  \citet{davis+14} explored the impact on SF due to a changing $X_{\mathrm{CO}}$ in the \atlas\ sample, arguing that $X_{\mathrm{CO}}$ variations driven by metallicity or gas density fluctuations between galaxies are unlikely to have a significant impact on SFR extrapolations and SFE estimates.  

In addition to the effects of ISM properties, galaxy dynamics may also influence the $X_{\mathrm{CO}}$ factor.  \citet{davis+14} found that the CO in the \atlas\ ETGs generally resides in the rising part of galactic rotation curve, indicating that much of the molecular gas in nearby ETGs is more centrally-concentrated compared to spirals.  Some studies have reported evidence that $X_{\mathrm{CO}}$ is lower in the central bulges of spiral galaxies (e.g., \citealt{sodroski+95, meier+04, strong+04, sandstrom+13}), however other studies have contradicted this finding (e.g., \citealt{leroy+13}).  We therefore cannot rule-out the possibility that the high CO-radio ratios in our sample are caused by a systematic overestimation of the H$_2$ masses due to a $X_{\mathrm{CO}}$ conversion factor that is lower than the canonical value.  Further studies of the influence of galaxy dynamics on $X_{\mathrm{CO}}$ in ETGs will be necessary to settle this issue.

\subsubsection{Low Star Formation Efficiency}
\label{sec:SFE}
\citet{martig+09, martig+13} presented hydrodynamical simulations suggesting that the kinematic conditions characteristic of galaxy bulges and ETGs, such as high stellar velocity dispersions, can render molecular gas discs too stable to fragment into clumps and form stars efficiently.  These studies concluded that this so-called ``morphological quenching" may be more pronounced in lower-mass molecular gas discs, whereas the SFEs of larger molecular gas reservoirs should be less affected.  The dynamical processes behind morphological quenching may also be responsible for decreased SFRs in the stellar bulges of spiral galaxies \citep{saintonge+12}, although we note that this remains a controversial issue \citep{leroy+13}.  Could reduced SFEs due to a process such as morphological quenching be responsible for the deficiency of radio continuum emission relative to the molecular gas mass discussed in Section~\ref{sec:radio_CO}?

\citet{davis+14} compared the Kennicutt-Schmidt (KS) relation \citep{kennicutt+98} of nearby spiral galaxies with that of the CO-detected \atlas\ ETGs.  They found that the ETGs had lower average SFR surface densities at a given molecular gas surface density compared to spirals, suggesting a decrease in the SFE of ETGs by a factor $\approx$2.5.  This is in agreement with recent simulations by\citet{martig+13} who predicted a decrease in the SFEs of ETGs by a similar amount.  

\begin{figure}
\includegraphics[clip=true, trim=0cm 0cm 2cm 2cm, scale=0.4]{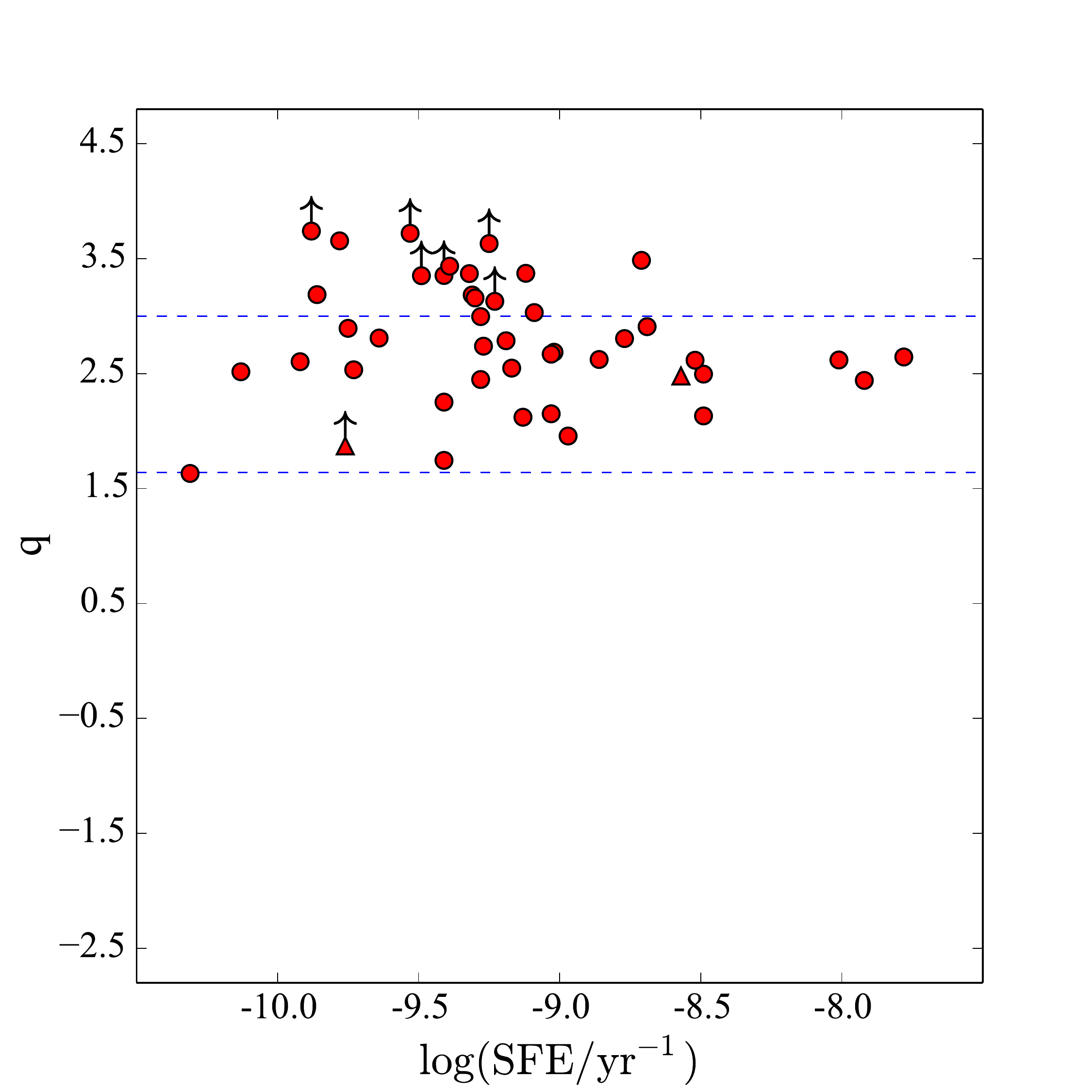}
\caption{Logarithmic FIR-radio ratio (q) of the \atlas\ galaxies with {\it IRAS} 60 and 100$\mu$m detections as a function of the star formation efficiency (SFE $\equiv$ SFR/M$_{\mathrm{gas}}$).  SFEs  were estimated from data provided in \citet{davis+14}.  
All objects shown in this figure represent the \atlas\ IRAM single-dish CO detections \citep{young+11}.  
Circles represent fast rotators while triangles represent slow rotators \citep{emsellem+11}.  The upper and lower dotted blue lines denote the classic divisions between sources with excess FIR (q $>$ 3.00) and radio (q $<$ 1.64) emission, respectively \citep{yun+01}.}
\label{fig:q_SFE}
\end{figure}

In Figure~\ref{fig:q_SFE}, we show the relationship between the FIR-radio ratio and SFE for the 44 CO-detected and candidate star-forming \atlas\ galaxies with {\it IRAS} FIR detections.  The SFE is defined here as SFR/$M_{\mathrm{gas}}$, where the SFR and $M_{\mathrm{gas}}$ are in units of M$_{\odot}$ yr$^{-1}$ and M$_{\odot}$, respectively.  $M_{\mathrm{gas}}$ is the total cold gas mass and includes gas in both the atomic (H{\tt I}) and molecular (H$_2$) phases.  The SFR and total cold gas mass for each candidate star-forming ETG were taken from Table~1 of \citet{davis+14}.  When possible, we selected the SFR measurements based on a combination of 22$\,\mu$m and far-UV data.  If far-UV data were not available, we used the 22$\,\mu$m-derived SFRs instead for calculating the SFEs.

Most of the ETGs shown in Figure~\ref{fig:q_SFE} that are forming stars with $\log$(SFE/yr$^{-1}$) $> -$9.0 have q-values within the range for typical star-forming galaxies.  However, at lower SFEs, the number of ETGs with high a q-value increases.  We speculate that this may be due to lower SFEs in these systems.  However, we emphasize that our sample is small and the difference between the incidence of high q-values at $\log$(SFE/yr$^{-1}$) $> -$9.0 and $\log$(SFE/yr$^{-1}$) $< -$9.0 is not statistically significant.

\begin{figure}
\includegraphics[clip=true, trim=0cm 0cm 2cm 2cm, scale=0.4]{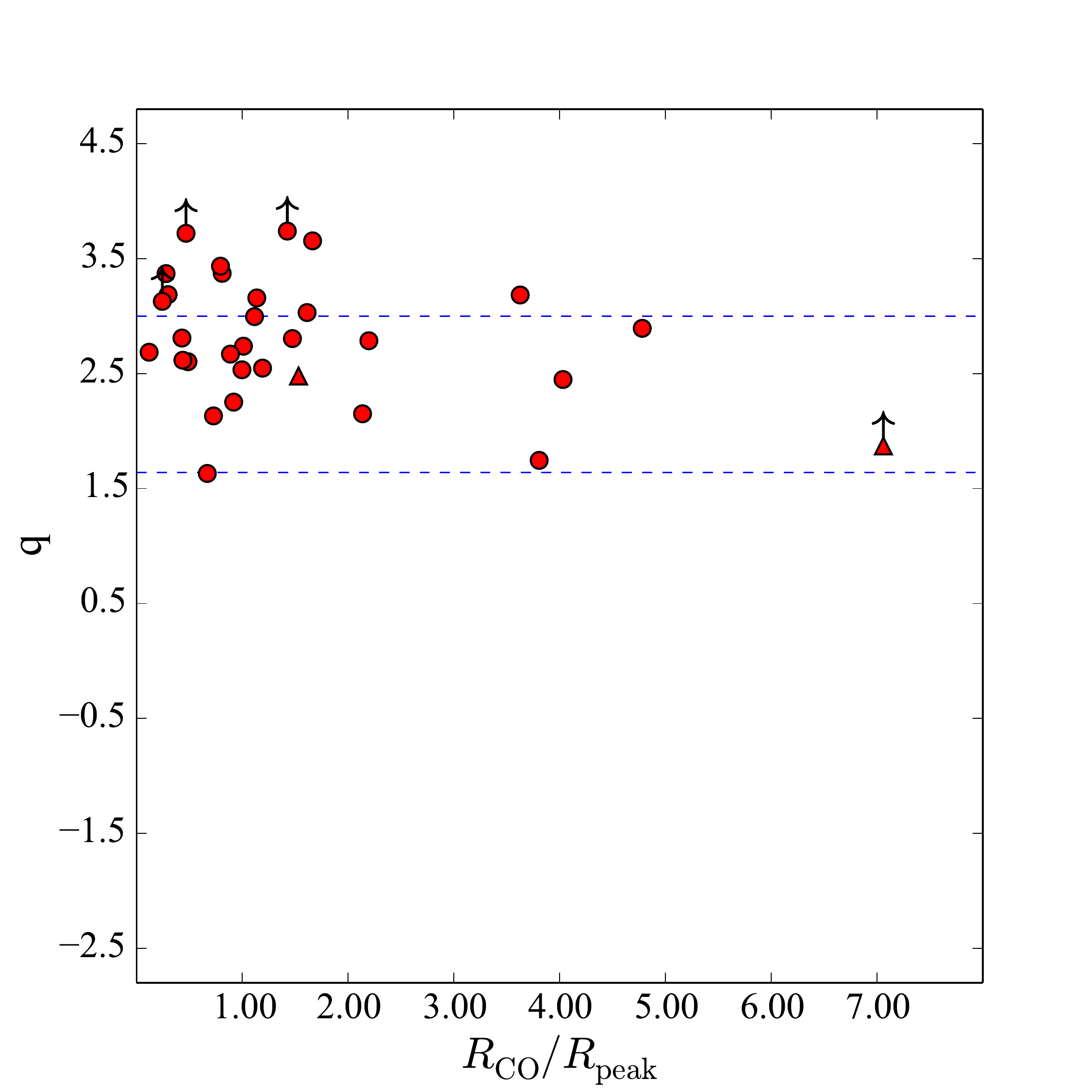}
\caption{Logarithmic FIR-radio ratio (q) of the \atlas\ galaxies with {\it IRAS} 60 and 100$\mu$m detections as a function of $R_{\mathrm{CO}}/R_{\mathrm{peak}}$, the ratio of the radius of the full extent of the molecular gas to that of the peak of the dynamically-modeled galactic rotation curve \citep{davis+11, davis+13, davis+14, cappellari+13}.  
All objects shown in this figure represent the \atlas\ IRAM single-dish CO detections \citep{young+11}.
Circles represent fast rotators while triangles represent slow rotators \citep{emsellem+11}.  The upper and lower dotted blue lines denote the classic divisions between sources with excess FIR (q $>$ 3.00) and radio (q $<$ 1.64) emission, respectively \citep{yun+01}.}
\label{fig:q_con}
\end{figure}

In Figure~\ref{fig:q_con}, we show the q-value as a function of the ratio of the radius of the full extent of the interferometrically-mapped molecular gas to that of the peak of the galactic rotation curve \citep{davis+14}.  Physically, this figure explores the dependence of q-value on the degree of central compactness of the molecular gas.  The rotation curve of each galaxy in the \atlas\ survey has been calculated based on dynamical models of the circular velocity.  Details of this calculation, and derived parameters such as the radius at which the rotation curve peaks for each galaxy ($R_{\mathrm{peak}}$), are provided in \citet{cappellari+13}.  Although the number of data points is small, Figure~\ref{fig:q_con} hints at the possibility that ETGs with more centrally concentrated reservoirs of molecular gas are more likely to also have higher q-values.  This would be consistent with the results of \citet{davis+14}, who found that the \atlas\ ETGs with the lowest SFEs had relatively compact distributions.  However, the difference between the incidence of FIR-excesses galaxies below and above $R_{\mathrm{CO}}/R_{\mathrm{peak}} = 2$ is less then 2$\sigma$, and is therefore not statistically significant.  

While decreased SFE may be responsible in part for the excess molecular gas relative to the radio continuum emission, it is a less viable explanation for the excess IR luminosity.  This is because a decreased SFE would be expected to lead to a reduction in both the radio and IR emission.  We discuss possible explanations for the excess IR emission in Section~\ref{sec:excess_IR}.

\subsection{Excess IR Emission}
\label{sec:excess_IR}

\begin{figure}
\includegraphics[clip=true, trim=0cm 0cm 2cm 2cm, scale=0.4]{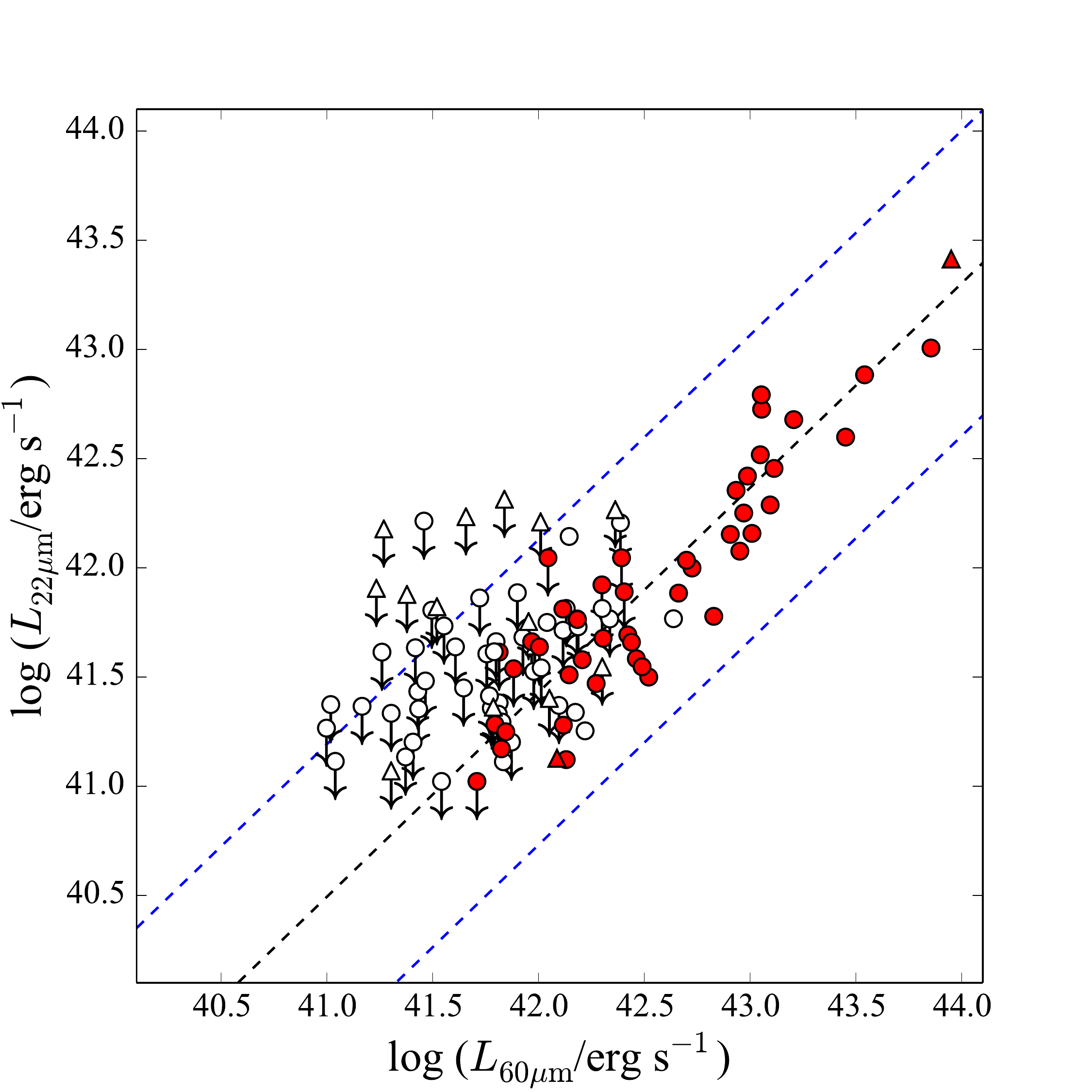}
\caption{60$\mu$m {\it IRAS} luminosity versus corrected 22$\mu$m {\it WISE} luminosity for the 106 \atlas\ ETGs with {\it IRAS} 60$\mu$m detections.  The 22$\mu$m luminosities have been corrected for the contribution due to evolved stars (Section~\ref{sec:MIR}).  Symbols filled in red represent the \atlas\ IRAM single-dish CO detections, while white symbols represent CO upper limits \citep{young+11}.  Circles represent fast rotators while triangles represent slow rotators \citep{emsellem+11}.  The dashed black line shows a 
linear fit between the 22 and 60$\mu$m luminosities for the CO-detected ETGs, and the blue dashed lines denote factors of 5 above and below it. 
Downward-pointing arrows denote corrected 22$\mu$m luminosities that are less than the intrinsic scatter of Equation~1 from \citet{davis+14}.}
\label{fig:WISE_IRAS}
\end{figure}

\subsubsection{Evolved Stars}
\label{sec:evolved_stars}
IR-based SFRs in ETGs may be contaminated by cool IR ``cirrus" emission and/or evolved stars, particularly in the MIR regime.  
IR cirrus emission at MIR and FIR wavelengths is produced by dust that has been heated by the interstellar radiation field.  Since the interstellar radiation field is driven by the evolved stellar population of a galaxy, the contribution to the IR emission by IR cirrus should depend on the stellar luminosity.  Thus, we argue that the correction applied to the MIR luminosities to account for contamination by an older stellar population in Section~\ref{sec:MIR} should effectively remove the cirrus component as well.

If any residual contamination is substantial, one might expect the IR-radio ratio to depend on the average age of the underlying stellar population.  We test this possibility in Figure~\ref{fig:q_SSP}.  This figure shows the IR-radio ratio as a function of the single stellar population (SSP) age derived from models of the optical absorption line indices measured in IFS observations \citep{mcdermid+15}.  Figure~\ref{fig:q_SSP} shows that some of the galaxies with the highest q-values have relatively young SSP ages, suggesting that excess IR emission associated with stellar mass loss from evolved stars is not the primary cause of the high IR-radio ratios.  However, we note that the SSP ages considered here are luminosity weighted, and as a consequence, young stars may dominate the light even if they constitute a less significant fraction of the total stellar mass.  Additionally, we note that Figure~\ref{fig:q_SSP} represents the relationship between q-value and SSP age in a globally-averaged sense.  Given that both the bulk of the molecular gas and the youngest stars tend to be centrally-concentrated in ETGs \citep{alatalo+13, mcdermid+12}, a spatially-resolved study of the variations of the IR-radio ratio with SSP age may lead to a different conclusion.

\subsubsection{Active Nuclei}
\label{sec:AGNs}
Seyfert nuclei are known to heat dust in their surroundings that re-radiates at IR wavelengths (e.g., \citealt{Ramos_Almeida+11, aalto+12}).  Thus, contamination from AGNs could contribute to the IR emission in IR-excess ETGs.  However, since only two ETGs in the candidate FIR-excess category are classified as Seyferts (NGC3156 and NGC4324) based on the optical emission line diagnostics reported in \citet{nyland+16}, we do not expect AGN contamination in the IR to be significant in our sample.

It is possible that dust-enshrouded AGNs are present in some of the candidate IR-excess ETGs.  Half of them (9/18; 50 $\pm$ 11\%) have nuclear radio sources identified in sub-arcsecond resolution 5~GHz data \citep{nyland+16}.  However, without any constraint on the bolometric luminosities associated with these low-power AGNs, it is difficult to assess just how much dust heating they might be able to provide.  None of the candidate IR-excess ETGs has nuclear X-ray measurements or high-resolution IR data available in the literature.  Existing high-resolution MIR studies of low-luminosity AGNs \citep{mason+12, asmus+14}, which include several \atlas\ ETGs with a nuclear radio source but a normal or low q-value, indicate that the MIR emission in most of these sources is still strongly host-galaxy dominated, arguing against the heating of dust by AGNs as a likely explanation for the IR-excess ETGs.  

\subsection{Deficient Radio Continuum Emission}
\label{sec:deficient_RC}
A number of physical scenarios could be responsible for the relative shortfall of radio continuum emission, including nascent SF, resolved-out radio continuum emission, a bottom-heavy stellar IMF, enhanced cosmic ray escape, weak galaxy magnetic fields, and environmental effects.  We discuss each of these scenarios in the sub-sections that follow.

\subsubsection{Nascent Star Formation}
\label{sec:nascent_SF}

\begin{figure}
\includegraphics[clip=true, trim=0cm 0cm 2cm 2cm, scale=0.4]{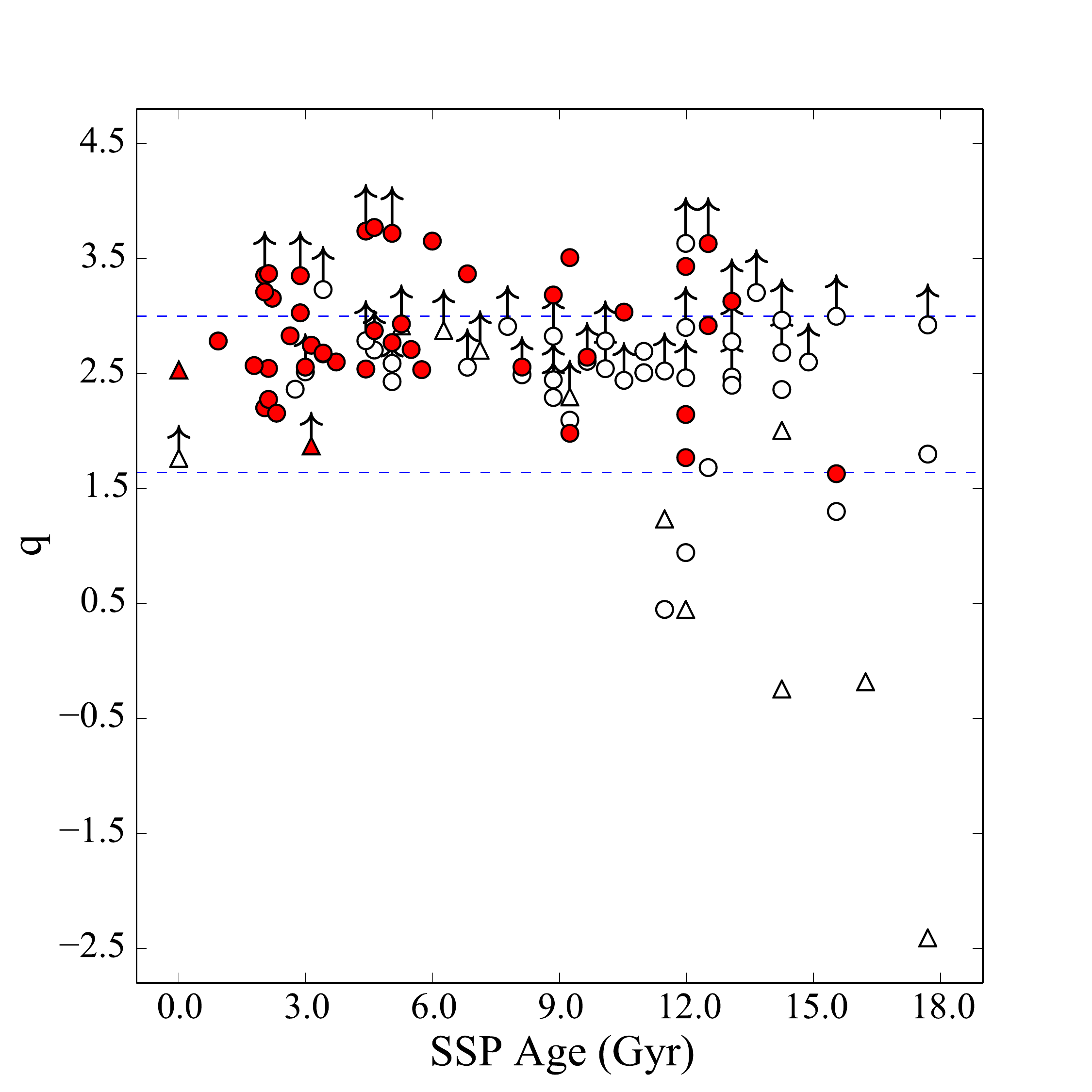}
\caption{Logarithmic FIR-radio ratio (q) of the \atlas\ galaxies with {\it IRAS} 60 and 100$\mu$m detections as a function of the single stellar population (SSP) age measured within one effective radius \citep{mcdermid+15}.  Symbols filled in red represent the \atlas\ IRAM single-dish CO detections, while white symbols represent CO upper limits \citep{young+11}.  Circles represent fast rotators while triangles represent slow rotators \citep{emsellem+11}.  The upper and lower dotted blue lines denote the classic divisions between sources with excess FIR (q $>$ 3.00) and radio (q $<$ 1.64) emission, respectively \citep{yun+01}.}
\label{fig:q_SSP}
\end{figure}

Could we have serendipitously caught some of the CO-detected ETGs in the \atlas\ sample at the cusp of a newly re-ignited episode of SF?  If nascent SF that only began a few Myrs ago were present, young stars would have had enough time to heat ambient dust and produce IR emission, but not necessarily enough time for significant numbers of supernovae to form.  This lack of supernova-driven cosmic rays would subsequently reduce the observed amount of radio continuum emission at 1.4~GHz relative to the IR emission.  Such a scenario has been suggested previously in the literature for disc-dominated galaxies with abnormally high IR-radio ratios (e.g., \citealt{roussel+03}).

However, the incidence of galaxies with q-values in the FIR-excess range is 40 times higher in the \atlas\ ETGs compared to that of \citet{yun+01}.  Therefore, it seems unlikely that young starbursts would be so much more common in ETGs compared to typical star-forming spirals.  Simple statistical considerations also argue against this scenario.  If we define the age of a nascent starburst to be less than 2 Myrs and assume the total length of the SF episode is similar to a typical orbital time of about 100~Myrs, then we would only expect to ``catch" about 2\% of the galaxies in this evolutionary state.  Thus, the expected detection rate of galaxies in a nascent SF state is at least an order of magnitude less than the fraction of ETGs that actually have deficient levels of radio continuum emission in our sample.

Figure~\ref{fig:q_SSP} further argues against the nascent SF possibility.  If nascent SF were a leading cause of the deficient radio continuum emission, we would expect the highest q-values to systematically correspond to the youngest SSP ages.  However, as already discussed in Section~\ref{sec:excess_IR}, there is no trend between q-value and SSP age.  Thus, we find that a dearth of cosmic rays due to a high incidence of exceptionally young SF in the \atlas\ ETGs is an unlikely explanation for the observed lack of radio continuum emission.

\subsubsection{Resolved-out Radio Emission}
\label{sec:resolved_out_radio}

\begin{figure*}
\includegraphics[clip=true, trim=0cm 0cm 1.5cm 1.75cm, scale=0.4]{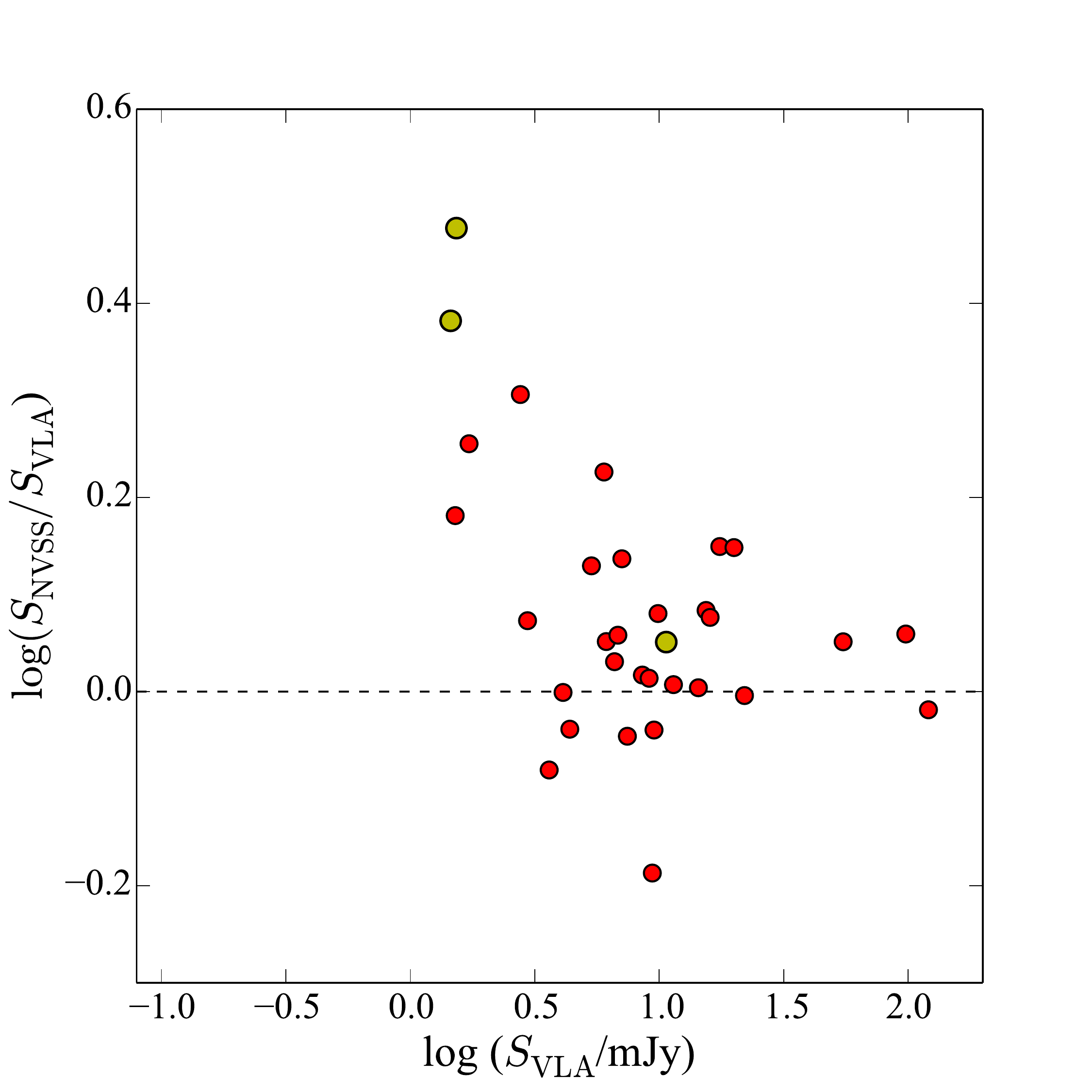}
\includegraphics[clip=true, trim=0cm 0cm 1.5cm 1.75cm, scale=0.4]{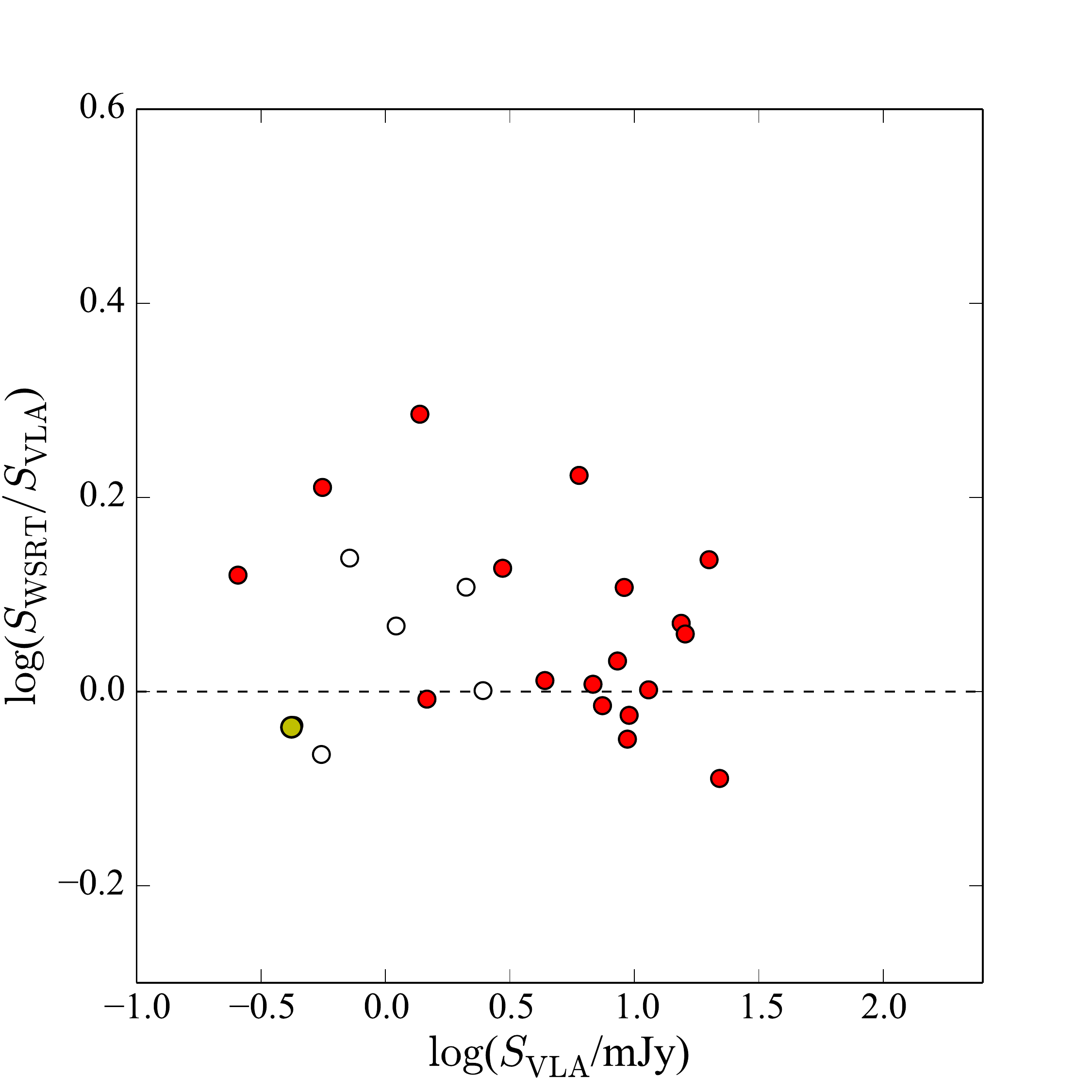}
\caption{{\bf Left:} Comparison between the higher-resolution ($\theta_{\mathrm{FWHM}} \approx 5^{\prime \prime}$) 1.4~GHz VLA flux densities and lower-resolution ($\theta_{\mathrm{FWHM}} \approx 45^{\prime \prime}$) NVSS flux densities for the 32 \atlas\ ETGs detected in both.  The $y$-axis shows the logarithmic flux ratio of the NVSS flux density to the VLA flux density and the $x$-axis shows the log of the VLA flux density at 5$^{\prime \prime}$ spatial resolution.  All of these ETGs were also detected in the \atlas\ IRAM single-dish CO observations \citep{young+11}.
The three yellow symbols highlight the galaxies initially categorized as FIR-excess sources that have both NVSS and new higher-resolution VLA detections available.  The black dashed line shows the expected logarithmic flux density ratio between the NVSS and VLA data if there were a one-to-one correspondence between the two series of data.
{\bf Right:} Same as the left panel, except that the comparison is between the flux densities of the 25 ETGs that have both new, higher-resolution ($\theta_{\mathrm{FWHM}} \approx 5^{\prime \prime}$), 1.4~GHz VLA detections and 1.4~GHz detections at lower spatial resolution ($\theta_{\mathrm{FWHM}} \approx 35^{\prime \prime}$) from WSRT.  Symbols filled in red represent the \atlas\ IRAM single-dish CO detections, while white symbols represent CO upper limits \citep{young+11}.}
\label{fig:NVSS_WSRT_compare}
\end{figure*}

The spatial resolution of the 1.4~GHz VLA data used in our analysis is $\approx 5^{\prime \prime}$.  Given the shortest spacing of 0.21~km in the VLA B-configuration in which these data were observed, structures on scales as large as $\approx 120^{\prime \prime}$ may be imaged given sufficient sensitivity.  If the radio continuum emission in some of the sample ETGs is in fact dominated by larger-scale, low-surface-brightness emission, this emission could be resolved-out or fall below our surface brightness sensitivity.  As a consequence, the q-values of ETGs with radio continuum emission predominantly distributed over larger spatial scales would actually represent upper limits.  This could in turn cause the q-values of some of these ETGs to be ``artificially'' boosted into the FIR-excess regime.

To check whether the impact of resolved-out radio emission on the q-values is significant, in Figure~\ref{fig:NVSS_WSRT_compare} we compare our VLA flux densities at lower spatial resolution with measurements from NVSS and the WSRT.  The left panel of Figure~\ref{fig:NVSS_WSRT_compare} shows the comparison between the higher-resolution VLA ($\theta_{\mathrm{FWHM}} \approx 5^{\prime \prime}$) and lower-resolution NVSS ($\theta_{\mathrm{FWHM}} \approx 45^{\prime \prime}$) flux densities for the 32 \atlas\ ETGs detected in both series of 1.4~GHz observations.  As mentioned in Section~\ref{sec:previous_studies}, the median ratio between the NVSS and VLA flux densities is 1.13 (with a range of 0.65 to 14.38).  

The right panel of Figure~\ref{fig:NVSS_WSRT_compare} is similar to the left panel, except here the 5$^{\prime \prime}$-resolution 1.4~GHz VLA flux densities are compared to the 1.4~GHz WSRT flux densities at a spatial resolution of $\theta_{\mathrm{FWHM}} \approx 35^{\prime \prime}$.  The WSRT flux densities used in this panel of Figure~\ref{fig:NVSS_WSRT_compare} were measured from images of the line-free channels from the \atlas\ H{\tt I} observations presented \citet{serra+12}.  A detailed description of these data, including flux density measurements, will be presented in a future study.  There are 25 ETGs that are detected in both the 5$^{\prime \prime}$-resolution VLA data and the lower-resolution WSRT data.  The ratio between the WSRT and VLA flux densities ranges from 0.81 to 1.93, with a median of 1.08.  Thus, compared to the WSRT data, the higher-resolution VLA data typically recovers about 92\% of the emission in the WSRT maps.  

To test whether the exclusion of any large-scale radio emission is responsible for the high q-values in any individual cases, we re-calculate q using the lower-resolution NVSS and WSRT 1.4~GHz data.  Of the four candidate FIR-excess ETGs with both 5$^{\prime \prime}$-resolution and lower-resolution 1.4~GHz data (IC0719, NGC4694, NGC4526 and UGC09519), the q-values of IC0719 and NGC4694 decrease when q is calculated using the lower-resolution radio data.  The new q values of these two ETGs based on the lower-resolution radio flux densities are now consistent with the range expected for ``normal star-forming" galaxies ($1.64 <$ q $< 3.00$).   Thus, it appears that the radio continuum emission associated with SF in IC0719 and NGC4694 is indeed much more extended than the spatial scales on which SF is actually occurring, perhaps similar to the situation in nearby starburst galaxies such as M82 (e.g., \citealt{seaquist+91}).  Since these two galaxies are not particularly unusual in other respects such as distance, CO flux density, angular size, or SFR, a future investigation into whether the spatially-extended synchrotron emission reflects increased cosmic ray diffusion length scales or unusual magnetic field configurations would be interesting.  

We conclude that incorporating radio emission on larger scales is important to avoid false identifications of FIR-excess galaxies.  An ideal means of fully addressing this issue would be to obtain sensitive, lower-resolution VLA data at 1.4~GHz in the C and D configurations for comparison with the existing higher-resolution data from the VLA B configuration.  However, we emphasize that some of the \atlas\ ETGs stubbornly remain in the FIR-excess category even when lower-resolution data are used to calculate the q-value.  The q-values of NGC4526 and UGC09519 remain high even when the radio flux density is integrated over much larger spatial scales.  Thus, some of the ETGs in our sample appear to be genuinely radio deficient.  

\subsubsection{Bottom-Heavy Stellar IMF}
\label{sec:IMF}
The stellar IMF has long been regarded as a ``universal" parameter (e.g., \citealt{bastian+10}).  
However, recent studies have reported that disc-dominated galaxies are best characterized by a Kroupa \citep{kroupa+03} IMF with a substantial fraction of high-mass stars, while massive ETGs are better characterized by a ``bottom heavy" IMF dominated by low- and intermediate-mass stars.  These studies have argued that the IMF varies systematically as a function of galaxy parameters such as mass-to-light ratio (M/L), total stellar mass, stellar velocity dispersion, bulge fraction, and metallicity (e.g., \citealt{cappellari+12, vandokkum+12, dutton+12, lasker+13, posacki+14}).  

Since only stars with $M_{\mathrm{star}} \gtrsim 8$ M$_{\odot}$ will ultimately end their lives as supernovae \citep{condon+92}, a more bottom-heavy stellar IMF in massive ETGs would reduce overall supernova rates in these galaxies.  Thus, a SFR tracer dominated by supernova-driven emission, such as centimeter-wave radio continuum observations, would naturally underestimate the SFR compared to both the molecular gas mass and the IR luminosity.  Given that all stars with masses in the range of 0.5 to 8~M$_{\odot}$ pass through the AGB phase of stellar evolution in which they produce circumstellar dust that may re-radiate at IR wavelengths \citep{marigo+08}, a bottom heavy IMF dominated by stellar masses within this range would also be consistent with deficient radio continuum emission relative to the IR (e.g., \citealt{condon+91a}).  However, we emphasize that the bottom-heavy IMF scenario for ETGs with IMFs dominated by stars with $M_{\mathrm{star}} < 0.5$ M$_{\odot}$ cannot explain the high q-values.

\begin{figure}
\includegraphics[clip=true, trim=0cm 0cm 2cm 2cm, scale=0.4]{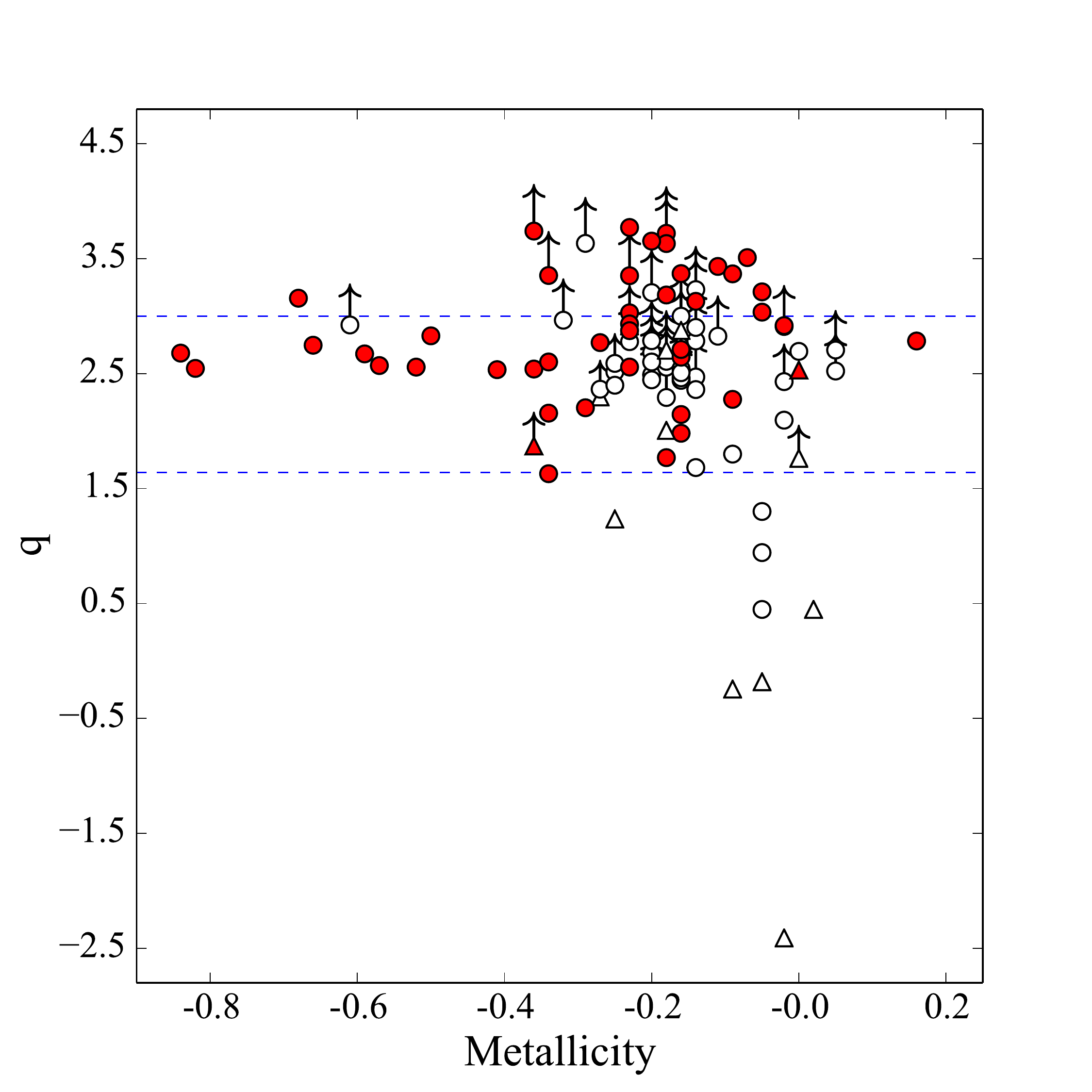}
\caption{The logarithmic FIR-radio ratio (q) for the \atlas\ galaxies with IRAS 60 and 100$\mu$m detections as a function of the metallicity, [Z/H], measured at one effective radius \citep{mcdermid+12}.  Symbols filled in red represent the \atlas\ IRAM single-dish CO detections, while white symbols represent CO upper limits \citep{young+11}.  Circles represent fast rotators while triangles represent slow rotators \citep{emsellem+11}.  The upper and lower dotted blue lines denote the classic divisions between sources with excess FIR (q $>$ 3.00) and radio (q $<$ 1.64) emission, respectively \citep{yun+01}.}
\label{fig:q_metallicity}
\end{figure}

We now consider the relationship between q-value and IMF using metallicity as a proxy.  The stellar IMF becomes systematically heavier with increasing metallicity \citep{smith+12, martin-navarro+15},  and could manifest itself as a tendency for higher-metallicity galaxies to be characterized by higher q-values due to the predominance of lower-mass stars.  In Figure~\ref{fig:q_metallicity}, we show the q-value as a function of metallicity.  Only a few of the lower-metallicity galaxies (e.g., [Z/H] $< -0.45$) have high q-values.  As argued in \citet{mcdermid+15}, many of the lowest-metallicity \atlas\ galaxies may have recently accreted new supplies of cold, low-metallicity gas.  It is interesting to note that none of these lower-metallicity ETGs have extreme q-values.  The ETGs with the highest q-values seem to have higher, near-solar metallicities.  This could be an indication that SF in higher-metallicity environments in ETGs has less radio continuum emission associated with it, possibly due to a more bottom-heavy stellar IMF.  

\subsubsection{Cosmic Ray Escape}
\label{sec:CR_losses}
More rapid/efficient cosmic ray escape would lead to a reduction in the observed radio continuum emission.  Unfortunately, robust estimates of cosmic ray diffusion rates require knowledge of many physical parameters, such as magnetic field strengths and cosmic ray production rates, that are poorly constrained at the present time, especially in bulge-dominated galaxies.  This has even resulted in conflicting predictions in the literature \citep{condon+92}.  Some theoretical studies have concluded that strong magnetic fields tend to drive cosmic rays away from their host galaxies more quickly (e.g., \citealt{chi+90}), while others have reported the opposite of this effect (e.g., \citealt{lerche+80}).

Cosmic ray escape via diffusion or convection may become significant in galaxies with low luminosities and/or masses \citep{yun+01, bell+03b, boyle+07, lacki+10a}.  \citet{yun+01} pointed out that galaxies with $L_{60\,\mu\mathrm{m}} \lesssim 10^9$ L$_{\odot}$ tend to have high q-values, and suggested that these could be lower-mass galaxies in which cosmic ray escape is more important compared to higher-mass systems.  Bell defined low-luminosity galaxies, which tend to have higher q-values, as those with $L \lesssim 0.01\, L_{*}$.  Converting the optical luminosities to stellar mass using the relation from \citet{bell+03} suggests that these low-luminosity galaxies have stellar masses of $\log(M_{*}/\mathrm{M}_{\odot}) < 10.8$.  

\begin{figure}
\includegraphics[clip=true, trim=0cm 0cm 2cm 2cm, scale=0.4]{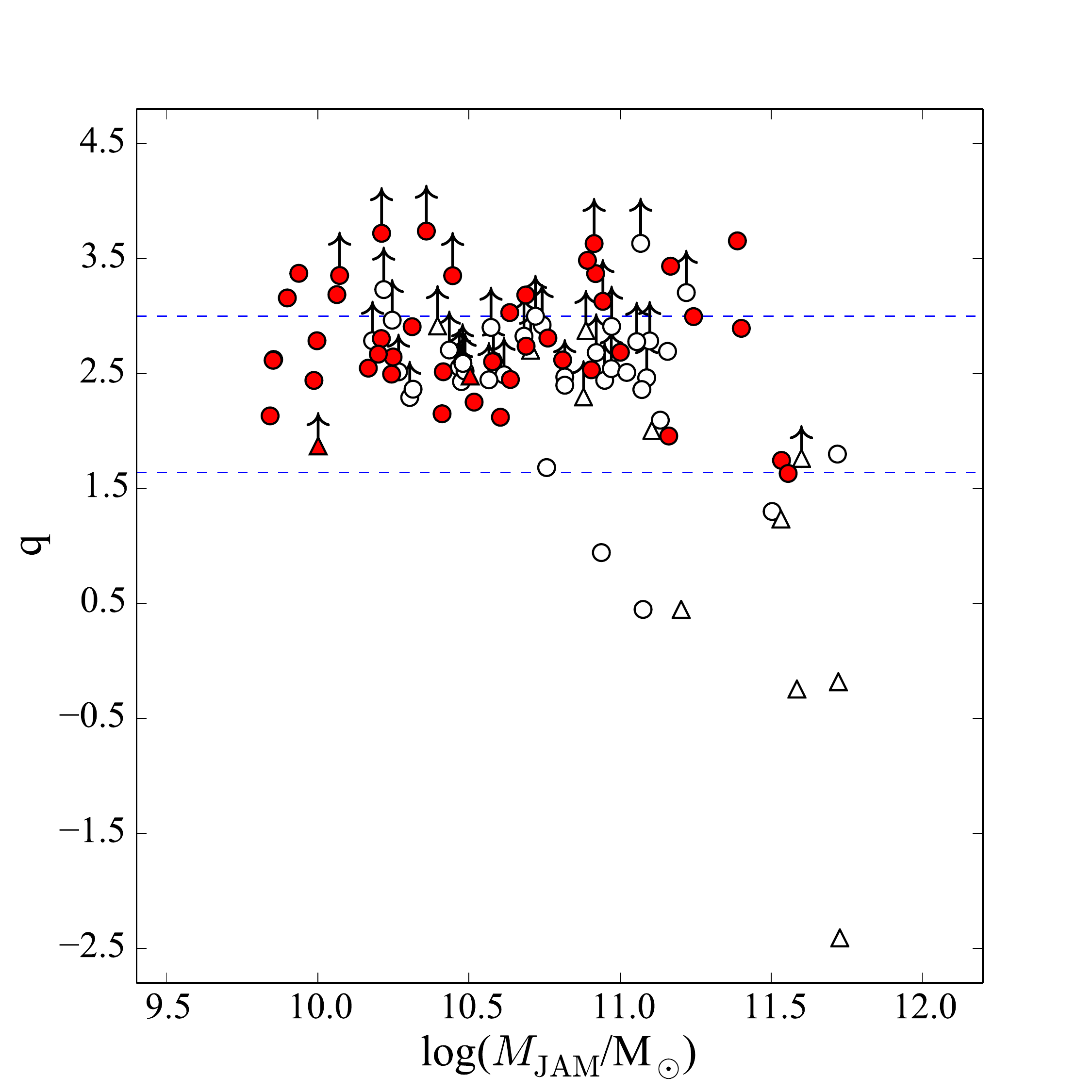}
\caption{Logarithmic FIR-radio ratio (q) of the \atlas\ galaxies with {\it IRAS} 60 and 100$\mu$m detections as a function of their dynamical mass, $M_{\mathrm{JAM}}$ \citep{cappellari+13}.  Symbols filled in red represent the \atlas\ IRAM single-dish CO detections, while white symbols represent CO upper limits \citep{young+11}.  Circles represent fast rotators while triangles represent slow rotators \citep{emsellem+11}.  The upper and lower dotted blue lines denote the classic divisions between sources with excess FIR (q $>$ 3.00) and radio (q $<$ 1.64) emission, respectively \citep{yun+01}.}
\label{fig:q_Mjam}
\end{figure}

We show the distribution of q-values as a function of dynamical mass for the \atlas\ sample in Figure~\ref{fig:q_Mjam}.  The \atlas\ stellar masses, $M_{\mathrm{JAM}}$, are based on dynamical models that account for variations in the stellar M/L due to both age and metallicity, as well as systematic variations in the IMF \citep{cappellari+12, cappellari+13}.  Figure~\ref{fig:q_Mjam} shows that the radio-excess galaxies that likely house AGNs tend to have high dynamical masses, as expected.  However, among the CO-rich ETGs likely to harbour SF, there is no strong dependence of the FIR-radio ratio on galaxy mass.  High q-values are present in some of the gas-rich ETGs with the highest dynamical masses in the sample, well above the ``low-mass" galaxy definition suggested in \citet{bell+03b}.  This suggests that cosmic ray escape due to low galaxy mass is not a dominant cause of the high q-values in our sample.

\subsubsection{Weak Magnetic Fields}
\label{sec:B_fields}
In addition to the presence of cosmic rays produced by recent supernovae, the level of radio continuum emission is also directly proportional to magnetic field strength.  While the magnetic field properties of star-forming spiral galaxies have been studied detail in the literature (e.g., \citealt{beck+13, wiegert+15, heesen+16}), the magnetic field properties of star-forming, bulge-dominated ETGs are essentially unknown.  Here, we estimate the minimum magnetic field strengths of the ETGs in our sample assuming near equipartition between the total cosmic ray particle and magnetic field energies.  We define the minimum magnetic field strength, $B_{\mathrm{min}}$, as follows:

\begin{equation}
\label{eq:bmin}
B_{\mathrm{min}}=2.3\,[(1+a)\,AL/V]^{2/7},
\end{equation}

\noindent where $a$ is the energy contribution of cosmic ray protons relative to that of electrons, $A$ is a constant\footnote{$A = C\, \frac{2\alpha + 2}{2\alpha + 1}\, \frac{\nu_{2}^{\alpha + 1/2} - \nu_{1}^{\alpha + 1/2}}{\nu_{2}^{\alpha + 1} - \nu_{1}^{\alpha + 1}}$, where $C$ is a constant of value 1.057 $\times$ 10$^{12}$ (g/cm)$^{3/4}$ s$^{-1}$, $\alpha$ is the radio spectral index, and $\nu_1$ and $\nu_2$ are the lower and upper frequencies of the radio spectrum, respectively.  We assume $\alpha = -0.8$, $\nu_1 = 10\,\mathrm{MHz}$, and $\nu_2 = 100\,\mathrm{GHz}$.}, $L$ is the radio luminosity, and $V$ is the volume of the synchrotron emitting region.  We assume a standard literature value of $a$ = 100 (e.g., \citealt{beck+01}), however, we note that the precise value of $a$, and the extent to which it varies among different galaxies or even among different environments within individual galaxies, is still poorly known.  For the volume, we assume disc-like geometries similar to those used in magnetic field studies of star-forming spiral galaxies (e.g., \citealt{tabatabaei+13}).  These disc volumes are calculated as $V = \pi(d/2)^2 z$, where $d$ is the major-axis diameter (Table~\ref{tab:gauss}) and $z$ is the scale height assumed here to be 1~kpc \citep{beck+13}.  NGC3182 and NGC3665 required special geometric considerations.  The 1.4~GHz emission of NGC3182 has a ring-like morphology, so an annular disc geometry was assumed.  For NGC3665, which has 1.4~GHz emission with an extended, narrow, jet-like morphology, we used a cylindrical geometry.

\begin{figure}
\includegraphics[clip=true, trim=0cm 0cm 2cm 2cm, scale=0.4]{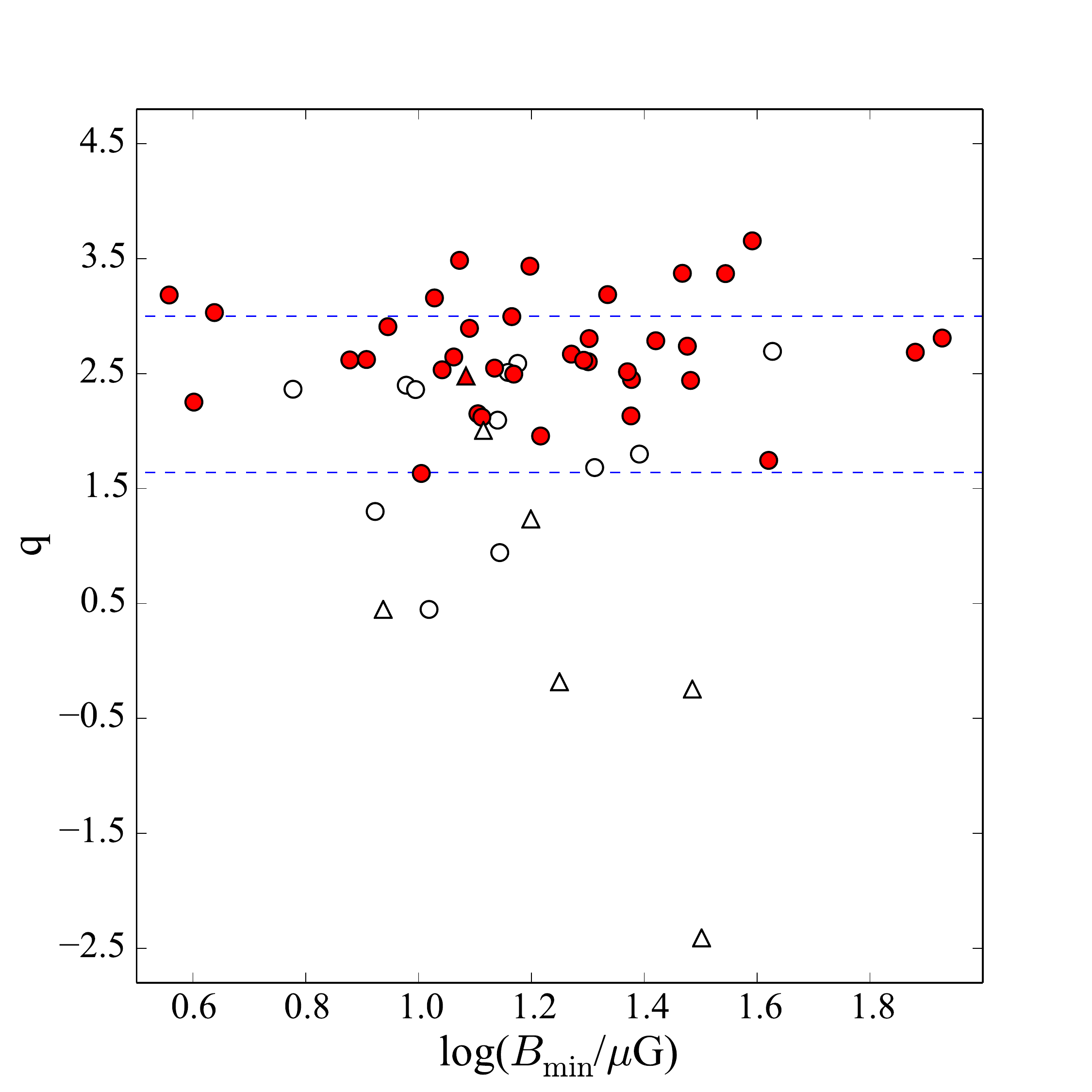}
\caption{The logarithmic FIR-radio ratio (q) of the \atlas\ galaxies with {\it IRAS} 60 and 100$\mu$m detections as a function of the minimum equipartition magnetic field.  Symbols filled in red represent the \atlas\ IRAM single-dish CO detections, while white symbols represent CO upper limits \citep{young+11}.  Circles represent fast rotators while triangles represent slow rotators \citep{emsellem+11}.  The upper and lower dotted blue lines denote the classic divisions between sources with excess FIR (q $>$ 3.00) and radio (q $<$ 1.64) emission, respectively \citep{yun+01}.  Only sources with radio detections are shown.}
\label{fig:q_B}
\end{figure}

We show the q-value as a function of the estimated (near) equipartition minimum magnetic field strength in Figure~\ref{fig:q_B}.  This figure shows no clear relationship between q-value and $B_{\mathrm{min}}$, however we emphasize that only sources with radio detections are shown.  The ETGs with the highest q-values have radio upper limits only.  It is therefore possible that weak magnetic fields in these ETGs are the dominant cause of the shortfall of radio continuum emission and subsequently high q-values.  
Figure~\ref{fig:q_B} also shows that the $B_{\mathrm{min}}$ values for the \atlas\ ETGs range from about 4 to 85$\,\mu$G, with a median magnetic field strength of about 15$\,\mu$G.  This is a factor of about 1.5 times above the average equipartition strength of the global magnetic field in the Milky Way and other similar spiral galaxies \citep{beck+01, beck+13}.  The galaxies with the strongest equipartition magnetic fields in our sample correspondingly have radio luminosities significantly higher than that of the Galactic centre.  In these galaxies, synchrotron emission associated with supermassive black hole accretion is likely contributing significantly to the 1.4~GHz flux density.  However, we emphasize that the $B_{\mathrm{min}}$ estimates for our sample ETGs carry a number of caveats.  The proton contribution to the particle energy budget compared to that of electrons ($a$), the cut-off frequencies of the radio continuum emission ($\nu_{1}$ and $\nu_{2}$), and the radio spectral index ($\alpha$) are not precisely known for these galaxies.  $B_{\mathrm{min}}$ is also fairly sensitive to changes in the radio source volume (e.g., if the diameter of the radio emitting region decreases by a factor of two, then $B_{\mathrm{min}}$ will increase by a factor of $\approx$1.8).  If larger-scale radio continuum emission is present but resolved-out in our observations for some ETGs, our $B_{\mathrm{min}}$ estimates would underestimate the true values.

We conclude that decreased magnetic fields are a plausible explanation for the high q-values in our sample.  However, additional observations of these galaxies over a broad range of frequencies and spatial scales, along with deep polarization measurements, would be necessary to verify that star-forming ETGs indeed have weaker magnetic fields than spirals and to quantify the magnitude of this effect on the q-values.

\subsubsection{Environment}
\label{sec:environment}

\begin{figure}
\includegraphics[clip=true, trim=0cm 0cm 2cm 2cm, scale=0.4]{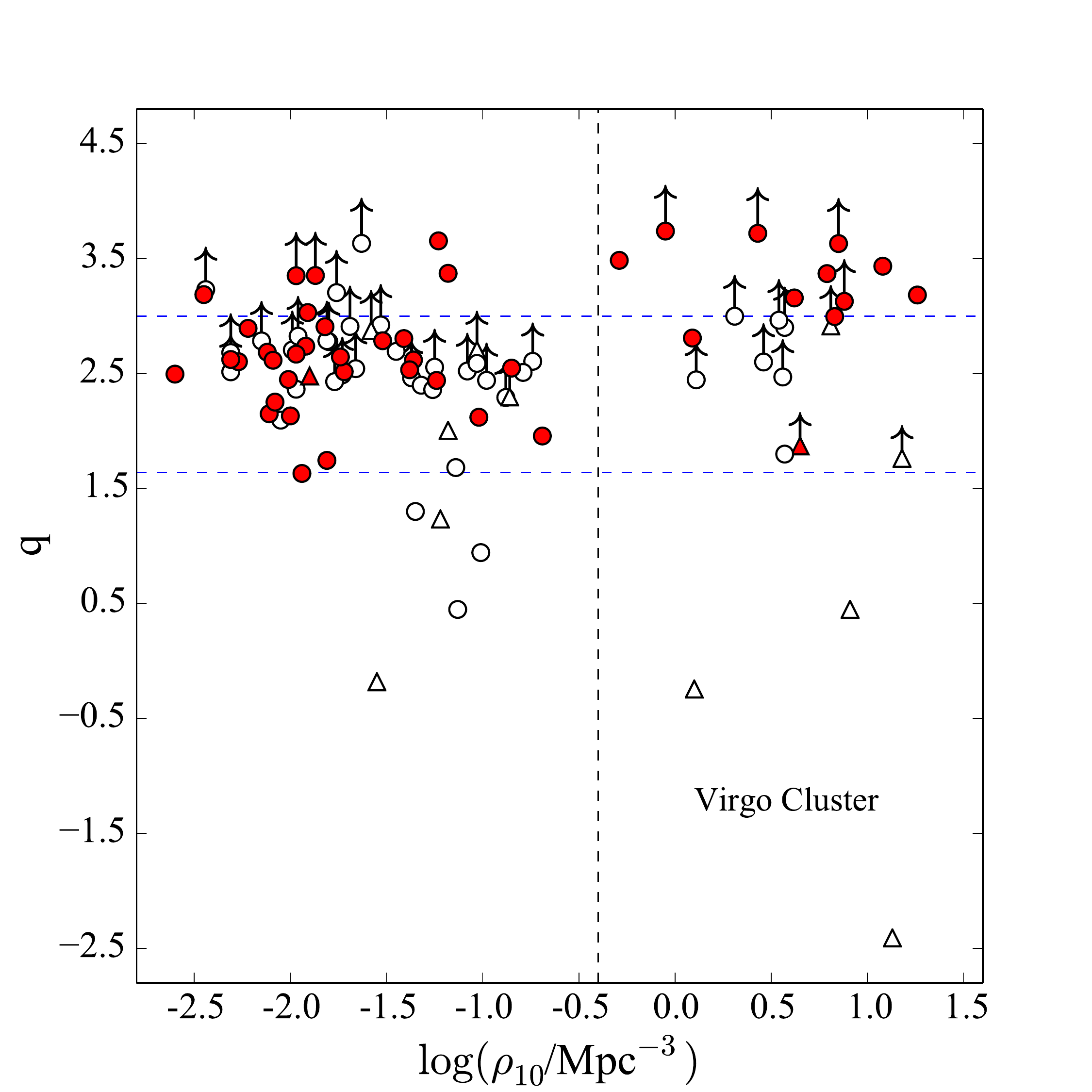}
\caption{Logarithmic FIR-radio ratio (q) of the \atlas\ galaxies with {\it IRAS} 60 and 100$\mu$m detections as a function of the mean volume density of galaxies within a sphere centered on each galaxy and containing the 10 nearest neighbours with $M_{\mathrm{K}} < -21$ ($\rho_{10}$; \citealt{cappellari+11b}).  The vertical dotted black line at $\log(\rho_{10}/\mathrm{Mpc}^{-3}) = -0.4$ separates Virgo and non-Virgo cluster members to the right and left, respectively.  Symbols filled in red represent the \atlas\ IRAM single-dish CO detections, while white symbols represent CO upper limits \citep{young+11}.  Circles represent fast rotators while triangles represent slow rotators \citep{emsellem+11}.  The upper and lower dotted blue lines denote the classic divisions between sources with excess FIR (q $>$ 3.00) and radio (q $<$ 1.64) emission, respectively \citep{yun+01}.}
\label{fig:q_density}
\end{figure}

\begin{figure*}
\includegraphics[clip=true, trim=0cm 0cm 1cm 1.5cm, scale=0.4]{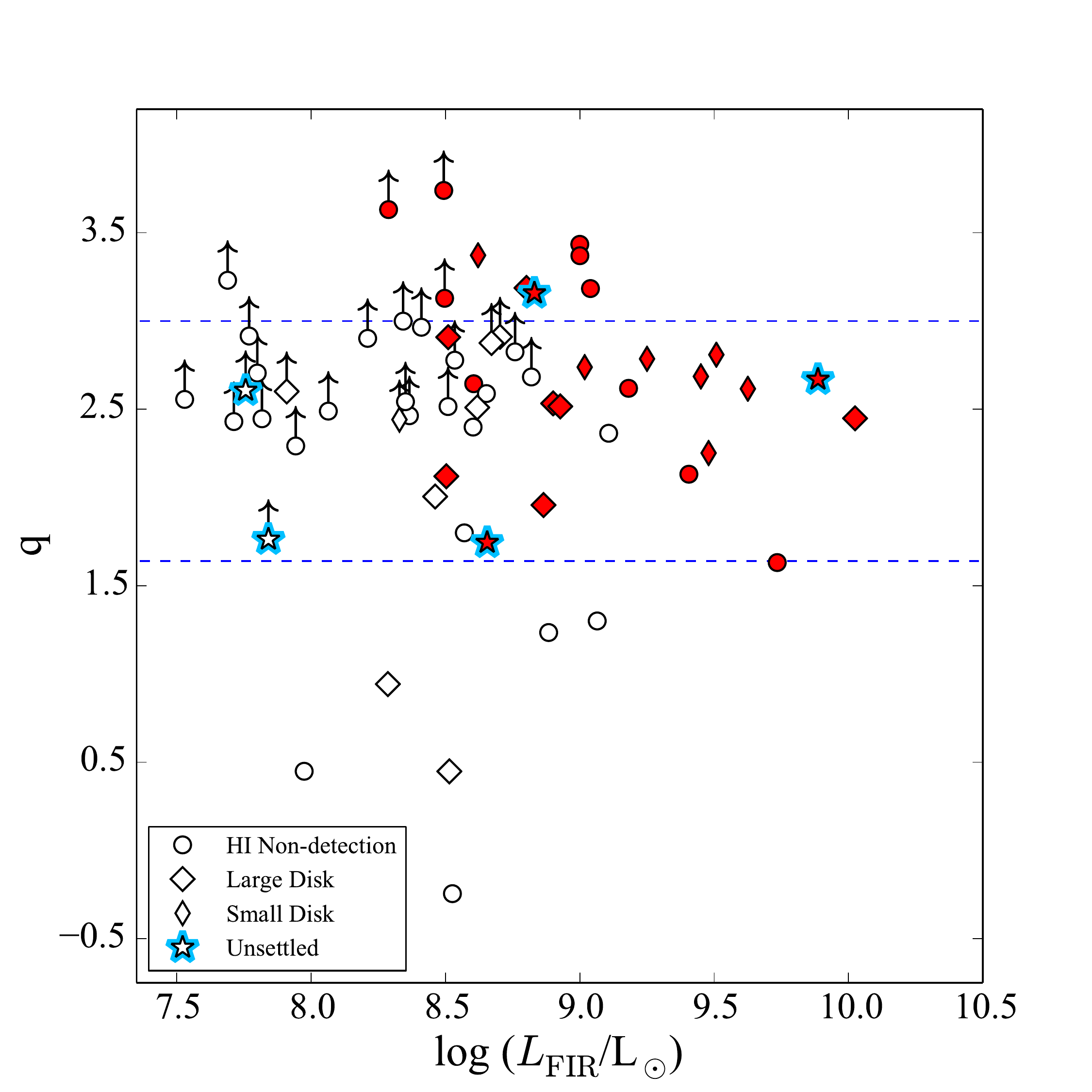}
\includegraphics[clip=true, trim=0cm 0cm 1cm 1.5cm, scale=0.4]{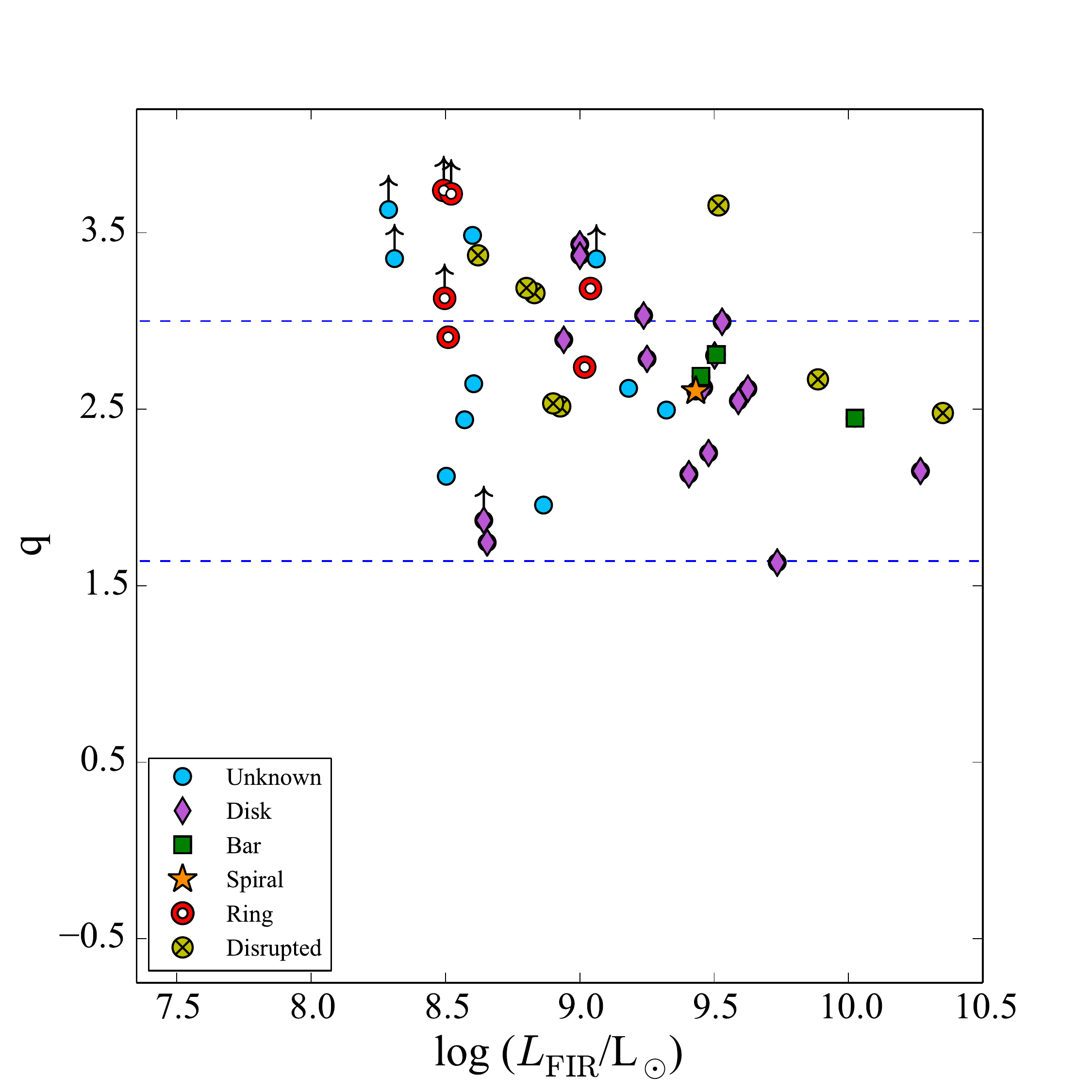}
\caption{Logarithmic FIR-radio ratio (q) as a function of the FIR luminosity for the \atlas\ galaxies with atomic and/or molecular gas morphological information.
The upper and lower dotted blue lines denote the classic divisions between sources with excess FIR (q $>$ 3.00) and radio (q $<$ 1.64) emission, respectively \citep{yun+01}.  
{\bf Left:} Symbols correspond to the H{\tt I} morphologies \citep{serra+12}, defined in the legend in the bottom-left corner of the figure.  Symbols representing ETGs with unsettled H{\tt I} morphologies are highlighted in light blue.  Symbols filled in red represent the \atlas\ IRAM single-dish CO detections, while white symbols represent CO upper limits  \citep{young+11}.
{\bf Right:} Same as the left panel, except the symbols represent the interferometric CO morphologies \citep{alatalo+13} as defined in the legend in the bottom-left corner of the figure.}
\label{fig:q_gas_morphology}
\end{figure*}

Galaxies residing in densely populated environments may suffer from gravitational interactions with the cluster potential or ``harassment" by other cluster galaxies, processes that could strip away loosely bound gas or even cosmic ray electrons \citep{moore+98, murphy+09}.  Cluster galaxies are also susceptible to ``ram pressure stripping'' \citep{gunn+72, vollmer+01}, in which gas is dislodged from galaxies as they travel through the hot intracluster medium (ICM).
Since the \atlas\ sample includes 58 (22\%) Virgo cluster members, we can study the effect of the cluster environment on a number of SF-related properties.  
\citet{serra+12} found that membership in the Virgo cluster has a strong impact on the detection rate and morphology of H{\tt I} in the \atlas\ survey, with a decreased H{\tt I} detection rate within the cluster.  That study also reported that among \atlas\ ETGs in the Virgo cluster with H{\tt I} detections, disturbed H{\tt I} morphologies are common.  

Virgo cluster ETGs also have, on average, older mass- and luminosity-weighted stellar population ages compared to field ETGs, even after controlling for galaxy mass \citep{mcdermid+15}.  This is an indication that SF histories are truncated earlier in Virgo cluster ETGs, and that their lower SFRs are long lived.  An additional clue that the ISM contents of Virgo cluster ETGs are different is their boosted $^{13}$CO/$^{12}$CO ratios relative to field ETGs, which may be due to preferential stripping of low-density molecular gas and/or the increased mid-plane pressure exerted on Virgo cluster galaxies by the ICM \citep{crocker+12, alatalo+15}.
The decreased detection rate of H{\tt I}, increased degree of central mass concentration of molecular gas, truncated SF histories, and boosted $^{13}$CO/$^{12}$CO ratios of Virgo cluster ETGs are all expected consequences of ram pressure stripping \citep{vollmer+01, tonnesen+09}.  Thus, ram pressure stripping is likely prevalent in at least some Virgo cluster \atlas\ ETGs, and it may therefore be an important process in shaping the residual SF in ETGs residing in cluster environments.

\subsubsection{Local Galaxy Density}
In Figure~\ref{fig:q_density} we show the q-value as a function of the local galaxy volume density to test whether a dense cluster environment has an effect on the FIR-radio ratio.  Although this figure shows no clear relationship, we note that galaxies with high q-values populate environments with both high and low local galaxy densities, but there are very few galaxies in the Virgo cluster with q-values consistent with normal SF.  In lower density environments, on the other hand, ETGs tend to have more moderate q-values.  

The lack of Virgo cluster ETGs with moderate q-values could be due to tidal interactions and/or ram pressure stripping that has reduced the level of radio emission associated with SF.  This would be in contrast to the results of previous studies of the FIR-radio ratio in ram pressure-stripped spiral galaxies, which have reported enhanced radio luminosities presumably due to the compression of magnetic fields via ram pressure and/or the thermal pressure of the ambient ICM \citep{miller+01} and cosmic ray re-acceleration in shock regions \citep{reddy+04, murphy+09}.  However, direct comparison of ram pressure stripping effects between spirals and ETGs may not be straightforward, and is further complicated by the fact that spirals have resided in the Virgo cluster for less time than the ETGs, and are thus not yet virialized in the cluster potential.

An alternative explanation for the deficit of Virgo ETGs that follow the radio-IR correlation is that the FIR emission is boosted by collisional dust heating due to ICM X-rays.  However, prior studies of the FIR-radio relation in clusters have failed to find evidence that such dust heating plays a significant role in generating excess emission at FIR wavelengths (e.g., \citealt{reddy+04}), so we find this scenario unlikely.  

\subsubsection{Ram Pressure Stripping}
We study the effects of ram pressure stripping in Figure~\ref{fig:q_gas_morphology}.  In the left panel of this figure, the q-value is shown as a function of the FIR luminosity for the 62 \atlas\ ETGs with both FIR and interferometric H{\tt I} data \citep{serra+12}.  Symbols are coded to represent the various neutral gas morphologies defined in \citet{serra+12}.  This figure shows no relationship between q-value and unsettled H{\tt I} morphologies, although this could be due to the small number of galaxies (5/62) with this particular H{\tt I} morphology.  Instead, disc-like H{\tt I} morphologies, as well as H{\tt I} non-detections, are more prevalent among the high-q-value sources.  Of the 22 galaxies with q-values in the normal range, 14 (64 $\pm$ 10\%) of them contain H{\tt I} distributed in a disc.  On the other hand, 24/33 (73 $\pm$ 8\%) of the ETGs with q $>$ 3.00, or q-values that are lower limits, lack any detectable H{\tt I} emission.  The lack of H{\tt I} in galaxies with high q-values could be a result of ram pressure stripping.  However, we note that these ETGs do not necessarily reside in dense environments, and additional studies will therefore be necessary to verify or refute this claim.

In the right panel of Figure~\ref{fig:q_gas_morphology}, we show the q-value as a function of FIR luminosity to study the relationship between the q-value and the molecular gas morphology for the 34 ETGs that have interferometric CARMA maps \citep{alatalo+13}.  We find no clear pattern between the CO morphology and q-value.  Of the seven galaxies identified as having disrupted molecular gas morphologies in \citet{alatalo+13}, three have q-values consistent with normal SF while four are characterized by q $> 3.00$.  Thus, we do not find compelling evidence that ETGs with signs of a recent gravitational disruption in their molecular gas distributions are more likely to have high FIR-radio ratios.  

\subsection{Origin of the CO-radio and IR-radio Relations?}
\label{sec:origins}

\subsubsection{Radio-deficient ETGs}
We now review the plausibility of explanations that could conceivably cause both the high CO-radio and IR-radio ratios seen in some \atlas\ ETGs.  Although variations in $X_{\mathrm{CO}}$ and decreased SFE could be responsible for the high CO-radio ratios, they cannot explain the high IR-radio ratios.  Thus, we find these explanations unlikely to be dominant factors in the deficient radio continuum emission, though it is possible that the high CO-radio and IR-radio ratios are caused by different mechanisms.

Since both the FIR-radio and MIR-radio ratios tend to be high for ETGs with lower luminosities, systematic effects in the IR datasets (e.g., confusion noise at low IR luminosities and contamination from dust associated with evolved stars) are likely not the root of the observed trends.  Thus, we conclude that the high q-values in some ETGs are likely the result of a genuine deficiency in the level of radio continuum emission that especially affects the lower IR luminosity and H$_2$ mass ETGs in our study.  

In some cases, the apparent deficient radio emission is due to resolved-out emission that could not be imaged by our $\theta_{\mathrm{FWHM}} = 5^{\prime \prime}$ resolution 1.4~GHz data.  However, the radio deficiency does persist in some ETGs even when data much more sensitive to extended, low-surface-brightness emission are included.  Although the radio continuum emission does appear to be genuinely suppressed in some ETGs, including those with substantial reservoirs of molecular gas, the underlying cause remains unclear.  Some scenarios, such as nascent SF, are highly unlikely.  We consider the following possibilities to be the most plausible at this time: i) weak magnetic fields, ii) ram pressure stripping of cosmic ray electrons/gravitational harassment in dense environments, and iii) bottom-heavy stellar IMFs.  Further studies of the relationship between radio continuum emission and other SF tracers will be necessary to improve our understanding of how SF proceeds in ETGs.

\subsubsection{ETGs that Follow the Radio-IR Relation}
\label{sec:normal_alternative}
While some nearby ETGs are deficient in their radio continuum emission compared to the IR, we note that many of the \atlas\ galaxies, particularly those with the highest molecular gas masses, do follow the radio-IR correlation.  In these systems, the radio-IR correlation likely originates from SF as it does in spirals.  However, a substantial fraction of the radio emission in some of the ETGs that lie on the radio-IR relation could have an AGN rather than a SF origin.  Previous studies of the radio-IR relation in low-luminosity AGNs have indeed shown that, unlike more powerful, radio-loud AGNs that show clear radio excesses when placed on the radio-IR correlation, many of these systems have q-values within the scatter of normal star-forming galaxies \citep{obric+06, moric+10, mauch+07, nyland+16}.

There is evidence that some \atlas\ ETGs with normal q-values may be dominated by AGN rather than SF emission at radio and IR wavelengths.  Some of the 1.4 to 5~GHz spectral index estimates reported in Table~\ref{tab:summary_all} are flat (i.e., $\alpha > -0.5$), a possible indication of self-absorbed synchrotron emission associated with an active nucleus \citep{condon+92}.  However, these spectral indices are based on observations taken a few decades apart in time and at very different sensitivities, and we regard these crude estimates as highly uncertain.  

An example of an \atlas\ ETG with strong multiwavelength evidence for the presence of an AGN that is characterized by a normal q-value is NGC1266 \citep{nyland+13, nyland+16}.  In this galaxy, the majority of the radio continuum emission is associated with kpc-scale radio lobes that may be interacting with the ISM of the host galaxy, yet its q-value of 2.15 is consistent with normal star-forming galaxies on the radio-IR relation.  In other systems with evidence for radio AGN emission that lack extended jets/lobes and also follow the radio-IR relation, such as NGC5273 \citep{nyland+16}, some portion of the radio continuum emission could even originate from coronal outflows from accretion discs, as recently suggested by \citet{wong+16}.  

\section{Summary and Conclusions}
\label{sec:summary}
We have presented new, sensitive 1.4~GHz VLA observations of the kpc-scale radio continuum emission in 72 ETGs from the volume- and magnitude-limited \atlas\ survey.  Combined with data from FIRST, we have studied the 1.4~GHz properties of 97\% of the \atlas\ ETGs.  We detected radio continuum emission in 71\% of our new 1.4~GHz VLA observations on scales ranging from $\approx$200 to 900~pc in compact sources to as large as 18~kpc in the most extended source.  For the majority of the ETGs in our sample, the 1.4~GHz emission has a morphology that is similar in appearance to the discs of radio emission associated with SF in spiral galaxies.  In at least two cases, the radio morphology is characterized by extended jets, and is clearly associated with an active nucleus rather than SF.    

We compared these radio data with existing molecular gas and IR observations to study the CO-radio and IR-radio relations in the largest sample of nearby ETGs to date.  
The main conclusions from this study are as follows:\\

\begin{enumerate}
\renewcommand{\theenumi}{(\arabic{enumi})}

\item The most molecular gas-rich \atlas\ ETGs have radio luminosities consistent with expectations from radio-SFR calibrations and SFRs derived from molecular gas masses \citep{gao+04, murphy+11}.  The gas-rich ETGs in our sample also follow the radio-IR correlation.  These ETGs may be in the process of efficiently forming stars, and SF likely proceeds in a manner similar to that in typical star-forming spiral galaxies.  The radio-IR relation in these systems likely arises from SF, but for some sources harboring low-luminosity radio AGNs, the correlation may be driven by AGN activity.

\item ETGs with lower H$_2$ masses tend to emit less radio continuum emission than expected based on standard H$_2$-SFR relations.  This population of ETGs is also characterized by high IR-radio ratios compared to ``normal" star-forming galaxies.  Correlations between the radio continuum and IR emission are similar for both FIR and MIR emission.  High q-values persist in the MIR even after correction for the contribution to the 22$\,\mu$m emission made by an underlying dusty, evolved stellar population.

\item The incidence of high q-values is much higher in this sample than in previous studies of the IR-radio relation in samples dominated by late-type galaxies.  About 19\% of our sample ETGs have high q-values and are candidate FIR-excess sources.  Considering \atlas\ ETGs with only upper limits the level of radio continuum emission, this fraction may even be as high as $\approx$50\%.

\item By comparing to lower-resolution archival radio data, we conclude that the amount of large-scale radio emission that would have been resolved-out by our higher-resolution data is modest.  While there are some ETGs in our study that have normal star-forming q-values when measurements are made using the lower-resolution radio data, the high q-values persist in other ETGs even when data more sensitive to extended, low-surface-brightness emission are included.

\item The high q-values in our sample tend to occur at low-IR luminosities but are not associated with low dynamical mass or metallicity.  This is in contrast to previous studies, which were dominated by late-type star-forming galaxies.

\item Possible explanations that could explain both the high CO-radio and IR-radio ratios in our sample of ETGs include bottom-heavy IMFs, weak magnetic fields, and a higher prevalence of environmental effects leading to enhanced cosmic ray electron escape compared to spirals.  

\end{enumerate}

Although our data indicate that some ETGs are deficient in their overall radio continuum emission compared to their CO and IR emission, further studies are needed to verify the underlying cause.  
Improved estimates of SF rates, SF efficiencies, ISM conditions, and galactic magnetic fields in ETGs will also help improve our understanding and interpretation of the correlations discussed in this work.  Examples of future studies include spectral energy distribution modeling, deep high-resolution imaging of denser molecular gas species with the Atacama Large Millimeter Array, and deep radio continuum polarization studies capable of tracing the strength and structure of the weak magnetic fields of nearby ETGs.

\section*{Acknowledgments}
We thank the anonymous referee for the many thoughtful comments that have strengthened this work.  The National Radio Astronomy Observatory is a facility of the National Science Foundation operated under cooperative agreement by Associated Universities, Inc.  This publication makes use of data products from the Wide-field Infrared Survey Explorer, which is a joint project of the University of California, Los Angeles, and the Jet Propulsion Laboratory/California Institute of Technology, funded by the National Aeronautics and Space Administration.  We also used the NASA/IPAC Extragalactic Database (NED), which is operated by the Jet Propulsion Laboratory, California Institute of Technology, under contract with the National Aeronautics and Space Administration.  Funding for this project was provided in part by National Science Foundation grant 1109803.  The research leading to these results has also received funding from the European Research Council under the European Union's Seventh Framework Programme (FP/2007-2013) / ERC Advanced Grant RADIOLIFE-320745.  KA is supported through Hubble Fellowship grant \hbox{\#HST-HF2-51352.001} awarded by the Space Telescope Science Institute, which is operated by the Association of Universities for Research in Astronomy, Inc., for NASA, under contract NAS5-26555.

{\it Facilities:} NRAO
\\

\footnotesize{\bibliographystyle{mnras}
\bibliography{a3d_sf_v20}}

\clearpage
\appendix

\section{Data Tables}

\begin{table*}
\begin{minipage}{13.75cm}
\caption{1.4~GHz VLA Image Properties} 
\label{tab:radio_parms}
\begin{tabular*}{13.75cm}{lccccccc}
\hline
\hline
Galaxy & D & Virgo & F/S & RMS & $S_\mathrm{peak}$ & $S_\mathrm{int}$ & log($L$) \\
  & (Mpc) &   &  & ($\mu$Jy beam$^{-1}$) & (mJy beam$^{-1}$) & (mJy) & (W Hz$^{-1}$) \\
(1) & (2) & (3) & (4) & (5) & (6) &  (7) & (8) \\
\hline
IC0676                     & 24.6 &    0 &   F &     43 &          3.44 $\pm$      0.04 &       6.78     $\pm$      0.23 &       20.69 \\ 
IC0719$^{\dagger}$         & 29.4 &    0 &   F &     28 &          0.18 $\pm$      0.02 &       2.15     $\pm$      0.28 &       20.35 \\ 
IC1024                     & 24.2 &    0 &   F &     67 &          3.57 $\pm$      0.03 &      17.52     $\pm$      0.56 &       21.09 \\ 
NGC0474                    & 30.9 &    0 &   F &     40 &  $<$     0.20                 &    \nodata                     &   $<$ 19.36 \\ 
NGC0509                    & 32.3 &    0 &   F &     28 &  $<$     0.14                 &    \nodata                     &   $<$ 19.24 \\ 
NGC0516                    & 34.7 &    0 &   F &     27 &  $<$     0.14                 &    \nodata                     &   $<$ 19.29 \\ 
NGC0524                    & 23.3 &    0 &   F &     29 &          1.36 $\pm$      0.02 &       1.63     $\pm$      0.07 &       20.02 \\ 
NGC0525                    & 30.7 &    0 &   F &     29 &  $<$     0.14                 &    \nodata                     &   $<$ 19.21 \\ 
NGC0680                    & 37.5 &    0 &   F &     27 &          0.99 $\pm$      0.02 &       1.12     $\pm$      0.05 &       20.28 \\ 
NGC0770                    & 36.7 &    0 &   F &     34 &  $<$     0.17                 &    \nodata                     &   $<$ 19.44 \\ 
NGC0821                    & 23.4 &    0 &   F &     31 &  $<$     0.15                 &    \nodata                     &   $<$ 19.01 \\ 
NGC1023                    & 11.1 &    0 &   F &     36 &          0.23 $\pm$      0.03 &       0.56     $\pm$      0.11 &       18.92 \\ 
$^{\star}$NGC1222              & 33.3 &    0 &   S &     70 &         16.24 $\pm$      0.02 &      48.64     $\pm$      1.46 &       21.81 \\ 
NGC1266                    & 29.9 &    0 &   F &     70 &         62.52 $\pm$      0.03 &     106.60     $\pm$      3.20 &       22.06 \\ 
NGC2685$^{\dagger}$        & 16.7 &    0 &   F &     29 &          1.05 $\pm$      0.01 &      44.91     $\pm$      1.40 &       21.18 \\ 
NGC2764                    & 39.6 &    0 &   F &     40 &          2.44 $\pm$      0.03 &      15.13     $\pm$      0.51 &       21.45 \\ 
NGC2768                    & 21.8 &    0 &   F &     42 &         13.48 $\pm$      0.02 &      13.65     $\pm$      0.41 &       20.89 \\ 
NGC2824                    & 40.7 &    0 &   F &     42 &          6.95 $\pm$      0.04 &    \nodata                     &       21.14 \\ 
NGC2852                    & 28.5 &    0 &   F &     35 &          0.71 $\pm$      0.03 &    \nodata                     &       19.84 \\ 
NGC3032                    & 21.4 &    0 &   F &     39 &          0.78 $\pm$      0.03 &       5.42     $\pm$      0.27 &       20.47 \\ 
NGC3073                    & 32.8 &    0 &   F &     52 &  $<$     0.26                 &    \nodata                     &   $<$ 19.52 \\ 
NGC3182$^{\dagger}$        & 21.8 &    0 &   F &     30 &          0.13 $\pm$      0.01 &       4.29     $\pm$      0.47 &       20.39 \\ 
NGC3193                    & 34.0 &    0 &   F &     30 &          0.24 $\pm$      0.03 &       0.48     $\pm$      0.08 &       19.82 \\ 
NGC3156                    & 33.1 &    0 &   F &     32 &  $<$     0.16                 &    \nodata                     &   $<$ 19.32 \\ 
NGC3245                    & 20.3 &    0 &   F &     33 &          6.28 $\pm$      0.03 &       7.05     $\pm$      0.22 &       20.54 \\ 
NGC3489                    & 11.7 &    0 &   F &     35 &          0.43 $\pm$      0.03 &       0.84     $\pm$      0.10 &       19.14 \\ 
NGC3599                    & 19.8 &    0 &   F &     40 &  $<$     0.20                 &    \nodata                     &   $<$ 18.97 \\ 
NGC3605                    & 20.1 &    0 &   F &     27 &  $<$     0.14                 &    \nodata                     &   $<$ 18.81 \\ 
NGC3607                    & 22.2 &    0 &   F &     28 &          4.37 $\pm$      0.02 &       5.47     $\pm$      0.17 &       20.51 \\ 
NGC3608                    & 22.3 &    0 &   S &     27 &          0.33 $\pm$      0.03 &    \nodata                     &       19.29 \\ 
$^{\star}$NGC3619$^{\dagger}$  & 26.8 &    0 &   F &     36 &          1.03 $\pm$      0.03 &       3.00     $\pm$      0.14 &       20.41 \\ 
NGC3626                    & 19.5 &    0 &   F &     40 &          2.92 $\pm$      0.03 &       4.55     $\pm$      0.15 &       20.32 \\ 
NGC3648                    & 31.9 &    0 &   F &     30 &          0.36 $\pm$      0.02 &       0.54     $\pm$      0.06 &       19.82 \\ 
$^{\star}$NGC3665$^{\dagger}$  & 33.1 &    0 &   F &     40 &         11.99 $\pm$      0.02 &      88.46     $\pm$      2.66 &       22.06 \\ 
NGC3945$^{\dagger}$        & 23.2 &    0 &   F &     34 &          1.63 $\pm$      0.03 &    \nodata                     &       20.02 \\ 
NGC4036                    & 24.6 &    0 &   F &     50 &          8.96 $\pm$      0.03 &      10.76     $\pm$      0.33 &       20.89 \\ 
NGC4111                    & 14.6 &    0 &   F &     48 &          4.55 $\pm$      0.05 &       7.69     $\pm$      0.26 &       20.29 \\ 
NGC4119                    & 16.5 &    1 &   F &     35 &  $<$     0.17                 &    \nodata                     &   $<$ 18.76 \\ 
NGC4150                   & 13.4 &    0 &   F &     29 &          0.66 $\pm$      0.02 &       0.82     $\pm$      0.05 &       19.25 \\ 
NGC4203                    & 14.7 &    0 &   F &     78 &          8.49 $\pm$      0.06 &    \nodata                     &       20.34 \\ 
NGC4324                    & 16.5 &    1 &   F &     39 &  $<$     0.20                 &    \nodata                     &   $<$ 18.80 \\ 
NGC4429                    & 16.5 &    1 &   F &     40 &          0.48 $\pm$      0.03 &       1.12     $\pm$      0.10 &       19.56 \\ 
NGC4459                    & 16.1 &    1 &   F &     40 &          1.14 $\pm$      0.03 &       1.39     $\pm$      0.07 &       19.63 \\ 
NGC4526                    & 16.4 &    1 &   F &     30 &          1.89 $\pm$      0.02 &       9.75     $\pm$      0.33 &       20.50 \\ 
NGC4550                    & 15.5 &    1 &   S &     50 &  $<$     0.25                 &    \nodata                     &   $<$ 18.86 \\ 
NGC4551                    & 16.1 &    1 &   F &     44 &  $<$     0.22                 &    \nodata                     &   $<$ 18.83 \\ 
NGC4564                    & 15.8 &    1 &   F &     31 &  $<$     0.15                 &    \nodata                     &   $<$ 18.67 \\ 
NGC4596                    & 16.5 &    1 &   F &     28 &  $<$     0.14                 &    \nodata                     &   $<$ 18.66 \\ 
NGC4643                    & 16.5 &    1 &   F &     29 &          0.24 $\pm$      0.03 &       0.41     $\pm$      0.07 &       19.13 \\ 
NGC4684                    & 13.1 &    0 &   F &     45 &          3.51 $\pm$      0.05 &       5.31     $\pm$      0.19 &       20.04 \\ 
NGC4694                    & 16.5 &    1 &   F &     35 &          0.90 $\pm$      0.03 &       1.44     $\pm$      0.07 &       19.67 \\ 
NGC4697                    & 11.4 &    0 &   F &     38 &  $<$     0.19                 &    \nodata                     &   $<$ 18.47 \\ 
NGC4710                    & 16.5 &    1 &   F &     28 &          2.63 $\pm$      0.02 &      13.30     $\pm$      0.42 &       20.64 \\ 
NGC4753                    & 22.9 &    0 &   F &     40 &          0.33 $\pm$      0.03 &       1.14     $\pm$      0.12 &       19.85 \\ 
NGC5173                    & 38.4 &    0 &   F &     32 &          1.24 $\pm$      0.03 &    \nodata                     &       20.34 \\ 
NGC5273$^{\flat}$                    & 16.1 &    0 &   F & \nodata &       \nodata                 &    \nodata                     &       \nodata \\ 
NGC5379                    & 30.0 &    0 &   F &     30 &          0.46 $\pm$      0.03 &    \nodata                     &       19.69 \\ 
$^{\star}$NGC5866$^{\dagger}$  & 14.9 &    0 &   F &     40 &         12.14 $\pm$      0.04 &      14.14     $\pm$      0.43 &       20.57 \\ 
NGC6014$^{\dagger}$        & 35.8 &    0 &   F &     38 &          2.75 $\pm$      0.03 &       3.63     $\pm$      0.12 &       20.75 \\ 
\end{tabular*}
\end{minipage} 
\end{table*}

\begin{table*}
\begin{minipage}{13.75cm}
\contcaption{} 
\begin{tabular*}{13.75cm}{lccccccc}
\hline
\hline
Galaxy & D & Virgo  & F/S &  RMS & $S_\mathrm{peak}$ & $S_\mathrm{int}$ & log($L$) \\
  & (Mpc) &   &  & ($\mu$Jy beam$^{-1}$) & (mJy) & (mJy) & (W Hz$^{-1}$) \\
(1) & (2) & (3) & (4) & (5) & (6) &  (7) & (8) \\
\hline
NGC6547                    & 40.8 &    0 &   F &     37 &          1.74 $\pm$      0.03 &       2.48     $\pm$      0.10 &       20.69 \\ 
NGC6798                    & 37.5 &    0 &   F &     29 &          0.25 $\pm$      0.03 &    \nodata                     &       19.62 \\ 
NGC7457                    & 23.2 &    0 &   S &     29 &  $<$     0.14                 &    \nodata                     &   $<$ 18.97 \\ 
NGC7454                    & 12.9 &    0 &   F &     27 &  $<$     0.14                 &    \nodata                     &   $<$ 18.43 \\ 
NGC7465                    & 29.3 &    0 &   F &     32 &          7.98 $\pm$      0.03 &      13.38     $\pm$      0.41 &       21.14 \\ 
PGC016060                  & 37.8 &    0 &   F &     60 &  $<$     0.30                 &    \nodata                     &   $<$ 19.71 \\ 
PGC029321                  & 40.9 &    0 &   F &     45 &          9.04 $\pm$      0.04 &    \nodata                     &       21.26 \\ 
PGC056772                  & 39.5 &    0 &   F &     28 &          2.18 $\pm$      0.03 &       3.14     $\pm$      0.11 &       20.77 \\ 
PGC058114                  & 23.8 &    0 &   F &     30 &          6.12 $\pm$      0.03 &       8.83     $\pm$      0.27 &       20.78 \\ 
PGC061468                  & 36.2 &    0 &   F &     36 &  $<$     0.18                 &    \nodata                     &   $<$ 19.45 \\ 
UGC05408                   & 45.8 &    0 &   F &     41 &          2.58 $\pm$      0.03 &       3.78     $\pm$      0.13 &       20.98 \\ 
UGC06176                   & 40.1 &    0 &   F &     29 &          4.81 $\pm$      0.03 &       6.17     $\pm$      0.19 &       21.07 \\ 
UGC09519                   & 27.6 &    0 &   F &     27 &          0.25 $\pm$      0.02 &       0.45     $\pm$      0.06 &       19.61 \\ 
\hline
\hline
\end{tabular*}
 
\medskip
{\bf Notes.} Column 1: galaxy name.  Column 2: official \atlas\ distance \citep{cappellari+11}.  Column 3: Virgo membership.  Column 4: kinematic class \citep{emsellem+11} of either fast rotator (F) or slow rotator (S).  Column 5: average RMS noise in the image.  Column 6:  peak flux density.  Column 7: integrated flux density.  Note that measurements of the integrated flux density are only given for sources that were resolved by JMFIT.  Column 8: log of the 1.4~GHz radio luminosity.  When an integrated flux density is given, $\log(L)$ is based on the integrated flux density.  If only a peak flux density is given (either a measurement or an upper limit), then $\log(L)$ is based on the peak flux density.

\medskip
$^{\star}$ Extended source not well-represented by a single two-dimensional Gaussian model.  The peak and integrated flux densities were calculated by drawing an aperture at the 3$\sigma$ level around the source in the CASA Viewer and then using the IMSTAT task to determine the flux density.
 
\medskip
$^{\dagger}$ Multi-component source.  The integrated flux density refers to the sum of all components.  See Table~\ref{tab:multi_flux} for information on individual components.

\medskip
$^{\flat}$ The NGC5273 dataset was of poor quality and the resultant flux density measurements were deemed unreliable.  This galaxy is, however, detected robustly in the FIRST survey and is therefore included in the analysis in this work as a ``detection."  

\end{minipage} 
\end{table*}

\begin{sidewaystable*}
\centering
\begin{minipage}{20.25cm}
\caption{Spatial Parameters of 1.4~GHz Detections 
\label{tab:gauss}}
\begin{tabular*}{20.25cm}{lcccccccc}
\hline
\hline
Galaxy & Morph. & R.A. & DEC.  & Beam & B.P.A. & $\theta_{\mathrm{M}} \times \theta_{\mathrm{m}}$ & P.A. & $M \times m$ \\
      & & (J2000) & (J2000) & (arcsec) & (deg) & (arcsec) & (deg) & (kpc) \\
 (1) & (2) & (3) & (4) & (5) & (6) & (7) & (8) & (9)\\
\hline
IC0676                     & R  & 11:12:39.764 &  09:03:23.19 & 3.68 $\times$ 3.06 &   18.04 &       4.39  $\pm$  0.13 $\times$  2.25 $\pm$ 0.14 & 164.95     $\pm$  2.65 &       0.52  $\times$ 0.27 \\ 
IC0719$^{\dagger}$         & R  & 11:40:18.411 &  09:00:34.36 & 6.31 $\times$ 4.16 &  -64.62 &      29.96  $\pm$  4.76 $\times$  8.21 $\pm$ 2.05 &  49.74     $\pm$  4.88 &       4.27  $\times$ 1.17 \\ 
IC1024                     & R  & 14:31:27.142 &  03:00:30.94 & 6.32 $\times$ 4.15 &  -52.95 &      13.17  $\pm$  0.17 $\times$  6.88 $\pm$ 0.15 &  27.68     $\pm$  1.04 &       1.55  $\times$ 0.81 \\ 
NGC0524                    & R  & 01:24:47.737 &  09:32:20.02 & 5.56 $\times$ 3.76 &  -53.95 &       2.31  $\pm$  0.41 $\times$  1.48 $\pm$ 0.61 &  70.08     $\pm$ 25.00 &       0.26  $\times$ 0.17 \\ 
NGC0680                    & R  & 01:49:47.297 &  21:58:15.06 & 5.16 $\times$ 3.80 &  -65.91 &       1.74  $\pm$  0.47 $\times$  1.07 $\pm$ 0.83 &  86.66     $\pm$ 38.48 &       0.32  $\times$ 0.19 \\ 
NGC1023                    & R  & 02:40:23.860 &  39:03:47.20 & 5.17 $\times$ 3.79 &  -66.08 &      13.26  $\pm$  2.79 $\times$  0.00 $\pm$ 0.00 &  70.50     $\pm$  2.96 &       0.71  $\times$ 0.00 \\ 
$^{\star}$NGC1222              & R  & 03:08:56.786 & -02:57:18.40 & 4.96 $\times$ 3.85 &  -68.11 &      24.99              $\times$ 19.77            &        \nodata       &       4.03  $\times$ 3.19 \\ 
NGC1266                    & R  &  03:16:0.739 & -02:25:39.21 & 4.95 $\times$ 3.85 &  -68.43 &       4.92  $\pm$  0.01 $\times$  2.15 $\pm$ 0.01 & 174.44     $\pm$  0.14 &       0.71  $\times$ 0.31 \\ 
NGC2685$^{\dagger}$        & R  & 08:55:25.186 &  58:44:42.47 & 4.05 $\times$ 3.85 &   24.36 &      95.24  $\pm$  1.04 $\times$  5.19 $\pm$ 0.08 & 117.10     $\pm$  0.06 &       7.71  $\times$ 0.42 \\ 
NGC2764                    & R  & 09:08:17.526 &  21:26:35.88 & 3.88 $\times$ 3.57 &   79.60 &      13.78  $\pm$  0.27 $\times$  5.79 $\pm$ 0.18 &  28.96     $\pm$  1.10 &       2.65  $\times$ 1.11 \\ 
NGC2768                    & R  & 09:11:37.413 &  60:02:14.91 & 3.85 $\times$ 3.73 &   65.61 &       0.55  $\pm$  0.11 $\times$  0.25 $\pm$ 0.19 &  87.33     $\pm$ 16.89 &       0.06  $\times$ 0.03 \\ 
NGC2824                    & U  &  09:19:2.231 &  26:16:11.85 & 3.91 $\times$ 3.46 &   62.14 &  $<$  2.36                                        &        \nodata       &  $<$  0.47                \\ 
NGC2852                    & U  & 09:23:14.637 &  40:09:49.76 & 5.23 $\times$ 4.14 &  -30.76 &  $<$  2.30                                        &        \nodata       &  $<$  0.32                \\ 
NGC3032                    & R  &  09:52:8.169 &  29:14:11.46 & 5.77 $\times$ 3.28 &   30.77 &      12.10  $\pm$  0.74 $\times$  9.60 $\pm$ 0.70 &  42.38     $\pm$ 13.45 &       1.26  $\times$ 1.00 \\ 
NGC3182$^{\dagger}$        & R  & 10:19:32.749 &  58:12:28.65 & 4.36 $\times$ 3.11 &  -70.13 &      34.51  $\pm$  4.64 $\times$ 13.25 $\pm$ 1.97 & 151.68     $\pm$  5.23 &       3.65  $\times$ 1.40 \\ 
NGC3193                    & R  & 10:18:24.896 &  21:53:38.51 & 5.35 $\times$ 3.16 &   72.76 &       6.50  $\pm$  2.02 $\times$  2.95 $\pm$ 2.46 & 153.57     $\pm$ 30.05 &       1.07  $\times$ 0.49 \\ 
NGC3245                    & R  & 10:27:18.377 &  28:30:26.60 & 4.36 $\times$ 3.12 &  -66.76 &       1.76  $\pm$  0.13 $\times$  0.98 $\pm$ 0.29 &   0.92     $\pm$ 80.34 &       0.17  $\times$ 0.10 \\ 
NGC3489                    & R  & 11:00:18.532 &   13:54:4.51 & 5.26 $\times$ 3.80 &   62.20 &       5.96  $\pm$  1.36 $\times$  2.73 $\pm$ 1.47 &  66.77     $\pm$ 18.03 &       0.34  $\times$ 0.15 \\ 
NGC3607                    & R  & 11:16:54.677 &   18:03:6.43 & 5.55 $\times$ 3.78 &  -87.13 &       2.04  $\pm$  0.09 $\times$  1.51 $\pm$ 0.13 & 118.13     $\pm$  8.48 &       0.22  $\times$ 0.16 \\ 
NGC3608                    & U  & 11:16:58.947 &  18:08:55.19 & 5.46 $\times$ 3.64 &  -81.04 &  $<$  2.68                                        &        \nodata       &  $<$  0.29                \\ 
$^{\star}$NGC3619$^{\dagger}$  & R  & 11:19:21.586 &  57:45:27.83 & 5.75 $\times$ 3.01 &   80.71 &      13.57              $\times$  9.18            &        \nodata       &       1.76  $\times$ 1.19 \\ 
NGC3626                    & R  &  11:20:3.810 &  18:21:24.54 & 4.74 $\times$ 3.16 &  -80.60 &       4.82  $\pm$  0.14 $\times$  2.66 $\pm$ 0.11 & 107.13     $\pm$  2.33 &       0.46  $\times$ 0.25 \\ 
NGC3648                    & R  & 11:22:31.448 &  39:52:37.01 & 6.02 $\times$ 4.02 &  -71.00 &       4.90  $\pm$  1.09 $\times$  0.44 $\pm$ 1.34 & 143.52     $\pm$ 16.66 &       0.76  $\times$ 0.07 \\ 
$^{\star}$NGC3665$^{\dagger}$  & R  & 11:24:43.662 &  38:45:46.13 & 8.37 $\times$ 3.42 &   51.92 &     112.76              $\times$ 20.12            &        \nodata       &      18.09  $\times$ 3.23 \\ 
NGC3945$^{\dagger}$        & U  & 11:53:13.624 &  60:40:32.15 & 5.49 $\times$ 3.59 &   62.63 &  $<$  1.59                                        &        \nodata       &  $<$  0.18                \\ 
NGC4036                    & R  & 12:01:26.656 &  61:53:44.03 & 6.11 $\times$ 3.10 &   43.70 &       3.61  $\pm$  0.15 $\times$  1.76 $\pm$ 0.07 &  67.04     $\pm$  2.53 &       0.43  $\times$ 0.21 \\ 
NGC4111                    & R  &  12:07:3.146 &  43:03:56.24 & 4.40 $\times$ 3.58 &  -81.89 &       3.43  $\pm$  0.10 $\times$  2.62 $\pm$ 0.10 &  60.88     $\pm$  5.07 &       0.24  $\times$ 0.19 \\ 
NGC4150                    & R  & 12:10:33.656 &   30:24:5.80 & 3.92 $\times$ 3.37 &   44.45 &       2.44  $\pm$  0.70 $\times$  1.59 $\pm$ 1.19 &  28.58     $\pm$ 47.71 &       0.16  $\times$ 0.10 \\ 
NGC4203                    & U  &  12:15:5.055 &  33:11:50.34 & 3.92 $\times$ 3.37 &   44.42 &  $<$  0.85                                        &        \nodata       &  $<$  0.06                \\ 
NGC4429                    & R  & 12:27:26.482 &  11:06:27.67 & 3.91 $\times$ 3.37 &   44.50 &       8.68  $\pm$  1.17 $\times$  3.62 $\pm$ 0.71 & 105.47     $\pm$  7.29 &       0.69  $\times$ 0.29 \\ 
NGC4459                    & R  &  12:29:0.027 &  13:58:42.53 & 5.42 $\times$ 3.15 &   86.60 &       2.51  $\pm$  0.63 $\times$  1.64 $\pm$ 1.08 &  69.42     $\pm$ 36.22 &       0.20  $\times$ 0.13 \\ 
\hline
\hline
\end{tabular*}
\end{minipage} 
\end{sidewaystable*}
 
\begin{sidewaystable*}
\centering
\begin{minipage}{20.25cm}
\contcaption{} 
\begin{tabular*}{20.25cm}{lcccccccc}
\hline
\hline
Galaxy & Morph. & R.A. & DEC.  & Beam & B.P.A. & $\theta_{\mathrm{M}} \times \theta_{\mathrm{m}}$ & P.A. & $M \times m$ \\
      & & (J2000) & (J2000) & (arcsec) & (deg) & (arcsec) & (deg) & (kpc) \\
 (1) & (2) & (3) & (4) & (5) & (6) & (7) & (8) & (9)\\
\hline
NGC4526                    & R  &  12:34:2.994 &  07:41:58.03 & 5.85 $\times$ 3.99 &  -71.56 &      12.14  $\pm$  0.22 $\times$  3.13 $\pm$ 0.13 & 110.82     $\pm$  0.62 &       0.97  $\times$ 0.25 \\ 
NGC4643                    & R  & 12:43:20.133 &  01:58:41.83 & 5.77 $\times$ 3.84 &  -85.22 &       8.47  $\pm$  2.19 $\times$  1.48 $\pm$ 1.27 &  41.53     $\pm$  9.63 &       0.68  $\times$ 0.12 \\ 
NGC4684                    & R  & 12:47:17.522 & -02:43:38.38 & 6.45 $\times$ 3.80 &  -77.47 &       3.67  $\pm$  0.16 $\times$  1.88 $\pm$ 0.19 &  11.63     $\pm$  4.20 &       0.23  $\times$ 0.12 \\ 
NGC4694                    & R  & 12:48:14.994 &   10:59:2.12 & 4.43 $\times$ 4.26 &   63.35 &       5.74  $\pm$  0.41 $\times$  2.02 $\pm$ 0.39 & 113.85     $\pm$  4.01 &       0.46  $\times$ 0.16 \\ 
NGC4710                    & R  & 12:49:38.821 &  15:09:56.32 & 9.92 $\times$ 3.45 &   57.78 &      10.96  $\pm$  0.13 $\times$  3.55 $\pm$ 0.07 &  27.03     $\pm$  0.46 &       0.88  $\times$ 0.28 \\ 
NGC4753                    & R  & 12:52:21.907 & -01:11:58.60 & 3.73 $\times$ 3.53 &   64.43 &      14.68  $\pm$  1.70 $\times$  2.23 $\pm$ 1.65 &  97.95     $\pm$  3.42 &       1.63  $\times$ 0.25 \\ 
NGC5173                    & U  & 13:28:25.271 &  46:35:29.81 & 6.79 $\times$ 3.39 &   50.66 &  $<$  2.92                                        &        \nodata       &  $<$  0.54                \\ 
NGC5379                    & U  & 13:55:34.343 &  59:44:34.17 & 3.68 $\times$ 2.64 &  -29.64 &  $<$  3.04                                        &        \nodata       &  $<$  0.44                \\ 
$^{\star}$NGC5866$^{\dagger}$  & R  & 15:06:29.491 &  55:45:47.62 & 6.16 $\times$ 4.08 &  -54.85 &      65.63              $\times$ 10.21            &        \nodata       &       4.74  $\times$ 0.74 \\ 
NGC6014$^{\dagger}$        & R  & 15:55:57.389 &  05:55:54.98 & 6.57 $\times$ 3.73 &  -68.92 &       3.14  $\pm$  0.29 $\times$  2.70 $\pm$ 0.36 & 169.26     $\pm$ 56.73 &       0.54  $\times$ 0.47 \\ 
NGC6547                    & R  & 18:05:10.806 &  25:13:40.71 & 5.18 $\times$ 3.94 &  -73.00 &       5.78  $\pm$  0.68 $\times$  0.00 $\pm$ 0.75 &  73.16     $\pm$  7.97 &       1.14  $\times$ 0.00 \\ 
NGC6798                    & U  &  19:24:3.133 &  53:37:29.78 & 3.70 $\times$ 3.16 & -177.60 &  $<$  2.81                                        &        \nodata       &  $<$  0.51                \\ 
NGC7465                    & R  &  23:02:0.973 &  15:57:53.16 & 3.70 $\times$ 3.14 &   33.77 &       5.11  $\pm$  0.07 $\times$  4.01 $\pm$ 0.04 &  75.91     $\pm$  2.25 &       0.73  $\times$ 0.57 \\ 
PGC029321                  & U  & 10:05:51.178 &  12:57:40.45 & 3.69 $\times$ 3.14 &   33.80 &  $<$  1.29                                        &        \nodata       &  $<$  0.26                \\ 
PGC056772                  & R  & 16:02:11.606 &   07:05:9.79 & 3.54 $\times$ 3.11 & -166.44 &       3.42  $\pm$  0.24 $\times$  2.20 $\pm$ 0.21 &  23.05     $\pm$  8.50 &       0.65  $\times$ 0.42 \\ 
PGC058114                  & R  &  16:26:4.235 &  02:54:23.82 & 3.58 $\times$ 3.11 &   -5.81 &       2.67  $\pm$  0.08 $\times$  2.10 $\pm$ 0.10 &  79.27     $\pm$  6.10 &       0.31  $\times$ 0.24 \\ 
UGC05408                   & R  & 10:03:51.896 &  59:26:10.48 & 6.83 $\times$ 3.27 &   44.89 &       3.02  $\pm$  0.20 $\times$  2.62 $\pm$ 0.10 &  82.00     $\pm$ 28.46 &       0.67  $\times$ 0.58 \\ 
UGC06176                   & R  & 11:07:24.674 &  21:39:25.46 & 4.18 $\times$ 3.54 &   -7.27 &       2.34  $\pm$  0.08 $\times$  1.56 $\pm$ 0.10 &  31.29     $\pm$  5.14 &       0.45  $\times$ 0.30 \\ 
UGC09519                   & R  & 14:46:21.106 &  34:22:13.73 & 5.61 $\times$ 3.99 &  -70.15 &       4.65  $\pm$  0.93 $\times$  1.05 $\pm$ 1.20 & 107.28     $\pm$ 12.18 &       0.62  $\times$ 0.14 \\ 
\hline
\hline
\end{tabular*}
 
\medskip
{\bf Notes.}  Column 1: galaxy name.  Column 2: radio morphology based on the output of the JMFIT task in {\sc AIPS}.  R = resolved and U = unresolved.  Column 3: right ascension of the emission at the peak flux density.  For sources with multiple components denoted by a $\dagger$ symbol, the position listed is that of the component closest to the optical nuclear position.  The format is sexagesimal and the epoch is J2000.  The positional uncertainty of each image is 0.1$^{\prime \prime}$ and is dominated by the positional uncertainty of the phase reference calibrator.  Column 4: declination of the central position of the emission, determined in the same manner as the right ascension in Column 3.  Column 5: angular dimensions of the synthesized beam (major $\times$ minor axis).  Column 6: synthesized beam position angle, measured anti clockwise from north.  Column 7: angular dimensions of the emission (major $\times$ minor axis).  If JMFIT was only able to deconvolve the major axis of the source, then the minor axis extent is given as 0.00.  The errors are from JMFIT and are only given if the emission was successfully deconvolved in at least one dimension and categorized as resolved.  For non-Gaussian sources, source dimensions were determined using the CASA Viewer and no error is reported.  Column 8: position angle of the emission from JMFIT.  For non-Gaussian, inherently complex sources, no position angle is reported. Column 9: linear dimensions of the emission (major $\times$ minor axis) in physical units.

\medskip
$^{\star}$ Extended source not well-represented by a single two-dimensional Gaussian model.  The peak and integrated flux densities were calculated by drawing an aperture at the 3$\sigma$ level around the source in the CASA Viewer and then using the IMSTAT task to determine the flux density.
 
\medskip
$^{\dagger}$ Multi-component source.  The integrated flux density refers to the sum of all components.  See Table~\ref{tab:multi_flux} for information on all components.  

\end{minipage} 
\end{sidewaystable*}

\begin{table*}
\begin{minipage}{14.5cm}
\caption{Image Properties of Sources with Multiple Components}
\label{tab:multi_flux}
\begin{tabular*}{14.5cm}{ccccccccc}
\hline
\hline
Galaxy & Component & R.A. & DEC. & $S_{\mathrm{peak}}$ & $S_{\mathrm{int}}$ & log($L$) \\
               &   & (J2000) & (J2000) & (mJy beam$^{-1}$) &  (mJy)  &  (W Hz$^{-1}$)  \\
 (1) & (2) & (3) & (4) & (5) & (6) & (7)  \\
\hline 
IC0719      &   $^{\star}$Central source & 11:40:18.588 & 09:00:36.77 &   0.20   $\pm$    0.03 &     0.37   $\pm$      0.06       &  19.58 \\ 
                   &   $^{\star}$Southern source & 11:40:18.154 & 09:00:31.71 &   0.20   $\pm$    0.03 &     0.32   $\pm$      0.05       &  19.52 \\ 
                   &   $^{\star}$Northern source & 11:40:18.862 & 09:00:41.68 &   0.17   $\pm$    0.03 &  \nodata                         &  19.25 \\ 
NGC2685     &   Northern Source & 08:55:33.694 & 58:44:8.43 &   0.33   $\pm$    0.02 &     0.61   $\pm$      0.05       &  19.31 \\ 
                       &  Southern Source & 08:55:34.477 & 58:44:4.09 &   0.15   $\pm$    0.02 &     1.08   $\pm$      0.15       &  19.56 \\ 
NGC3182     &   $^{\star}$Northern source & 10:19:33.043 & 58:12:25.17 &   0.23   $\pm$    0.03 &     0.52   $\pm$      0.16       &  19.47 \\ 
                       &   $^{\star}$Western source & 10:19:33.589 & 58:12:15.74 &   0.19   $\pm$    0.03 &     0.50   $\pm$      0.19       &  19.45 \\ 
                       &   $^{\star}$Eastern source & 10:19:32.474 & 58:12:20.16 &   0.16   $\pm$    0.03 &     0.41   $\pm$      0.18       &  19.37 \\ 
NGC3619     &   $^{\star}$Southern source & 11:19:21.755 & 57:45:25.90 &   1.13   $\pm$    0.05 &     1.43   $\pm$      0.12       &  20.09 \\ 
                       &   $^{\star}$Northern source & 11:19:21.476 & 57:45:29.14 &   0.89   $\pm$    0.05 &     1.18   $\pm$      0.11       &  20.01 \\ 
NGC3665     &   $^{\star}$Eastern jet  & 11:24:43.277 & 38:45:49.79 &  13.35   $\pm$    0.40 &    37.94   $\pm$      1.60       &  21.70 \\ 
                       &   $^{\star}$Western jet  & 11:24:44.221 & 38:45:40.84 &  13.38   $\pm$    0.40 &    38.24   $\pm$      1.40       &  21.70 \\ 
                       &   $^{\star}$Core & 11:24:43.012 & 38:45:52.23 &  12.21   $\pm$    0.37 &    18.72   $\pm$      0.64       &  21.39 \\ 
NGC3945     &   Northern source & 11:53:13.625 & 60:40:32.15 &   1.64   $\pm$    0.03 &  \nodata                         &  20.02 \\ 
                       &   Southern source & 11:53:13.473 & 60:40:21.17 &   0.42   $\pm$    0.03 &  \nodata                         &  19.43 \\ 
NGC5866     &   Central source & 15:06:29.491 & 55:45:47.62 &  12.14   $\pm$    0.04 &    14.14   $\pm$      0.43       &  20.57 \\ 
                       &   $^{\star}$Northern source & 15:06:34.984 & 55:45:20.19 &   0.64   $\pm$    0.04 &     4.28   $\pm$      1.50       &  20.06 \\ 
                       &   $^{\star}$Southern source & 15:06:23.306 & 55:46:29.15 &   0.95   $\pm$    0.05 &     2.26   $\pm$      1.08       &  19.78 \\ 
NGC6014     &   Central source & 15:55:57.389 & 05:55:54.98 &   2.75   $\pm$    0.03 &     3.63   $\pm$      0.12       &  20.75 \\ 
                       &   Northern source & 15:55:56.695 & 05:56:11.46 &   0.47   $\pm$    0.03 &  \nodata                         &  19.86 \\ 
\hline
\hline
\end{tabular*}
 
\medskip
{\bf Notes.} Column 1: galaxy name.  Column 2: radio component name.  Column 3: right ascension of the central position of the component as determined by IMFIT in CASA.  The format is sexagesimal and the epoch is J2000.  The positional uncertainty of each image is 0.1$^{\prime \prime}$ and is dominated by the positional uncertainty of the phase reference calibrator.  Column 4: declination of the central position of the emission, determined in the same manner as the right ascension in Column 3.  Column 5: peak flux density.  Column 6: integrated flux density.  Note that measurements of the integrated flux density are only given for sources that were resolved by JMFIT.  Column 7: log of the 1.4~GHz radio luminosity.  When an integrated flux density is given, $\log(L)$ is based on the integrated flux density.  If only a peak flux density is given (either a measurement or an upper limit), then $\log(L)$ is based on the peak flux density.

\medskip
$^{\star}$ Extended source not well-represented by a single two-dimensional Gaussian model.  The peak and integrated flux densities were calculated by drawing an aperture at the 3$\sigma$ level around the source in the CASA Viewer and then using the IMSTAT task to determine the flux density.

\end{minipage}
\end{table*}

\begin{table*}
\begin{minipage}{14cm}
\caption{Spatial Properties of Sources with Multiple Components}
\label{tab:multi_spatial}
\begin{tabular*}{14cm}{ccccccccc}
\hline
\hline
 Galaxy & Component & Morph. & $\theta_{\mathrm{M}} \times \theta_{\mathrm{m}}$ & P.A.  & $M \times m$ \\
               &                       &               & (arcsec)                                                     & (deg) & (kpc) \\
 (1) & (2) & (3) & (4) & (5) & (6)  \\
\hline 
IC0719      & $^{\star}$Central source  &    R &        10.67              $\times$  9.59            &        \nodata         &           1.52  $\times$    1.37 \\ 
                    & $^{\star}$Southern source  &    R &        13.55              $\times$  9.95            &        \nodata         &           1.93  $\times$    1.42 \\ 
                    & $^{\star}$Northern source  &    R &         8.49              $\times$  6.37            &        \nodata         &           1.21  $\times$    0.91 \\ 
NGC2685     & Northern Source  &    R &         4.34  $\pm$  0.76 $\times$  2.32 $\pm$ 0.92 &  75.75     $\pm$   18.24 &           0.35  $\times$    0.19 \\ 
                       & Southern Source  &    R &        18.54  $\pm$  3.07 $\times$  3.96 $\pm$ 1.04 & 105.15     $\pm$    3.79 &           1.50  $\times$    0.32 \\ 
NGC3182     & $^{\star}$Northern source  &    R &         8.78              $\times$  4.38            &        \nodata         &           0.93  $\times$    0.46 \\ 
                       & $^{\star}$Western source  &    R &         7.52              $\times$  3.28            &        \nodata         &           0.79  $\times$    0.35 \\ 
                       & $^{\star}$Eastern source  &    R &        11.11              $\times$  3.62            &        \nodata         &           1.17  $\times$    0.38 \\ 
NGC3619     & $^{\star}$Southern source  &    R &         9.22              $\times$  6.59            &        \nodata         &           1.20  $\times$    0.86 \\ 
                       & $^{\star}$Northern source  &    R &         9.83              $\times$  6.78            &        \nodata         &           1.28  $\times$    0.88 \\ 
NGC3665     & $^{\star}$Eastern jet   &    R &        28.78              $\times$ 14.23            &        \nodata         &           4.62  $\times$    2.28 \\ 
                       & $^{\star}$Western jet   &    R &        42.01              $\times$ 15.72            &        \nodata         &           6.74  $\times$    2.52 \\ 
                       & $^{\star}$Core  &    R &        20.12              $\times$  7.33            &        \nodata         &           3.23  $\times$    1.18 \\ 
NGC3945     & Northern source  &    U &    $<$  1.57                                        &        \nodata         &    $<$    0.18                   \\ 
                       & Southern source  &    U &    $<$  2.62                                        &        \nodata         &    $<$    0.29                   \\ 
NGC5866     & Central source  &    R &         2.69  $\pm$  0.08 $\times$  1.18 $\pm$ 0.07 & 116.29     $\pm$    2.20 &           0.19  $\times$    0.09 \\ 
                       & $^{\star}$Northern source  &    R &        25.95              $\times$ 10.47            &        \nodata         &           1.87  $\times$    0.76 \\ 
                       & $^{\star}$Southern source  &    R &        19.88              $\times$  6.97            &        \nodata         &           1.44  $\times$    0.50 \\ 
NGC6014     & Central source  &    R &         3.13  $\pm$  0.29 $\times$  2.67 $\pm$ 0.37 & 170.19     $\pm$   57.68 &           0.54  $\times$    0.46 \\ 
                       & Northern source  &    U &    $<$  3.56                                        &        \nodata         &    $<$    0.62                   \\ 
\hline
\hline
\end{tabular*}
 
\medskip
{\bf Notes.} Column 1: galaxy name.  Column 2: radio component name.  Column 3: radio morphology based on the output of the JMFIT task in {\sc AIPS}.  R = resolved and U = unresolved.  Column 4: angular dimensions of the emission (major $\times$ minor axis).  If JMFIT was only able to deconvolve the major axis of the source, then the minor axis extent is given as 0.00.  The errors are from JMFIT and are only given if the emission was successfully deconvolved in at least one dimension and categorized as resolved.  For non-Gaussian sources, source dimensions were determined using the CASA Viewer and no error is reported.  Column 5: position angle of the emission from JMFIT.  For non-Gaussian, inherently complex sources, no position angle is reported.  Column 6: linear dimensions of the emission (major $\times$ minor axis) in physical units.

\medskip
$^{\star}$ Extended source not well-represented by a single two-dimensional Gaussian model.  The peak and integrated flux densities were calculated by drawing an aperture at the 3$\sigma$ level around the source in the CASA Viewer and then using the IMSTAT task to determine the flux density.  

\end{minipage}
\end{table*}

\begin{sidewaystable*}
\centering
\begin{minipage}{22.5cm}
\caption{Summary of Radio, CO and IR Data
\label{tab:summary_all}}
\begin{tabular*}{22.5cm}{lccccccccccccccc}
\hline
\hline
Galaxy & $S_{\mathrm{NVSS}}$ &  $S_{\mathrm{FIRST}}$ &  $S_{\mathrm{VLA}}$ &  $S_{\mathrm{Wrobel}}$ &  $\alpha_{1.4\,\mathrm{GHz}}^{5\,\mathrm{GHz}}$ & $\log(M_{\mathrm{H}2})$ & $S_{60\, \mu\mathrm{m}}$ & $S_{100\, \mu\mathrm{m}}$ & q & $S_{22\, \mu\mathrm{m}}$ & $\log(L_{22\, \mu\mathrm{m}})$ & q$_{22\, \mu\mathrm{m}}$\\
              & (mJy) & (mJy) & (mJy) & (mJy) &   & (M$_{\odot}$)			& (Jy) 					& (Jy)				      &						 	 & (mJy) & (erg~s$^{-1}$)				& \\
(1) & (2) & (3) & (4)  & (5) & (6) & (7) & (8) & (9) & (10) & (11) & (12) & (13) \\
\hline 
IC0560$^{\flat}$       &    \nodata                     &           $<$         0.74                         &                \nodata                         &      \nodata                                     &      \nodata             &         $<$         7.67 &      \nodata &      \nodata &                  \nodata &        18.53 &                    41.26 &         $>$         1.24 \\ 
IC0598       &    \nodata                     &           $<$         0.67                         &                \nodata                         &      \nodata                                     &      \nodata             &         $<$         8.02 &         0.10 &         0.39 &         $>$         2.52 &         7.81 &         $<$        41.20 &                  \nodata \\ 
IC0676       &        9.7     $\pm$       0.5 &                       7.25      $\pm$         0.73 &                   6.78       $\pm$        0.31 &                     3.60       $\pm$        0.20 &                     1.73 &                     8.63 &         3.21 &         5.15 &                     2.80 &       544.46 &                    42.73 &                     1.89 \\
\hline
\hline
\end{tabular*}
\end{minipage} 
\end{sidewaystable*}
 
\begin{sidewaystable*}
\centering
\begin{minipage}{22.5cm}
\contcaption{} 
\begin{tabular*}{22.5cm}{lccccccccccccccc}
\hline
\hline
Galaxy & $S_{\mathrm{NVSS}}$ &  $S_{\mathrm{FIRST}}$ &  $S_{\mathrm{VLA}}$ &  $S_{\mathrm{Wrobel}}$ &  $\alpha_{1.4\,\mathrm{GHz}}^{5\,\mathrm{GHz}}$ & $\log(M_{\mathrm{H}2})$ & $S_{60\, \mu\mathrm{m}}$ & $S_{100\, \mu\mathrm{m}}$ & q & $S_{22\, \mu\mathrm{m}}$ & $\log(L_{22\, \mu\mathrm{m}})$ & q$_{22\, \mu\mathrm{m}}$\\
              & (mJy) & (mJy) & (mJy) & (mJy) &   & (M$_{\odot}$)			& (Jy) 					& (Jy)				      &						 	 & (mJy) & (erg~s$^{-1}$)				& \\
(1) & (2) & (3) & (4)  & (5) & (6) & (7) & (8) & (9) & (10) & (11) & (12) & (13) \\
\hline
NGC2880$^{\flat}$      &    \nodata                     &           $<$         0.70                         &                \nodata                         &      \nodata                                     &      \nodata             &         $<$         7.44 &         0.10 &         0.38 &         $>$         2.49 &         6.26 &         $<$        41.35 &                  \nodata \\ 
NGC2950      &    \nodata                     &           $<$         0.77                         &                \nodata                         &      \nodata                                     &      \nodata             &         $<$         7.12 &         0.16 &         0.20 &         $>$         2.43 &        37.84 &         $<$        41.33 &                  \nodata \\ 
NGC2962$^{\flat}$      &    \nodata                     &           $<$         0.75                         &                \nodata                         &      \nodata                                     &      \nodata             &         $<$         7.85 &         0.22 &         0.79 &         $>$         2.78 &        19.43 &         $<$        41.77 &                  \nodata \\
\hline
\hline
\end{tabular*}
\end{minipage} 
\end{sidewaystable*}
 
\begin{sidewaystable*}
\centering
\begin{minipage}{22.5cm}
\contcaption{} 
\begin{tabular*}{22.5cm}{lccccccccccccccc}
\hline
\hline
Galaxy & $S_{\mathrm{NVSS}}$ &  $S_{\mathrm{FIRST}}$ &  $S_{\mathrm{VLA}}$ &  $S_{\mathrm{Wrobel}}$ &  $\alpha_{1.4\,\mathrm{GHz}}^{5\,\mathrm{GHz}}$ & $\log(M_{\mathrm{H}2})$ & $S_{60\, \mu\mathrm{m}}$ & $S_{100\, \mu\mathrm{m}}$ & q & $S_{22\, \mu\mathrm{m}}$ & $\log(L_{22\, \mu\mathrm{m}})$ & q$_{22\, \mu\mathrm{m}}$\\
              & (mJy) & (mJy) & (mJy) & (mJy) &   & (M$_{\odot}$)			& (Jy) 					& (Jy)				      &						 	 & (mJy) & (erg~s$^{-1}$)				& \\
(1) & (2) & (3) & (4)  & (5) & (6) & (7) & (8) & (9) & (10) & (11) & (12) & (13) \\
\hline
NGC3658      &    \nodata                     &           $<$         0.67                         &                \nodata                         &         $<$         0.50                         &      \nodata             &         $<$         7.82 &         0.16 &         0.92 &         $>$         2.83 &        14.71 &         $<$        41.54 &                  \nodata \\ 
NGC3665      &      112.2     $\pm$       3.7 &                       8.89      $\pm$         0.89 &                  98.38       $\pm$        8.75 &      \nodata                                     &      \nodata             &                     8.91 &         1.91 &         7.53 &                     1.63 &       138.66 &                    42.29 &                     1.28 \\ 
NGC3674      &    \nodata                     &           $<$         0.71                         &                \nodata                         &         $<$         0.50                         &      \nodata             &         $<$         7.78 &         0.04 &         0.14 &                  \nodata &         1.32 &         $<$        41.45 &                  \nodata \\
\hline
\hline
\end{tabular*}
\end{minipage} 
\end{sidewaystable*}
 
\begin{sidewaystable*}
\centering
\begin{minipage}{23.5cm}
\contcaption{} 
\begin{tabular*}{23.5cm}{lccccccccccccccc}
\hline
\hline
Galaxy & $S_{\mathrm{NVSS}}$ &  $S_{\mathrm{FIRST}}$ &  $S_{\mathrm{VLA}}$ &  $S_{\mathrm{Wrobel}}$ &  $\alpha_{1.4\,\mathrm{GHz}}^{5\,\mathrm{GHz}}$ & $\log(M_{\mathrm{H}2})$ & $S_{60\, \mu\mathrm{m}}$ & $S_{100\, \mu\mathrm{m}}$ & q & $S_{22\, \mu\mathrm{m}}$ & $\log(L_{22\, \mu\mathrm{m}})$ & q$_{22\, \mu\mathrm{m}}$\\
              & (mJy) & (mJy) & (mJy) & (mJy) &   & (M$_{\odot}$)			& (Jy) 					& (Jy)				      &						 	 & (mJy) & (erg~s$^{-1}$)				& \\
(1) & (2) & (3) & (4)  & (5) & (6) & (7) & (8) & (9) & (10) & (11) & (12) & (13) \\
\hline
NGC4350      &    \nodata                     &           $<$         0.68                         &                \nodata                         &         $<$         0.50                         &      \nodata             &         $<$         7.18 &         0.36 &         1.09 &         $>$         3.00 &        35.81 &         $<$        41.41 &                  \nodata \\ 
NGC4365      &    \nodata                     &           $<$         0.72                         &                \nodata                         &         $<$         0.50                         &      \nodata             &         $<$         7.62 &         0.04 &         0.65 &                  \nodata &        62.00 &         $<$        42.25 &                  \nodata \\ 
NGC4371      &    \nodata                     &           $<$         0.74                         &                \nodata                         &         $<$         0.50                         &      \nodata             &         $<$         7.29 &         0.05 &         0.16 &                  \nodata &        30.36 &         $<$        41.54 &                  \nodata \\
\hline
\hline
\end{tabular*}
\end{minipage} 
\end{sidewaystable*}
 
\begin{sidewaystable*}
\centering
\begin{minipage}{22.5cm}
\contcaption{} 
\begin{tabular*}{22.5cm}{lccccccccccccccc}
\hline
\hline
Galaxy & $S_{\mathrm{NVSS}}$ &  $S_{\mathrm{FIRST}}$ &  $S_{\mathrm{VLA}}$ &  $S_{\mathrm{Wrobel}}$ &  $\alpha_{1.4\,\mathrm{GHz}}^{5\,\mathrm{GHz}}$ & $\log(M_{\mathrm{H}2})$ & $S_{60\, \mu\mathrm{m}}$ & $S_{100\, \mu\mathrm{m}}$ & q & $S_{22\, \mu\mathrm{m}}$ & $\log(L_{22\, \mu\mathrm{m}})$ & q$_{22\, \mu\mathrm{m}}$\\
              & (mJy) & (mJy) & (mJy) & (mJy) &   & (M$_{\odot}$)			& (Jy) 					& (Jy)				      &						 	 & (mJy) & (erg~s$^{-1}$)				& \\
(1) & (2) & (3) & (4)  & (5) & (6) & (7) & (8) & (9) & (10) & (11) & (12) & (13) \\
\hline
NGC4596$^{\dagger}$      &    \nodata                     &           $<$         0.72                         &         $<$       0.14                         &         $<$         0.50                         &         $<$         0.09 &                     7.31 &         0.40 &         0.75 &         $>$         3.63 &        41.61 &         $<$        41.61 &                  \nodata \\ 
NGC4608      &    \nodata                     &           $<$         0.74                         &                \nodata                         &         $<$         1.00                         &      \nodata             &         $<$         7.30 &         0.03 &         0.24 &                  \nodata &        12.12 &         $<$        41.34 &                  \nodata \\ 
NGC4612      &    \nodata                     &           $<$         0.72                         &                \nodata                         &         $<$         0.50                         &      \nodata             &         $<$         7.20 &         0.04 &         0.09 &                  \nodata &        14.16 &         $<$        41.18 &                  \nodata \\
\hline
\hline
\end{tabular*}
\end{minipage} 
\end{sidewaystable*}
 
\begin{sidewaystable*}
\centering
\begin{minipage}{22.5cm}
\contcaption{} 
\begin{tabular*}{22.5cm}{lccccccccccccccc}
\hline
\hline
Galaxy & $S_{\mathrm{NVSS}}$ &  $S_{\mathrm{FIRST}}$ &  $S_{\mathrm{VLA}}$ &  $S_{\mathrm{Wrobel}}$ &  $\alpha_{1.4\,\mathrm{GHz}}^{5\,\mathrm{GHz}}$ & $\log(M_{\mathrm{H}2})$ & $S_{60\, \mu\mathrm{m}}$ & $S_{100\, \mu\mathrm{m}}$ & q & $S_{22\, \mu\mathrm{m}}$ & $\log(L_{22\, \mu\mathrm{m}})$ & q$_{22\, \mu\mathrm{m}}$\\
              & (mJy) & (mJy) & (mJy) & (mJy) &   & (M$_{\odot}$)			& (Jy) 					& (Jy)				      &						 	 & (mJy) & (erg~s$^{-1}$)				& \\
(1) & (2) & (3) & (4)  & (5) & (6) & (7) & (8) & (9) & (10) & (11) & (12) & (13) \\
\hline
NGC5574      &    \nodata                     &           $<$         0.75                         &                \nodata                         &         $<$         0.50                         &      \nodata             &         $<$         7.51 &         0.04 &         0.50 &                  \nodata &         8.91 &         $<$        41.08 &                  \nodata \\ 
NGC5576      &    \nodata                     &           $<$         0.75                         &                \nodata                         &         $<$         0.50                         &      \nodata             &         $<$         7.60 &         0.09 &         0.21 &         $>$         2.30 &        20.04 &         $<$        41.82 &                  \nodata \\ 
NGC5582      &    \nodata                     &           $<$         0.69                         &                \nodata                         &         $<$         0.50                         &      \nodata             &         $<$         7.67 &         0.04 &         0.10 &                  \nodata &         5.31 &         $<$        41.47 &                  \nodata \\
\hline
\hline
\end{tabular*}
\end{minipage} 
\end{sidewaystable*}
 
\begin{sidewaystable*}
\centering
\begin{minipage}{22.5cm}
\contcaption{} 
\begin{tabular*}{22.5cm}{lccccccccccccccc}
\hline
\hline
Galaxy & $S_{\mathrm{NVSS}}$ &  $S_{\mathrm{FIRST}}$ &  $S_{\mathrm{VLA}}$ &  $S_{\mathrm{Wrobel}}$ &  $\alpha_{1.4\,\mathrm{GHz}}^{5\,\mathrm{GHz}}$ & $\log(M_{\mathrm{H}2})$ & $S_{60\, \mu\mathrm{m}}$ & $S_{100\, \mu\mathrm{m}}$ & q & $S_{22\, \mu\mathrm{m}}$ & $\log(L_{22\, \mu\mathrm{m}})$ & q$_{22\, \mu\mathrm{m}}$\\
              & (mJy) & (mJy) & (mJy) & (mJy) &   & (M$_{\odot}$)			& (Jy) 					& (Jy)				      &						 	 & (mJy) & (erg~s$^{-1}$)				& \\
(1) & (2) & (3) & (4)  & (5) & (6) & (7) & (8) & (9) & (10) & (11) & (12) & (13) \\
\hline
PGC051753    &    \nodata                     &           $<$         0.69                         &                \nodata                         &      \nodata                                     &      \nodata             &         $<$         7.92 &      \nodata &      \nodata &                  \nodata &         0.55 &         $<$        40.93 &                  \nodata \\ 
PGC054452    &    \nodata                     &           $<$         0.73                         &                \nodata                         &      \nodata                                     &      \nodata             &         $<$         7.73 &      \nodata &      \nodata &                  \nodata &         1.70 &         $<$        40.80 &                  \nodata \\ 
PGC056772    &        3.0     $\pm$       0.4 &                       3.30      $\pm$         0.33 &                   3.14       $\pm$        0.14 &      \nodata                                     &      \nodata             &                     8.19 &         0.92 &         0.99 &                     2.50 &        90.36 &                    42.35 &                     0.19 \\
\hline
\hline
\end{tabular*}
 
\medskip
{\bf Notes.} Column 1: galaxy name.  Column 2: 1.4~GHz flux density from the NVSS catalog.  The spatial resolution of these data is $\theta_{\mathrm{FWHM}} \approx 45^{\prime \prime}$ and their typical sensitivity is $\approx$ 0.5~mJy~beam$^{-1}$.  Column 3: 1.4~GHz flux density from the FIRST survey.  The spatial resolution of these data is $\theta_{\mathrm{FWHM}} \approx 5^{\prime \prime}$ and their typical sensitivity is $\approx 0.15$~mJy~beam$^{-1}$.  Column 4: 1.4~GHz flux density from this work.  Column 5: 5~GHz flux density from \citet{wrobel+91b}.  Column 6: Radio spectral index estimates or limits from 1.4~GHz (Column 4 or Column 3) to 5~GHz (Column 5) at near-matched spatial resolution.  A value is only reported if the source is detected in at least one of the two frequencies.  Column 7: IRAM single-dish $H_2$ mass from \citet{young+11}.  Column 8: {\it IRAS} 60~$\mu$m flux density.  Column 9: {\it IRAS} 100~$\mu$m flux density.  Column 10: logarithmic FIR-radio ratio.  Column 11: {\it WISE} 22~$\mu$m flux density.  Column 12: corrected {\it WISE} 22~$\mu$m luminosity (see Equation~1 in \citealt{davis+14}).  Column 13: logarithmic 22~$\mu$m-radio ratio.

\medskip
$^{\star}$ While NGC3648 is reported as a detection in \citet{wrobel+91a}, we believe that measurement is actually associated with a nearby background source at an angular separation of about 25$^{\prime \prime}$ from NGC3648.

\medskip
$^{\spadesuit}$ Source detected in the 1.4~GHz observations presented in this work with $\theta_{\mathrm{FWHM}} \approx 5^{\prime \prime}$, but not at 5~GHz with $\theta_{\mathrm{FWHM}} \approx 0.5^{\prime \prime}$ in \citet{nyland+16}.

\medskip
$^{\flat}$ Source detected at 5~GHz with $\theta_{\mathrm{FWHM}} \approx 0.5^{\prime \prime}$ in \citet{nyland+16}, but not in the 1.4~GHz data used in this work with $\theta_{\mathrm{FWHM}} \approx 5^{\prime \prime}$.

\medskip
$^{\dagger}$ Candidate FIR-excess source.  
 
\end{minipage} 
\end{sidewaystable*}

\clearpage
\section{Radio Continuum Maps}
\begin{figure*}
{\label{fig:sub:PGC056772}\includegraphics[clip=True, trim=0cm 0cm 0cm 0cm, scale=0.23]{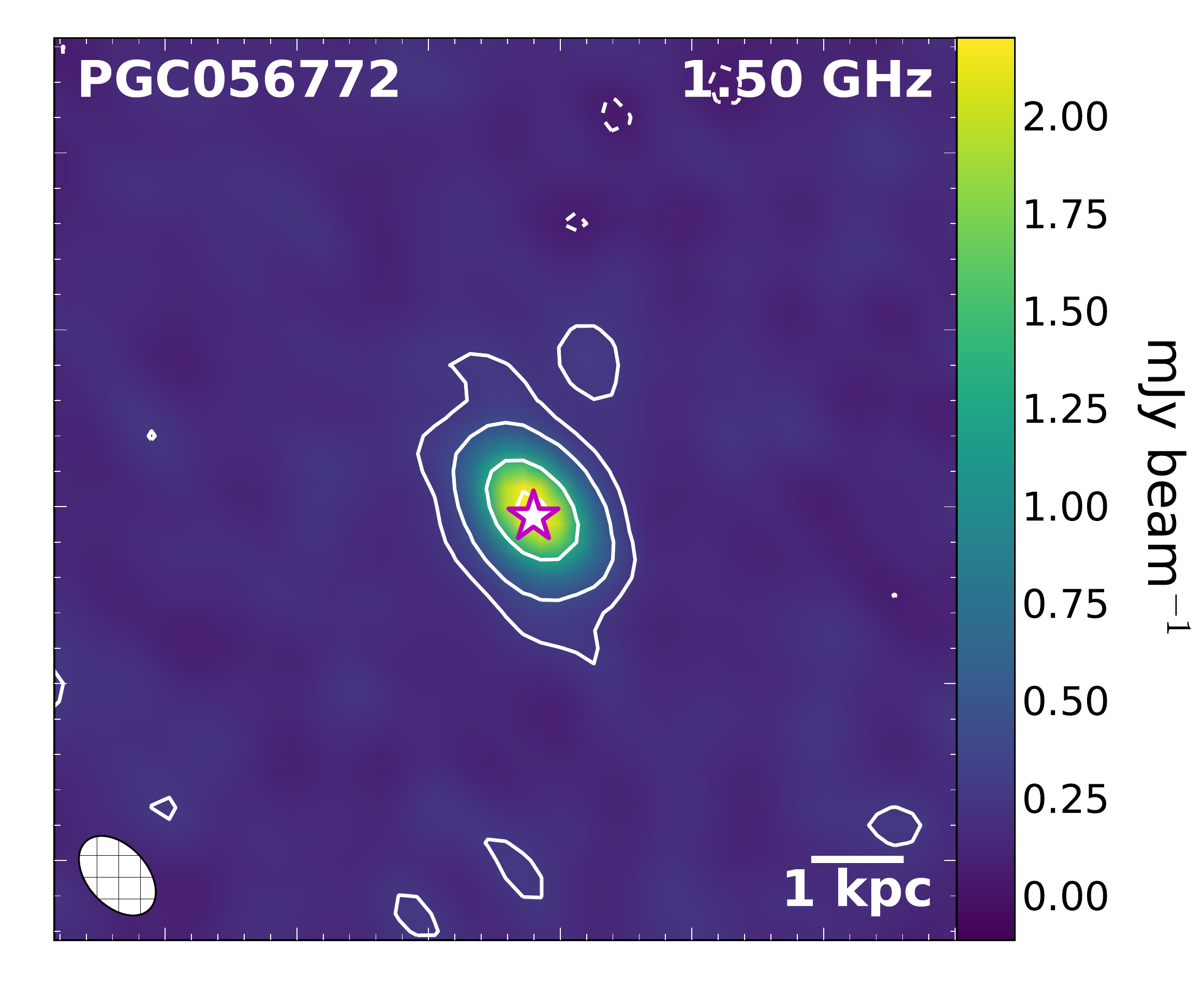}}
{\label{fig:sub:PGC058114}\includegraphics[clip=True, trim=0cm 0cm 0cm 0cm, scale=0.23]{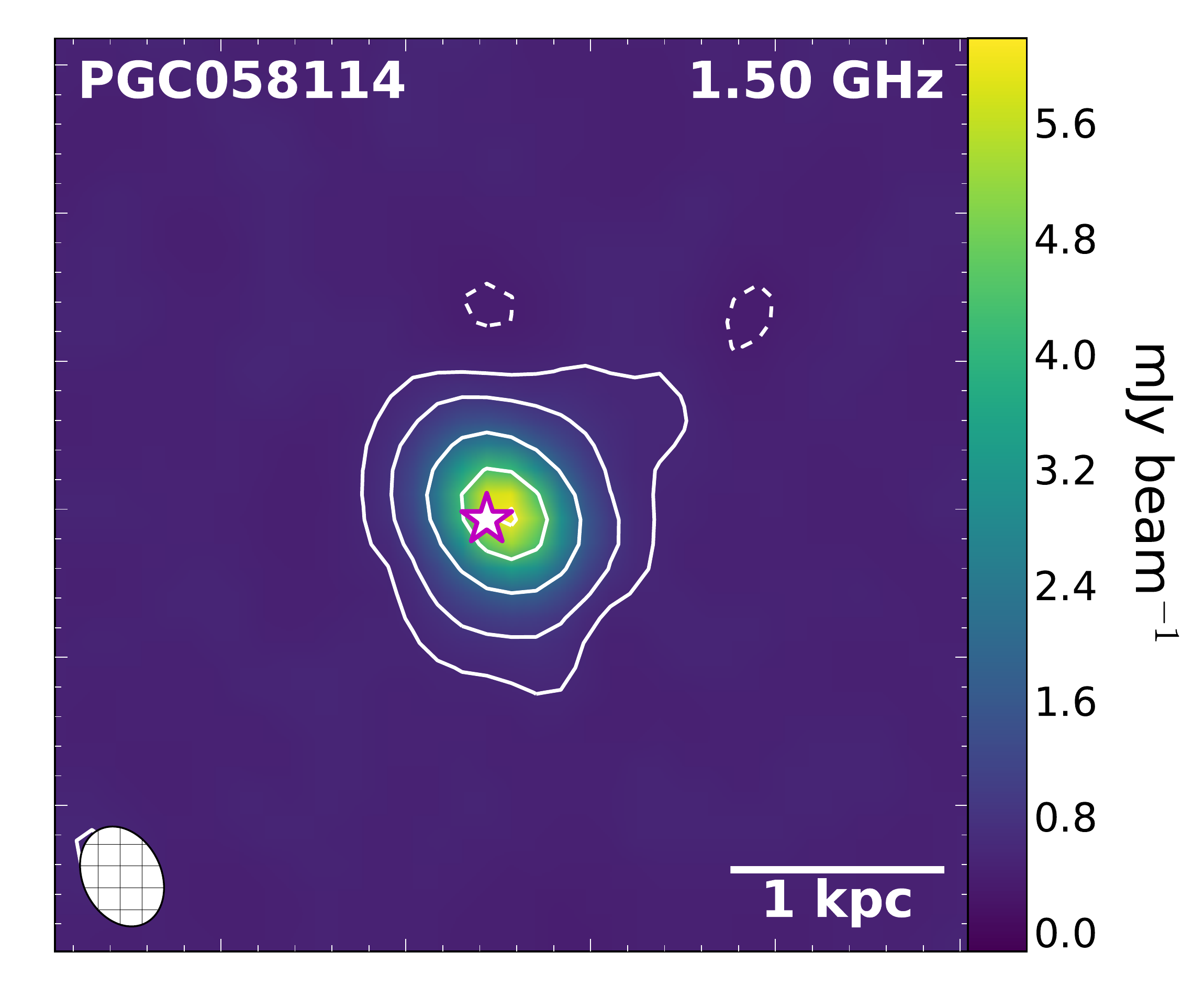}}
{\label{fig:sub:UGC05408}\includegraphics[clip=True, trim=0cm 0cm 0cm 0cm, scale=0.23]{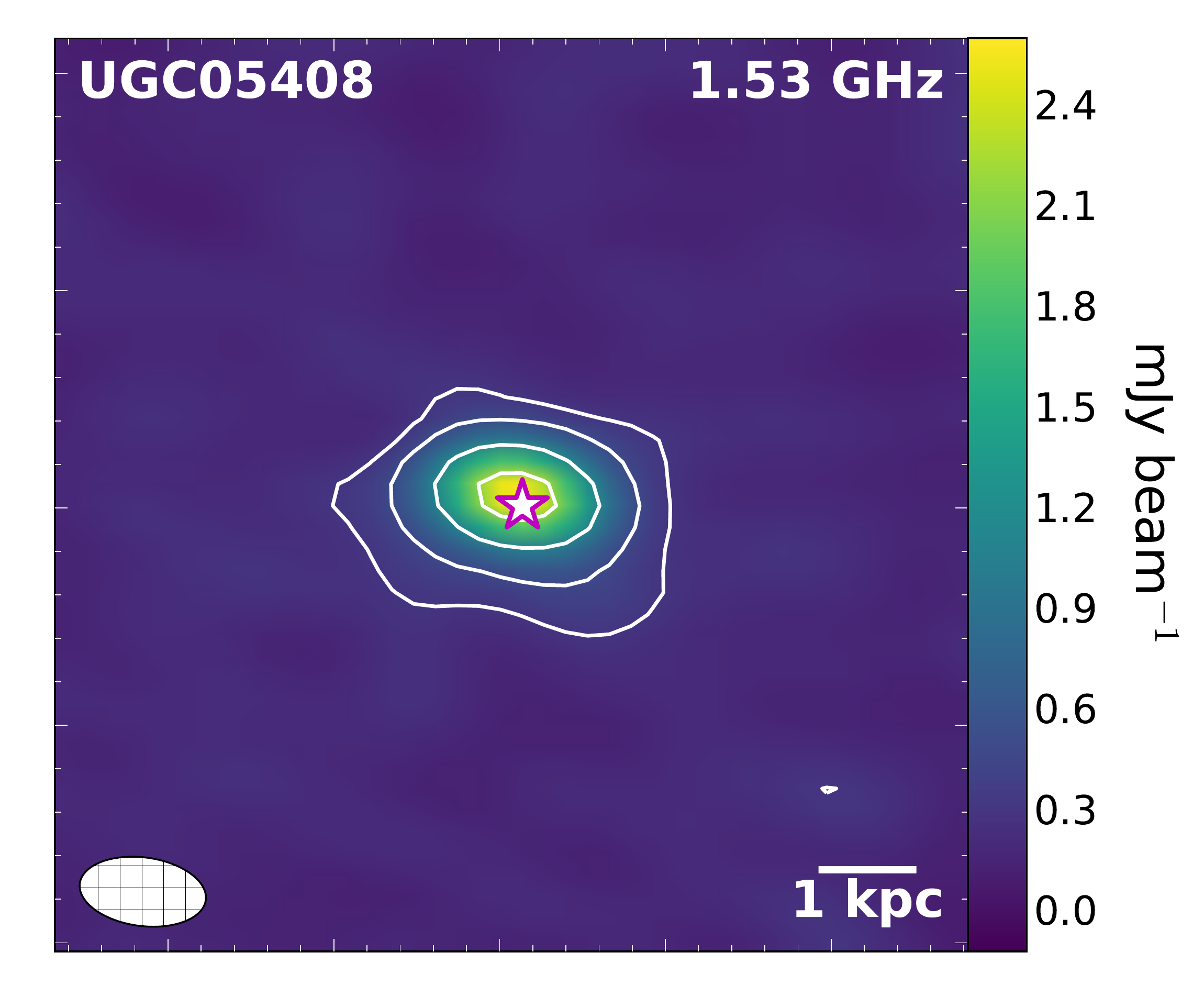}}
{\label{fig:sub:UGC06176}\includegraphics[clip=True, trim=0cm 0cm 0cm 0cm, scale=0.23]{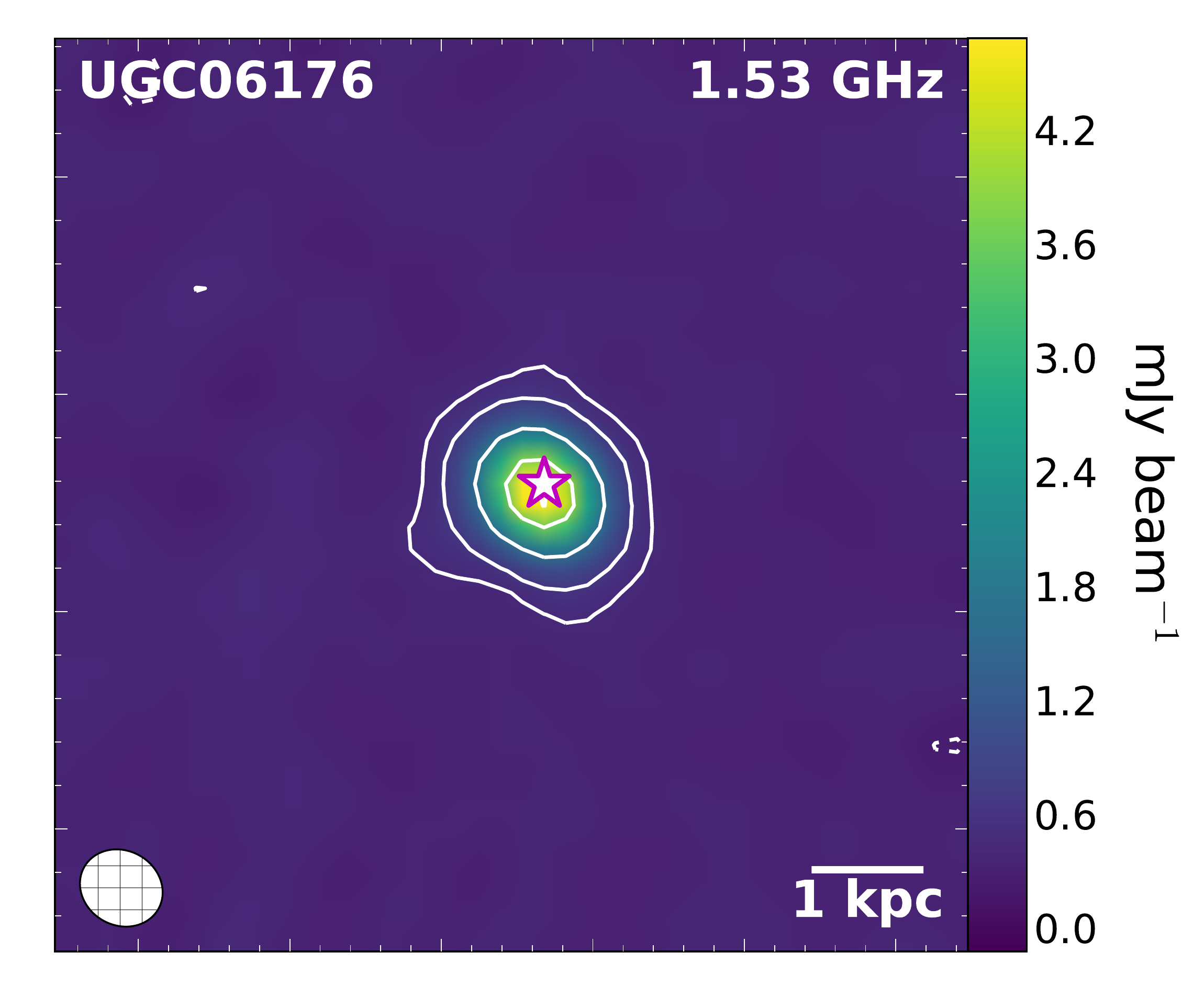}}
{\label{fig:sub:UGC09519}\includegraphics[clip=True, trim=0cm 0cm 0cm 0cm, scale=0.23]{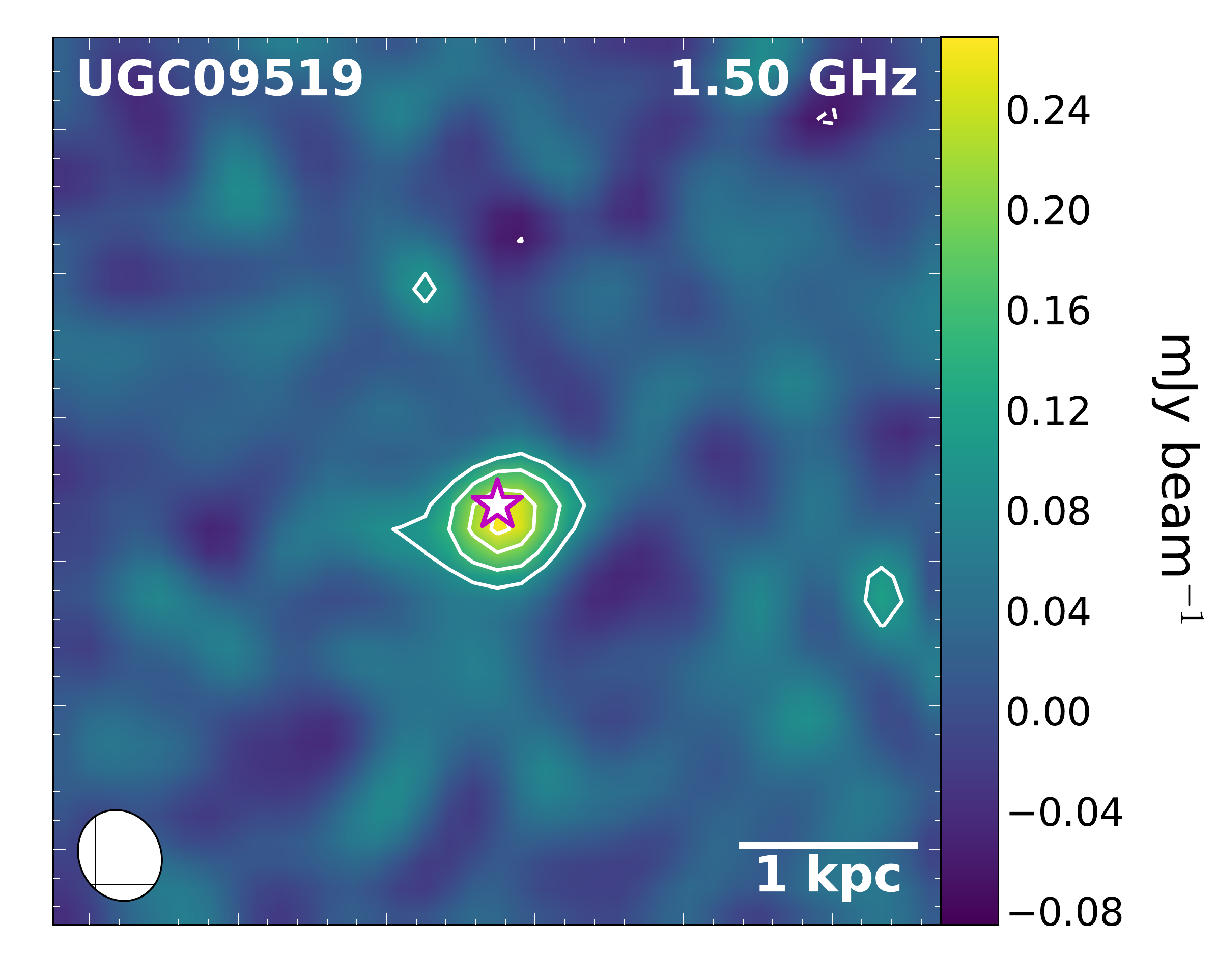}}
 
\caption{1.4~GHz continuum images with contours.  Negative contours are dashed.  The contour levels are spaced as multiples of the rms noise in each image.  Relative contour levels and rms noises are listed in Table~\ref{tab:contours}.  The synthesized beam is shown as a filled magenta ellipse in the lower left corner of each image.  In the upper right corner of each image the central observing frequency is shown.  A magenta star denotes the official optical position in the \atlas\ survey \citep{cappellari+11}.  A scale bar denoting a size of 1~kpc is shown in the lower left corner of each image.  We note that the bright component to the southwest of NGC3648 is most likely associated with a background source about 25$^{\prime \prime}$ away.}
\label{fig:radio_images}
\end{figure*}

\begin{figure*}
{\label{fig:sub:NGC4710}\includegraphics[clip=True, trim=0cm 0cm 0cm 0cm, scale=0.25]{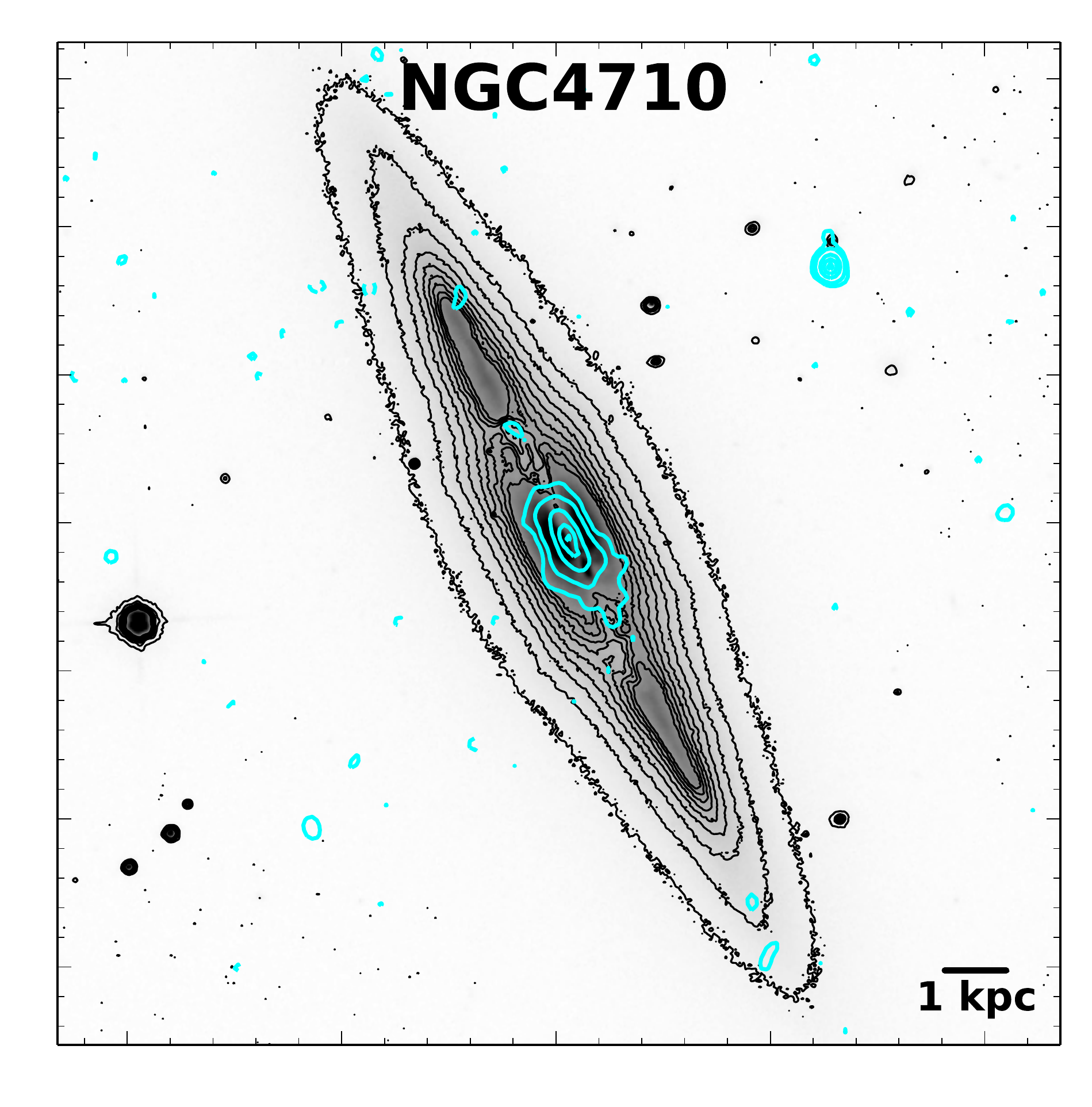}}
{\label{fig:sub:NGC4753}\includegraphics[clip=True, trim=0cm 0cm 0cm 0cm, scale=0.25]{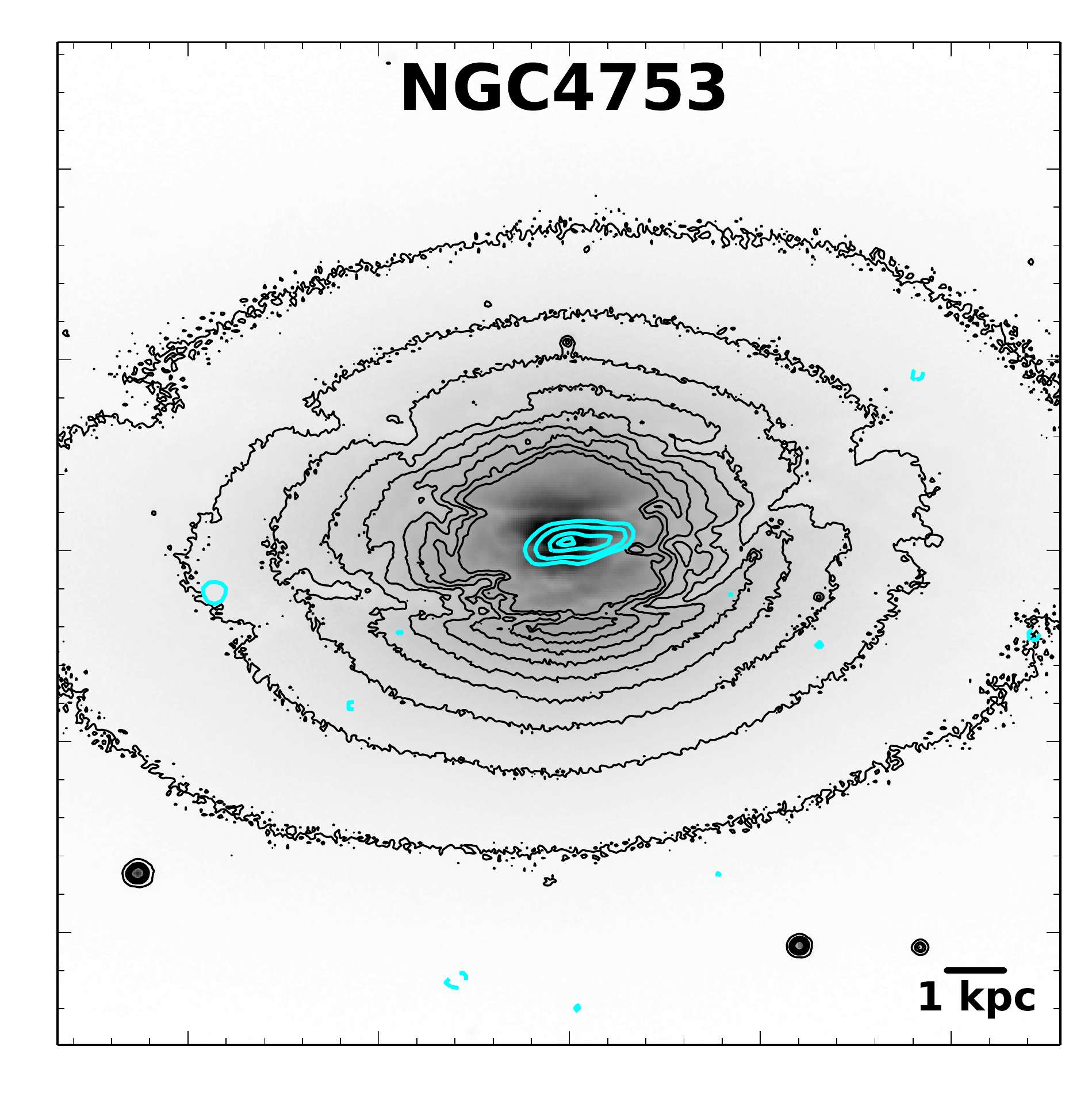}}
{\label{fig:sub:NGC5866}\includegraphics[clip=True, trim=0cm 0cm 0cm 0cm, scale=0.25]{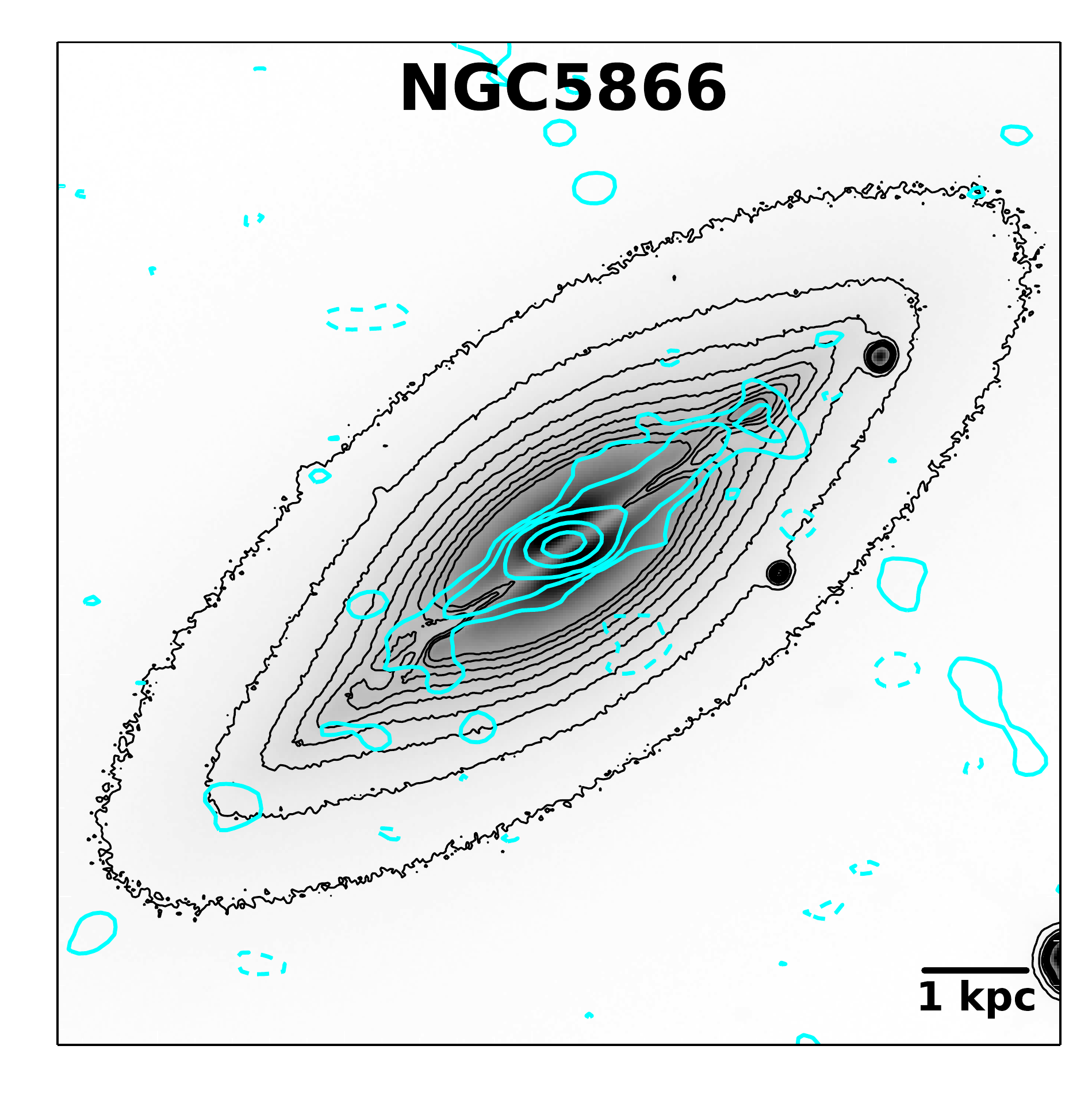}}
{\label{fig:sub:NGC6014}\includegraphics[clip=True, trim=0cm 0cm 0cm 0cm, scale=0.25]{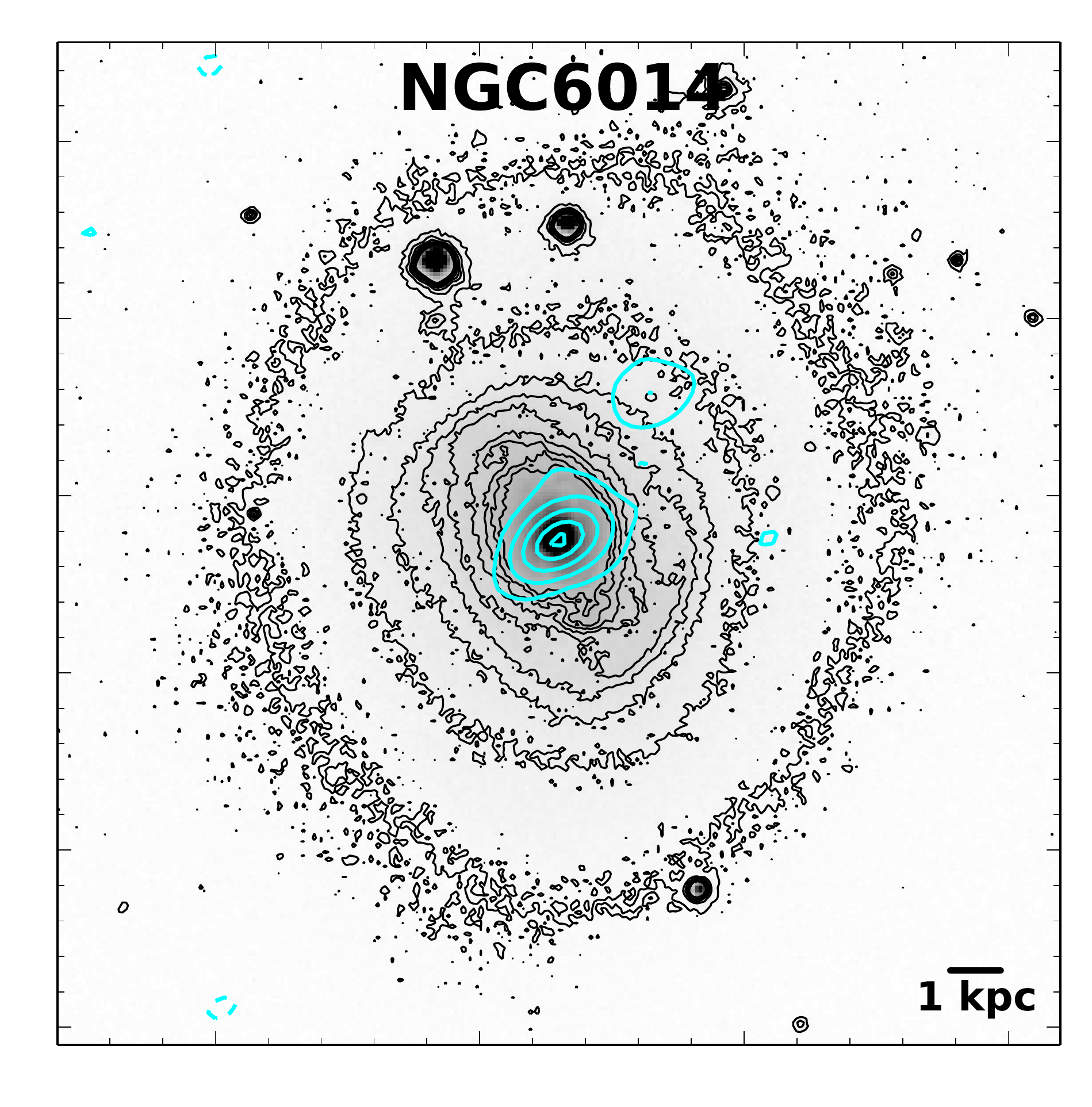}}
{\label{fig:sub:NGC7465}\includegraphics[clip=True, trim=0cm 0cm 0cm 0cm, scale=0.25]{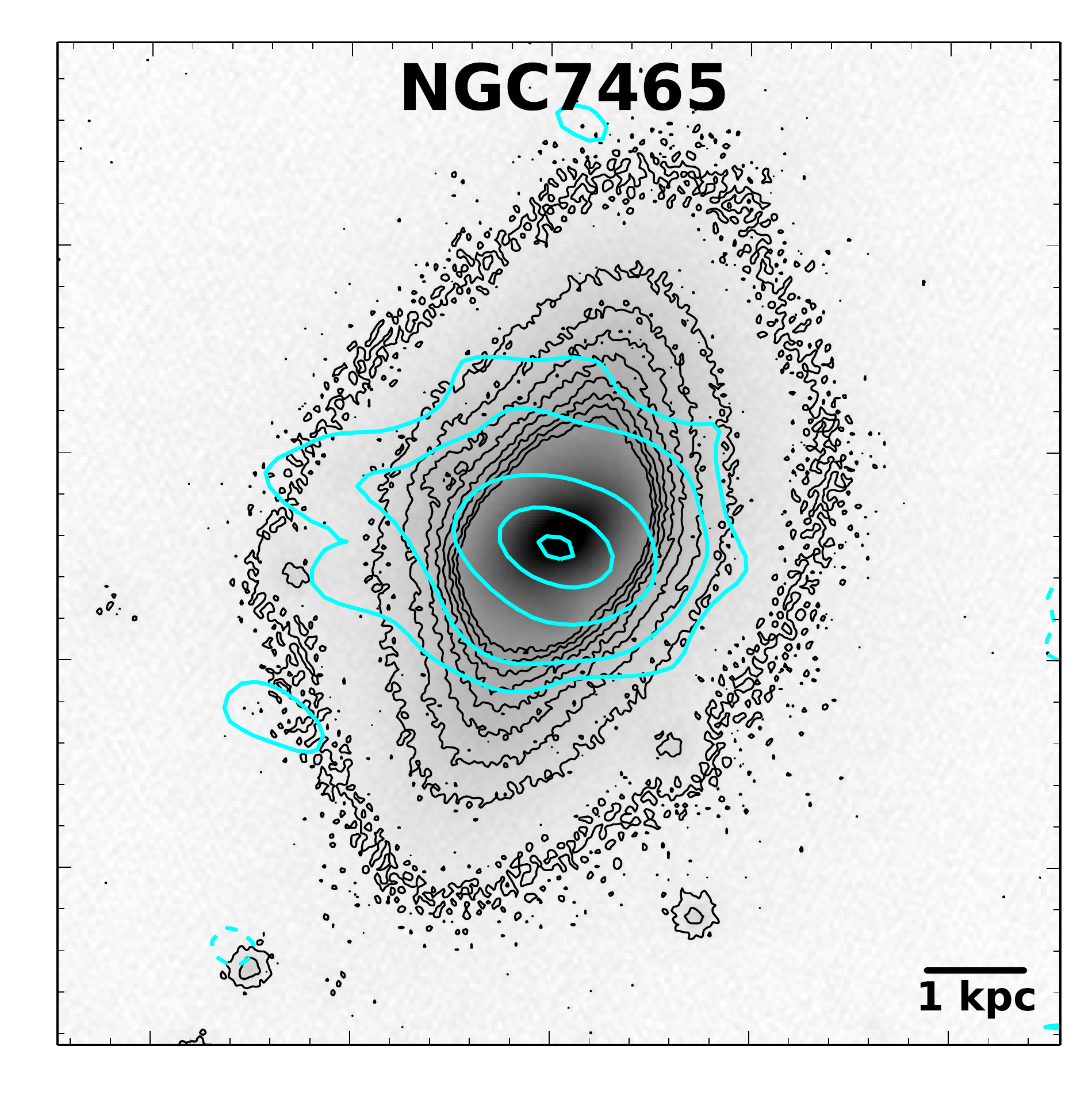}}
{\label{fig:sub:PGC058114}\includegraphics[clip=True, trim=0cm 0cm 0cm 0cm, scale=0.25]{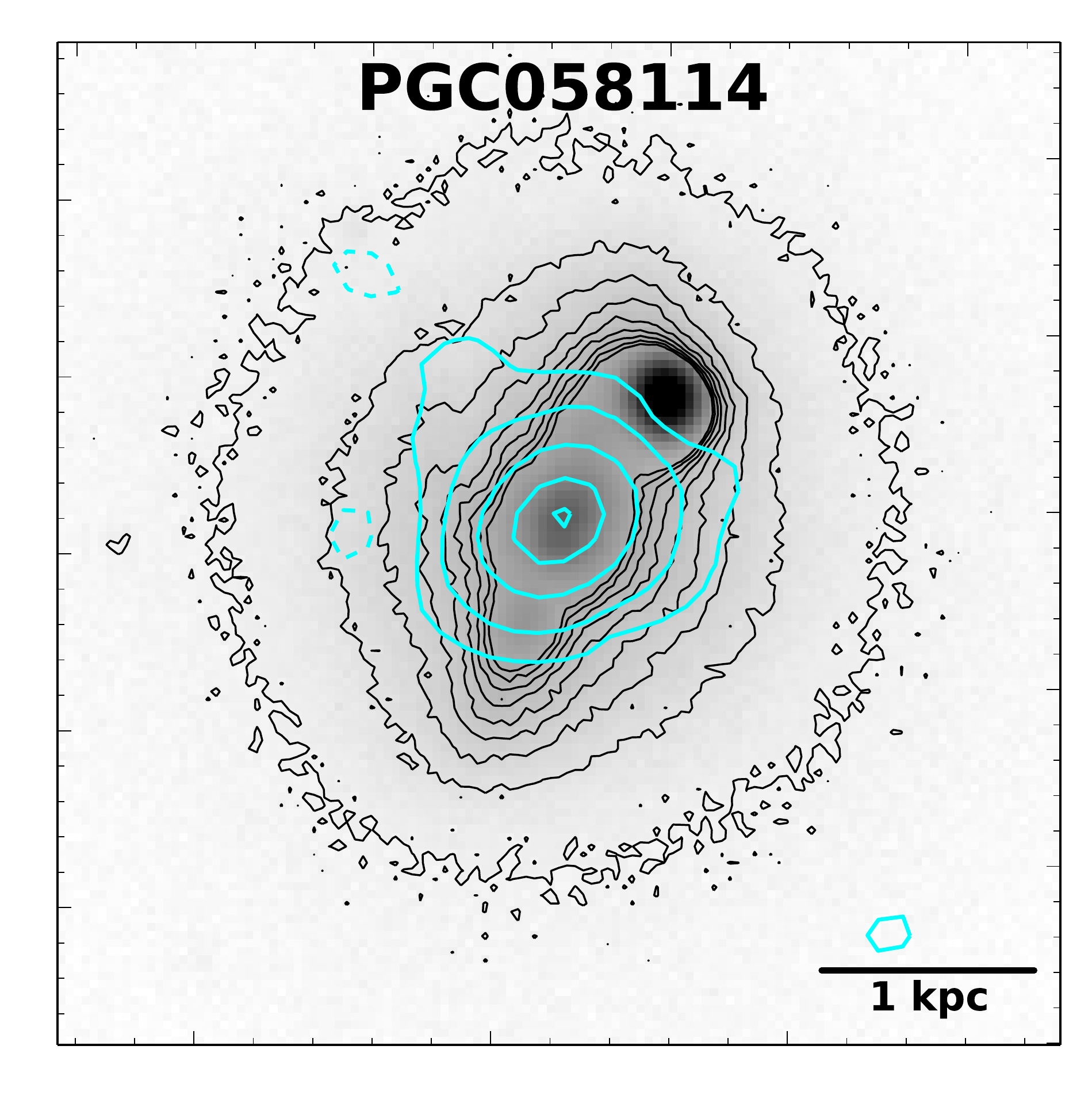}}
{\label{fig:sub:UGC05408}\includegraphics[clip=True, trim=0cm 0cm 0cm 0cm, scale=0.25]{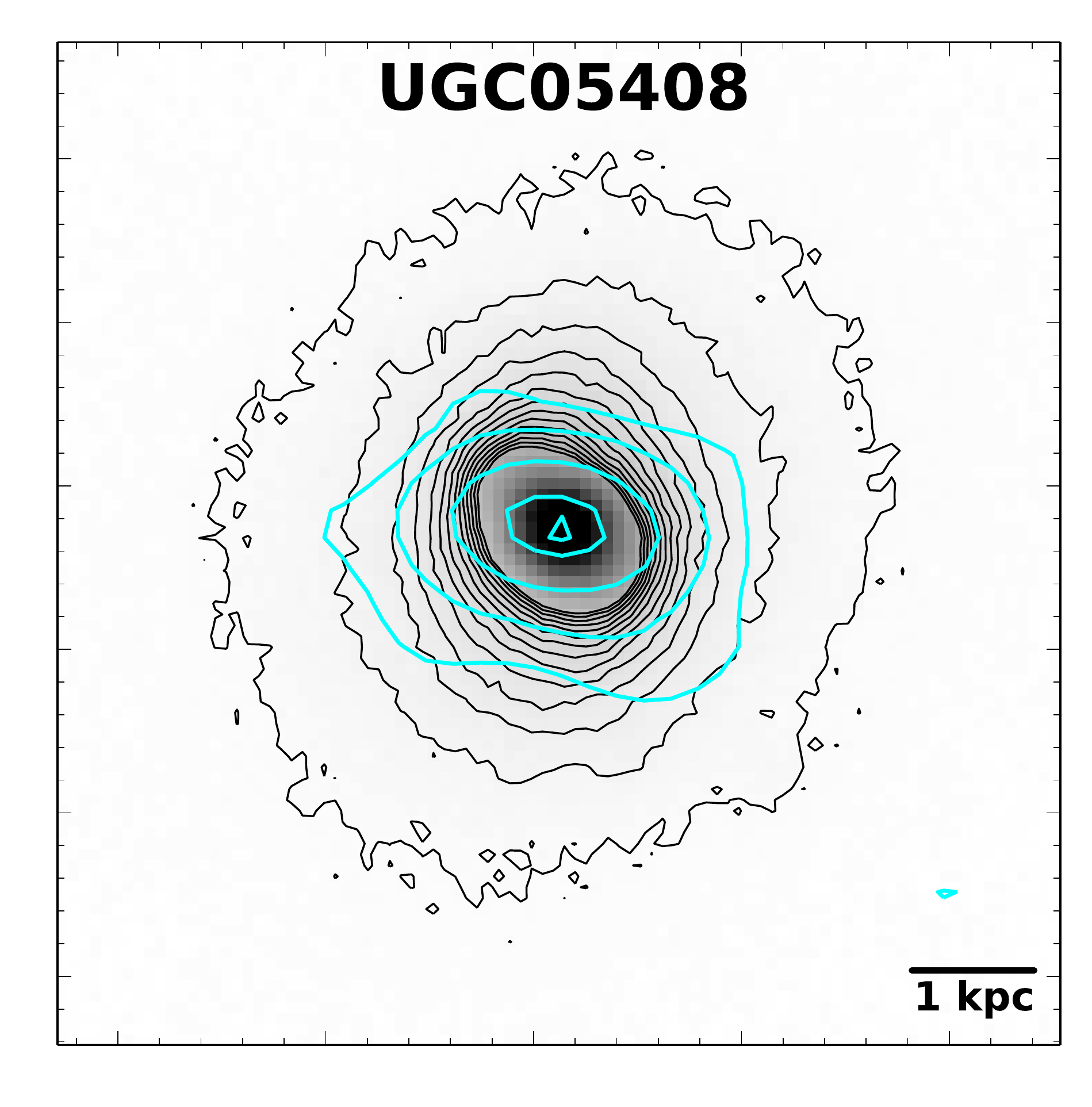}}
  
\caption{Optical r-band images (greyscale and black contours) with 1.4~GHz radio continuum contours overlaid in cyan for the 19 well-resolved radio sources from our new VLA observations.  The radio contour levels are the same as those shown in Figure~\ref{fig:radio_images} and listed in Table~\ref{tab:contours}.}
\label{fig:radio_overlays}
\end{figure*}

For each ETG included in our new 1.4~GHz VLA observations, we provide a map of the radio continuum emission with contours in Figure~\ref{fig:radio_images}.  The RMS noise level and relative contours of each detected ETG are listed in Table~\ref{tab:contours}.  Optical images with radio continuum contours are shown in Figure~\ref{fig:radio_overlays} for the 19 well-resolved sources.

\begin{figure*}
{\label{fig:sub:IC0676}\includegraphics[clip=True, trim=0cm 0cm 0cm 0cm, scale=0.23]{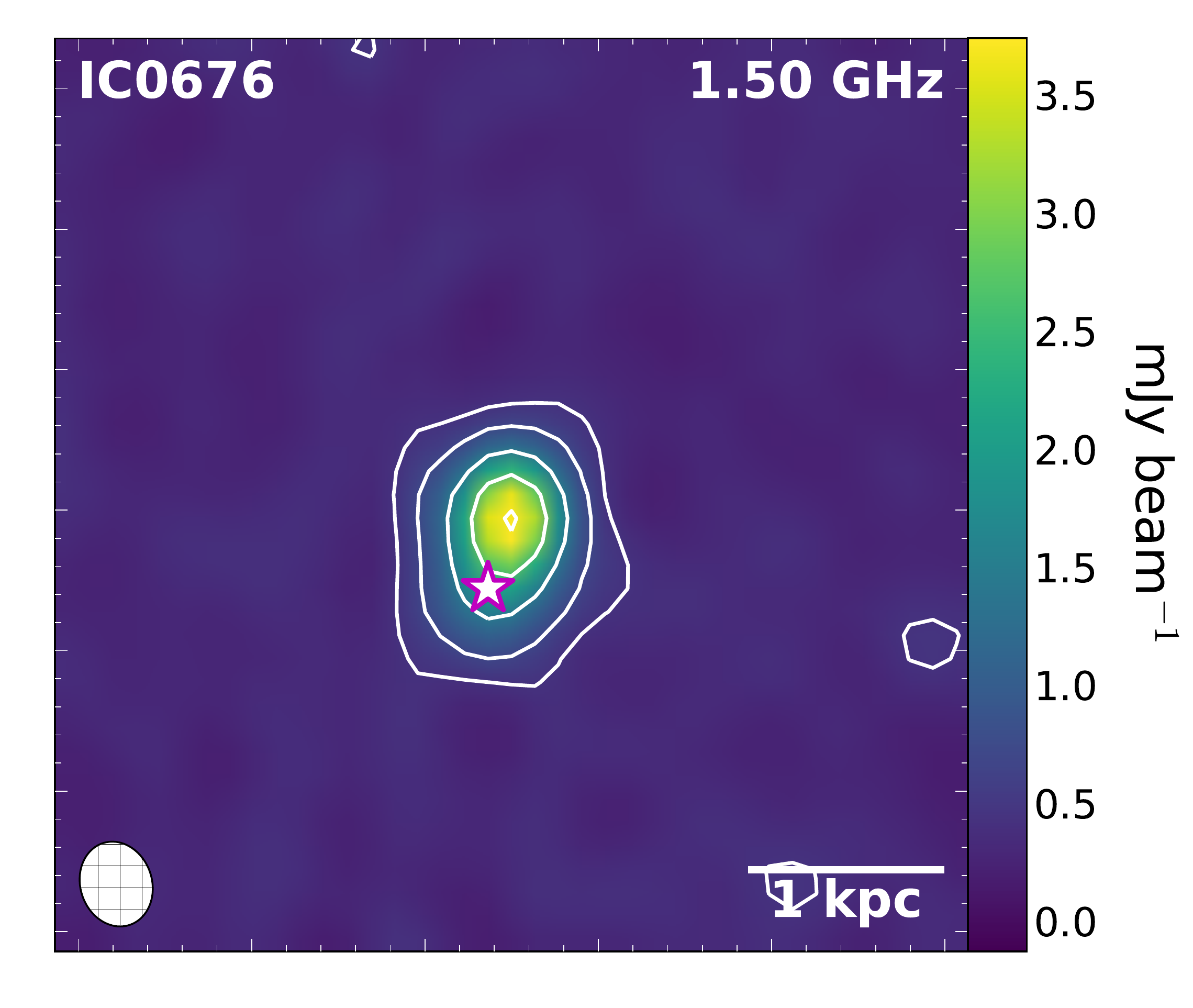}}
{\label{fig:sub:IC0719}\includegraphics[clip=True, trim=0cm 0cm 0cm 0cm, scale=0.23]{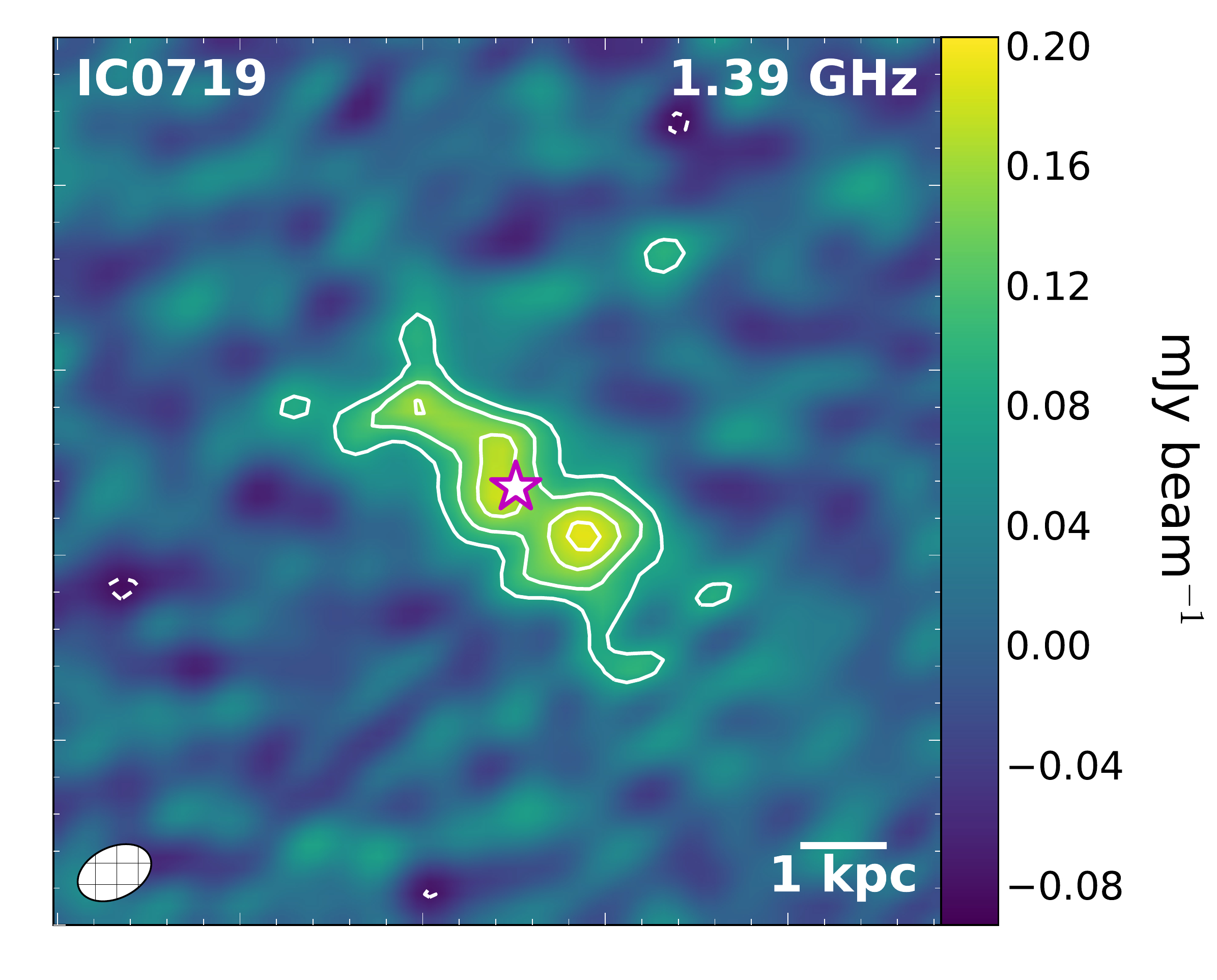}}
{\label{fig:sub:IC1024}\includegraphics[clip=True, trim=0cm 0cm 0cm 0cm, scale=0.23]{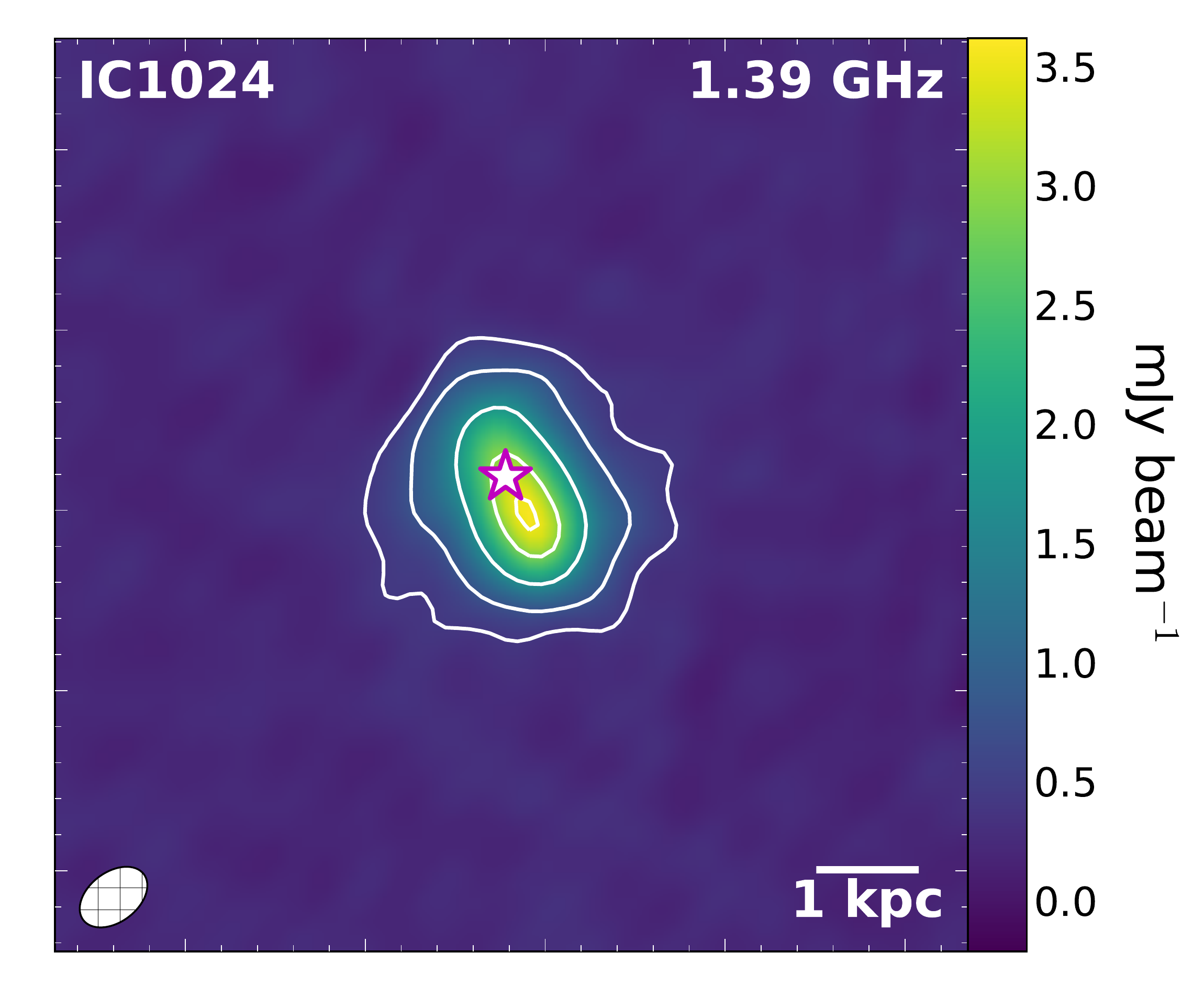}}
{\label{fig:sub:NGC0524}\includegraphics[clip=True, trim=0cm 0cm 0cm 0cm, scale=0.23]{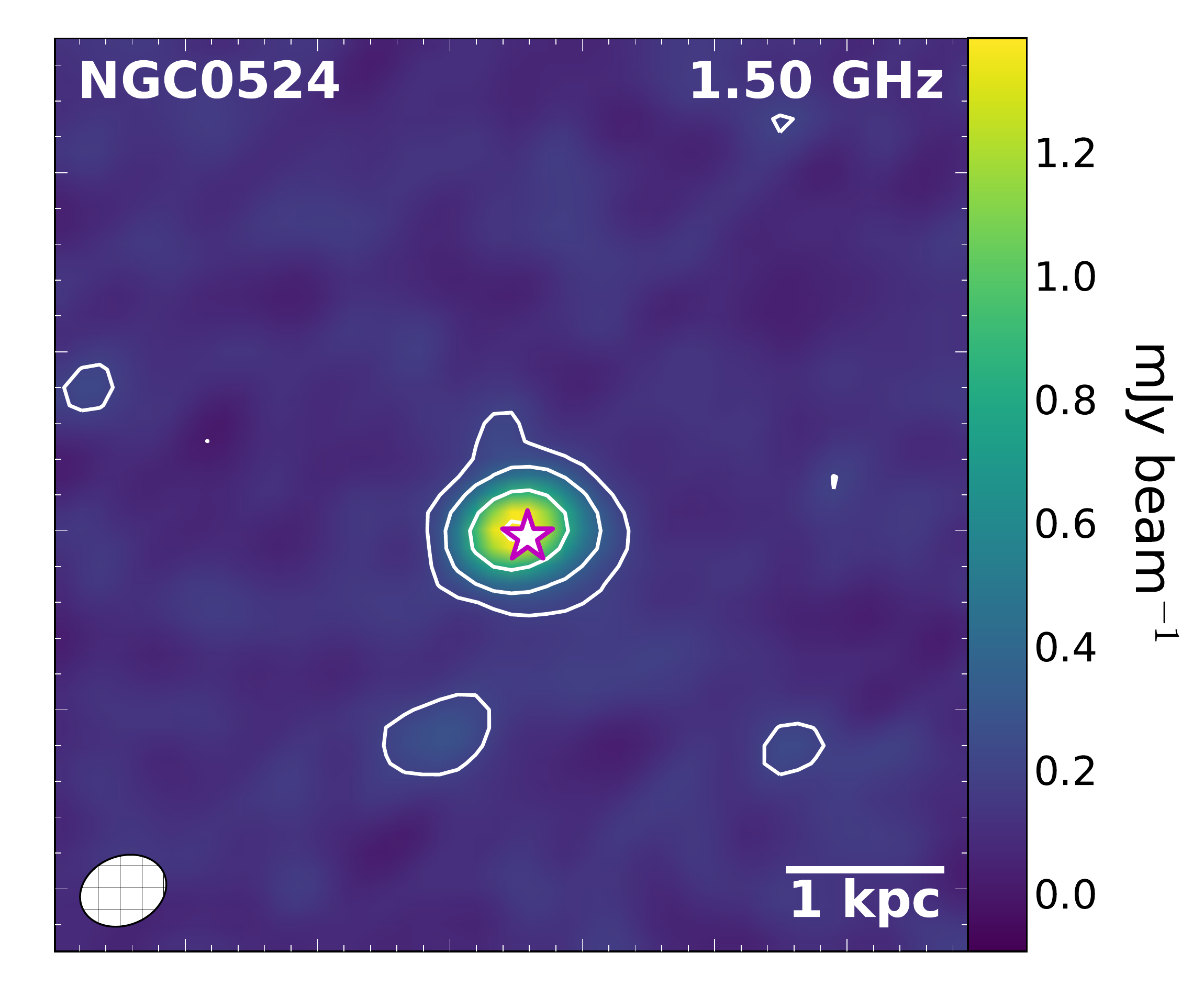}}
{\label{fig:sub:NGC0680}\includegraphics[clip=True, trim=0cm 0cm 0cm 0cm, scale=0.23]{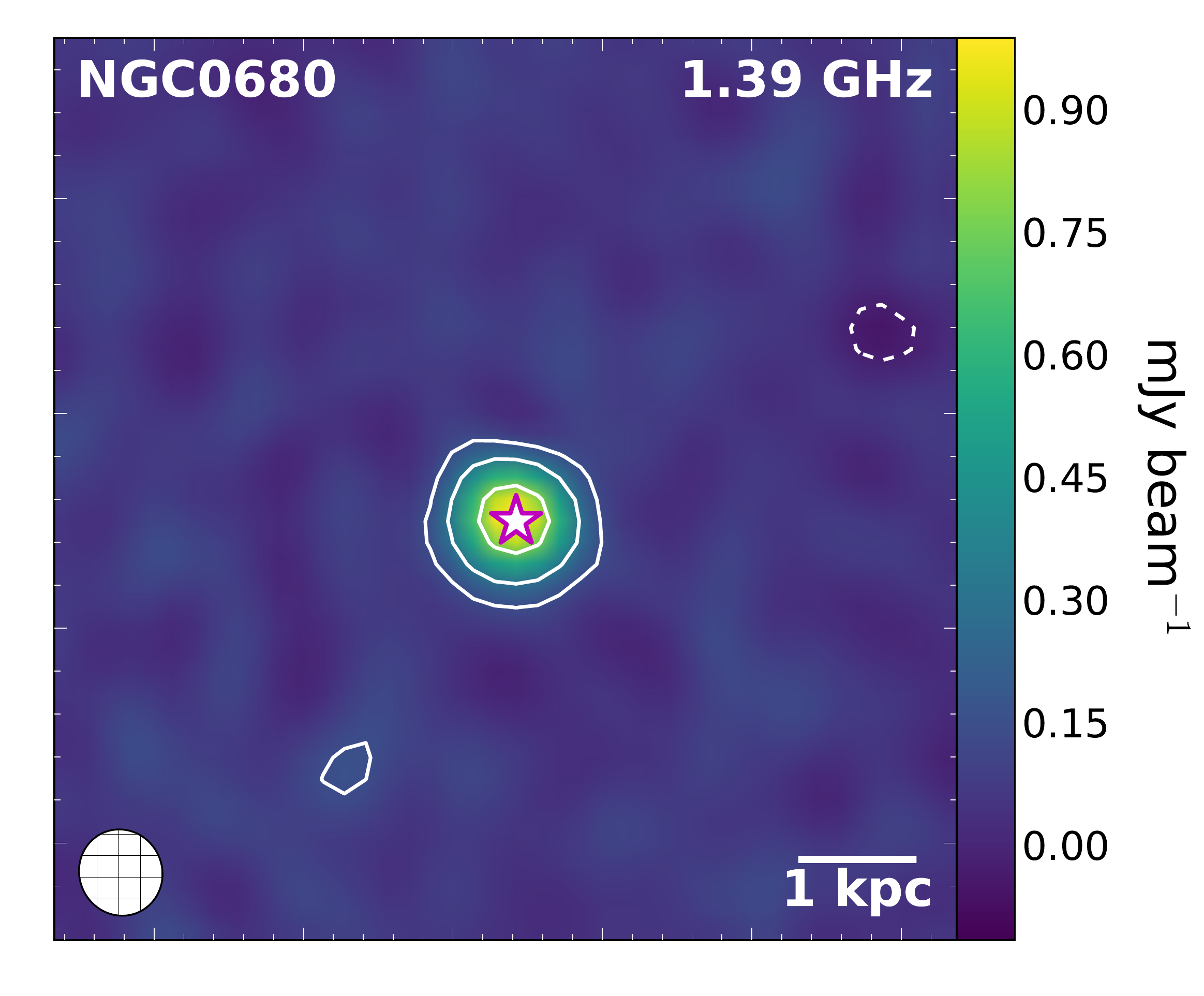}}
{\label{fig:sub:NGC1023}\includegraphics[clip=True, trim=0cm 0cm 0cm 0cm, scale=0.23]{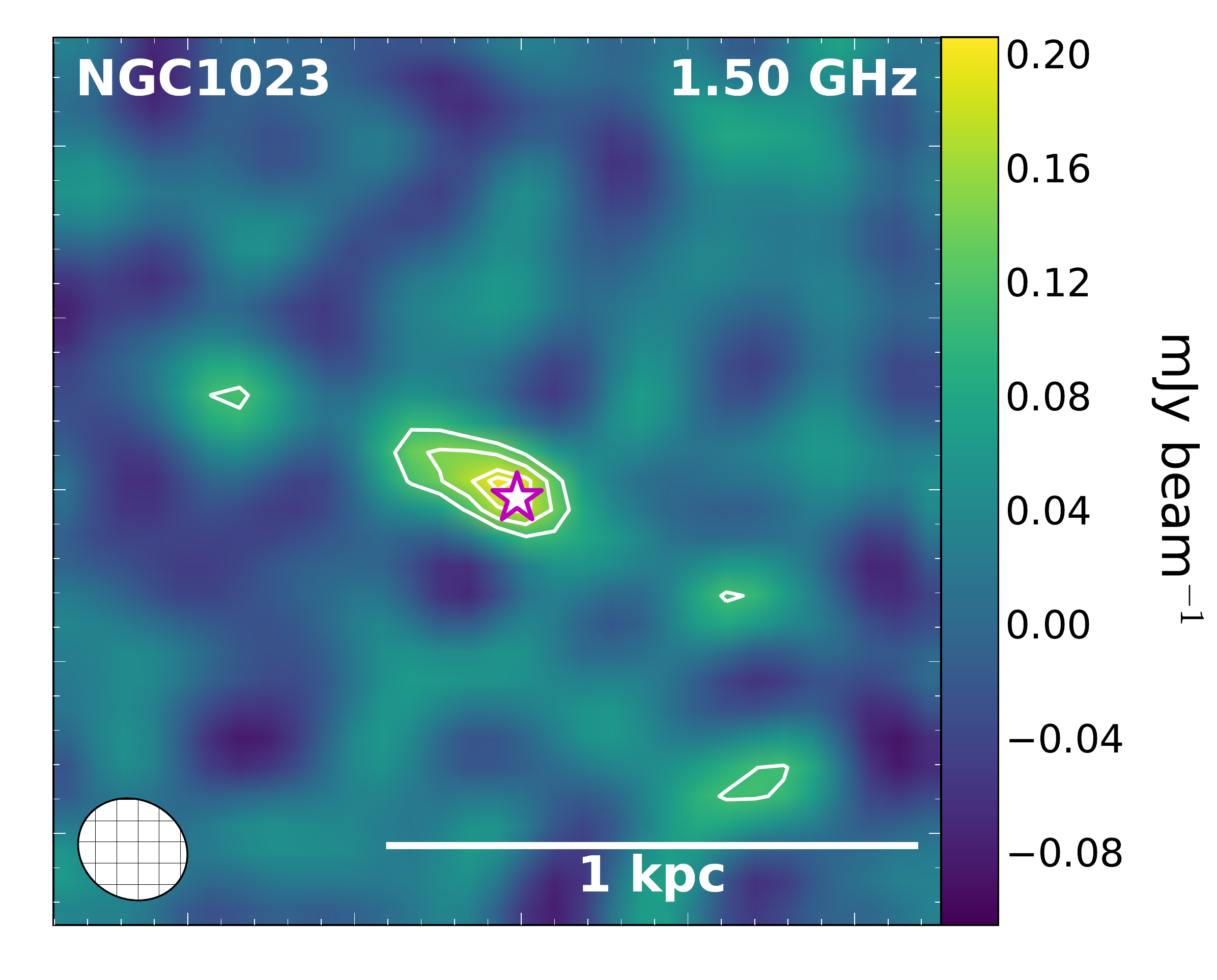}}
{\label{fig:sub:NGC1222}\includegraphics[clip=True, trim=0cm 0cm 0cm 0cm, scale=0.23]{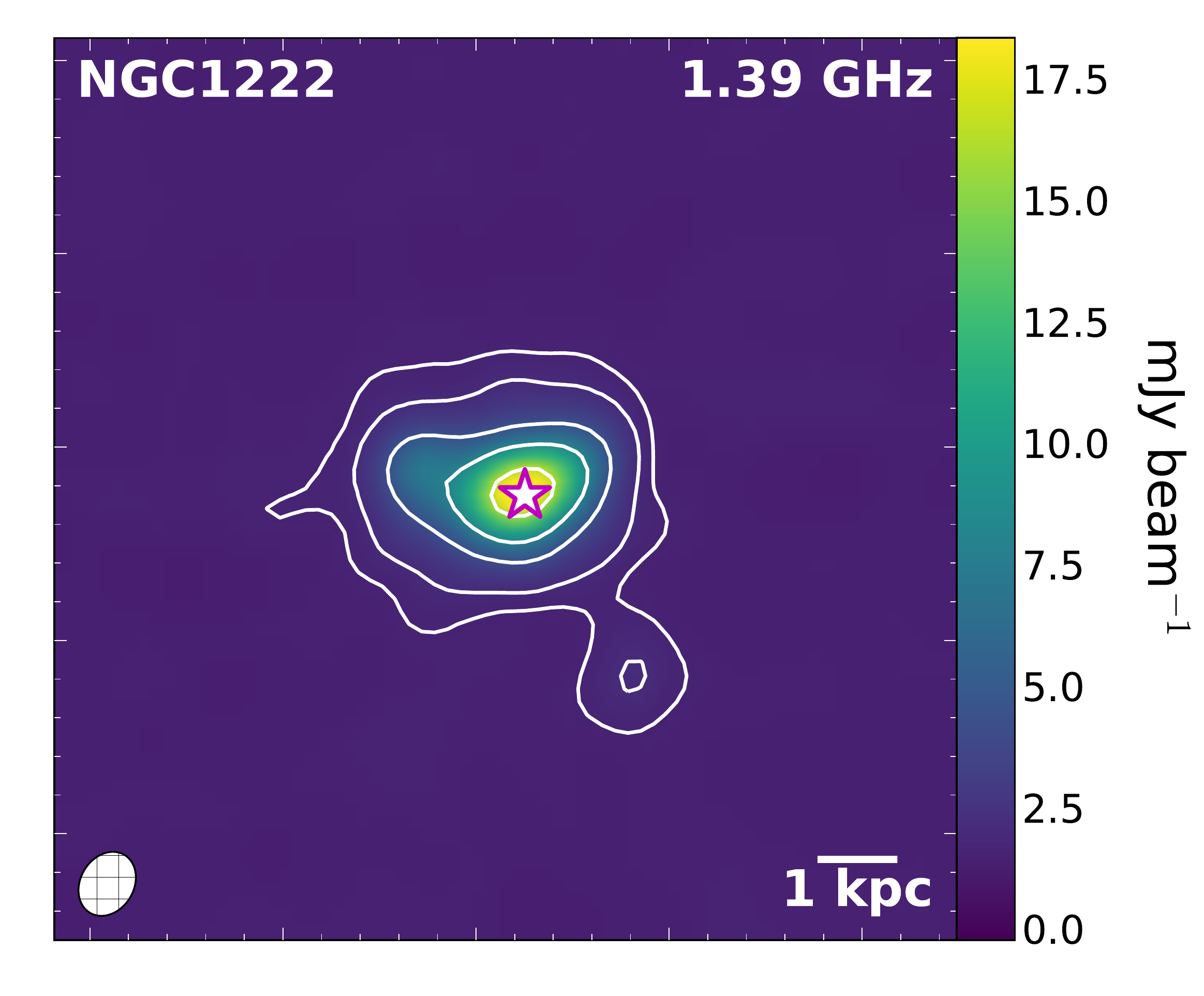}}
{\label{fig:sub:NGC1266}\includegraphics[clip=True, trim=0cm 0cm 0cm 0cm, scale=0.23]{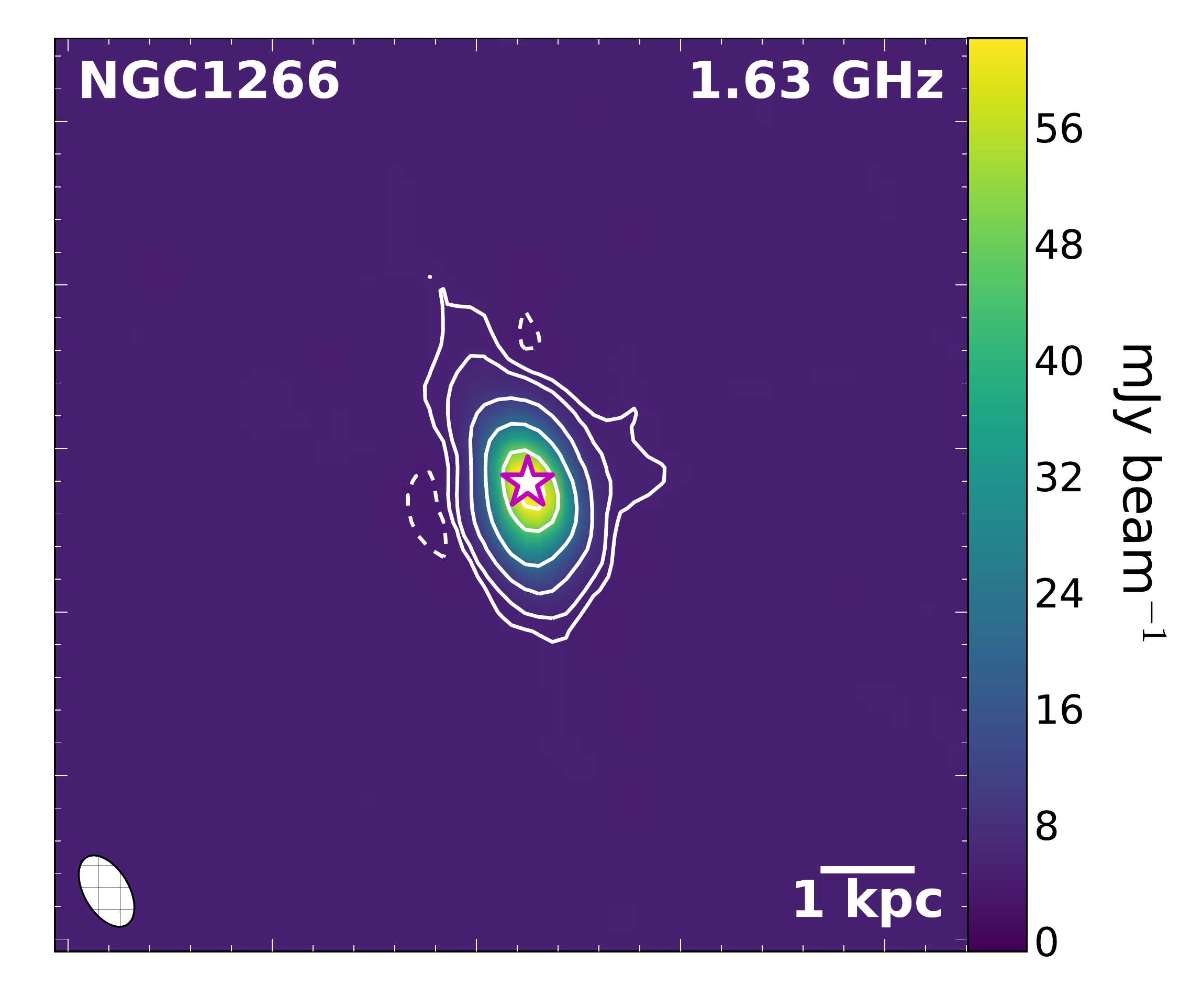}}
{\label{fig:sub:NGC2685}\includegraphics[clip=True, trim=0cm 0cm 0cm 0cm, scale=0.23]{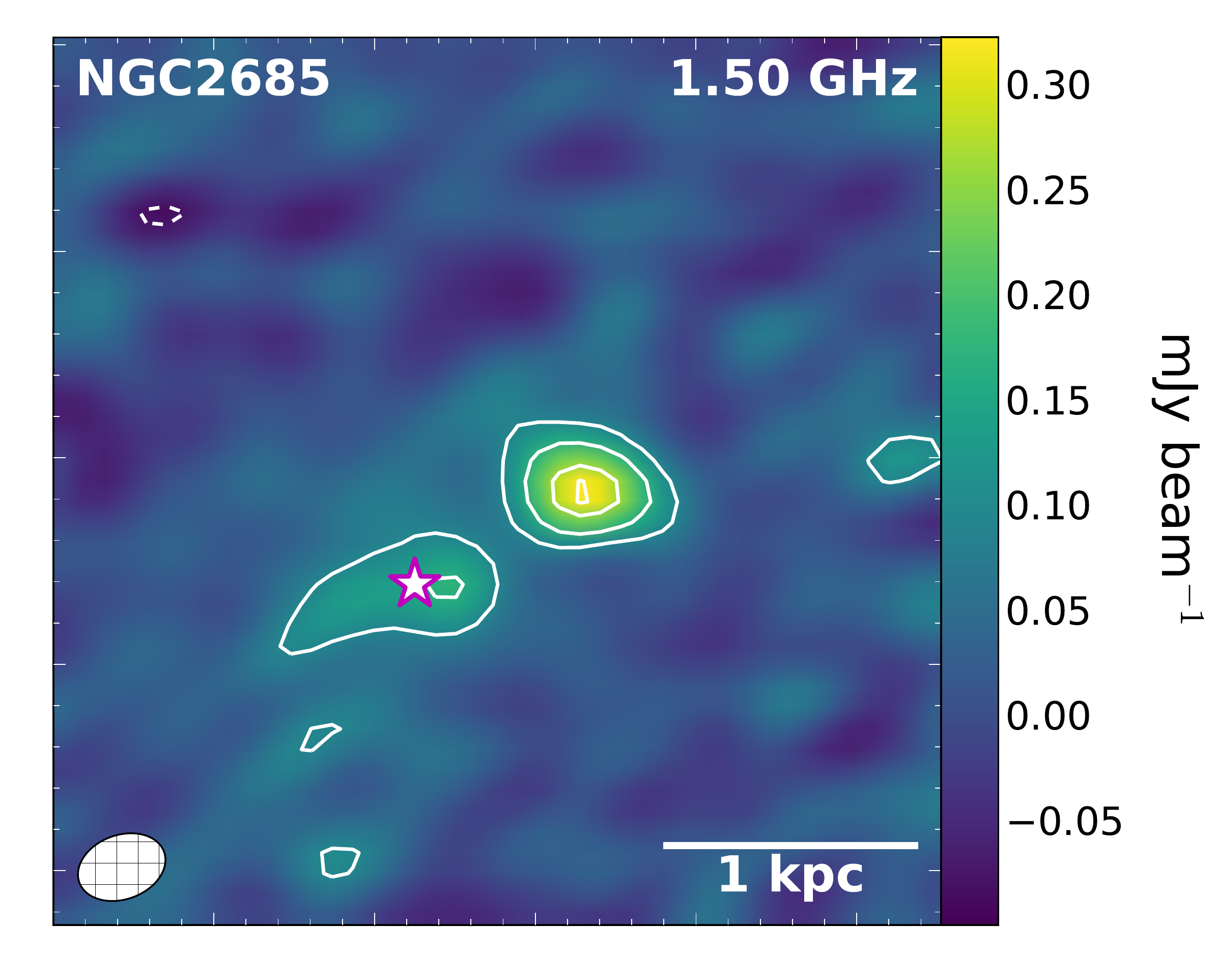}}
{\label{fig:sub:NGC2764}\includegraphics[clip=True, trim=0cm 0cm 0cm 0cm, scale=0.23]{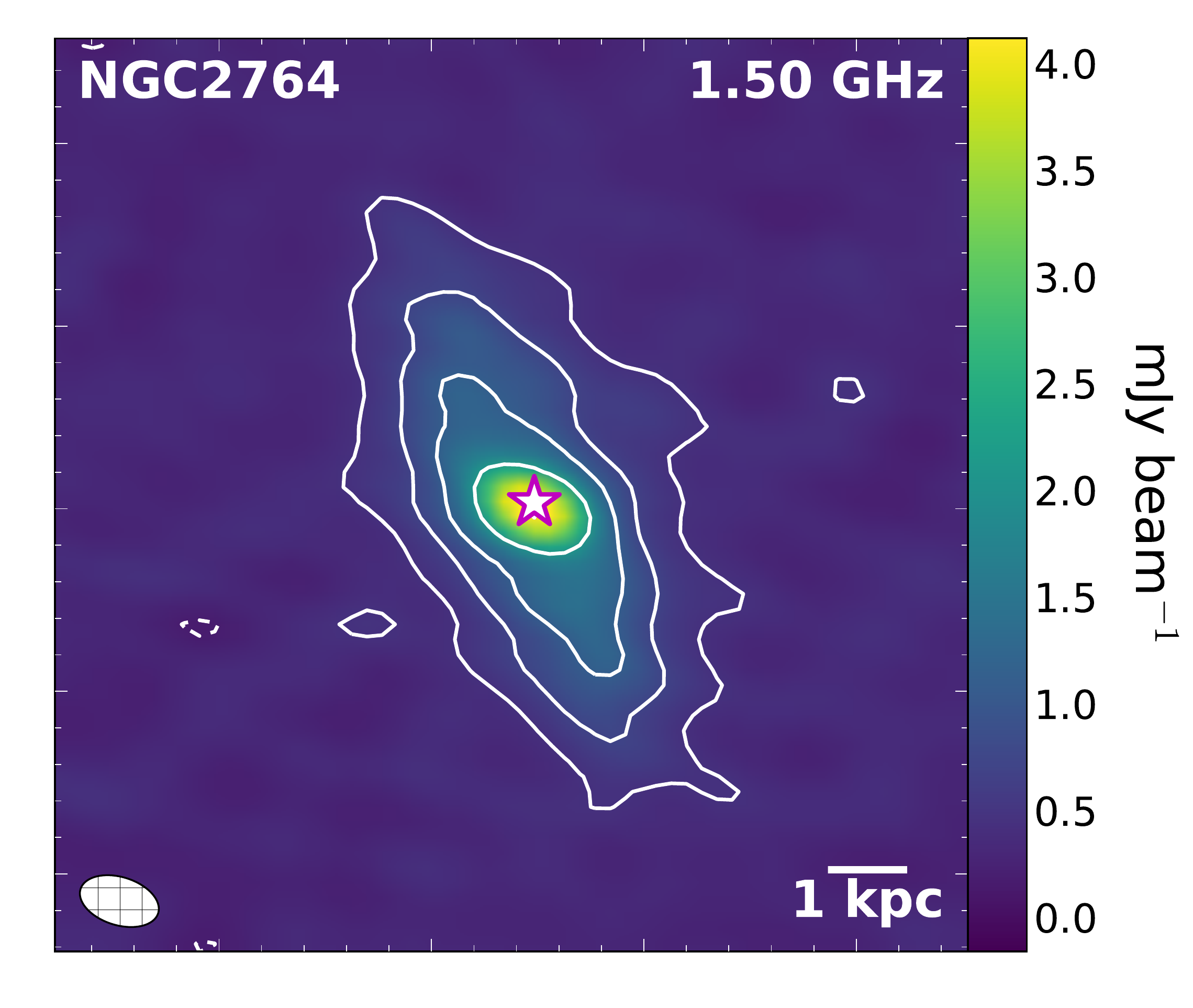}}
{\label{fig:sub:NGC2768}\includegraphics[clip=True, trim=0cm 0cm 0cm 0cm, scale=0.23]{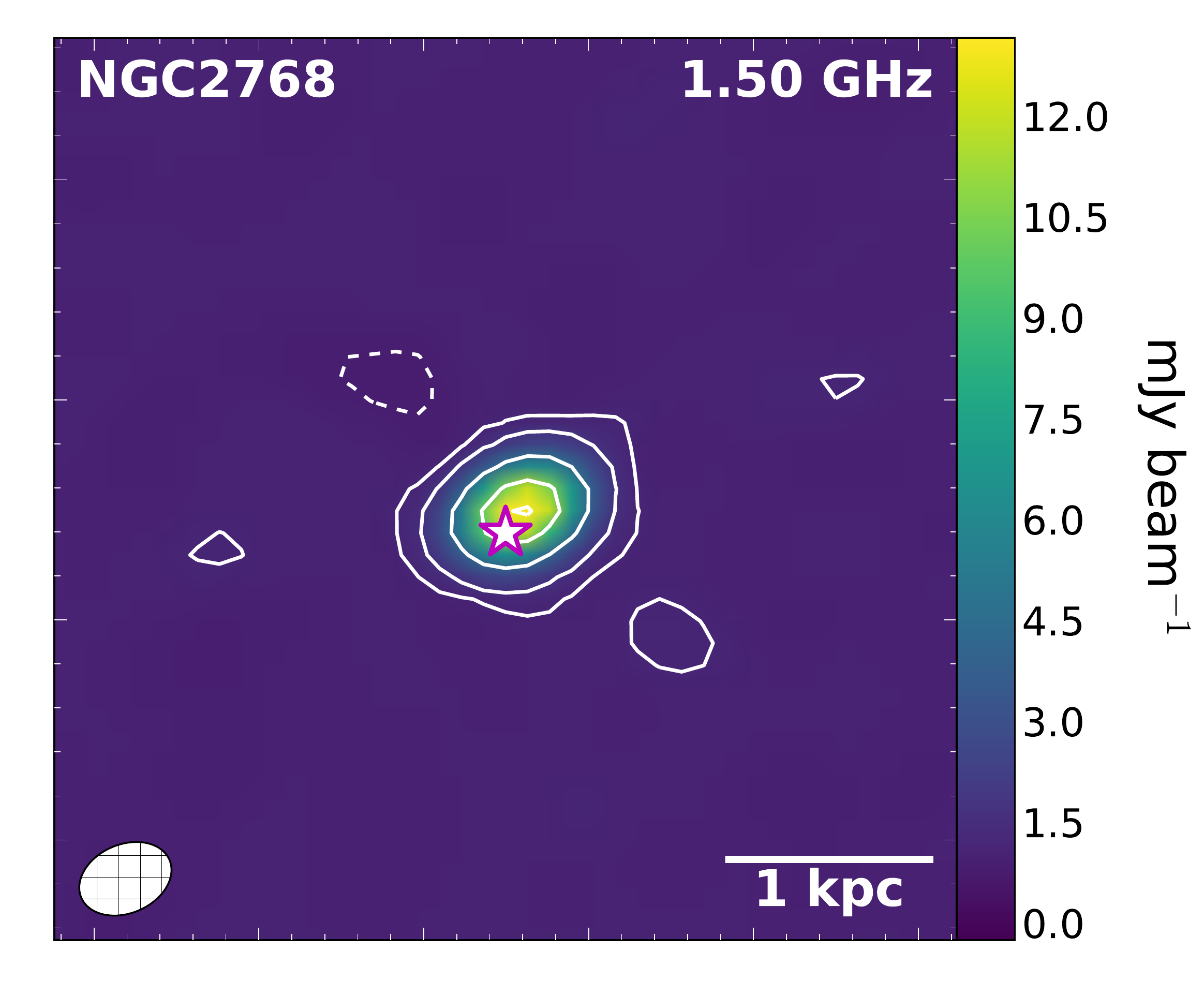}}
{\label{fig:sub:NGC2824}\includegraphics[clip=True, trim=0cm 0cm 0cm 0cm, scale=0.23]{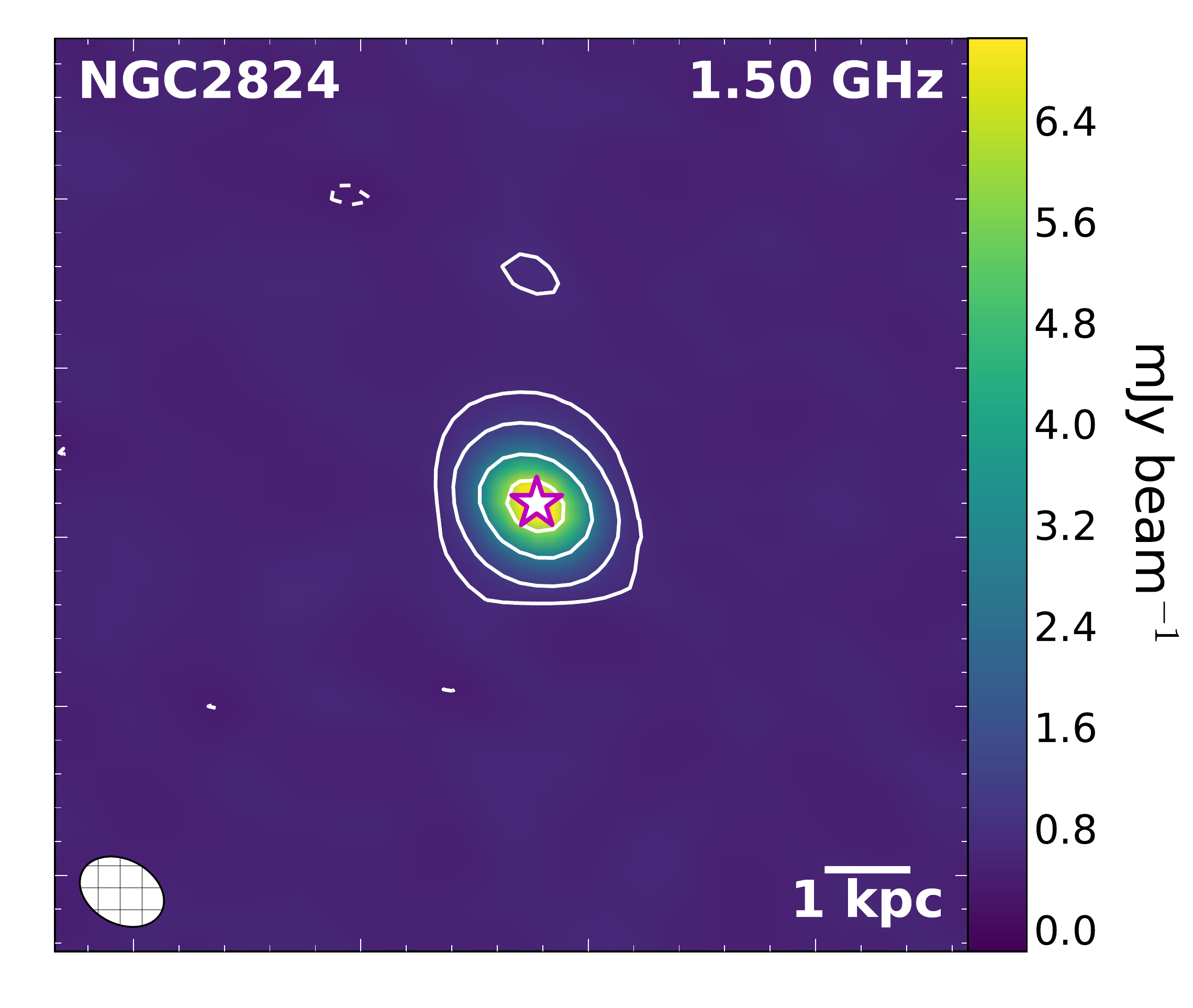}}
{\label{fig:sub:NGC2852}\includegraphics[clip=True, trim=0cm 0cm 0cm 0cm, scale=0.23]{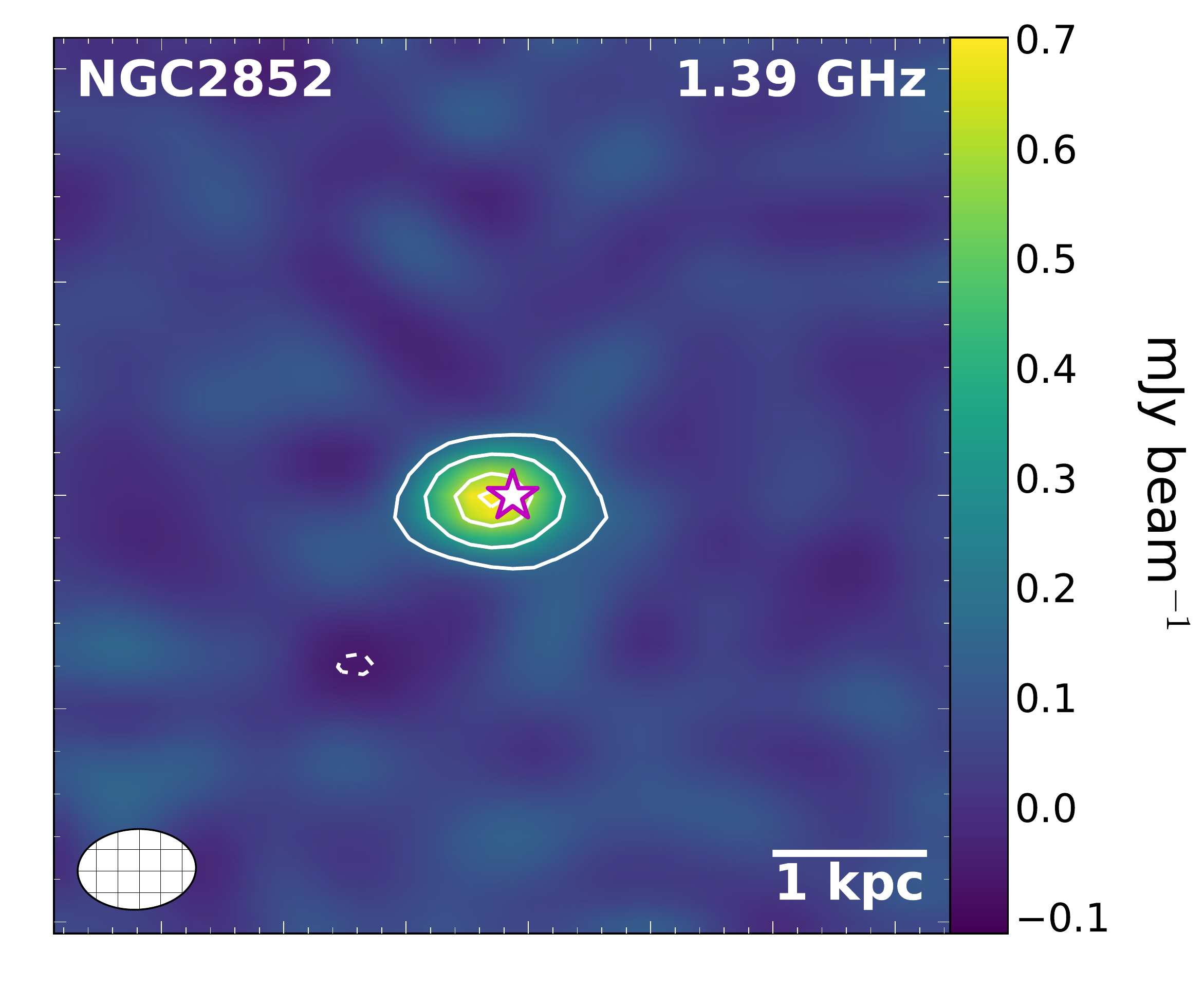}}
{\label{fig:sub:NGC3032}\includegraphics[clip=True, trim=0cm 0cm 0cm 0cm, scale=0.23]{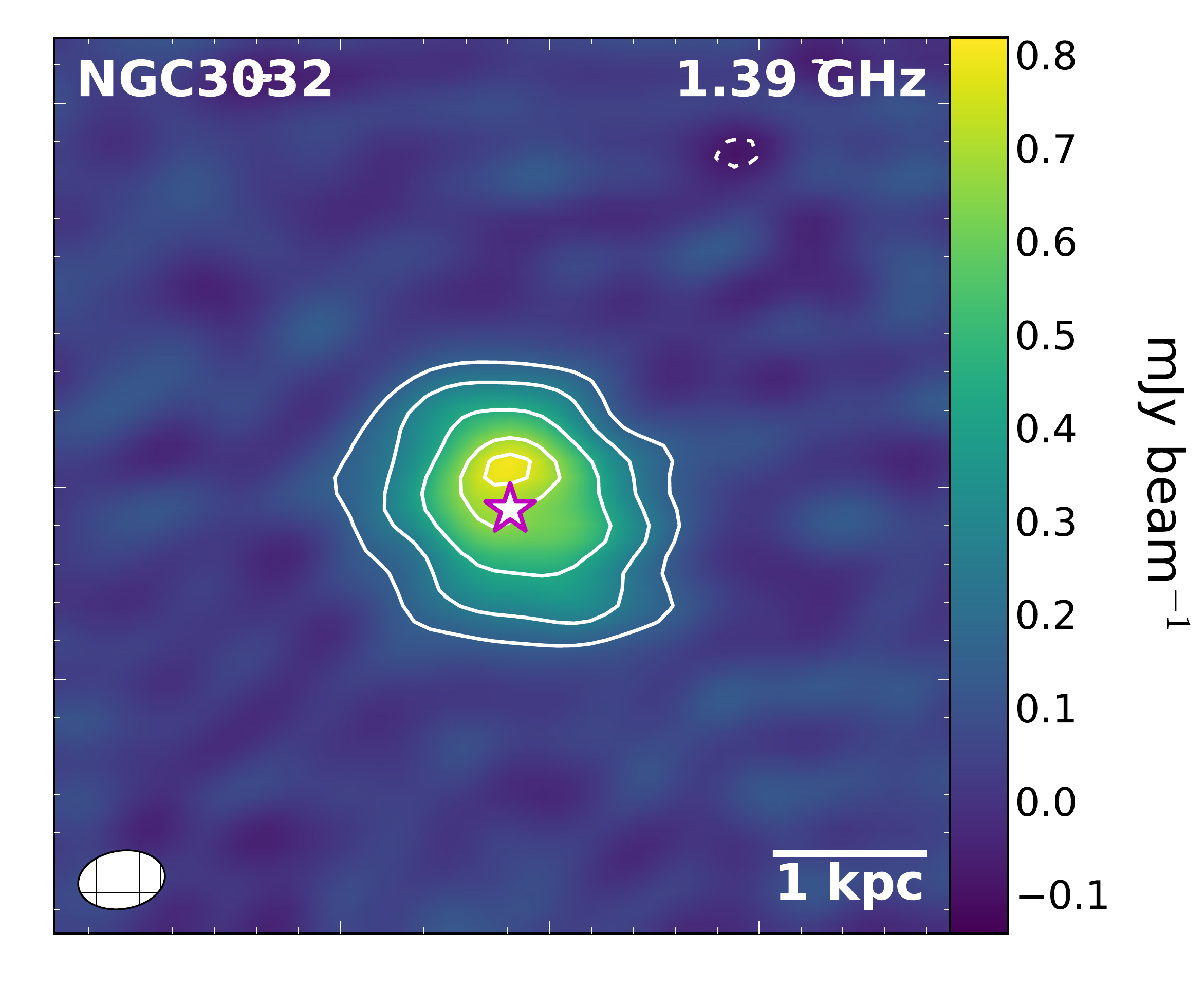}}
{\label{fig:sub:NGC3182}\includegraphics[clip=True, trim=0cm 0cm 0cm 0cm, scale=0.23]{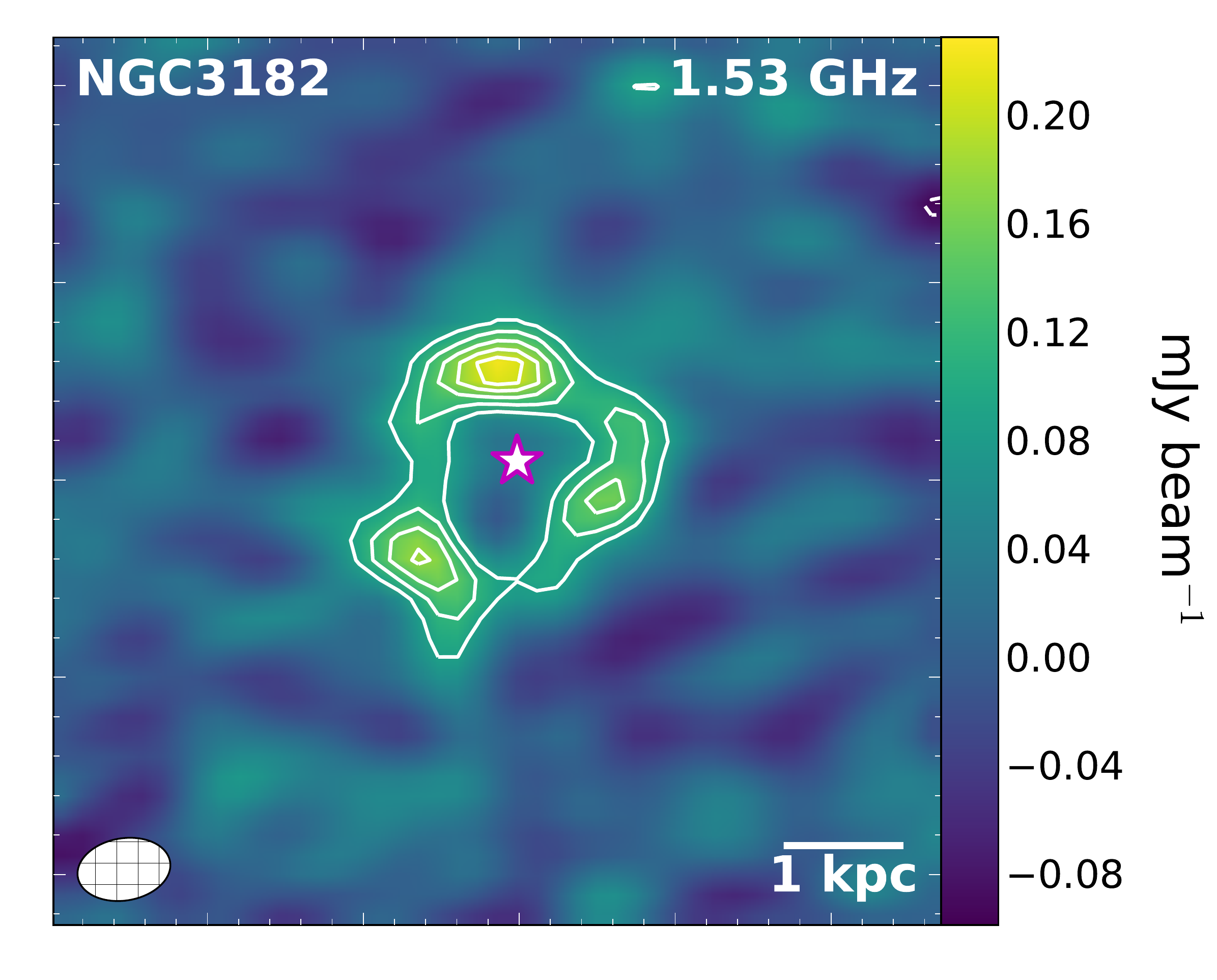}}
\end{figure*}
 
\begin{figure*}
{\label{fig:sub:NGC3193}\includegraphics[clip=True, trim=0cm 0cm 0cm 0cm, scale=0.23]{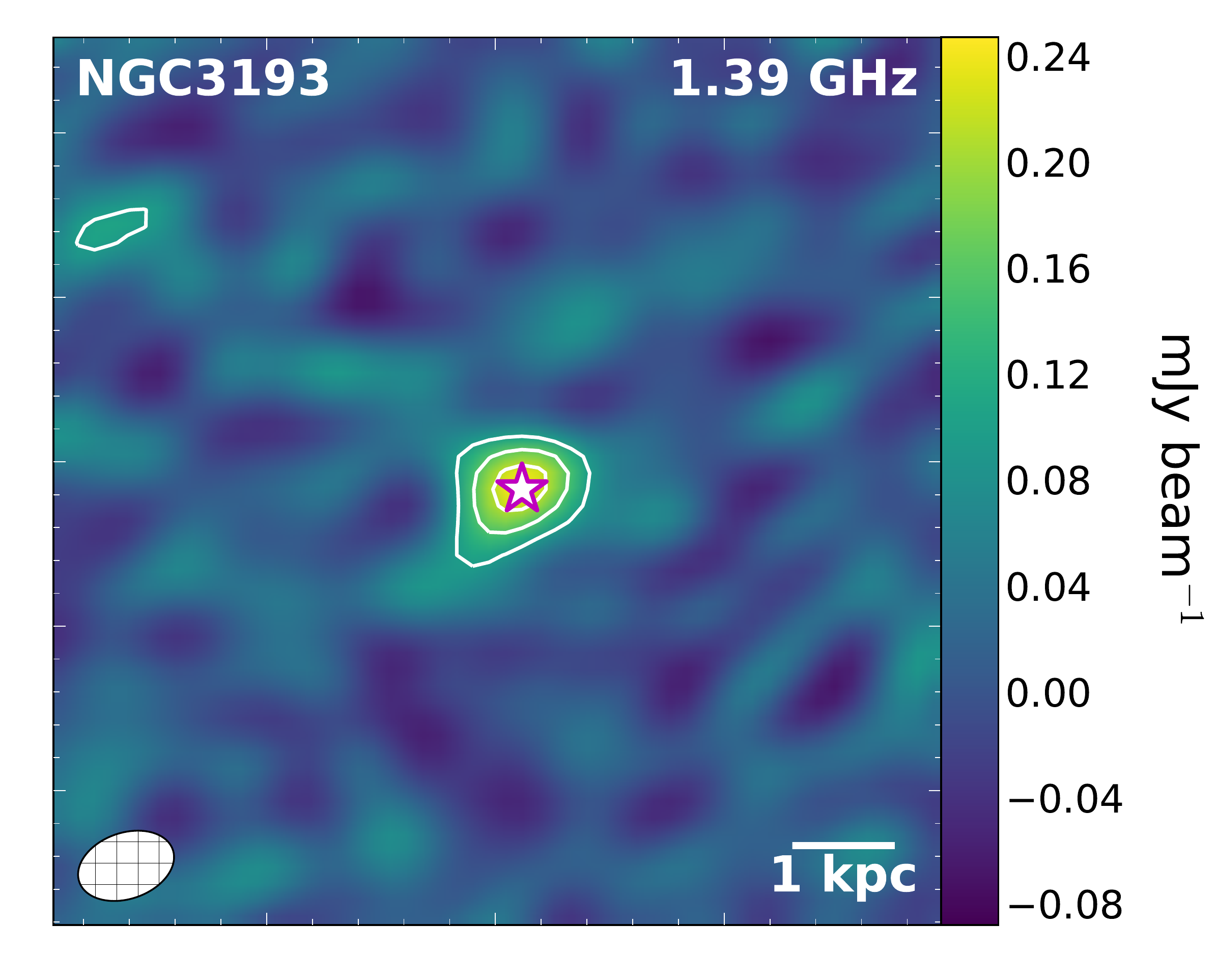}}
{\label{fig:sub:NGC3245}\includegraphics[clip=True, trim=0cm 0cm 0cm 0cm, scale=0.23]{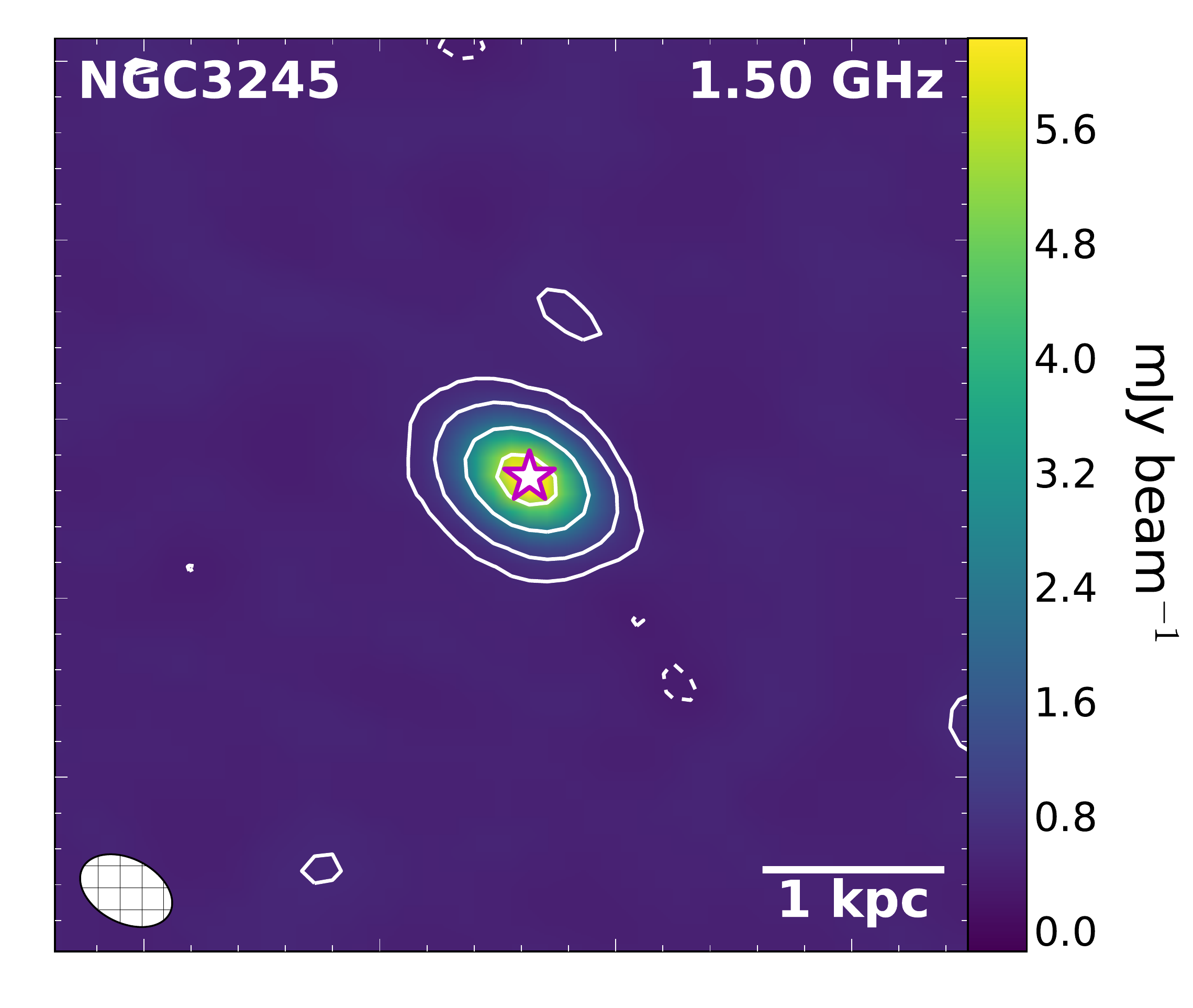}}
{\label{fig:sub:NGC3489}\includegraphics[clip=True, trim=0cm 0cm 0cm 0cm, scale=0.23]{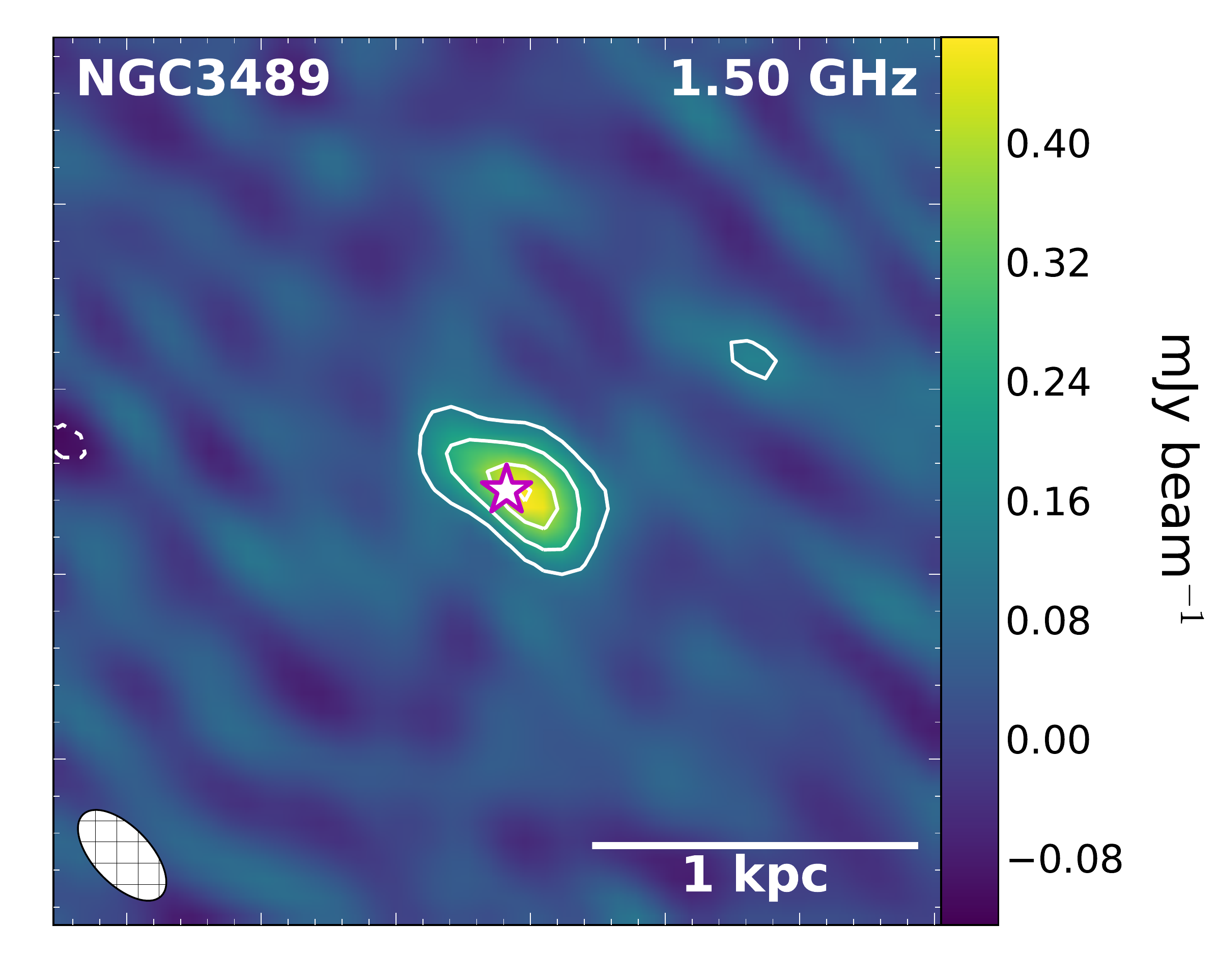}}
{\label{fig:sub:NGC3607}\includegraphics[clip=True, trim=0cm 0cm 0cm 0cm, scale=0.23]{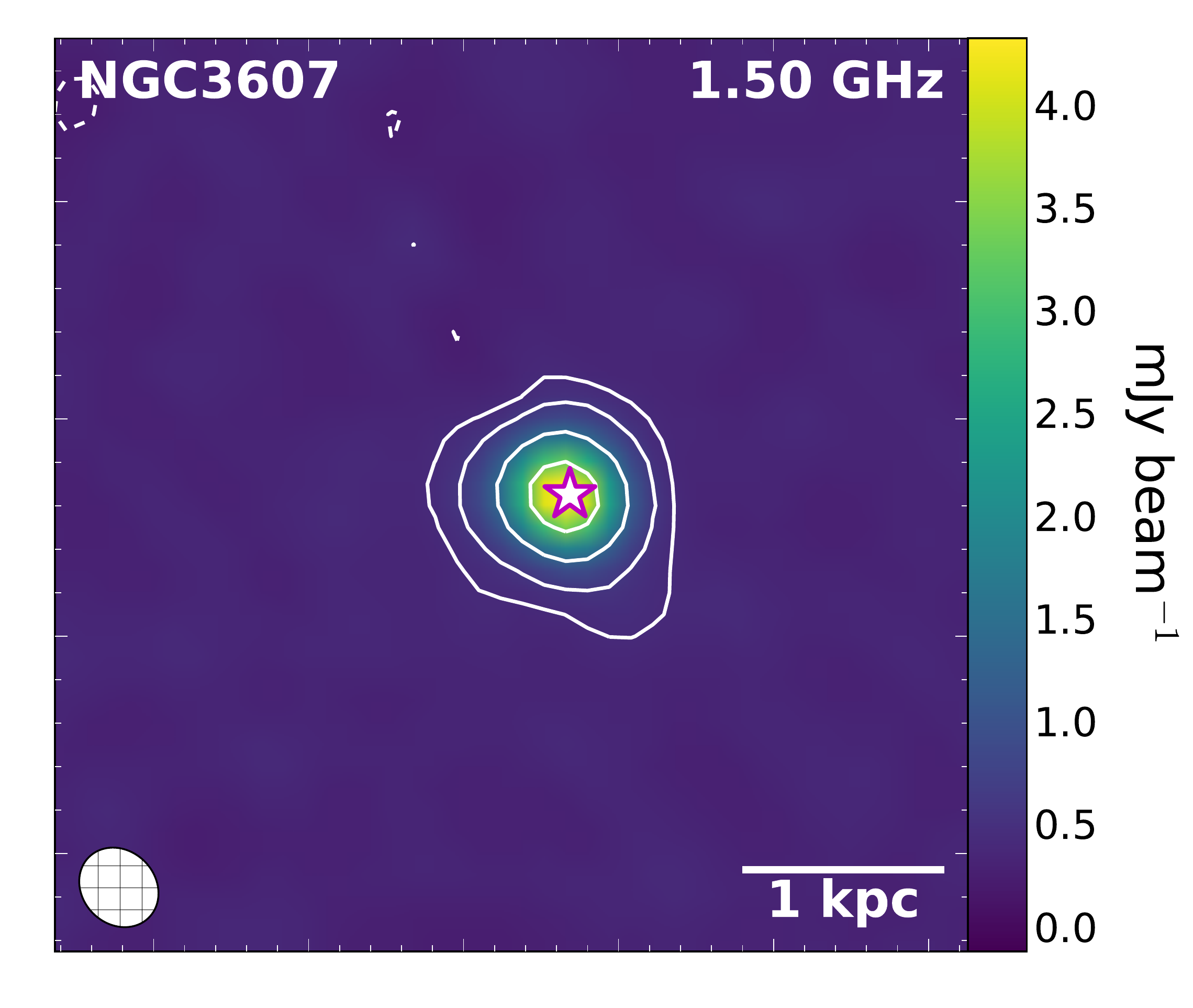}}
{\label{fig:sub:NGC3608}\includegraphics[clip=True, trim=0cm 0cm 0cm 0cm, scale=0.23]{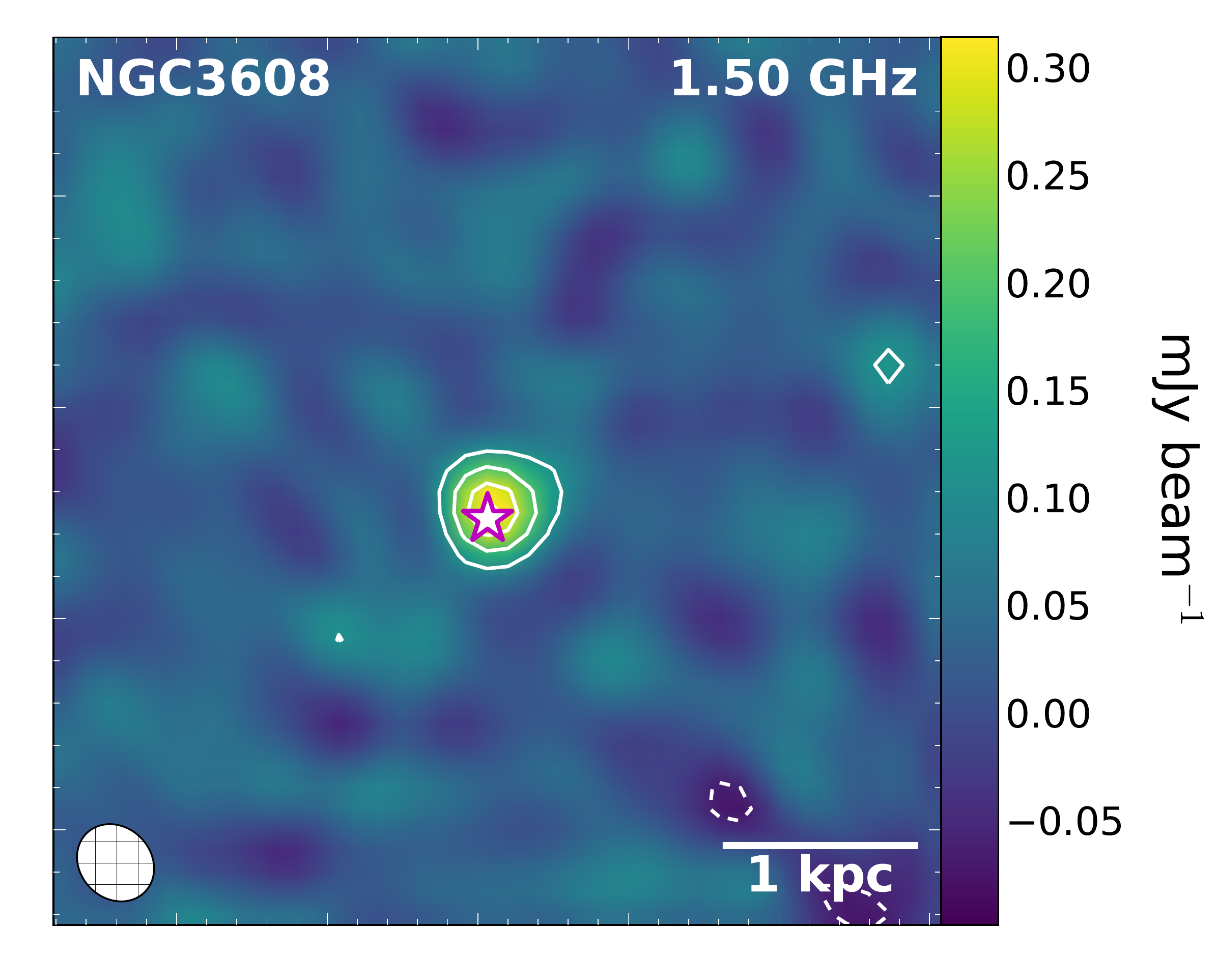}}
{\label{fig:sub:NGC3619}\includegraphics[clip=True, trim=0cm 0cm 0cm 0cm, scale=0.23]{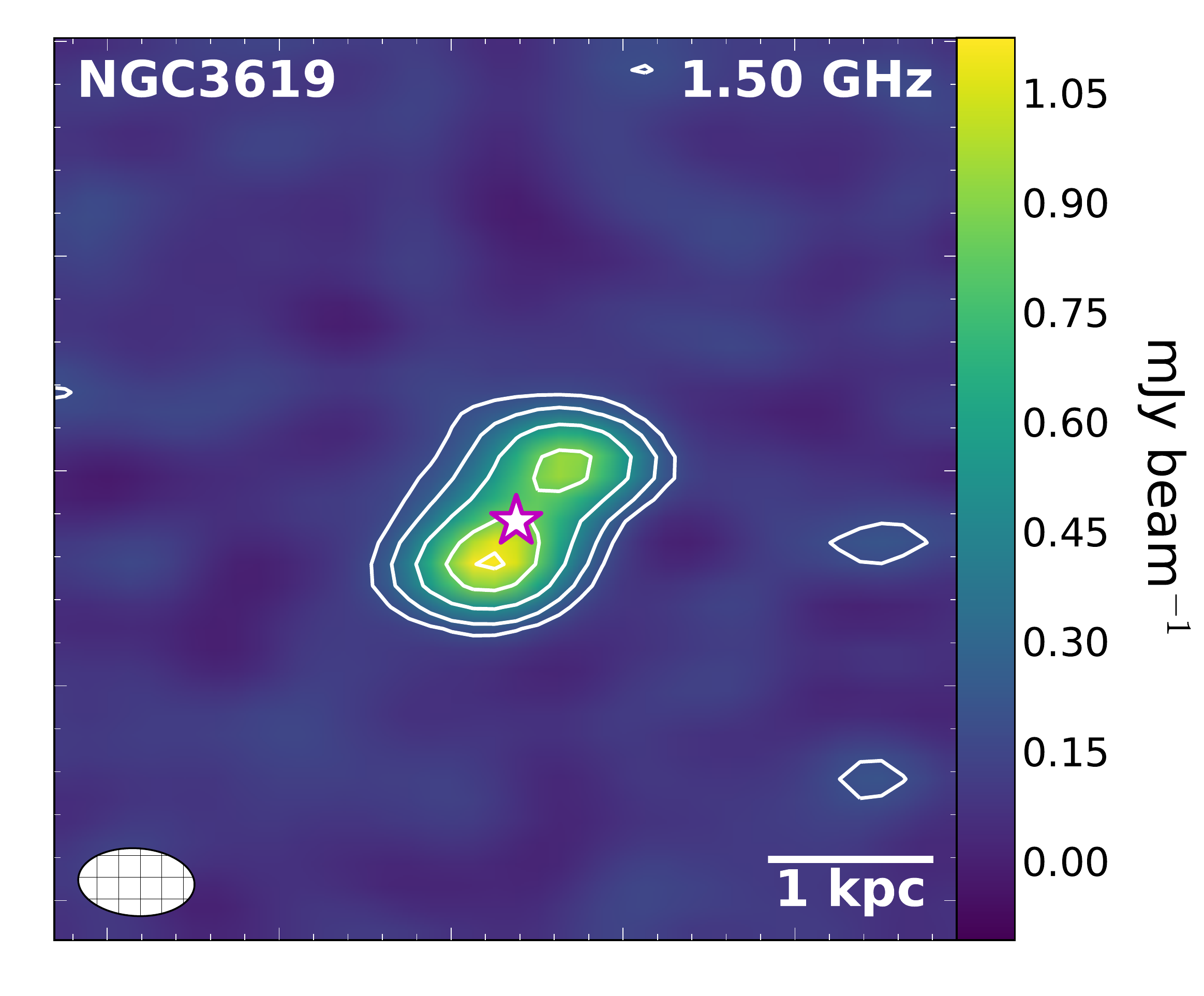}}
{\label{fig:sub:NGC3626}\includegraphics[clip=True, trim=0cm 0cm 0cm 0cm, scale=0.23]{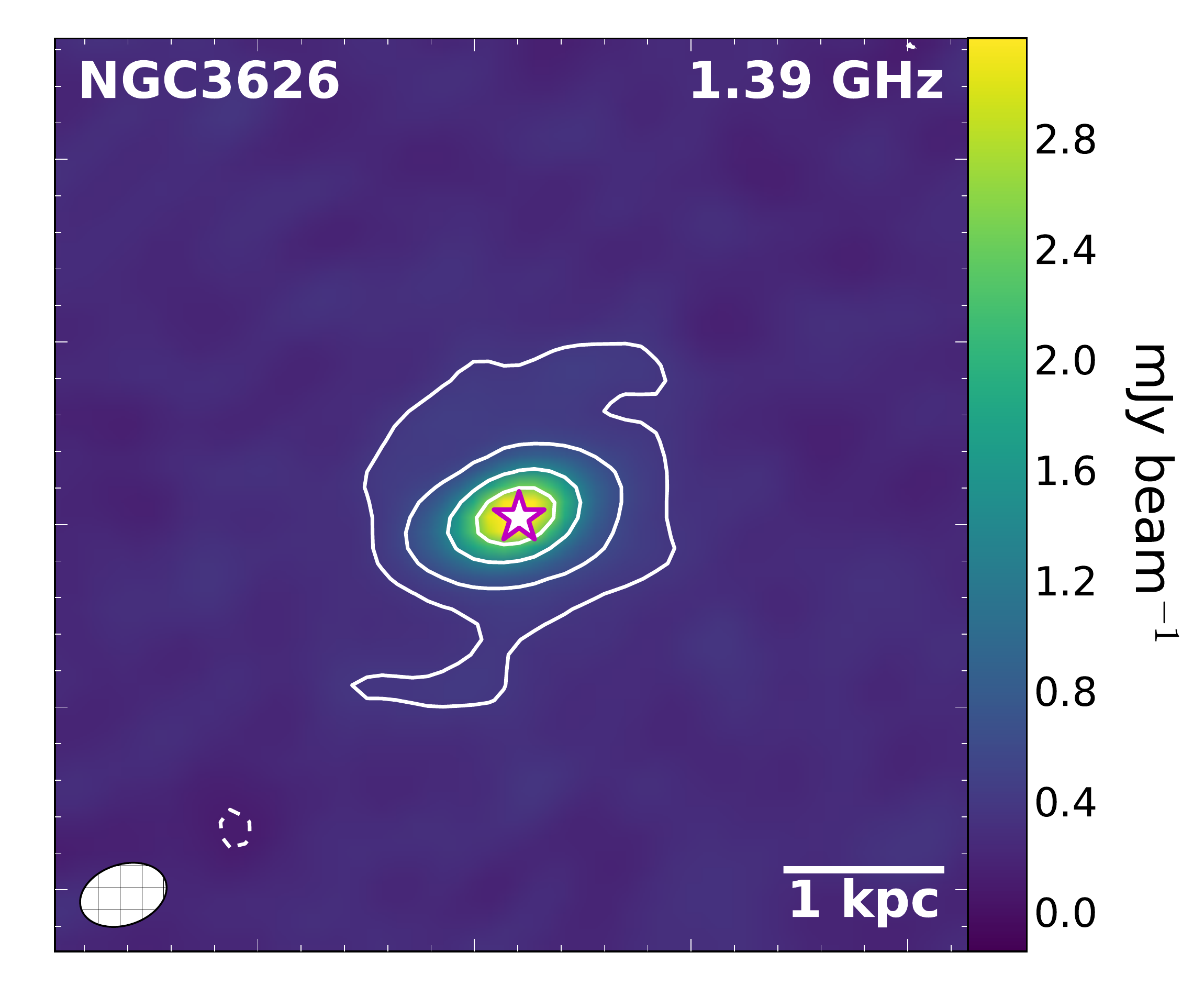}}
{\label{fig:sub:NGC3648}\includegraphics[clip=True, trim=0cm 0cm 0cm 0cm, scale=0.23]{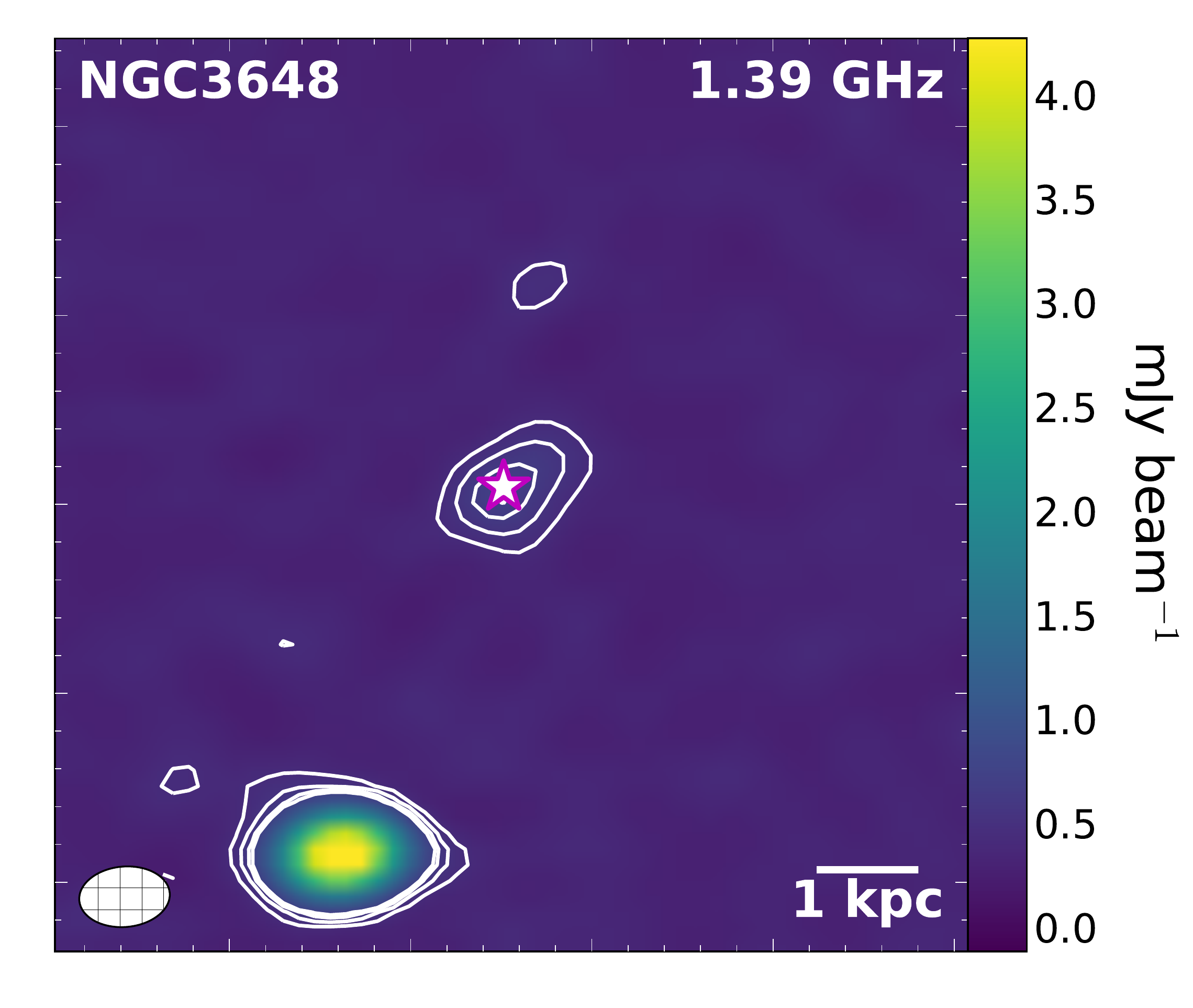}}
{\label{fig:sub:NGC3665}\includegraphics[clip=True, trim=0cm 0cm 0cm 0cm, scale=0.23]{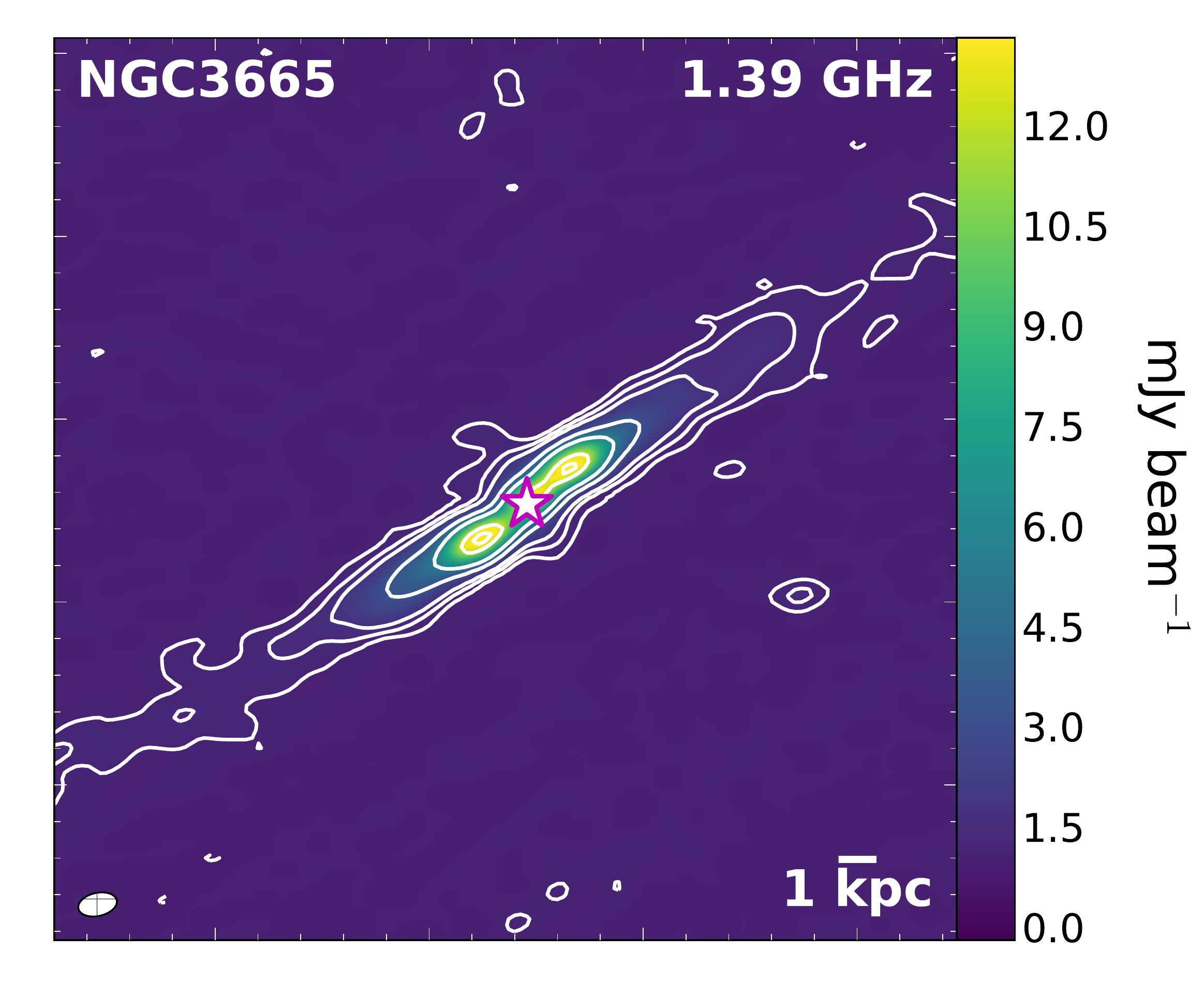}}
{\label{fig:sub:NGC3945}\includegraphics[clip=True, trim=0cm 0cm 0cm 0cm, scale=0.23]{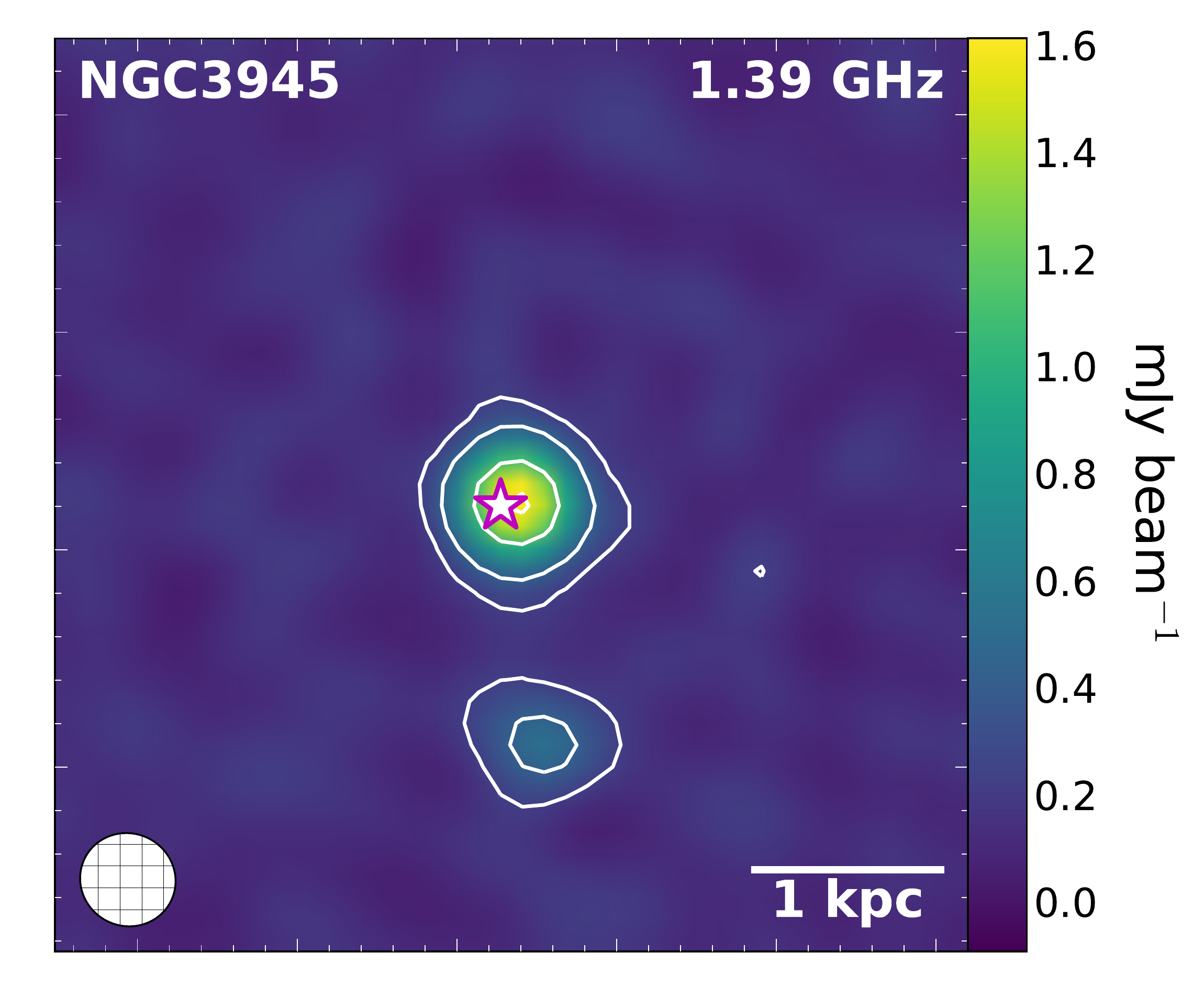}}
{\label{fig:sub:NGC4036}\includegraphics[clip=True, trim=0cm 0cm 0cm 0cm, scale=0.23]{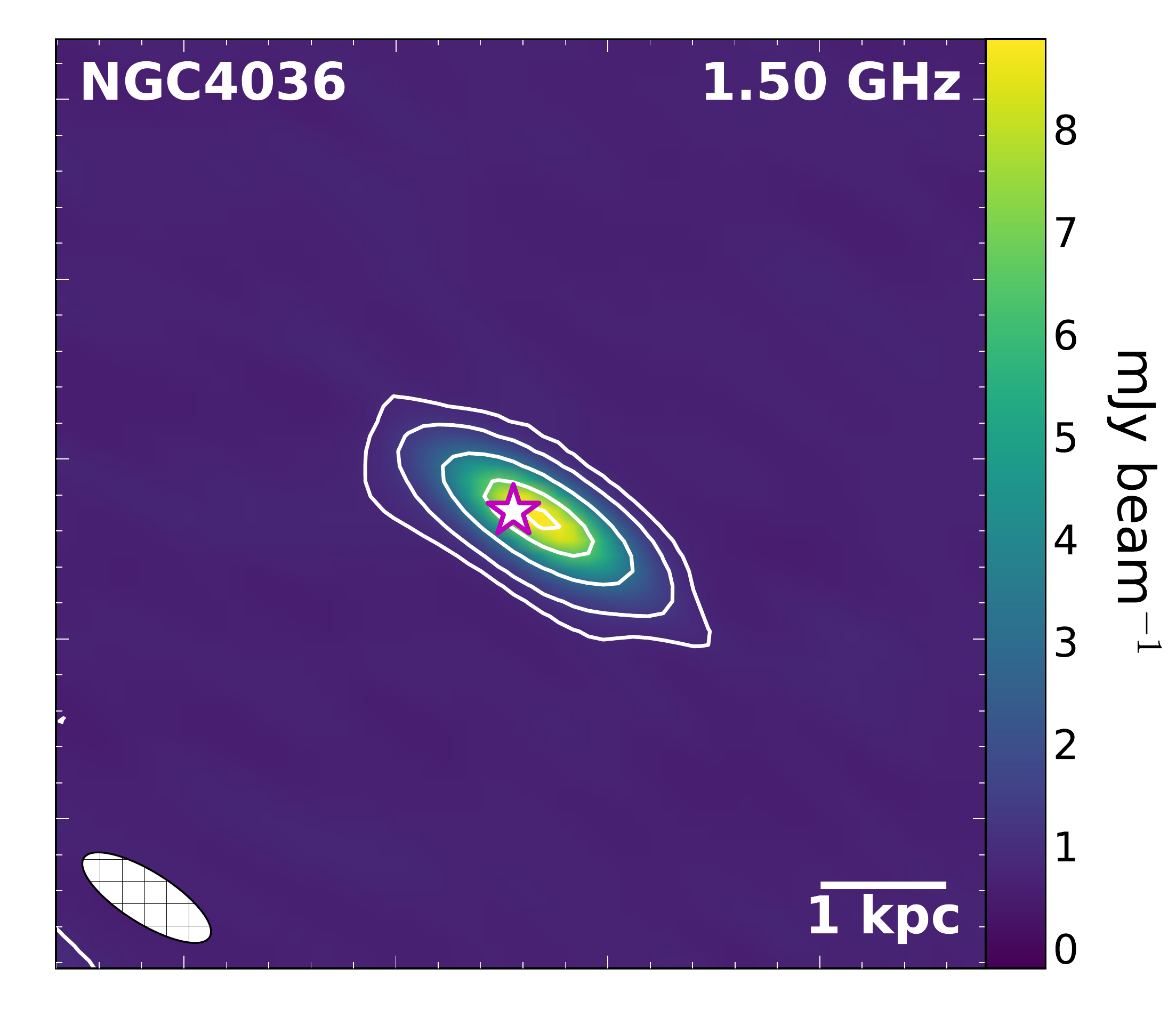}}
{\label{fig:sub:NGC4111}\includegraphics[clip=True, trim=0cm 0cm 0cm 0cm, scale=0.23]{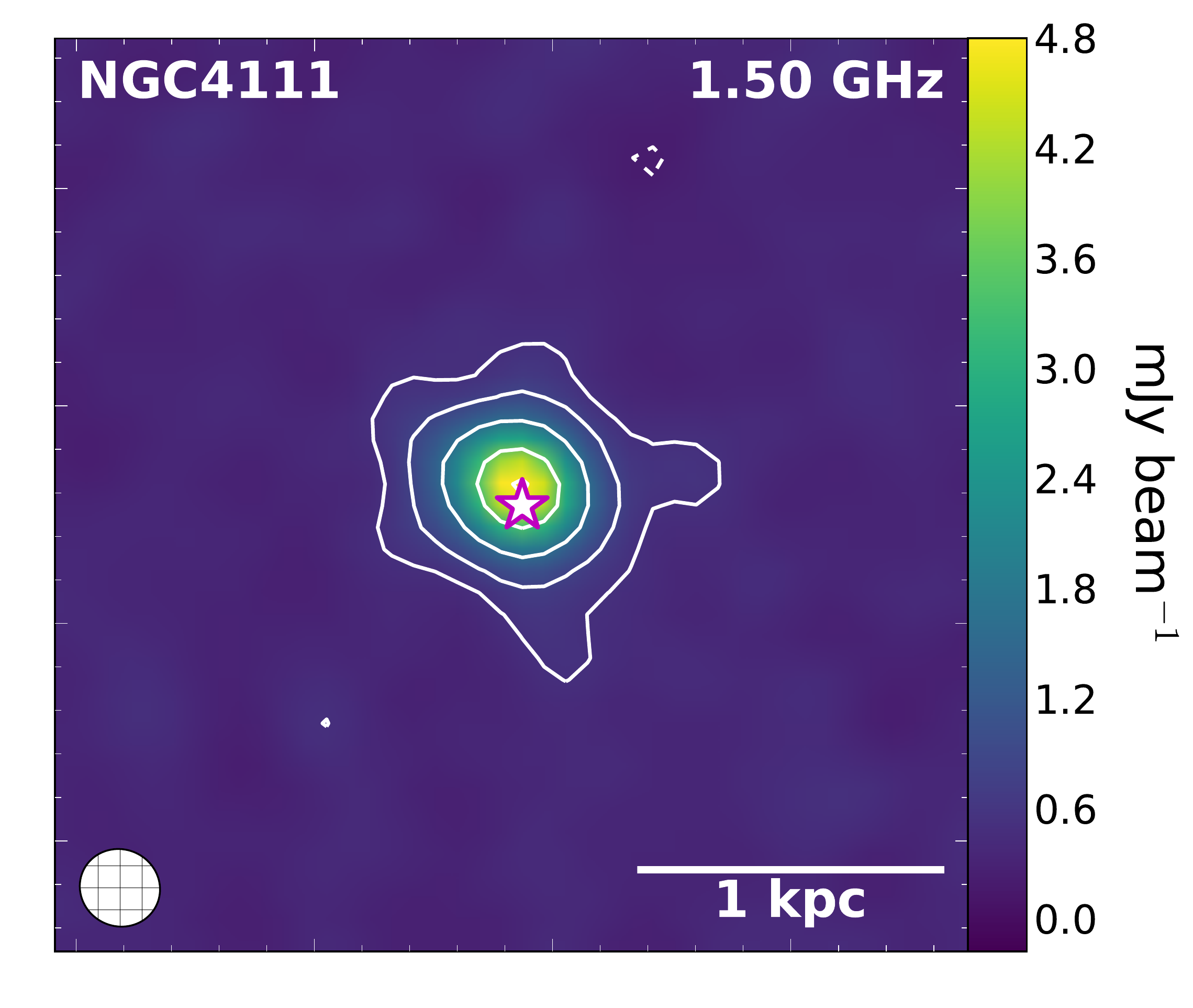}}
{\label{fig:sub:NGC4150}\includegraphics[clip=True, trim=0cm 0cm 0cm 0cm, scale=0.23]{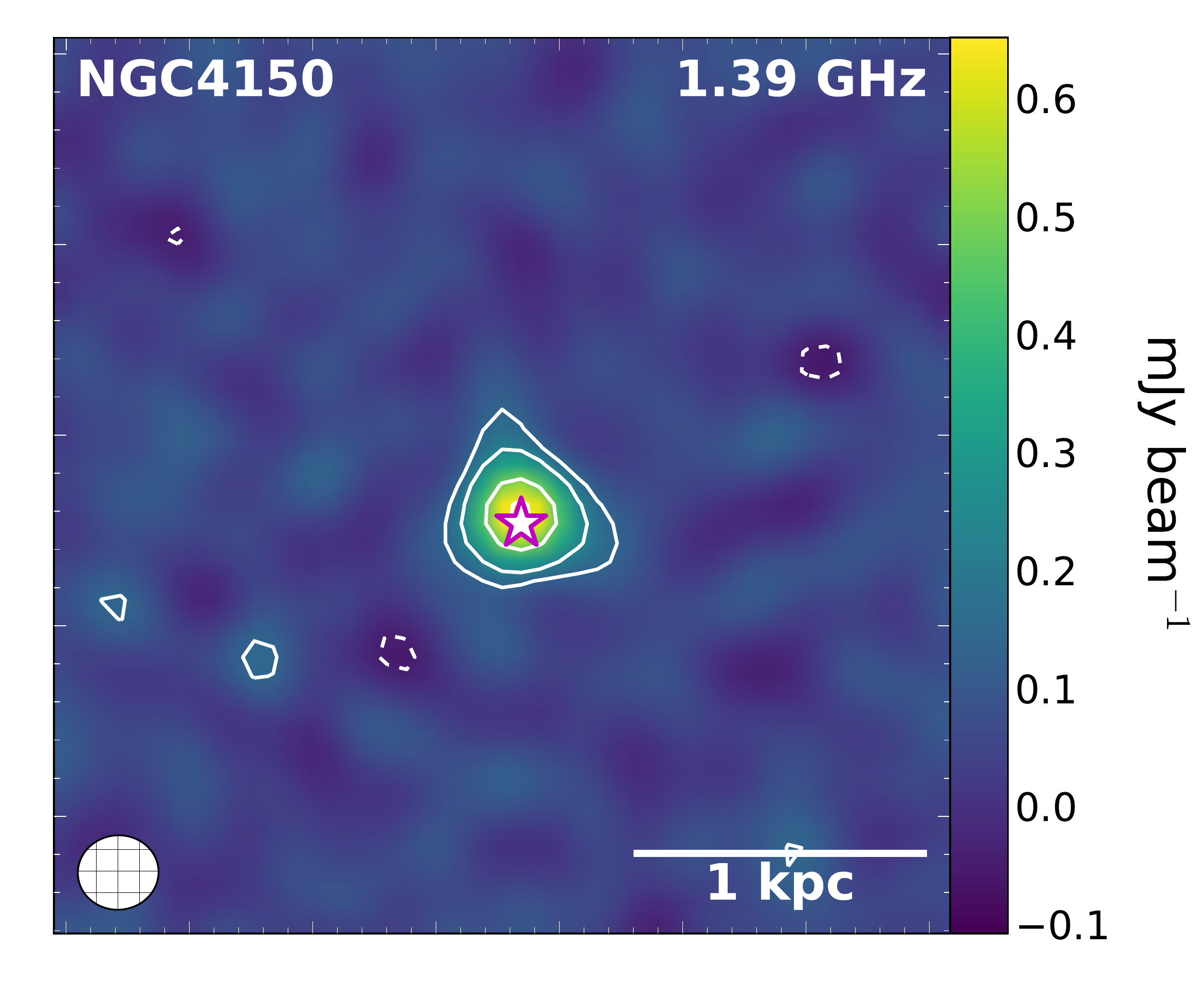}}
{\label{fig:sub:NGC4203}\includegraphics[clip=True, trim=0cm 0cm 0cm 0cm, scale=0.23]{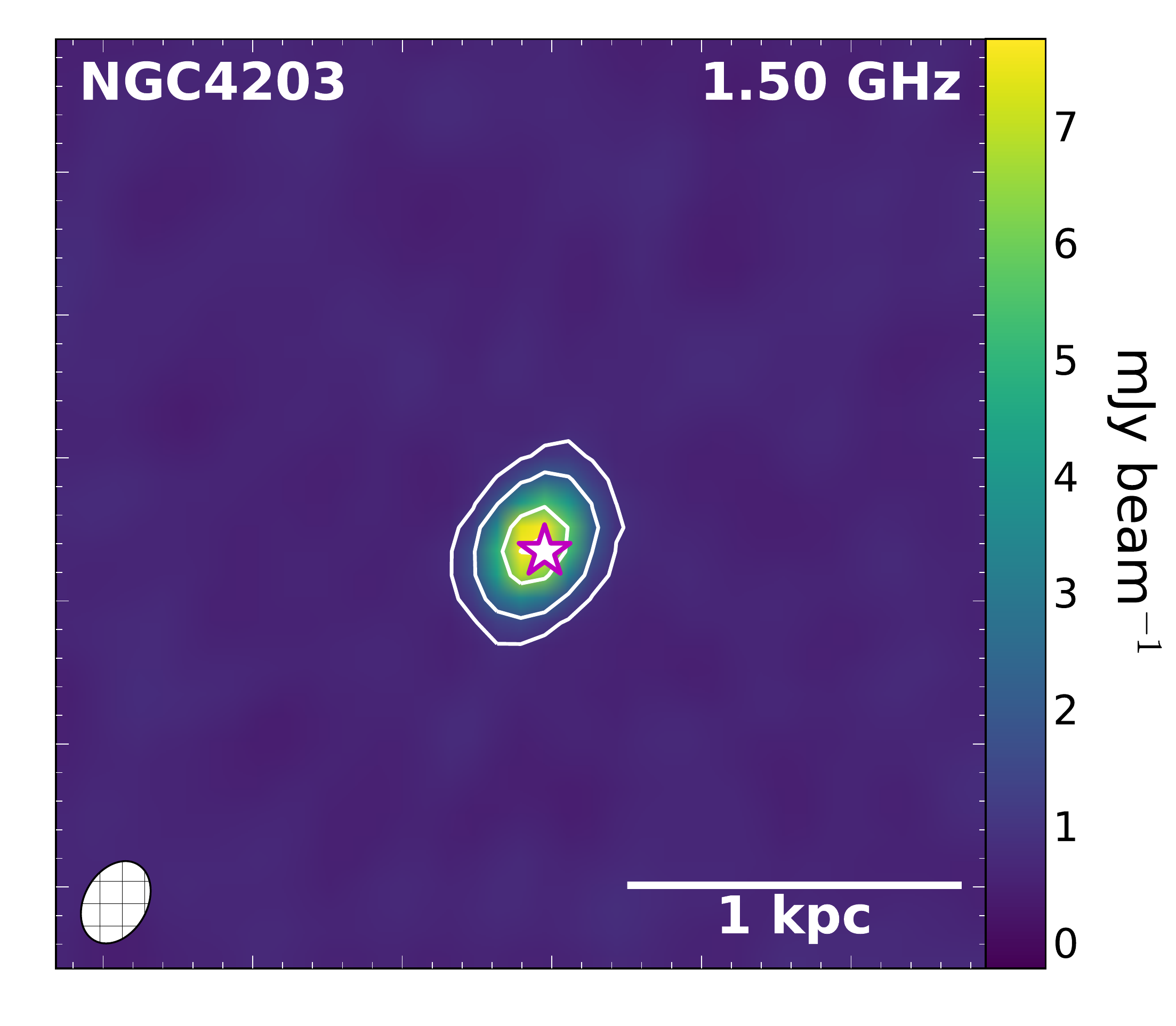}}
{\label{fig:sub:NGC4429}\includegraphics[clip=True, trim=0cm 0cm 0cm 0cm, scale=0.23]{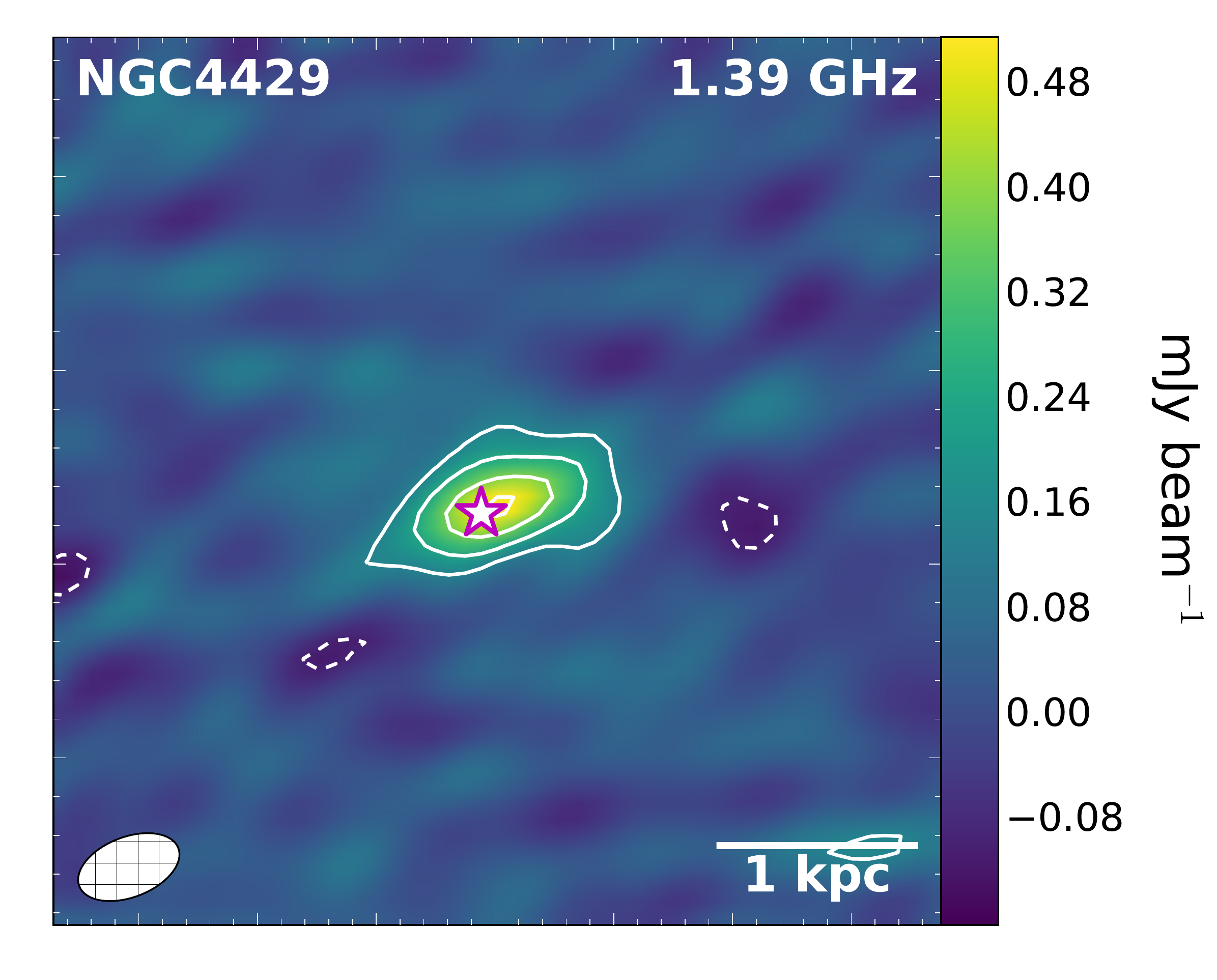}}
\end{figure*}
 
\begin{figure*}
{\label{fig:sub:NGC4459}\includegraphics[clip=True, trim=0cm 0cm 0cm 0cm, scale=0.23]{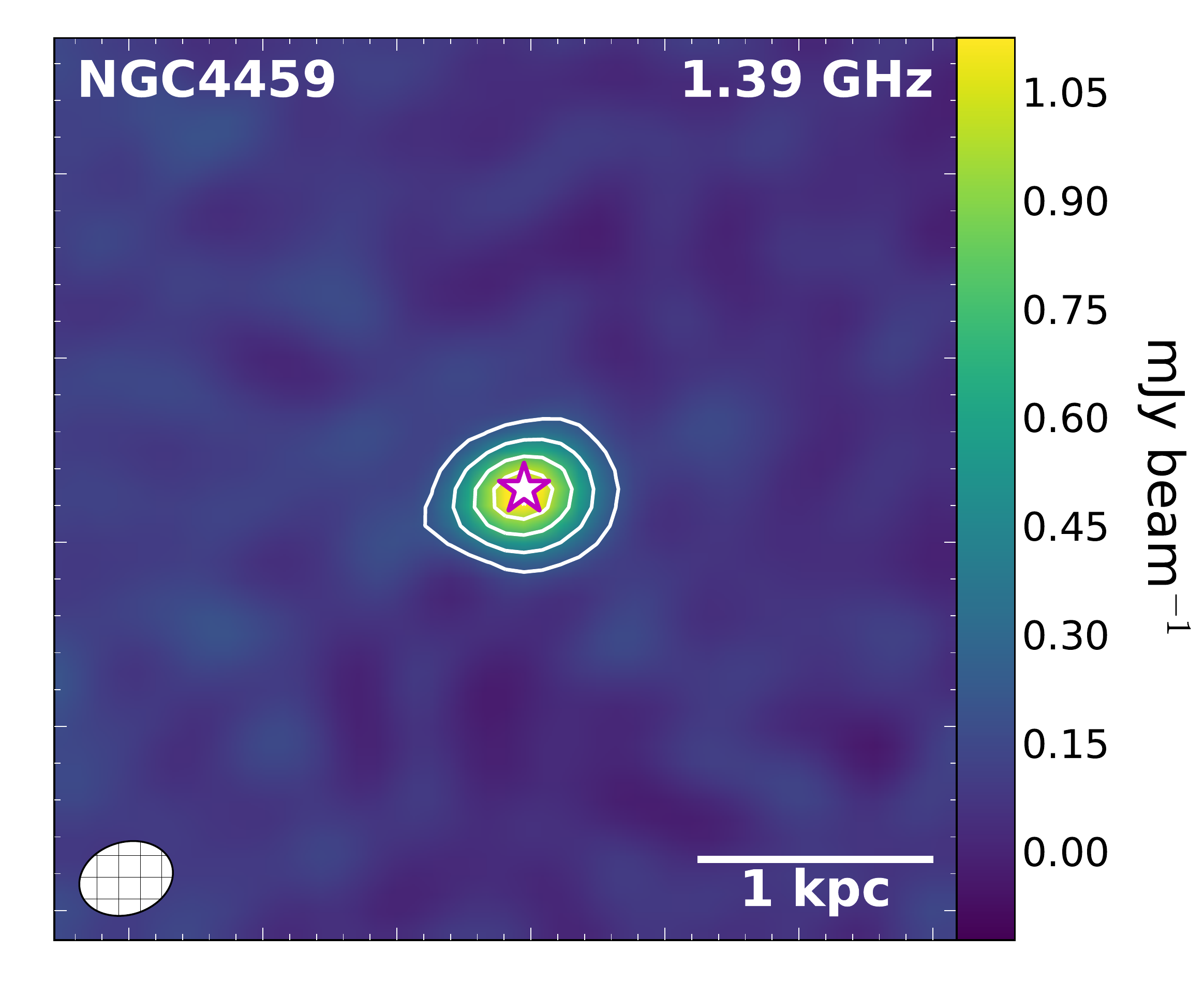}}
{\label{fig:sub:NGC4526}\includegraphics[clip=True, trim=0cm 0cm 0cm 0cm, scale=0.23]{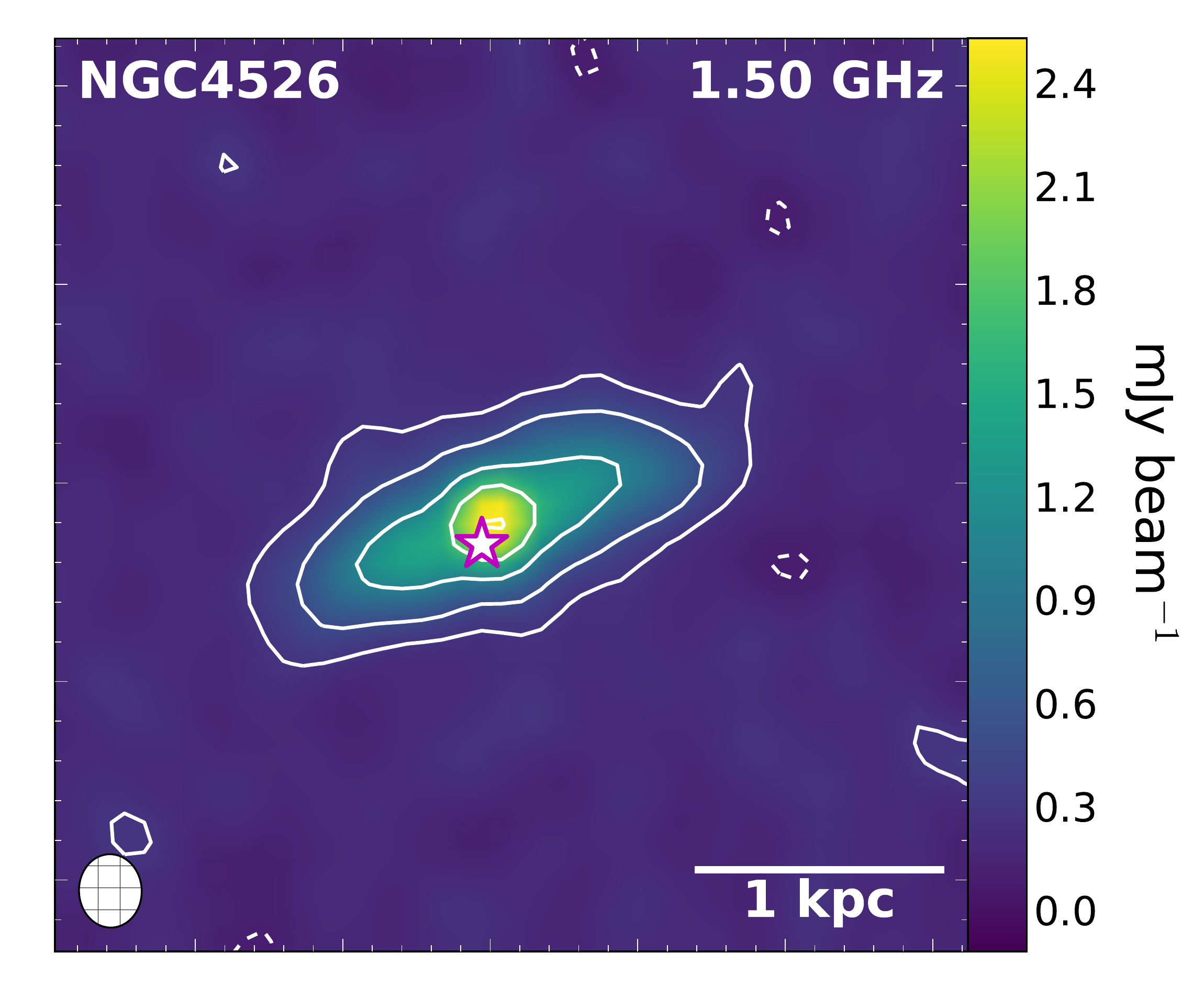}}
{\label{fig:sub:NGC4643}\includegraphics[clip=True, trim=0cm 0cm 0cm 0cm, scale=0.23]{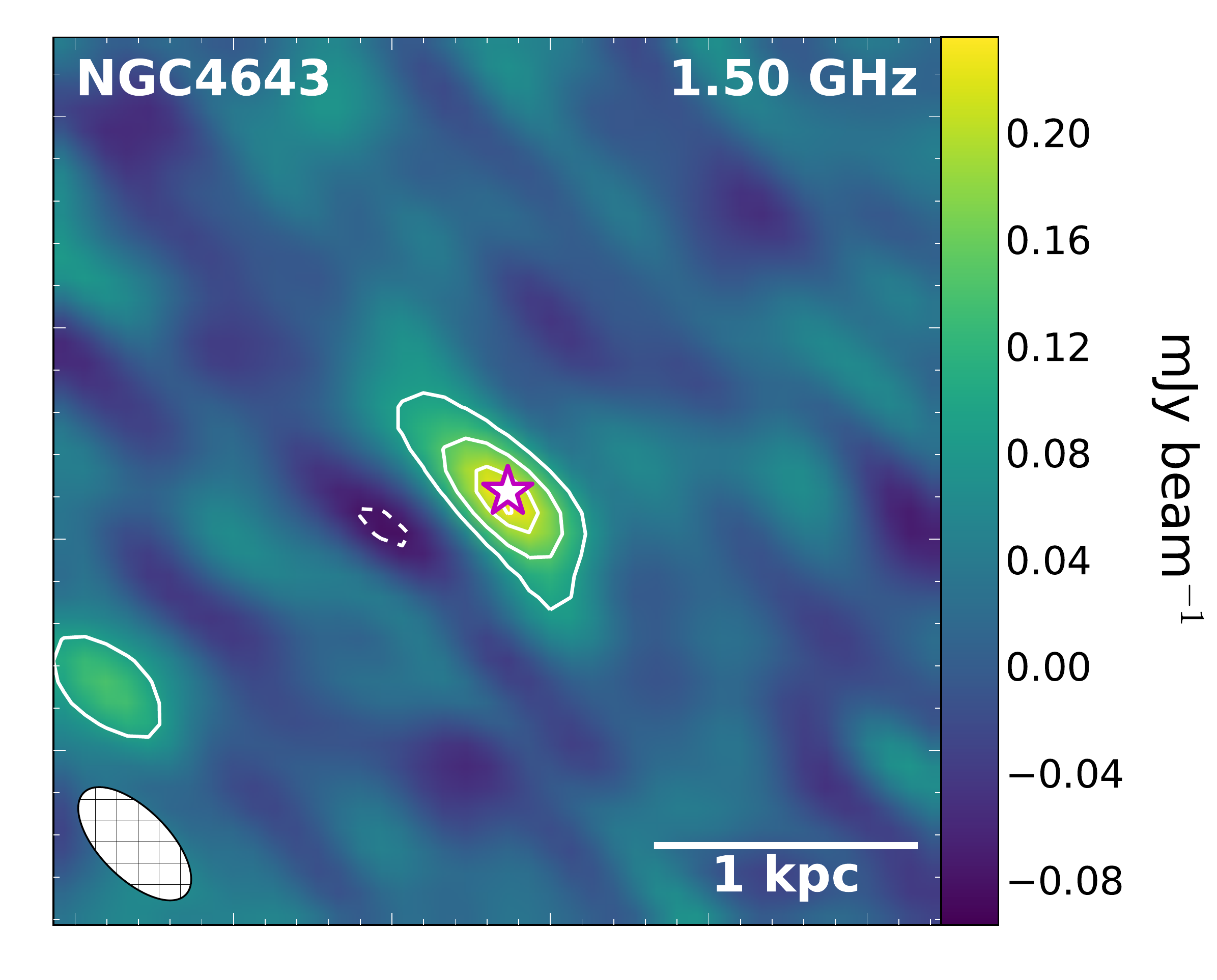}}
{\label{fig:sub:NGC4684}\includegraphics[clip=True, trim=0cm 0cm 0cm 0cm, scale=0.23]{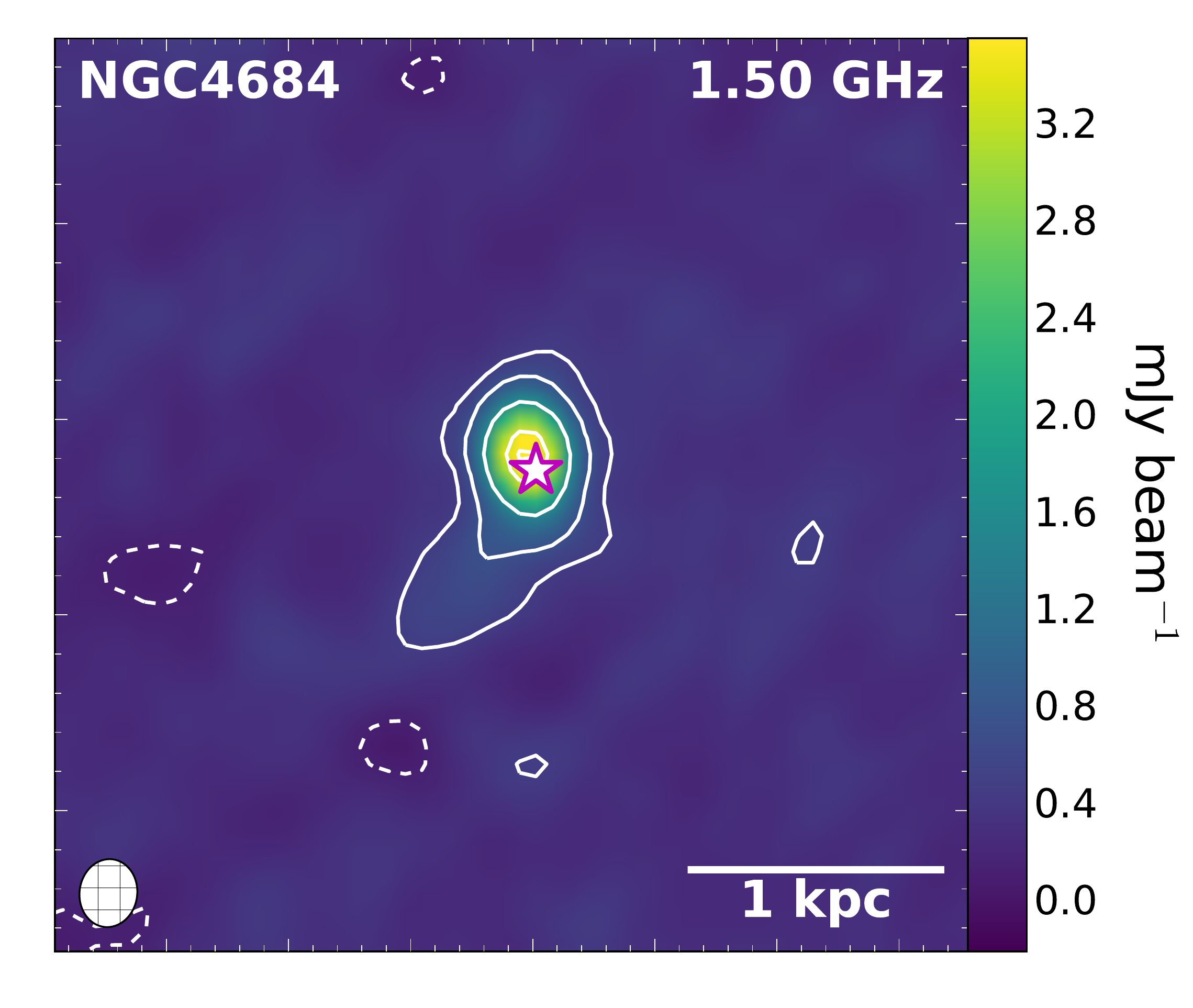}}
{\label{fig:sub:NGC4694}\includegraphics[clip=True, trim=0cm 0cm 0cm 0cm, scale=0.23]{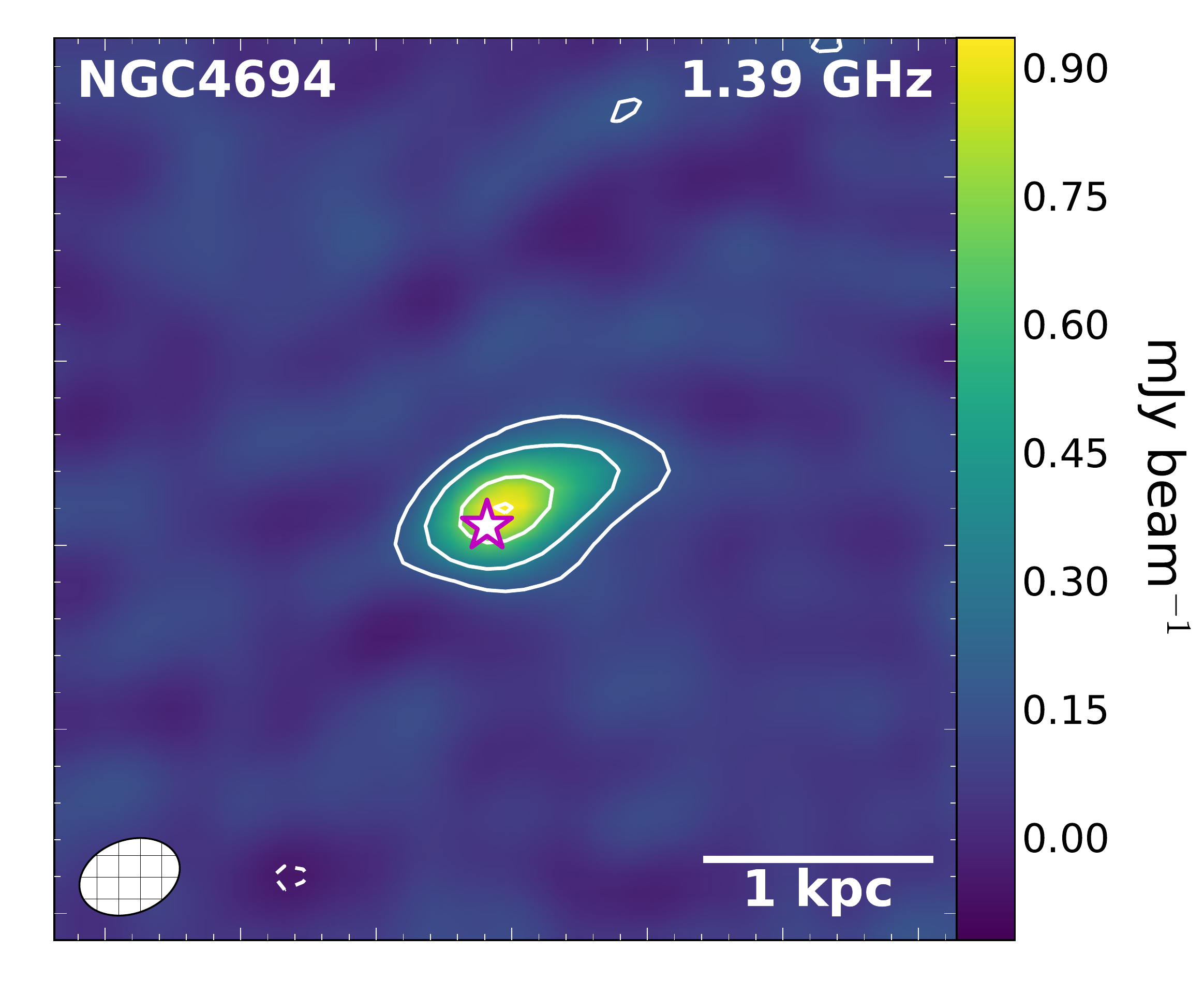}}
{\label{fig:sub:NGC4710}\includegraphics[clip=True, trim=0cm 0cm 0cm 0cm, scale=0.23]{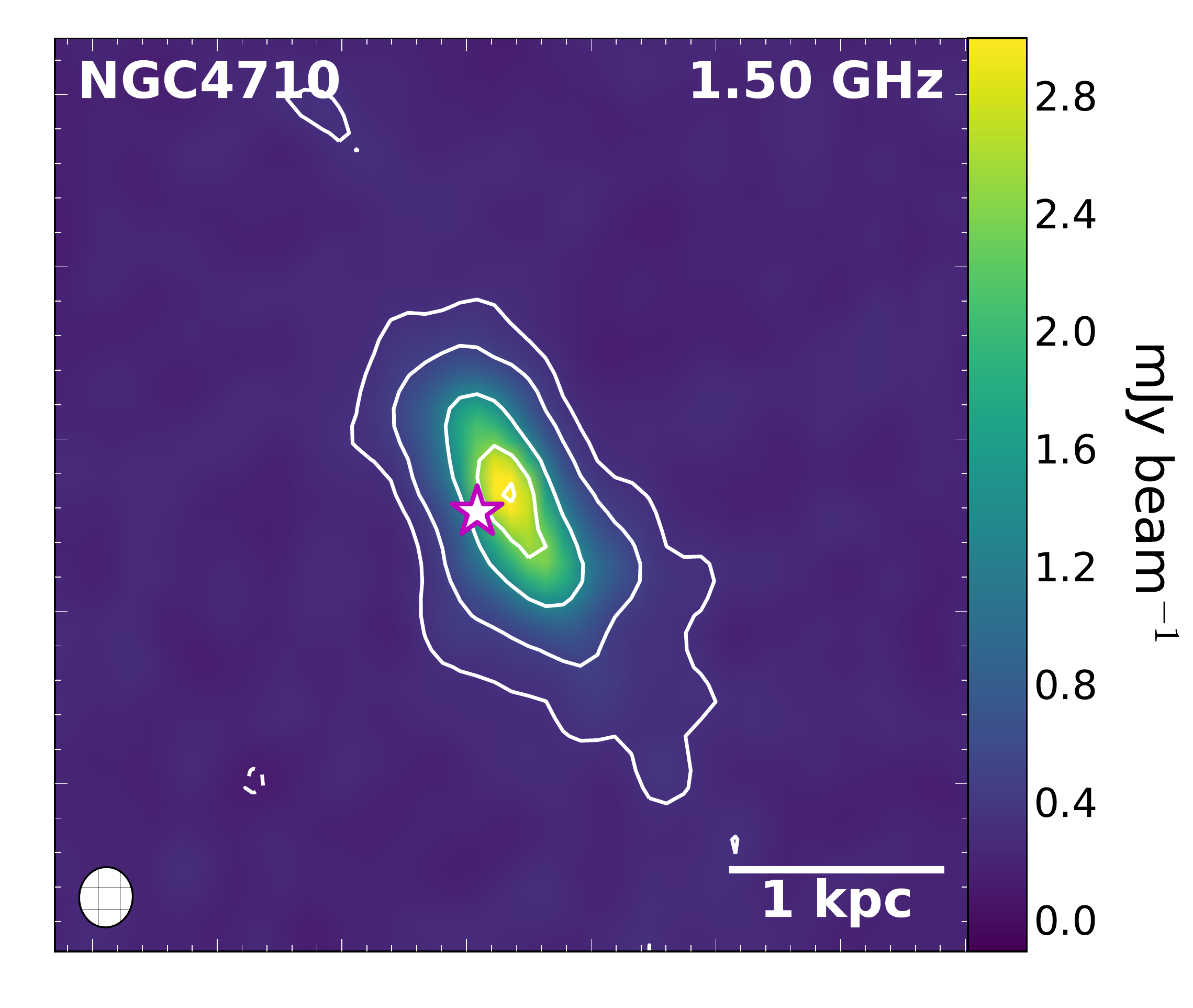}}
{\label{fig:sub:NGC4753}\includegraphics[clip=True, trim=0cm 0cm 0cm 0cm, scale=0.23]{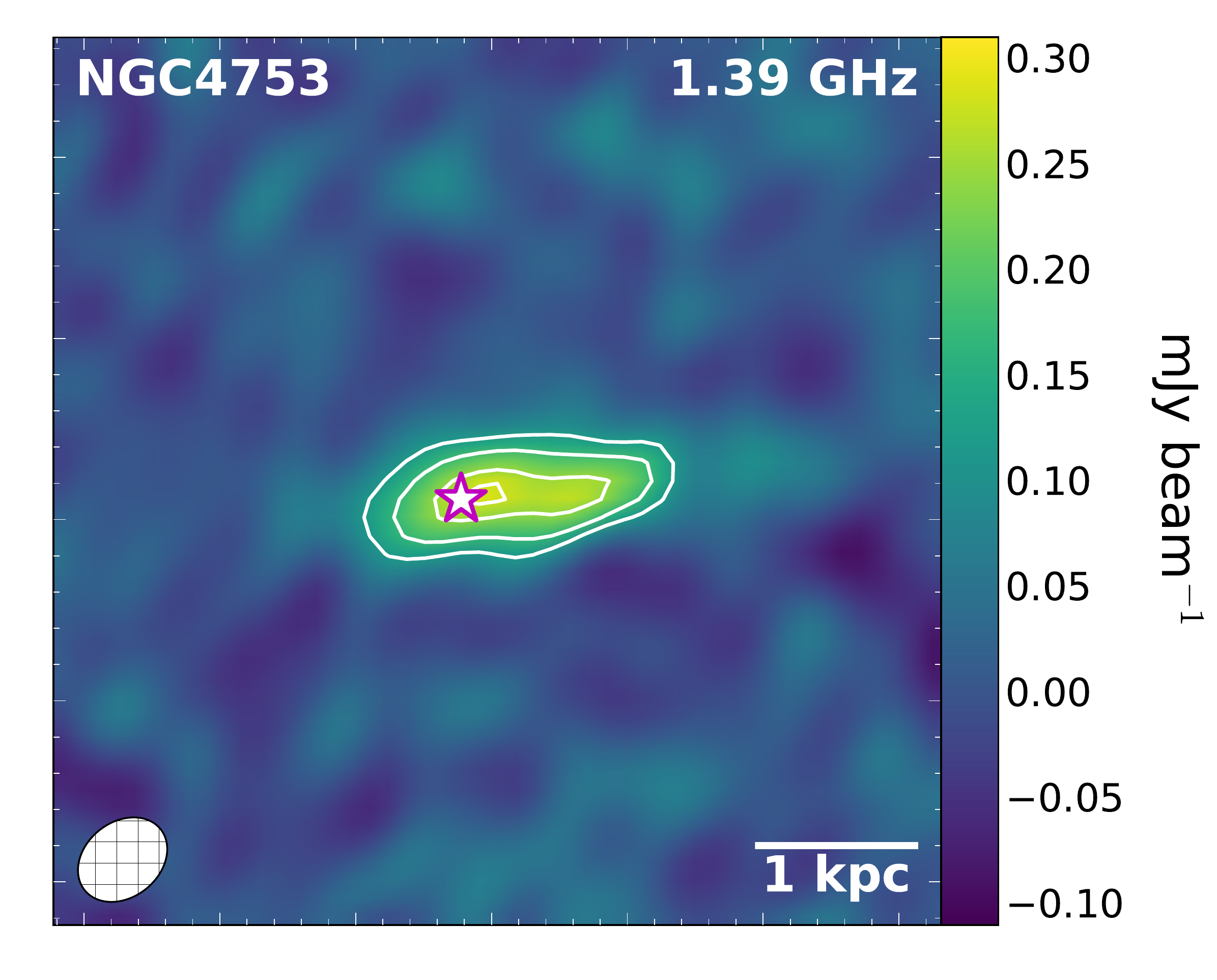}}
{\label{fig:sub:NGC5173}\includegraphics[clip=True, trim=0cm 0cm 0cm 0cm, scale=0.23]{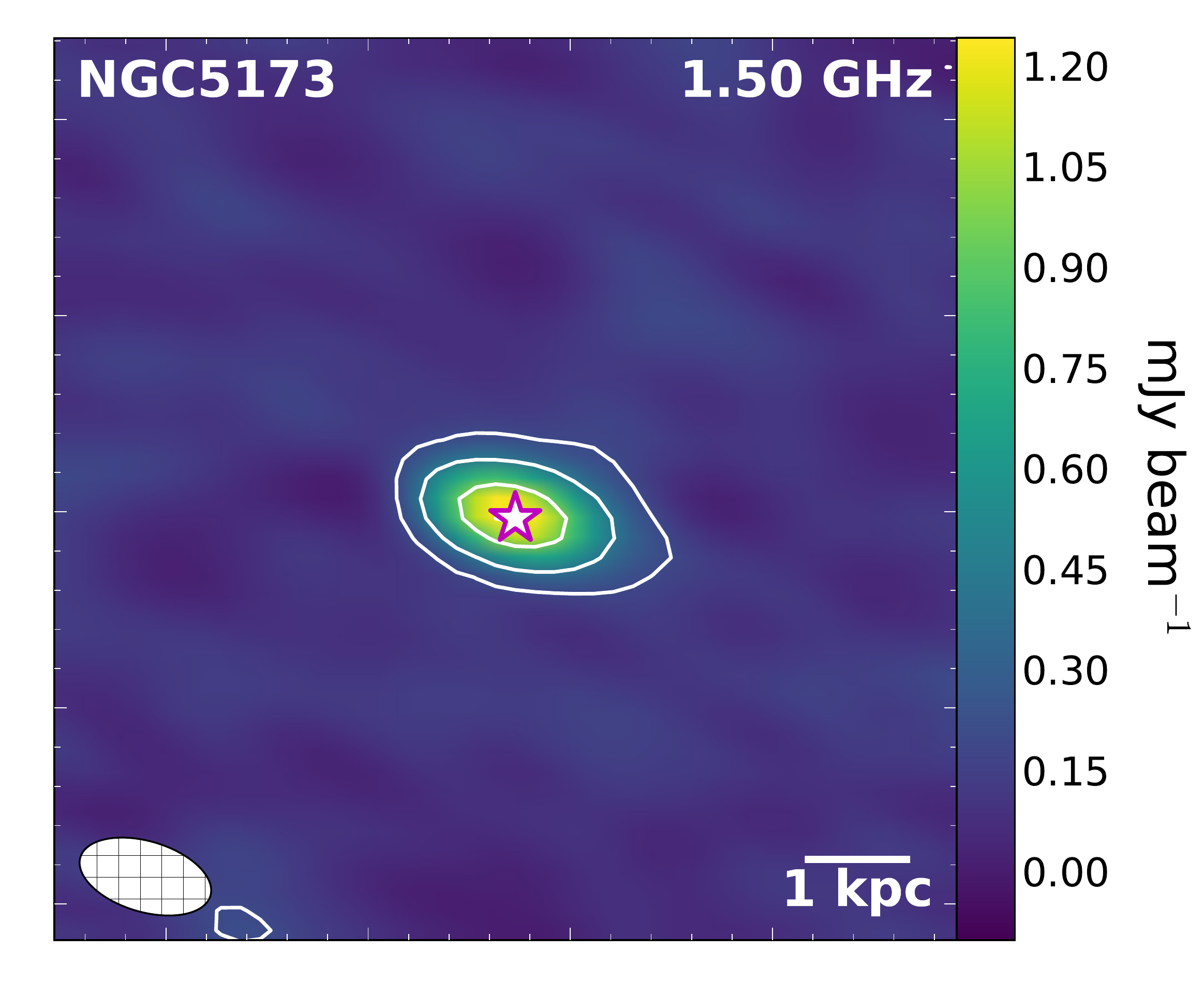}}
{\label{fig:sub:NGC5379}\includegraphics[clip=True, trim=0cm 0cm 0cm 0cm, scale=0.23]{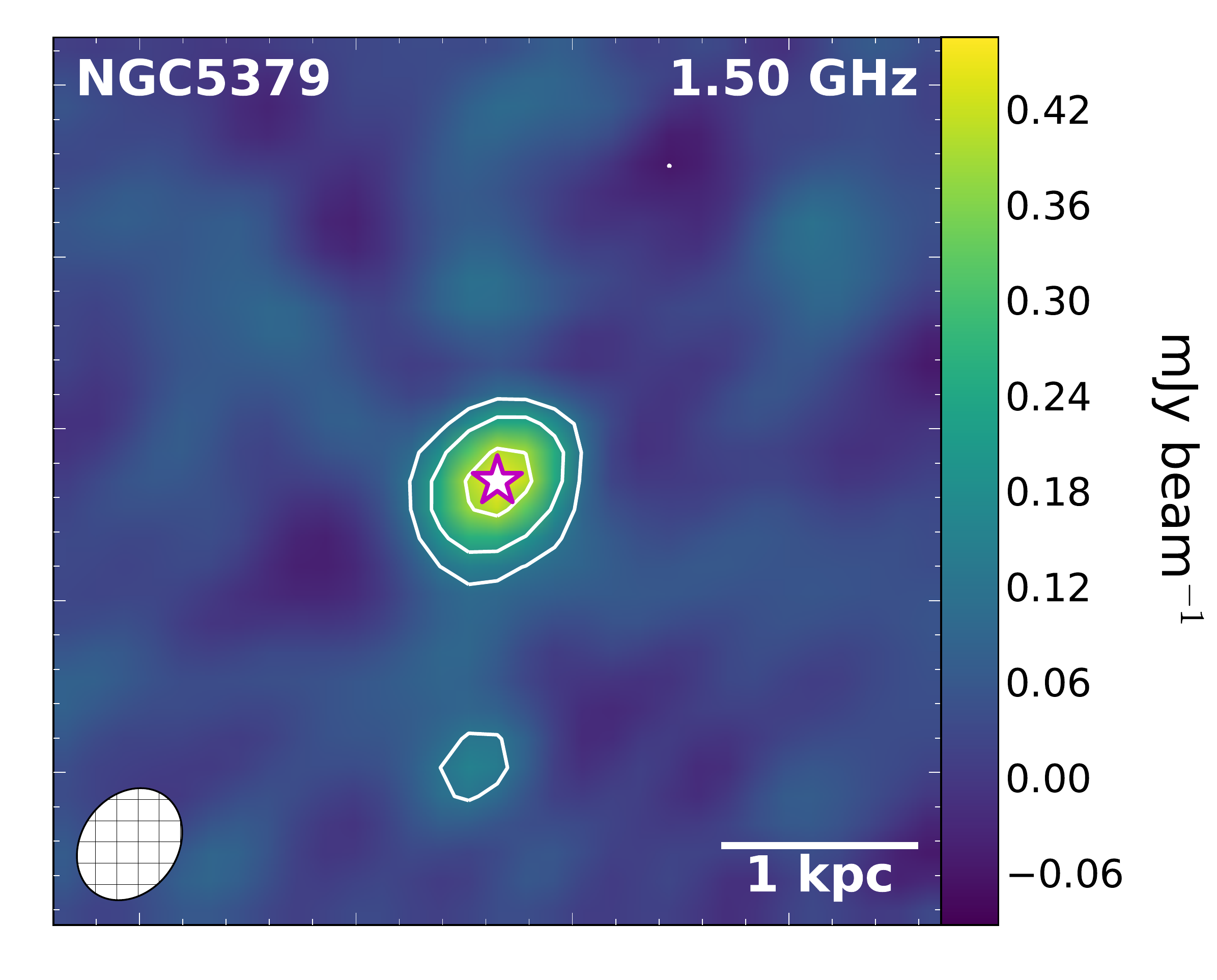}}
{\label{fig:sub:NGC5866}\includegraphics[clip=True, trim=0cm 0cm 0cm 0cm, scale=0.23]{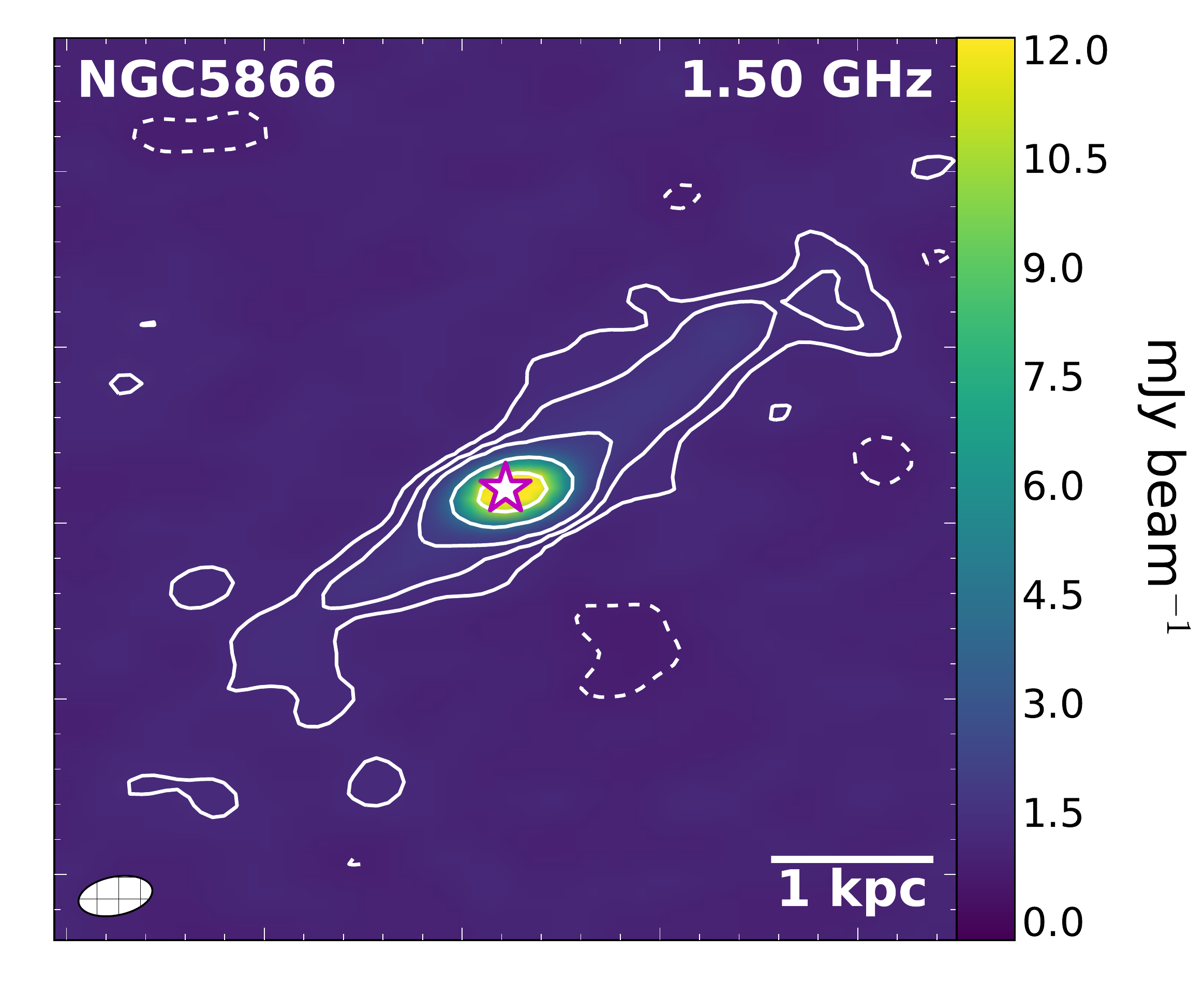}}
{\label{fig:sub:NGC6014}\includegraphics[clip=True, trim=0cm 0cm 0cm 0cm, scale=0.23]{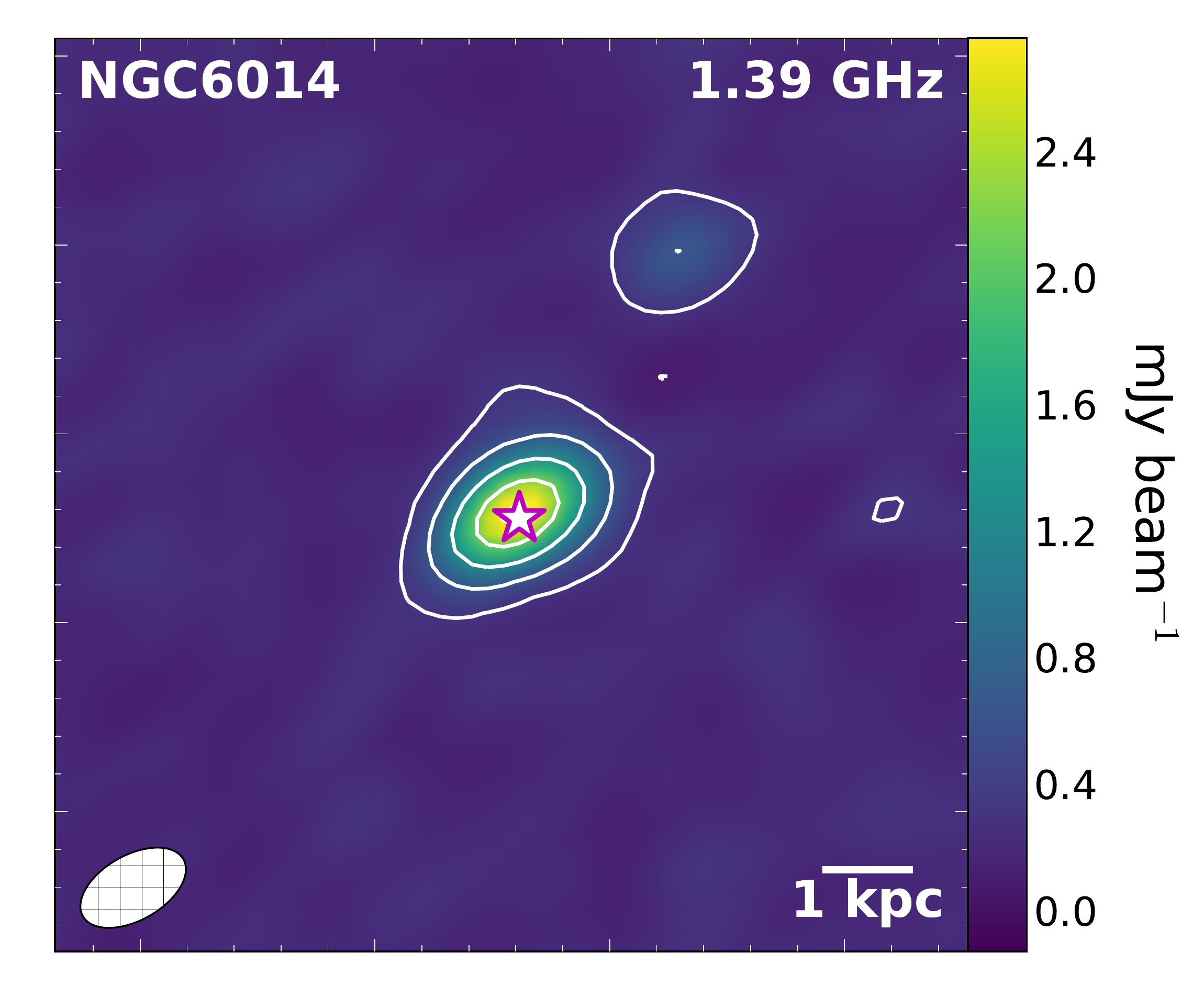}}
{\label{fig:sub:NGC6547}\includegraphics[clip=True, trim=0cm 0cm 0cm 0cm, scale=0.23]{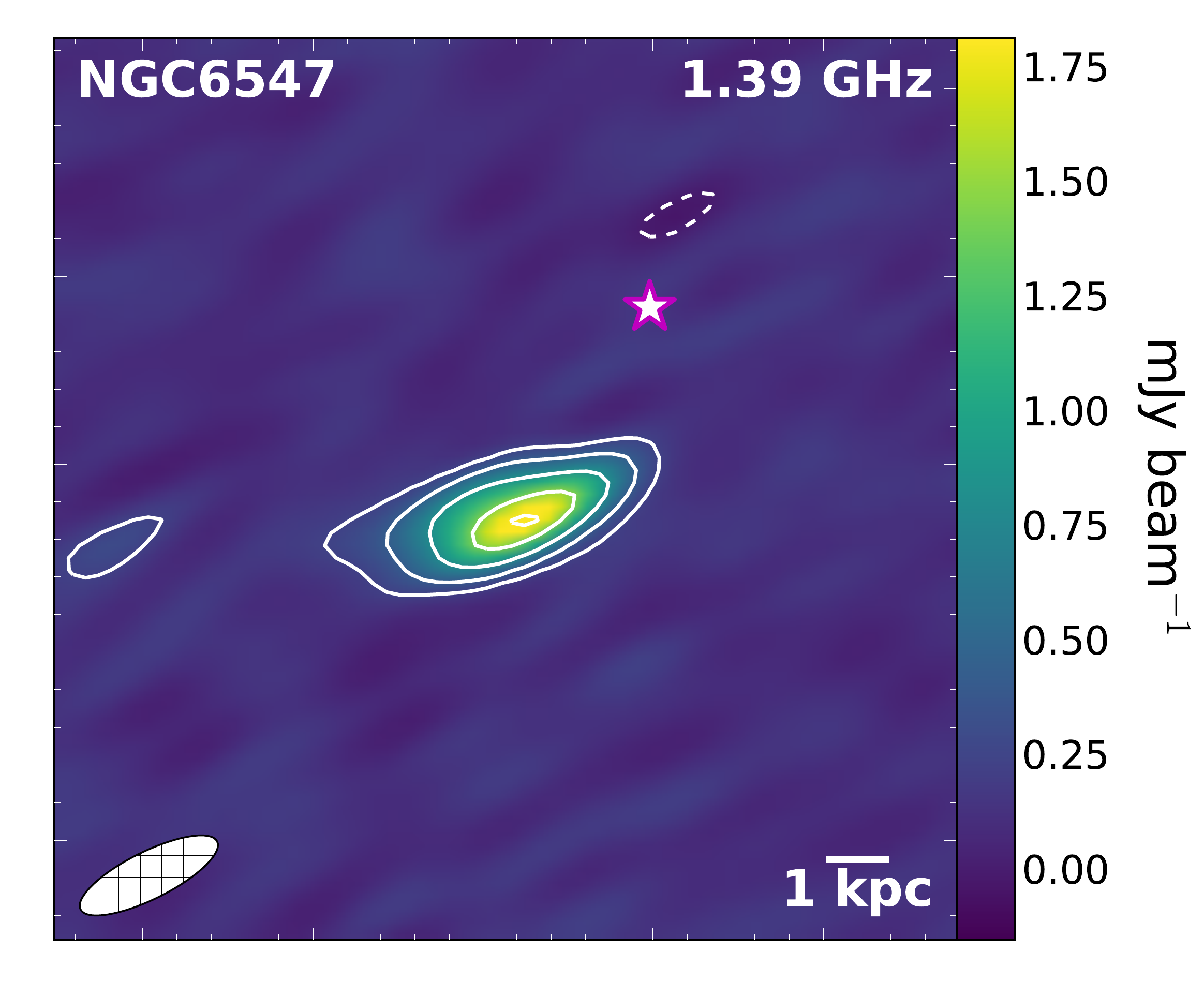}}
{\label{fig:sub:NGC6798}\includegraphics[clip=True, trim=0cm 0cm 0cm 0cm, scale=0.23]{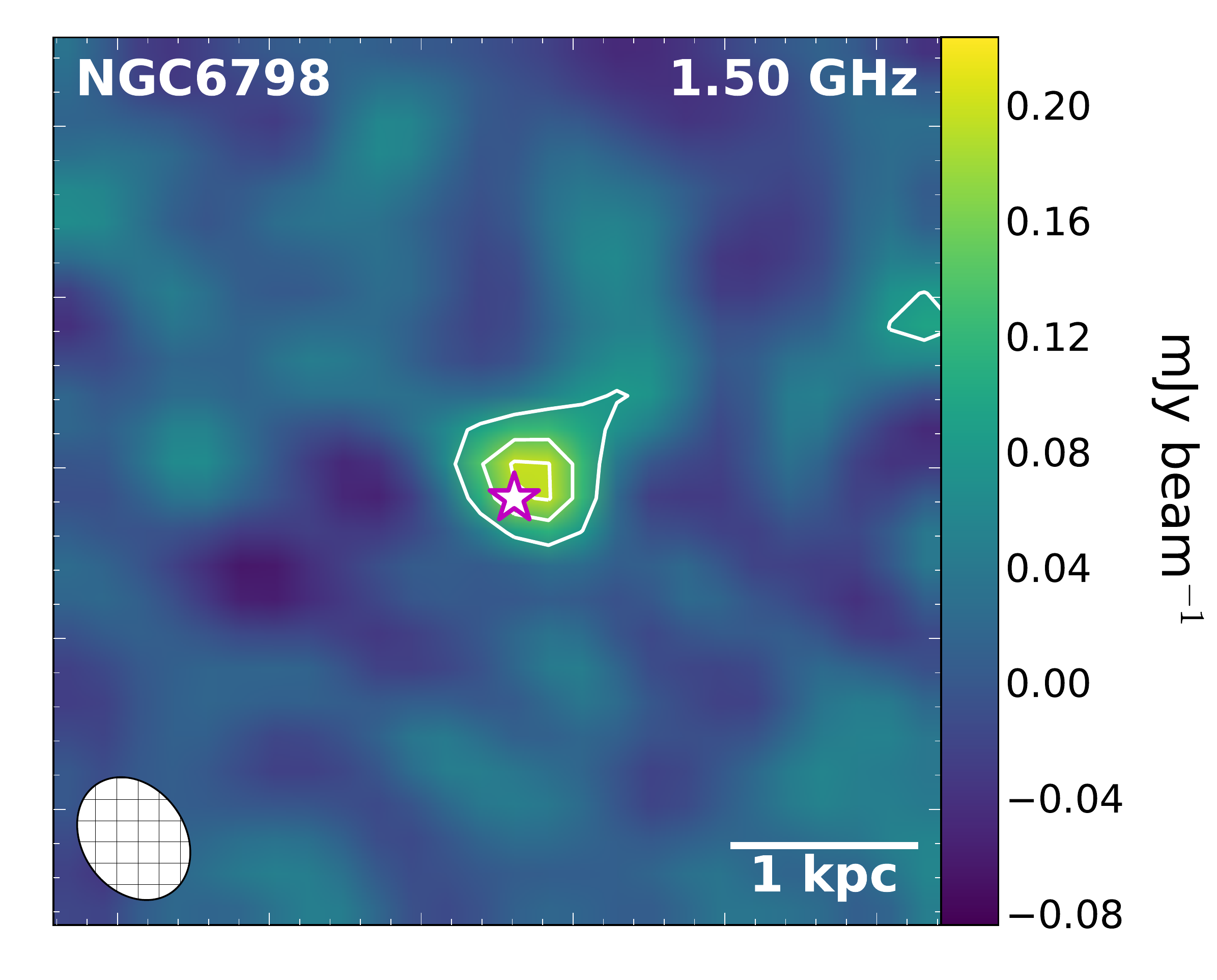}}
{\label{fig:sub:NGC7465}\includegraphics[clip=True, trim=0cm 0cm 0cm 0cm, scale=0.23]{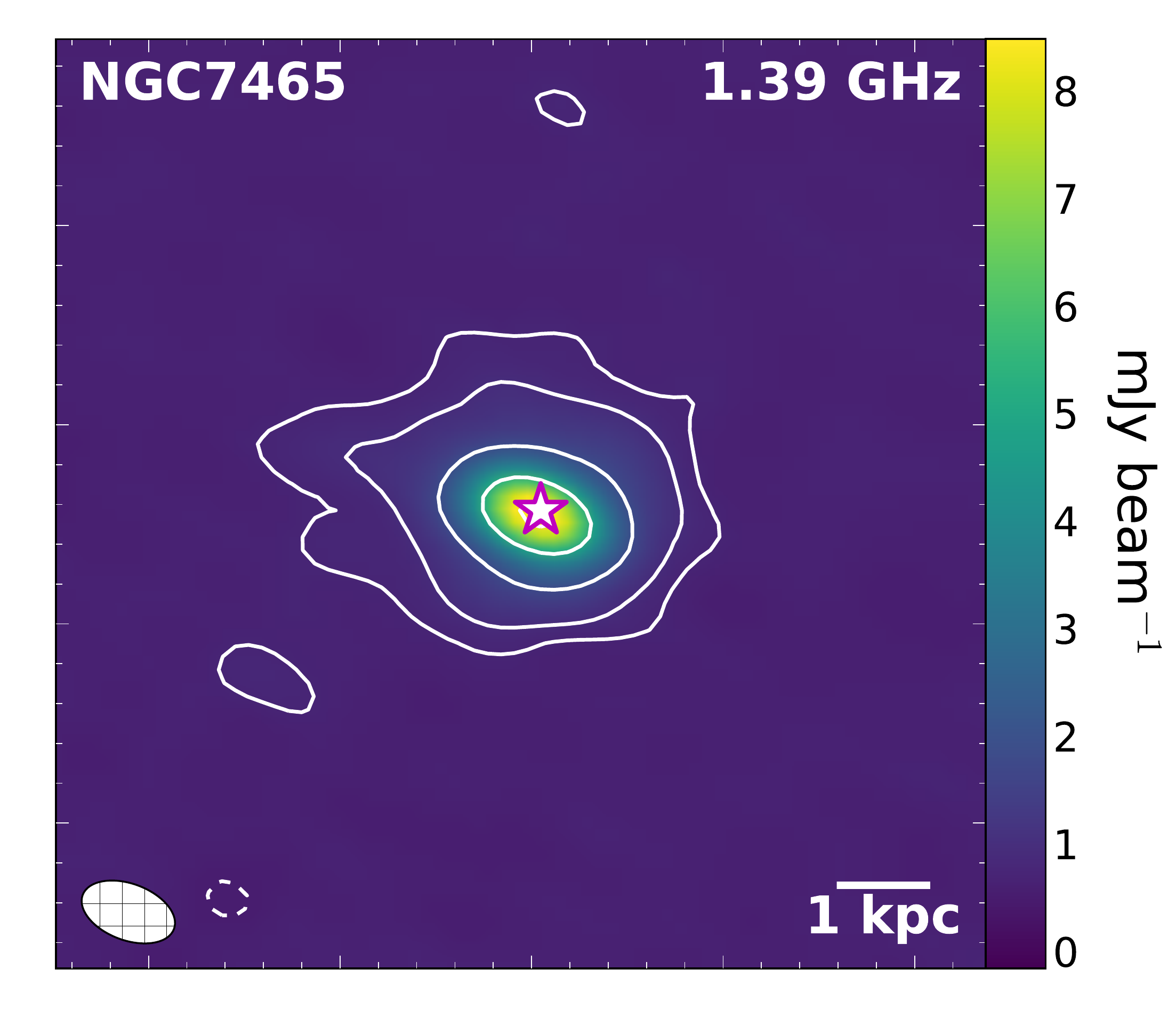}}
{\label{fig:sub:PGC029321}\includegraphics[clip=True, trim=0cm 0cm 0cm 0cm, scale=0.23]{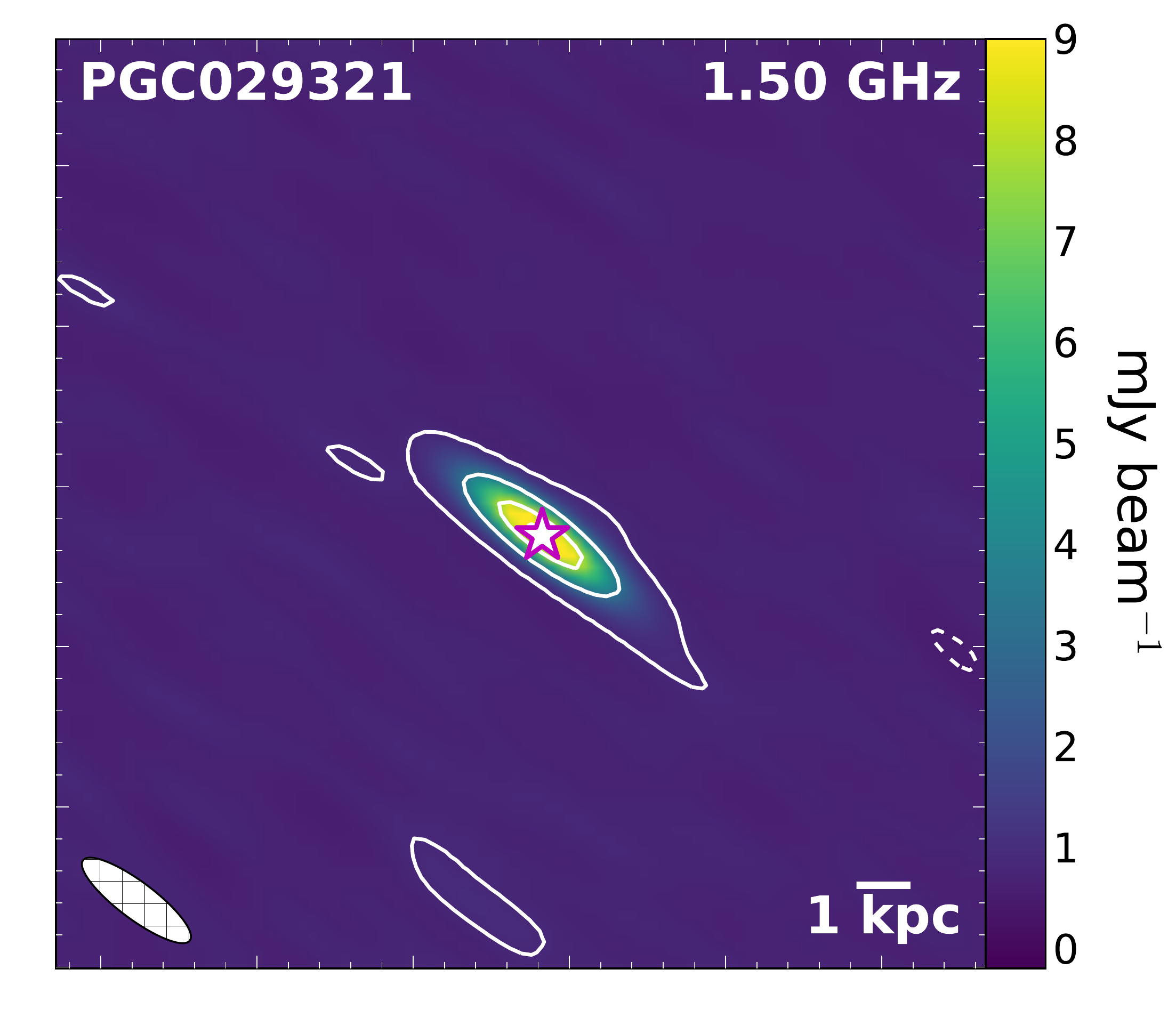}}
\end{figure*}
 
\begin{table*}
\begin{minipage}{8.6cm}
\caption{Relative Contour Levels in the 1.4~GHz Continuum Maps}
\label{tab:contours}
\begin{tabular*}{8.6cm}{lcr}
\hline
\hline
Galaxy      & rms & Relative Contours \\
                   & ($\mu$Jy beam$^{-1}$) &  \\
\hline 
IC0676     &     43 &                        [-3, 3, 9, 25, 50, 83]\\ 
IC0719     &     28 &                            [-3, 3, 4.5, 6, 7]\\ 
IC1024     &     67 &                        [-3, 3, 9, 24, 43, 53]\\ 
NGC0524    &     29 &                            [-3, 3, 8, 22, 44]\\ 
NGC0680    &     27 &                            [-3, 3, 9, 24, 35]\\ 
NGC1023    &     36 &                            [-3, 3, 4, 5, 5.5]\\ 
NGC1222    &     70 &                  [-3, 3, 9, 40, 90, 198, 258]\\ 
NGC1266    &     74 &                [-3, 3, 12, 60, 200, 500, 740]\\ 
NGC2685    &     29 &                           [-3, 3, 5.5, 9, 11]\\ 
NGC2764    &     40 &                       [-3, 3, 9, 18, 40, 100]\\ 
NGC2768    &     42 &                     [-3, 3, 15, 72, 192, 302]\\ 
NGC2824    &     42 &                     [-3, 3, 12, 50, 120, 164]\\ 
NGC2852    &     35 &                            [-3, 3, 8, 15, 19]\\ 
NGC3032    &     39 &                        [-3, 3, 6, 11, 17, 20]\\ 
NGC3182    &     30 &                           [-3, 3, 4, 5, 6, 7]\\ 
NGC3193    &     30 &                              [-3, 3, 5, 7, 8]\\ 
NGC3245    &     33 &                     [-3, 3, 14, 58, 140, 185]\\ 
NGC3489    &     35 &                            [-3, 3, 6, 10, 13]\\ 
NGC3607    &     28 &                      [-3, 3, 9, 36, 100, 150]\\ 
NGC3608    &     27 &                             [-3, 3, 6, 9, 11]\\ 
NGC3619    &     36 &                        [-3, 3, 6, 12, 22, 30]\\ 
NGC3626    &     40 &                        [-3, 3, 8, 25, 50, 76]\\ 
NGC3648    &     30 &                             [-3, 3, 6, 9, 11]\\ 
NGC3665    &     40 &             [-3, 3, 8, 16, 50, 100, 250, 316]\\ 
NGC3945    &     34 &                            [-3, 3, 9, 28, 45]\\ 
NGC4036    &     50 &                     [-3, 3, 12, 48, 115, 170]\\ 
NGC4111    &     48 &                        [-3, 3, 9, 25, 60, 96]\\ 
NGC4150    &     29 &                            [-3, 3, 6, 14, 21]\\ 
NGC4203    &     78 &                           [-3, 3, 15, 60, 96]\\ 
NGC4429    &     40 &                         [-3, 3, 5.5, 9, 12.5]\\ 
NGC4459    &     40 &                        [-3, 3, 8, 15, 22, 27]\\ 
NGC4526    &     30 &                       [-3, 3, 10, 28, 50, 82]\\ 
NGC4643    &     29 &                              [-3, 3, 5, 7, 8]\\ 
NGC4684    &     45 &                        [-3, 3, 8, 24, 60, 77]\\ 
NGC4694    &     35 &                            [-3, 3, 8, 18, 26]\\ 
NGC4710    &     28 &                       [-3, 3, 9, 36, 72, 103]\\ 
NGC4753    &     40 &                        [-3, 3, 4.5, 6.5, 7.5]\\ 
NGC5173    &     32 &                           [-3, 3, 10, 25, 37]\\ 
NGC5273    &     20 &                           [-3, 3, 5, 8, 9.75]\\ 
NGC5379    &     30 &                            [-3, 3, 6, 12, 15]\\ 
NGC5866    &     44 &                  [-3, 3, 6, 12, 52, 160, 282]\\ 
NGC6014    &     38 &                       [-3, 3, 12, 28, 50, 70]\\ 
NGC6547    &     37 &                        [-3, 3, 8, 18, 34, 47]\\ 
NGC6798    &     29 &                             [-3, 3, 5.5, 7.5]\\ 
NGC7465    &     32 &                      [-3, 3, 9, 36, 130, 250]\\ 
PGC029321  &     45 &                         [-3, 3, 48, 130, 190]\\ 
PGC056772  &     28 &                            [-3, 3, 9, 36, 72]\\ 
PGC058114  &     30 &                     [-3, 3, 12, 50, 128, 198]\\ 
UGC05408   &     41 &                        [-3, 3, 8, 24, 51, 62]\\ 
UGC06176   &     29 &                     [-3, 3, 10, 42, 110, 158]\\ 
UGC09519   &     27 &                          [-3, 3, 5, 7.5, 9.5]\\ 
\hline
\hline
\end{tabular*}

\end{minipage} 
\end{table*}

\begin{figure*}
{\label{fig:sub:IC0719}\includegraphics[clip=True, trim=0cm 0cm 0cm 0cm, scale=0.25]{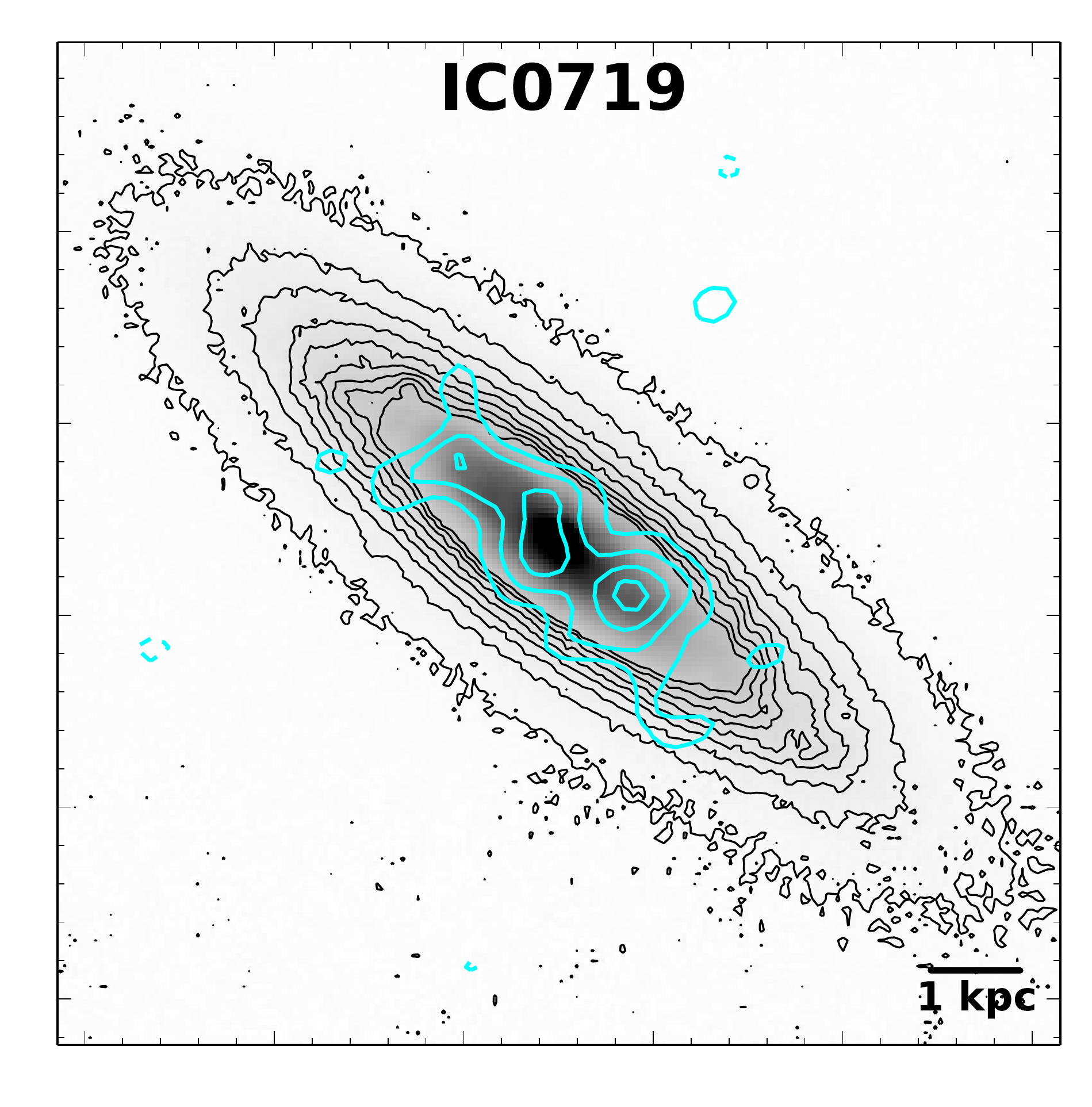}}
{\label{fig:sub:NGC1222}\includegraphics[clip=True, trim=0cm 0cm 0cm 0cm, scale=0.25]{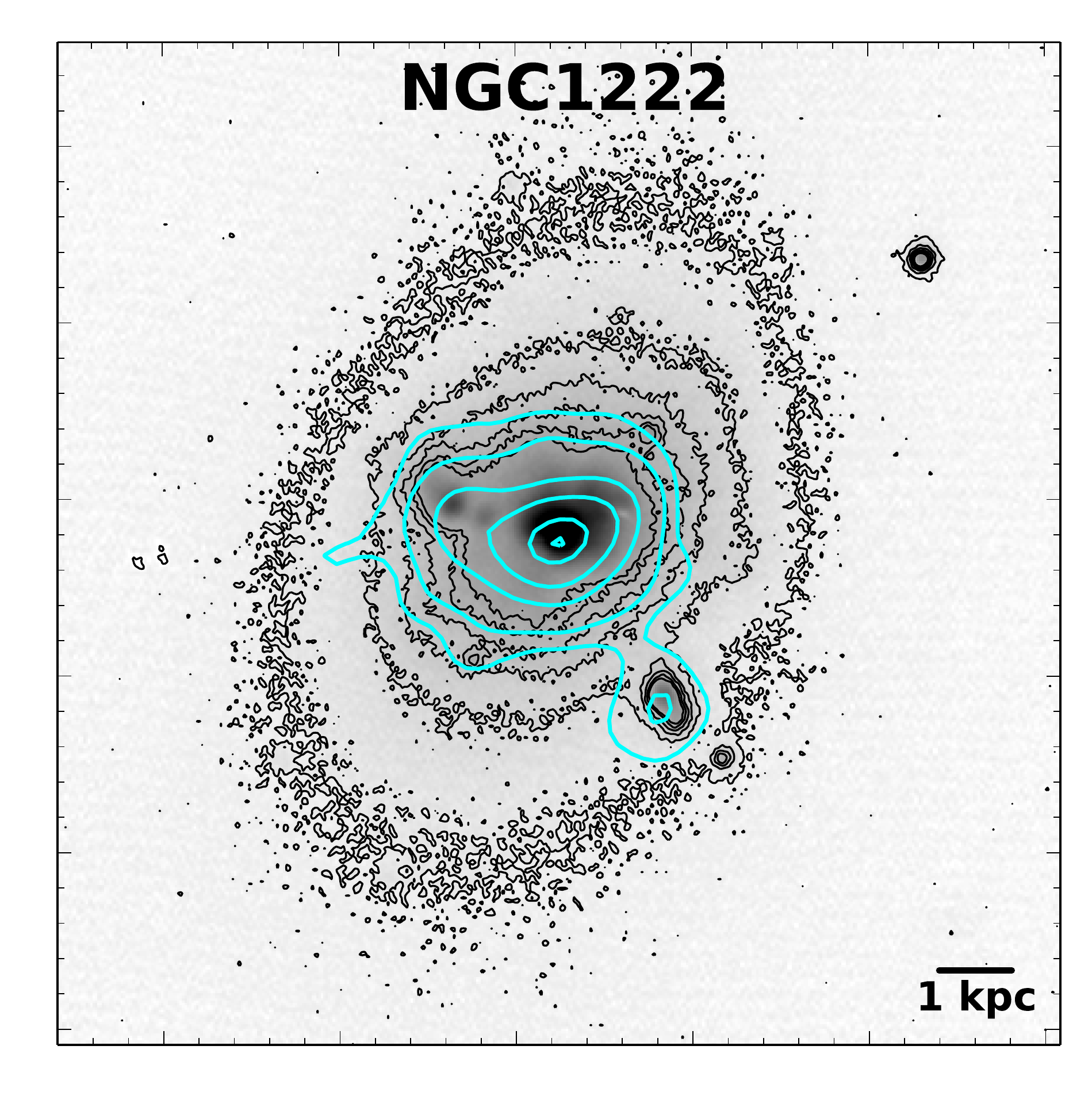}}
{\label{fig:sub:NGC2685}\includegraphics[clip=True, trim=0cm 0cm 0cm 0cm, scale=0.25]{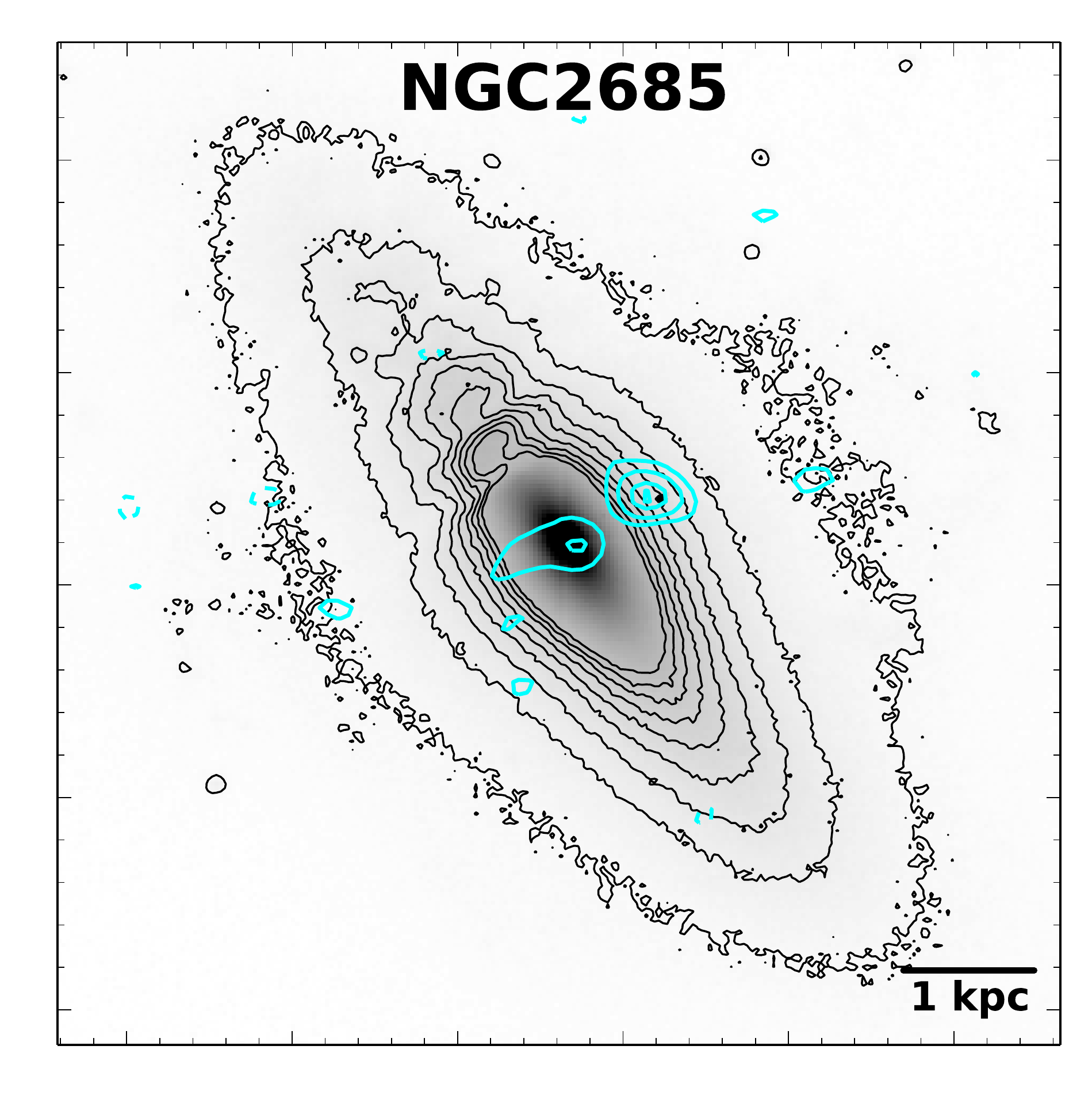}}
{\label{fig:sub:NGC2764}\includegraphics[clip=True, trim=0cm 0cm 0cm 0cm, scale=0.25]{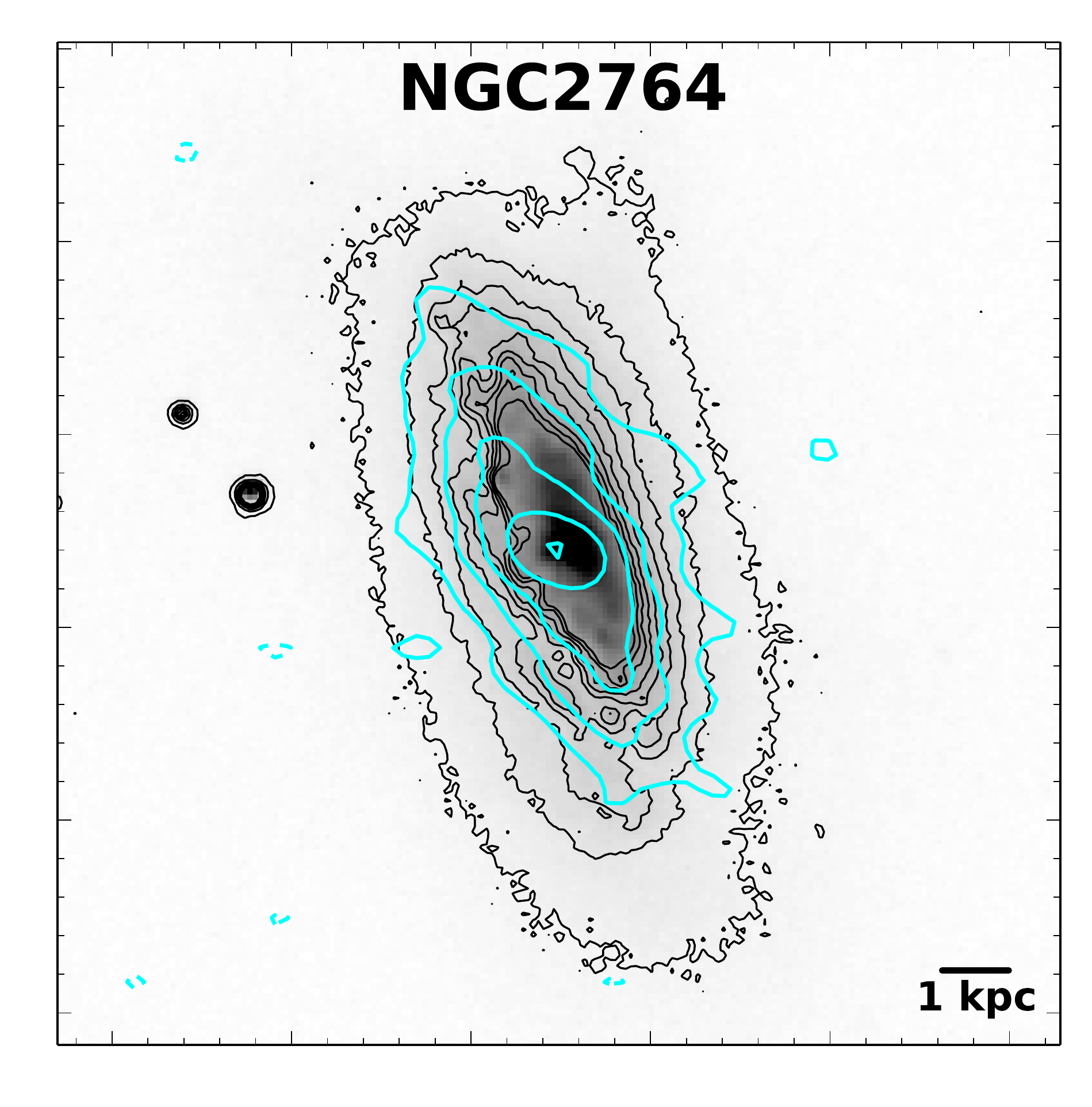}}
{\label{fig:sub:NGC3032}\includegraphics[clip=True, trim=0cm 0cm 0cm 0cm, scale=0.25]{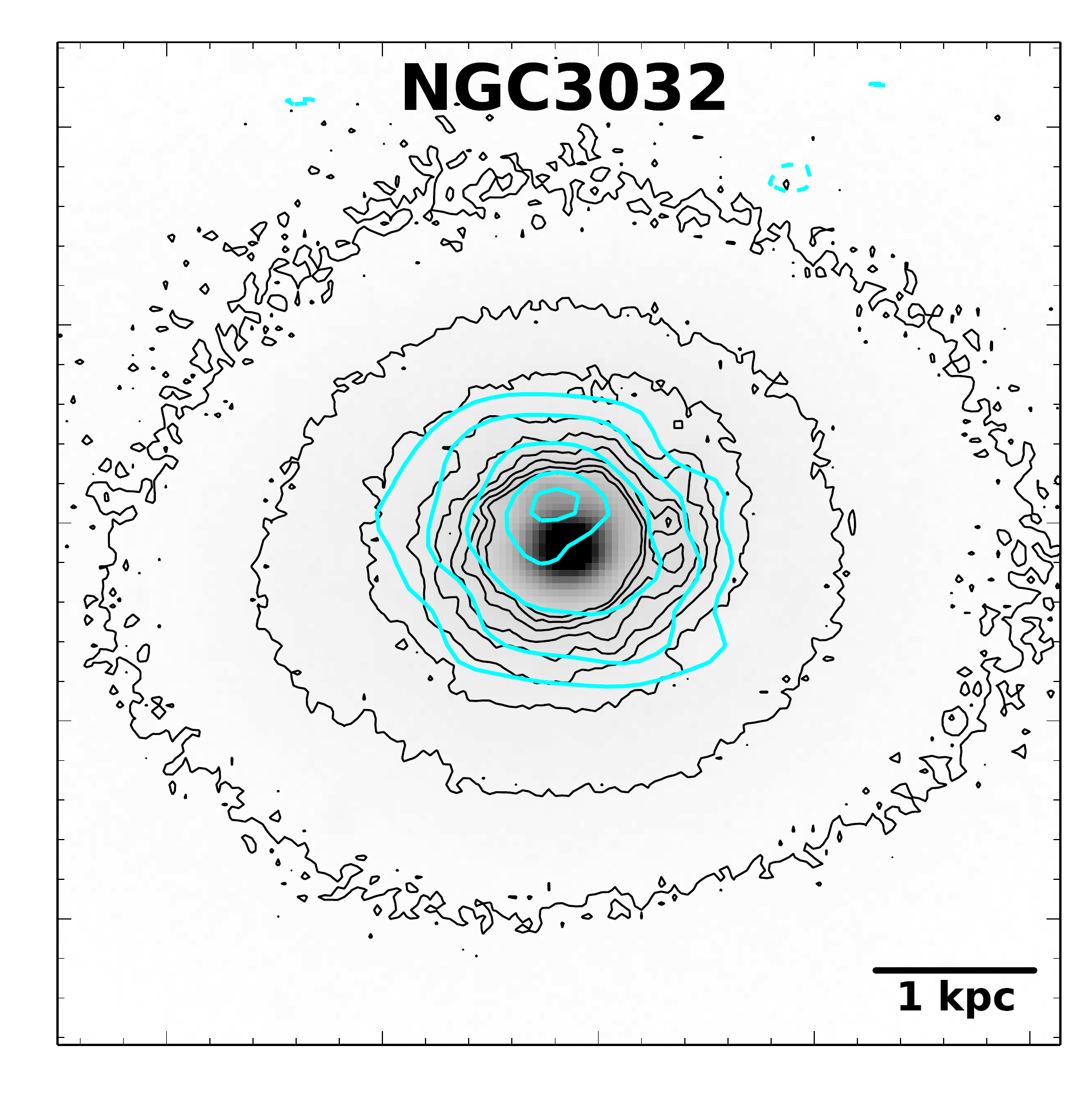}}
{\label{fig:sub:NGC3182}\includegraphics[clip=True, trim=0cm 0cm 0cm 0cm, scale=0.25]{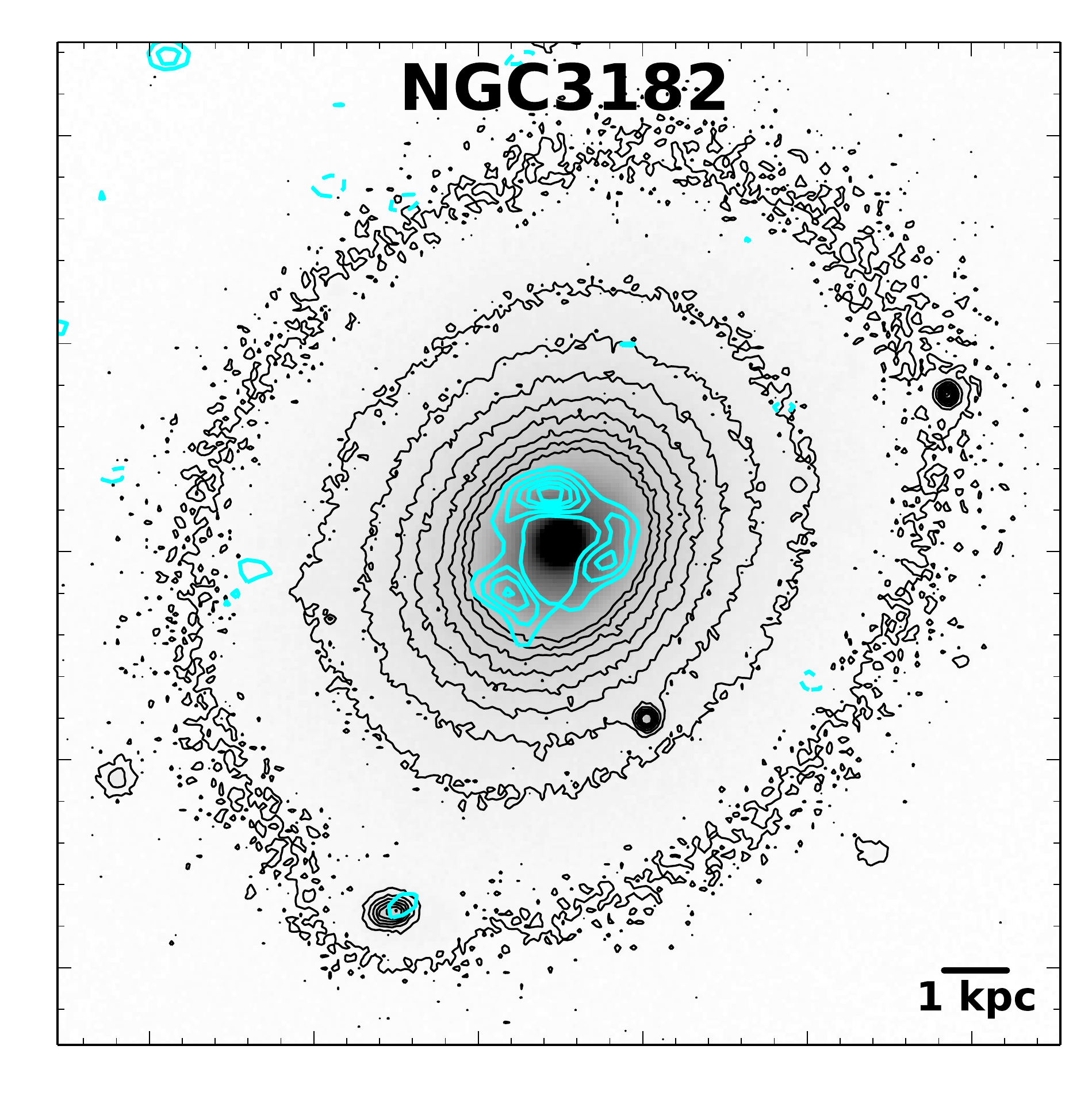}}
{\label{fig:sub:NGC3619}\includegraphics[clip=True, trim=0cm 0cm 0cm 0cm, scale=0.25]{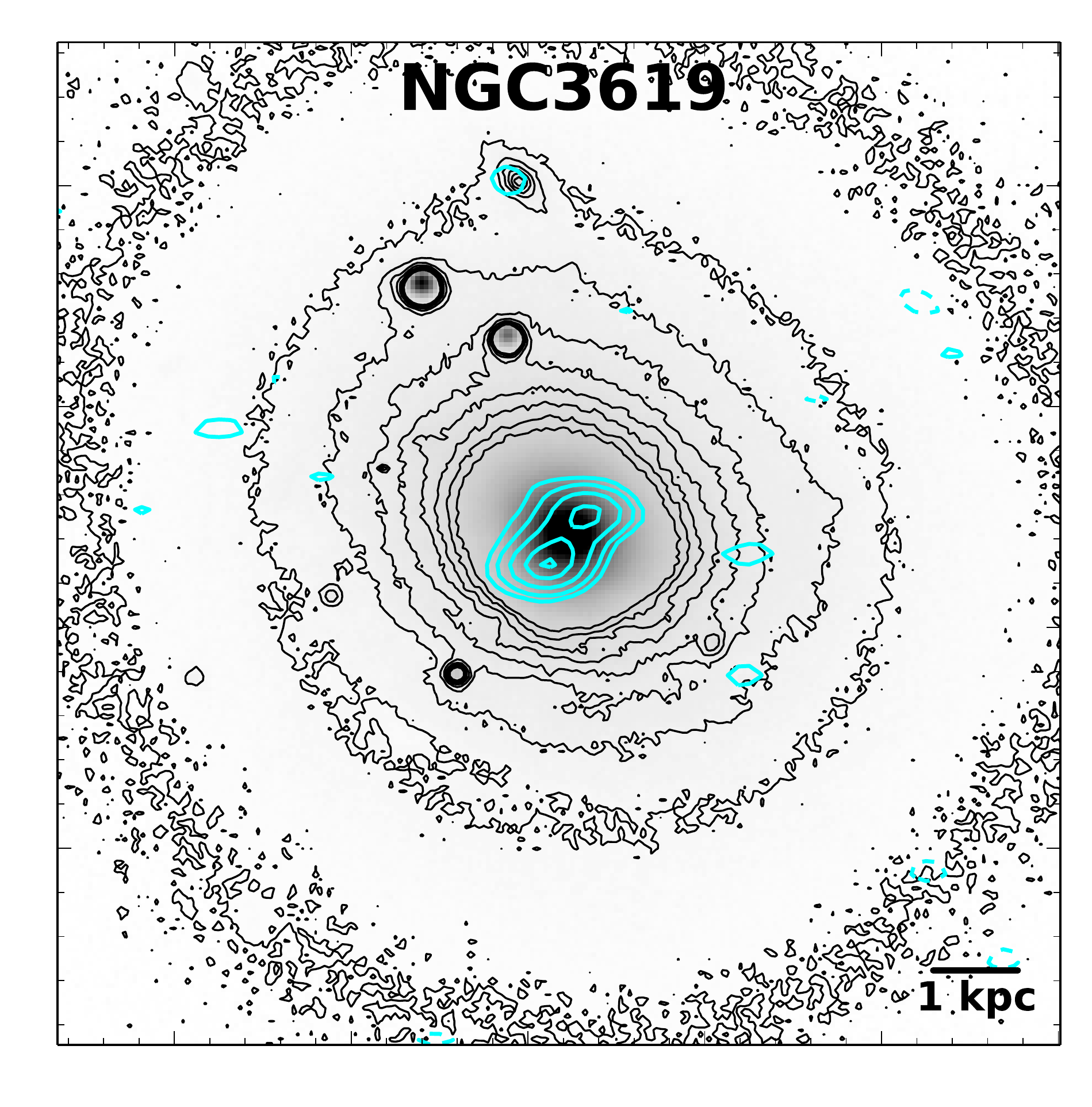}}
{\label{fig:sub:NGC3626}\includegraphics[clip=True, trim=0cm 0cm 0cm 0cm, scale=0.25]{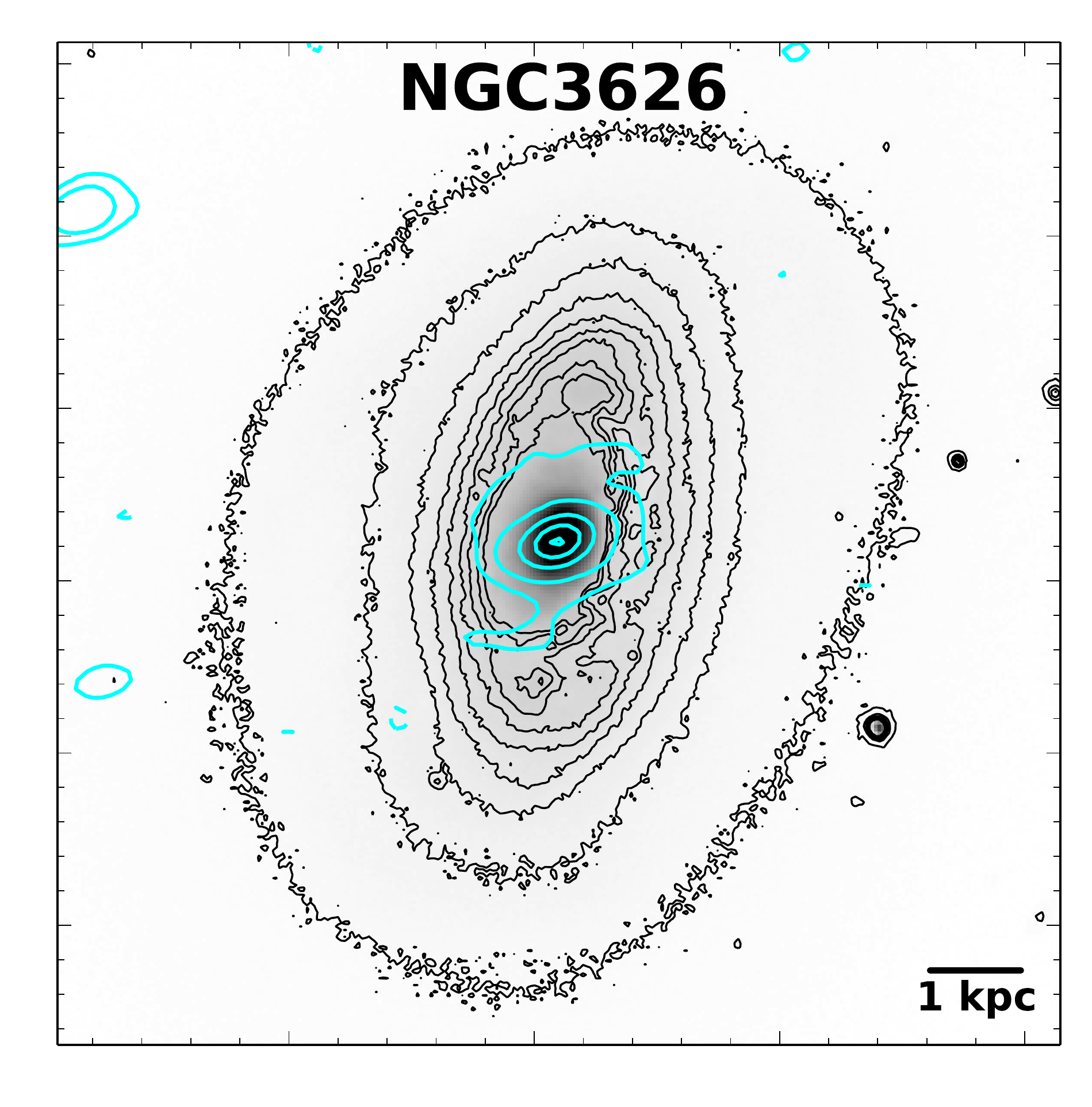}}
{\label{fig:sub:NGC3665}\includegraphics[clip=True, trim=0cm 0cm 0cm 0cm, scale=0.25]{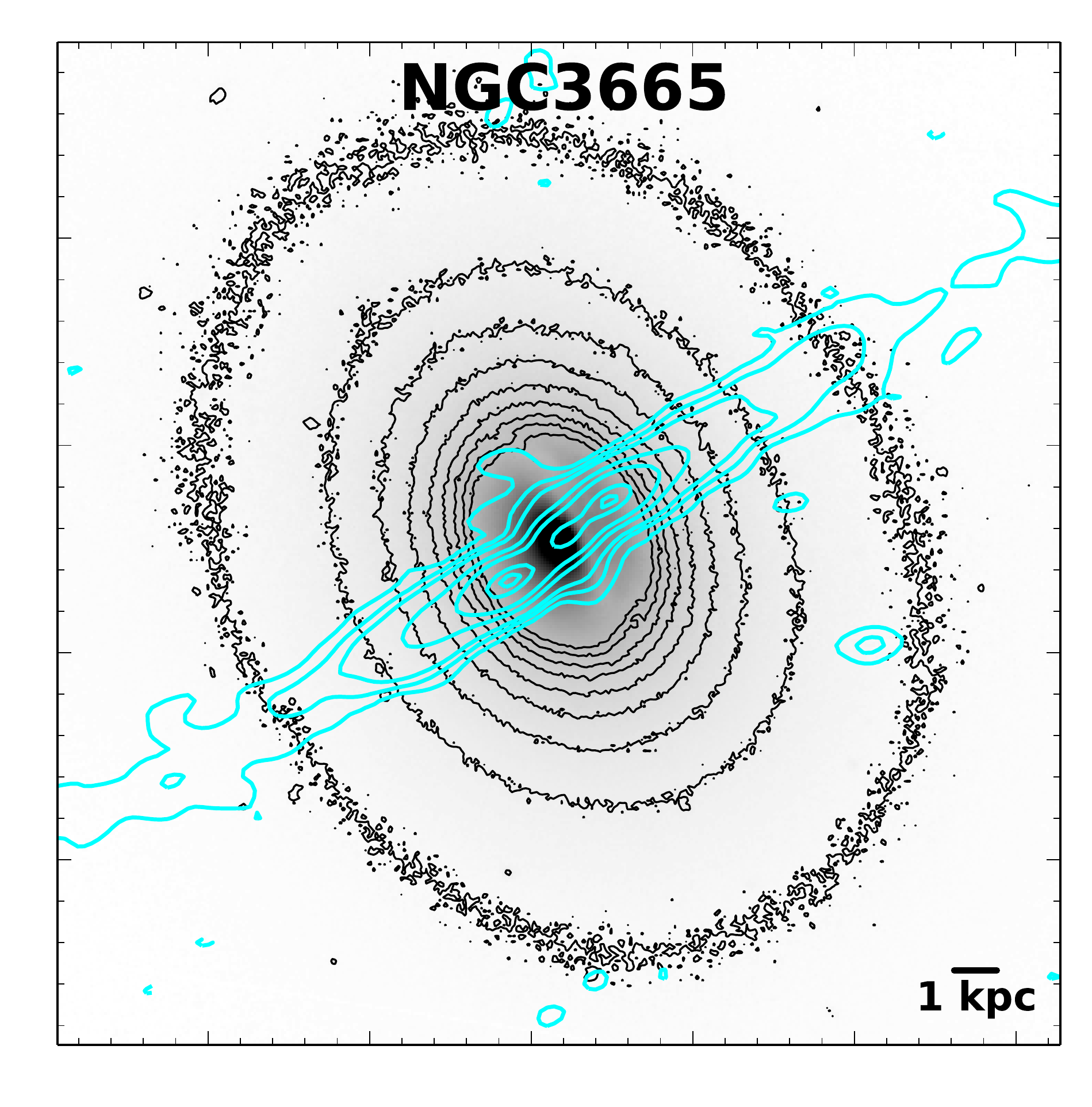}}
{\label{fig:sub:NGC3945}\includegraphics[clip=True, trim=0cm 0cm 0cm 0cm, scale=0.25]{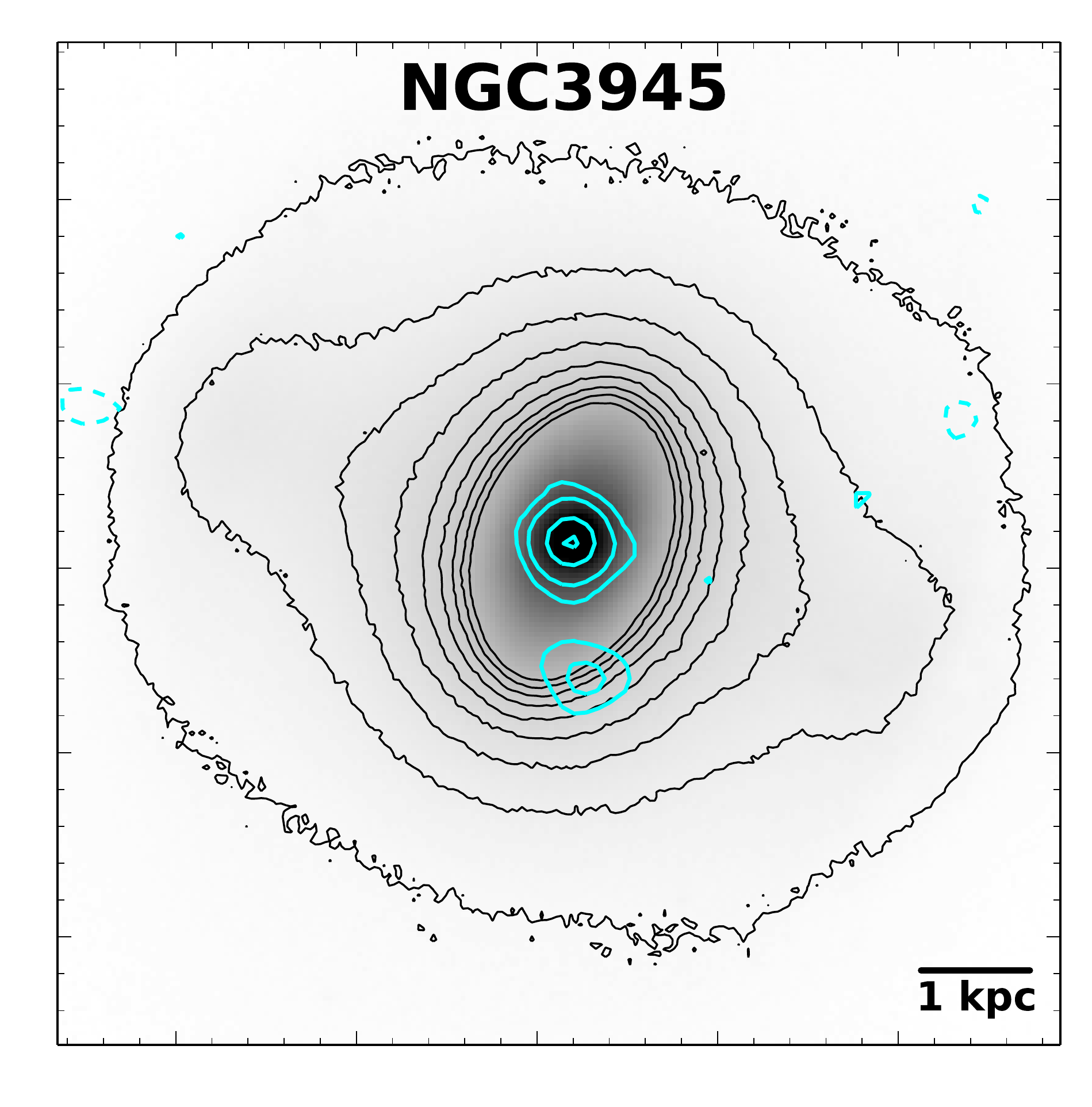}}
{\label{fig:sub:NGC4526}\includegraphics[clip=True, trim=0cm 0cm 0cm 0cm, scale=0.25]{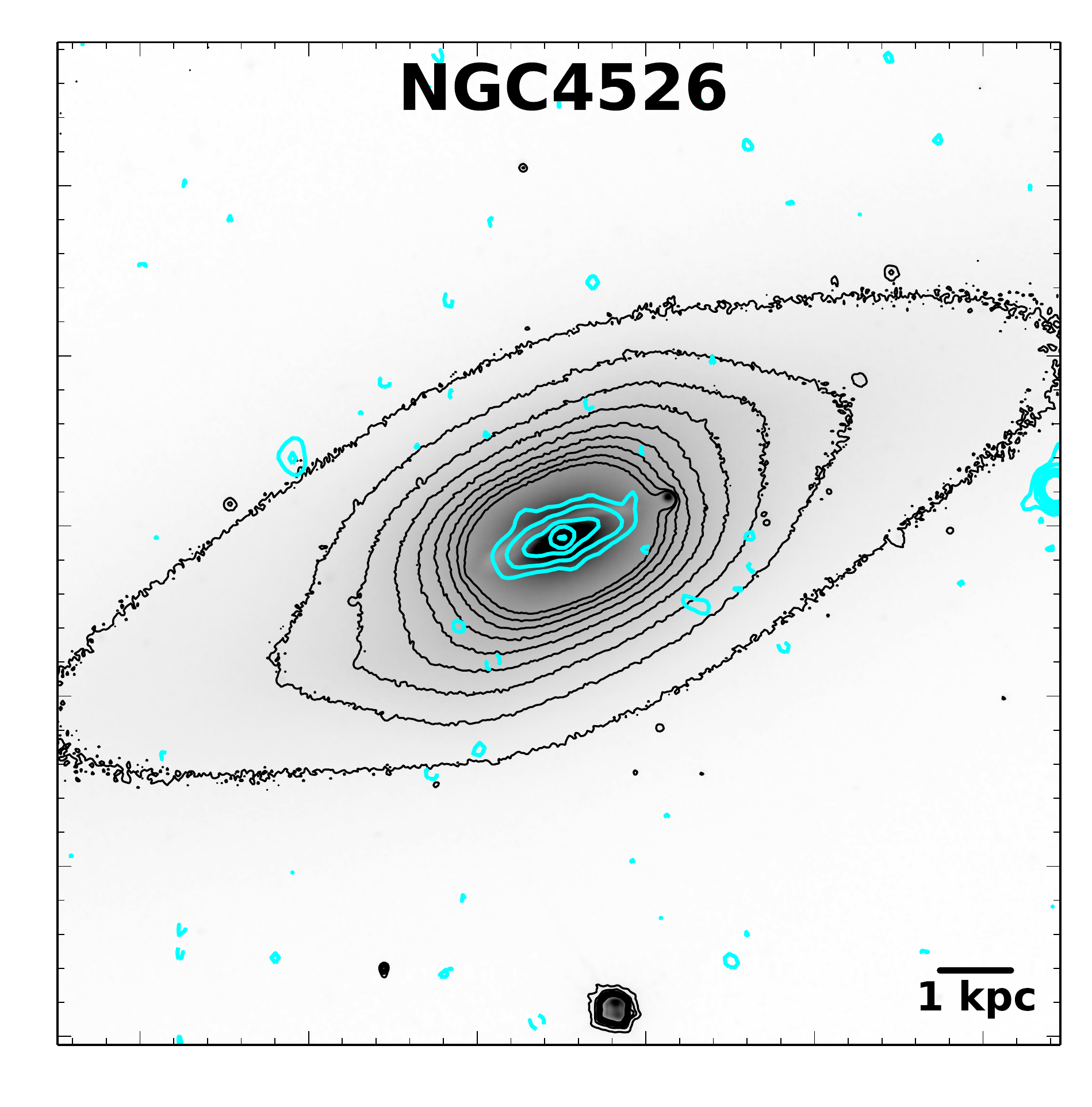}}
{\label{fig:sub:NGC4684}\includegraphics[clip=True, trim=0cm 0cm 0cm 0cm, scale=0.25]{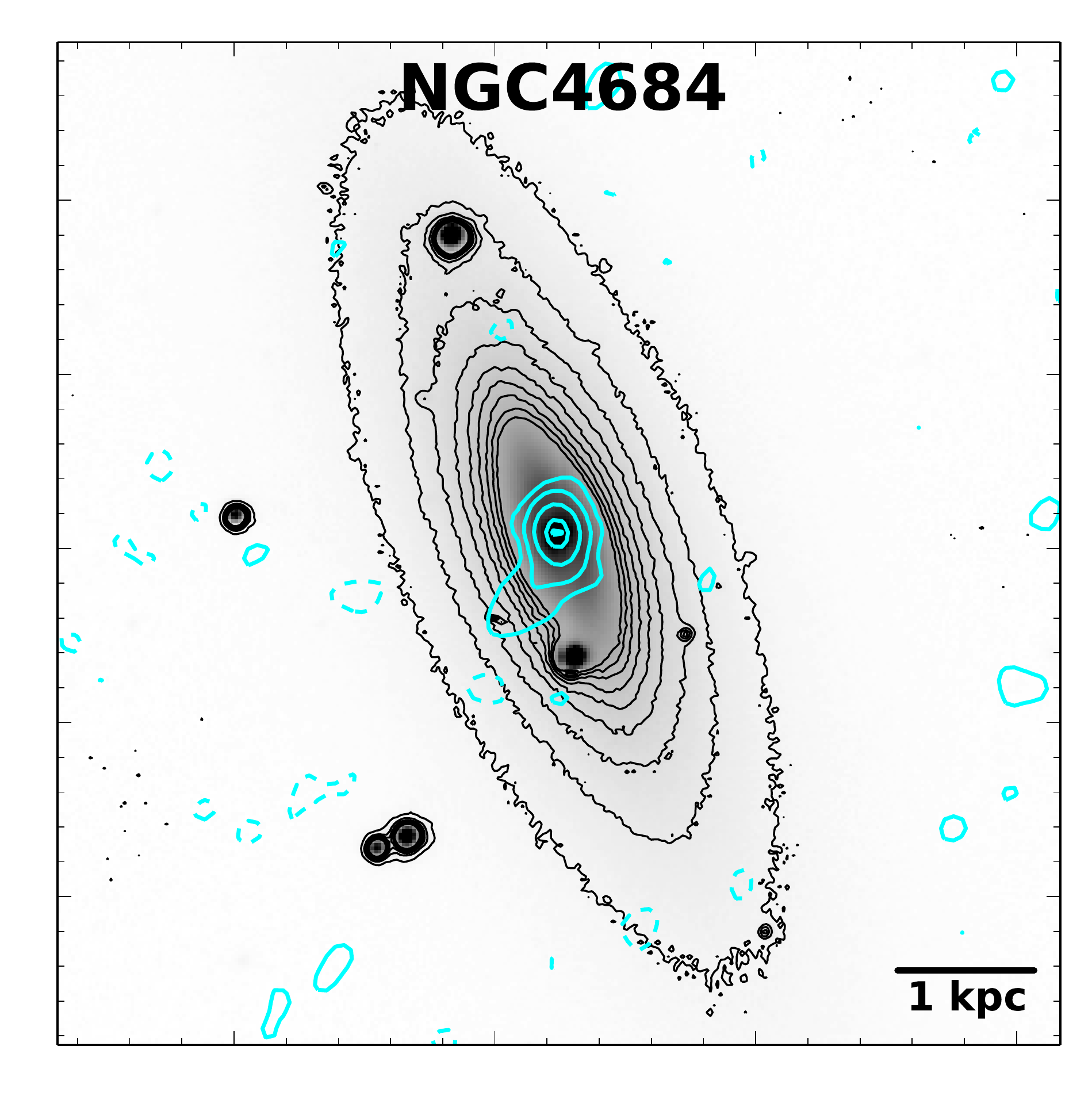}}
\end{figure*}
 
\bsp
\label{lastpage}
\end{document}